\documentclass[12pt,notitlepage]{report}

\pdfoutput=1

\usepackage{bibunits}
\usepackage{uwmthesis}
\usepackage{graphicx}
\usepackage{color}
\usepackage{amsmath}
\usepackage{amssymb}
\usepackage{amsfonts}
\usepackage{rotating}
\usepackage{tensor}

\DeclareGraphicsExtensions{.pdf,.png}

\newcommand{\ospsd}{\ensuremath{S_n\left(\left|f_{k}\right|\right)}}

\newcommand{\cqg}{{\it Class. Quant. Grav.}}

\newcommand{\prd}{{\it Phys. Rev. D}}

\begin{document}
\title{
Searching for Gravitational Radiation from Binary Black Hole MACHOs 
in the Galactic Halo
}
\author{\bf Duncan A. Brown}
\majorprof{Patrick Brady}
\submitdate{December 2004}
\degree{Doctor of Philosophy}
\program{Physics}
\copyrightyear{2004}
\majordept{Physics}
\havededicationtrue
\dedication{to\\ Mum and Dad}
\haveminorfalse
\copyrighttrue
\doctoratetrue
\figurespagetrue
\tablespagetrue

\Abstract{
The Laser Interferometer Gravitational Wave Observatory (LIGO) is one of a new
generation of detectors of gravitational radiation. The existence of
gravitational radiation was first predicted by Einstein in 1916, however
gravitational waves have not yet been directly observed.

One source of gravitation radiation is binary inspiral. Two compact bodies
orbiting each other, such as a pair of black holes, lose energy to
gravitational radiation. As the system loses energy the bodies spiral towards
each other. This causes their orbital speed and the amount of gravitational
radiation to increase, producing a characteristic ``chirp'' waveform in the
LIGO sensitive band.

In this thesis, matched filtering of LIGO science data is used to search for
low mass binary systems in the halo of dark matter surrounding the Milky Way.
Observations of gravitational microlensing events of stars in the Large
Magellanic Cloud suggest that some fraction of the dark matter in the halo may
be in the form of Massive Astrophysical Compact Halo Objects (MACHOs). It has
been proposed that low mass black holes formed in the early universe may be a
component of the MACHO population; some fraction of these black hole MACHOs
will be in binary systems and detectable by LIGO.

The inspiral from a MACHO binary composed of two 0.5 solar mass black holes
enters the LIGO sensitive band around 40 Hz.  The chirp signal increases in
amplitude and frequency, sweeping through the sensitive band to 4400 Hz in 140
seconds. By using evidence from microlensing events and theoretical
predictions of the population an upper limit is placed on the rate of black
hole MACHO inspirals in the galactic halo.
}
\beforepreface
\prefacesection{Preface}
The work presented in this thesis stems from my participation in the LIGO
Scientific Collaboration. 

\vspace*{0.5cm}
\noindent The upper limit on the rate of binary neutron 
star inspirals quoted in chapter \ref{ch:introduction} is based on

\vspace*{0.25cm}

\noindent B.~Abbott~{\it et~al.} (The LIGO Scientific Collaboration),
``Analysis of LIGO data for gravitational waves from binary neutron stars,''
\prd~{\bf 69} (2004) 122001.

\vspace*{0.5cm}

\noindent Chapter \ref{ch:pipeline} is based on material from

\vspace*{0.25cm}

\noindent Duncan A. Brown~{\it et~al.}, ``Searching for Gravitational Waves
from Binary Inspirals with LIGO,'' \cqg~{\bf 21}, S1625 (2004).

\vspace*{0.25cm}

\noindent and

\vspace*{0.25cm}

\noindent B. Abbott~{\it et~al.} (The LIGO Scientific Collaboration), ``Search
for binary neutron star coalescence in the Local Group,'' to be submitted to
\cqg

\vspace*{0.5cm}

\noindent Chapter \ref{ch:hardware} is based on 

\vspace*{0.25cm}

\noindent Duncan A. Brown (for the LIGO Scientific Collaboration), ``Testing
the LIGO inspiral analysis with hardware injections,'' \cqg~{\bf 21}, S797
(2004).

\vspace*{0.5cm}

\noindent Chapter \ref{ch:result} is based on

\vspace*{0.25cm}

\noindent B. Abbott~{\it et~al.} (The LIGO Scientific Collaboration), ``Search
for binary black hole MACHO coalescence in the Galactic Halo,'' to be
submitted to \prd

\prefacesection{Acknowledgments}

As a member of the LIGO Scientific Collaboration, I have been fortunate to
have benefited through advice from and discussions with many people. It would
not be possible to thank everyone who I have worked with over the past five
years without making these acknowledgments the longest chapter in this
dissertation, so I shall only attempt to thank those who I have interacted
with the most and hope that the others forgive me.

First and foremost, I would like to thank Patrick Brady for his constant
guidance and patience over the past five years as my advisor and my friend. I
have been fortunate to work with someone with the ability and integrity of
Patrick.  I hope that our collaboration can continue for many years.

I would also like to thank Jolien Creighton for his help and enthusiasm over
the past five years. It has been fun working with Jolien and I have learnt a
great deal from him.

I am grateful to Bruce Allen for suggesting the search for binary inspiral as
a research topic and his assistance with the scientific and computational
obstacles along the way. Thanks also to Gabriela Gonz\'{a}lez for patiently
answering my many stupid questions about the LIGO interferometers helping me
understand the data that I have been analyzing, and to Scott Koranda for
helping me get the data analyzed.

I would like to thank the members of my committee: Daniel Agterberg, John
Friedman and Leonard Parker for their careful reading of this dissertation and
helpful suggestions for its improvement.

I also would like to thank Warren Anderson, Teviet Creighton, Stephen
Fairhurst, Eirini Messaritaki, Ben Owen, Xavier Siemens and Alan Wiseman for
help, advice and pints of beer. I am also indebted to Axel's for stimulating
many useful discussions.

Thanks to Steve Nelson, Wyatt Osato and Quiana Robinson for their help in the
preparation of this dissertation and, of course, to Sue Arthur for making
everything run smoothly.

I could not have come this far without the constant love and support of my
parents, to whom this thesis is dedicated. Finally, I would like to thank
Emily Dobbins for all her love and understanding over the past two years.

\prefacesection{Conventions}
There are two possible sign conventions for the Fourier transform of a time
domain quantity $v(t)$. In this thesis, we define the Fourier transform
$\tilde{v}(f)$ of a $v(t)$ to be
\begin{equation*}
\tilde{v}(f)=\int_{-\infty}^\infty dt\,v(t)\, e^{- 2 \pi i f t}
\end{equation*}
and the inverse Fourier transform to be 
\begin{equation*}
v(t)=\int_{-\infty}^\infty df\,\tilde{v}(f)\, e^{2 \pi i f t}.
\end{equation*}
This convention differs from that used in some gravitational wave literature,
but is the adopted convention in the LIGO Scientific Collaboration.

\vspace{0.5cm}

\noindent The time-stamps of interferometer data are measured in Global
Positioning System (GPS) seconds: seconds since 00:00.00 UTC January 6, 1980
as measured by an atomic clock.

\vspace{0.5cm}

\noindent Astronomical distances are quoted in parsecs
\begin{equation*}
1\,\mathrm{pc} = 3.0856775807 \times 10^{16}\,\mathrm{m}
\end{equation*}
and masses in units of solar mass
\begin{equation*}
1\,\mathrm{M}_\odot = 1.98892 \times 10^{30}\,\mathrm{kg}.
\end{equation*}

\afterpreface

\Chapter{Introduction}
\label{ch:introduction}
One of the earliest predictions of the Theory of General Relativity was the
existence of gravitational waves. By writing the metric $g_{\mu\nu}$ as the sum
of the flat Minkowski metric $\eta_{\mu\nu}$ and a small perturbation $h_{\mu\nu}$, 
\begin{equation}
g_{\mu\nu} = \eta_{\mu\nu} + h_{\mu\nu},
\end{equation}
and considering bodies with negligible self-gravity, 
Einstein showed\cite{Einstein:1916}
\begin{quotation}
``that these $h_{\mu\nu}$ can be calculated in a manner analogous to that of
the retarded potentials of electrodynamics.''
\end{quotation}
It follows that
gravitational fields propagate at the speed of light.  In electrodynamics, the
lowest multipole moment that produces radiation is the electric dipole; there
is no electric monopole radiation due to the conservation of electric charge.
Similarly in General Relativity, the lowest multipole that produces
gravitational waves is the quadrupole moment. Radiation from the 
mass monopole, mass dipole and momentum dipole vanish due to conservation of
mass, momentum and angular momentum respectively. Einstein also derived the
\emph{quadrupole formula} for the gravitational wave field, which states that
the spacetime perturbation is proportional to the second time derivative of
the quadrupole moment of the source.  The strength of the gravitational waves
decreases as the inverse of the distance to the source.  We can estimate
this strength at a distance $r$ by noticing that the quadrupole moment
involves terms of dimension mass $\times$ length$^2$ and so the second time
derivative of the quadrupole moment is proportional to the kinetic energy of
the source associated with non-spherical motion $E^\mathrm{ns}_\mathrm{kin}$.
Using the quadrupole formula, which we will see in equation
(\ref{eq:quadrupole}), we then approximate the strength of the gravitational
waves as
\begin{equation}
h \sim \frac{G}{c^4}\frac{E^\mathrm{ns}_\mathrm{kin}}{r}.
\label{eq:strainest}
\end{equation}
The effect of a
gravitational wave is to cause the measured distance $L$ between two freely
falling bodies to change by a distance $\Delta L \sim h L$. 

Interferometers were suggested as a way of measuring the change in 
length between two test masses by Pirani in 1956\cite{Pirani:1956} and
the first working detector was constructed by Forward in
1971\cite{Forward:1971}. The fundamental designs of modern laser
interferometers were developed by Weiss\cite{Weiss:1972} and
Drever\cite{Drever:1980} in the 1970s. The principle upon which
interferometric detectors operate is to use laser light to measure the change
in distance between two mirrors as a gravitational wave passes through the
detector.  The sensitivity of an interferometer on the Earth is limited by
\emph{gravity gradient noise} at frequencies below $\sim
5$~Hz\cite{Saulson:1994}.  Any time changing distribution of matter near the
detector, for example compression waves in the Earth, cause fluctuations in
the local gravitational field. These fluctuations will cause the test masses
to move producing a spurious response in the interferometer which masks the
presence of gravitational waves. In fact, Earth based interferometers are
typically limited in sensitivity to frequencies above $\sim 10$~Hz due to the
seismic motion of the earth.


The canonical example of an astrophysical source of gravitational waves is the
Hulse-Taylor binary pulsar, PSR~$1913+16$\cite{1975ApJ...195L..51H}. This
system is composed of two neutron stars, each of mass $\sim 1.4\,M_\odot$,
with average separation and orbital velocity of $\sim10^9$~m and $\sim
10^5$ms$^{-1}$, respectively. The period of the orbit is $7.75$~hours and the
binary is at a distance from the earth of $\sim 6$~kpc.  Hulse and Taylor
observed that the orbital period of the binary is decreasing and that the rate
of orbital energy loss agrees with the expected loss of energy due to the
radiation of gravitational waves to within
$0.3\%$\cite{Taylor:1982,Taylor:1989}.   Since the quadrupole moment
of an equal mass binary is periodic at half the orbital period, we would
expect the frequency of the gravitational waves emitted to be twice
the orbital frequency.  Thus, the gravitational waves from PSR~$1913+16$ have
a frequency $f_\mathrm{GW} \sim 10^{-4}$~Hz that is outside the
sensitive band of earth based detectors.   Nevertheless,  the orbit
will continue to tighten by gravitational wave emission, and the two
neutron stars are expected to merge in about 300 million years;   in the last
several minutes prior to merger,  the gravitational wave frequency will sweep
upward from $\sim 10$ Hz reaching about 1500 Hz just before the merger.

It is worthwhile to estimate the strength of the gravitational waves from
a neutron star binary since it informs the target sensitivity for
modern interferometric detectors.   The non-spherical kinetic energy
of this system is 
\begin{equation}
E_\mathrm{kin}^\mathrm{ns} \sim 
1.4\,\mathrm{M}_\odot (\pi a / T)^2
\label{eq:ensht}
\end{equation}
where $T$ is the binary period and $a$ is the average separation.  The period,
separation and mass of a binary are related by Kepler's third law,
\begin{equation}
T^2 = \frac{4\pi^2}{GM}a^3,
\label{eq:kep3}
\end{equation}
where $M$ is the total mass of the binary.  Using equation
(\ref{eq:strainest}), Kepler's third law and the non-spherical kinetic energy
given in equation (\ref{eq:ensht}), we can estimate the strength of the waves
from a neutron star binary as
\begin{equation}
h \sim 10^{-20} \times 
\left(\frac{6.3\,\mathrm{kpc}}{r}\right)
\left(\frac{M}{2.8M_\odot}\right)^{5/3}
\left(\frac{T}{1\,\mathrm{s}}\right)^{-\frac{2}{3}}.
\label{eq:binaryest}
\end{equation}
When the orbital separation is $a \sim 10^5$~m,  the orbital period will be
$T\sim10^{-2}$~seconds and the gravitational wave strain will be 
$h \sim 10^{-19}$.

To date, four binary neutron star systems that will merge within a Hubble time
have been discovered.   By considering the time to merger, position and
efficiency of detecting such binary pulsar systems, the galactic merger rate
for inspirals can be estimated\cite{Phinney:1991ei}.  The latest estimates of
neutron star inspirals in the Milky Way are $8.3 \times 10^{-6}$~yr$^{-1}$. 
Extrapolating this rate to the neighboring Universe using the
blue-light luminosity gives an (optimistic) estimate of the rate at
$0.3$~yr$^{-1}$ within a distance of $\sim 20$~Mpc.  To measure the waves from
a neutron star binary at this distance,  we must construct
interferometers that are sensitive to gravitational waves of strength $h \sim
10^{-22}$. An overview of the theory and experimental techniques underlying
the generation and detection of gravitational waves from binary inspiral is
presented in chapter \ref{ch:inspiral}.

A world-wide network of gravitational wave interferometers has been
constructed that have the sensitivity necessary to detect the gravitational
waves from astrophysical sources. Among these is the Laser Interferometric
Gravitational Wave Observatory (LIGO)\cite{Barish:1999}. LIGO has completed
three science data taking runs. The first, referred to as S1, lasted for 17
days between August 23 and September 9, 2002; the second, S2, lasted for 59
days between February 14 and April 14, 2003; the third, S3, lasted for 70 days
between October 31, 2003 and January 9, 2004.  During the runs, all three LIGO
detectors were operated: two detectors at the LIGO Hanford observatory (LHO)
and one at the LIGO Livingston observatory (LLO).  The detectors are not yet
at their design sensitivity, but the detector sensitivity and amount of
usable data has improved between each data taking run. The noise level is low
enough that searches for coalescing compact neutron stars are worthwhile, and
since the start of S2, these searches are sensitive to extra-galactic sources.
Using the techniques of \emph{matched filtering} described in chapter
\ref{ch:findchirp} of this dissertation, the S1 binary neutron star search set
an upper limit of
\begin{equation}
\mathcal{R}_{90\%} < 1.7 \times 10^2 \;\textrm{per year per Milky Way Equivalent Galaxy (MWEG)}
\end{equation}
with no gravitational wave signals detected. Details of this analysis can be
found in \cite{LIGOS1iul}. 

In this dissertation, we are concerned with the
search for gravitational waves from a different class of compact binary
inspiral: those from binary black holes in the galactic halo. Observations of
the gravitational microlensing of stars in the Large Magellanic cloud suggest
that $\sim 20\%$ of the galactic halo consists of objects of mass $\sim
0.5\,M_\odot$ of unknown origin. In chapter \ref{ch:macho} we discuss a
proposal that these Massive Astrophysical Compact Halo Objects (MACHOs) may be
black holes formed in the early universe and that some fraction of them may be
in binaries whose inspiral is detectable by LIGO\cite{Nakamura:1997sm}.  The
upper bound on the rate of such binary black hole MACHO inspirals are
projected to be $R \sim 0.1$~yr$^{-1}$ for initial LIGO, much higher than the
binary neutron star rates discussed above. It should be noted however, that
while binary neutron stars have been observed, there is no direct
observational evidence of the existence of binary black hole MACHOs. Despite
this, the large projected rates make them a tempting source for LIGO. In
chapter \ref{ch:pipeline} we describe an \emph{analysis pipeline} that has
been used to search the LIGO S2 data for binary black hole MACHOs\footnote{The
same pipeline has also been used to search for binary neutron star inspiral in
the S2 data and the results of this search will be presented in
\cite{LIGOS2iul}.}. Chapter \ref{ch:hardware} describes how the search
techniques were tested on data from the gravitational wave interferometers.
Finally we present the result of the S2 binary black hole MACHO search in
chapter \ref{ch:result}. 

\Chapter{Gravitational Radiation from Binary Inspiral}
\label{ch:inspiral}

In this chapter we review some of the physics underlying the detection of
gravitational waves from binary inspiral.  In section~\ref{s:effect} we review
the effect of gravitational waves on a pair of freely falling particles in
order to introduce some of the concepts that we need to discuss the detection
of gravitational waves from binary inspiral.  For a detailed description of
gravitational waves, we refer to \cite{MTW73,Thorne:1982cv}.
Section~\ref{s:ifos} describes how a laser interferometer can be used to
measure this effect. The gravitational waveform produced by the inspiral of
two compact objects, such as neutron stars or black holes, are discussed in
section \ref{s:inspiralgw}. We also derive the waveform that will be used to
search for gravitational waves from binary inspiral events in the Universe.

\section{The Effect of Gravitational Waves on Freely Falling Particles}
\label{s:effect}

The 4-velocity $\vec{u}$ of a freely falling test particle satisfies the
geodesic equation\cite{Wald:1984}
\begin{equation}
\label{eq:geodessic}
\left(\nabla_{\vec{u}} \vec{u}\right)^\alpha =
\tensor{u}{^\alpha_{;\mu}}u^{\mu} = 0,
\end{equation}
where $;$ denotes the covariant derivative, that is,
\begin{equation}
\tensor{u}{^\alpha_{;\mu}}\tensor{u}{^{\mu}} = \left(\tensor{u}{^\alpha_{,\mu}} +
\Gamma^\alpha_{\mu\nu}u^\nu\right) u^\mu,
\end{equation}
where $\Gamma^\alpha_{\mu\nu}$ is the connection coefficient of the metric
$g_{\mu\nu}$ and $,\mu$ represents the standard partial derivative with
respect to the coordinate $x^\mu$. 

\begin{figure}[p]
\begin{center}
\includegraphics[width=\linewidth]{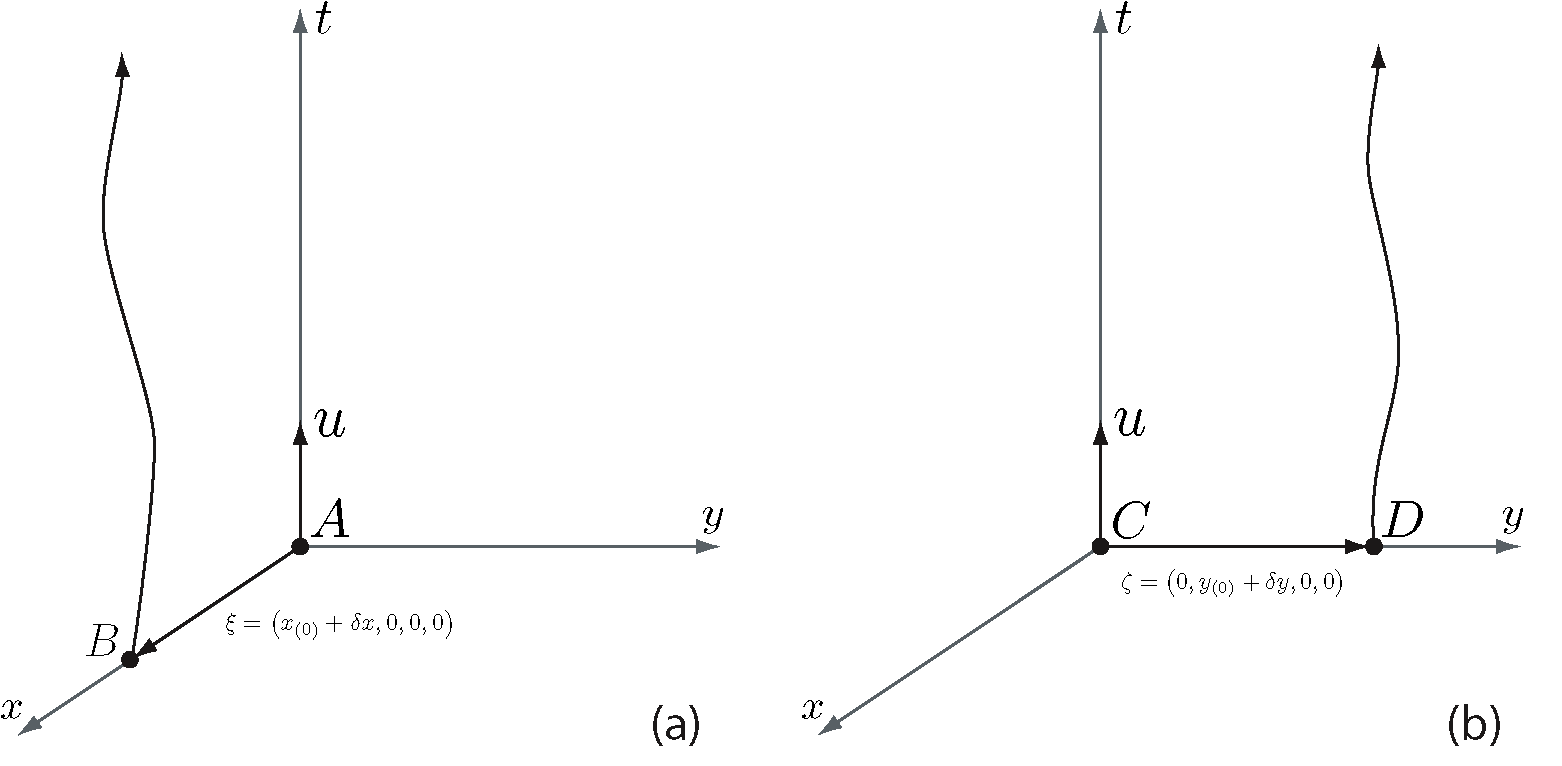}
\end{center}
\caption[Effect of a Gravitational Wave on Test Particles]{%
\label{f:particles}
The axes shown in (a) and (b) represent Local Lorentz frames for particles $A$
and $C$ respectively. The effect of a gravitational wave in these frame can be
described in terms of its effect on the vectors $\xi$ and $\zeta$ separating
the particles at the origin from particles $B$ and $D$.
}
\end{figure}

Consider two particles $A$ and $B$, as shown in figure \ref{f:particles}~(a),
with separation vector $\vec{\xi}$. The particles are initially at rest with
respect to each other, so
\begin{align}
\nabla_{\vec{u}} \vec{u} &= 0, \\
\vec{u} \cdot \vec{\xi} & = 0.
\end{align}
If the spacetime is curved, the second derivative of $\vec{\xi}$ along
$\vec{u}$ is non-zero; it is given by the equation of geodesic deviation
\begin{equation}
\nabla_{\vec{u}}\nabla_{\vec{u}} \vec{\xi} = -
R(\_,\vec{u},\vec{\xi},\vec{u}),
\end{equation}
where $R(\_,\vec{u},\vec{\xi},\vec{u})$ is the Riemann curvature tensor. 
If the spacetime is flat with weak gravitational waves propagating in it, we
can describe it by a metric 
\begin{equation}
g_{\mu\nu} = \eta_{\mu\nu} + h_{\mu\nu},
\end{equation}
where $h_{\mu\nu}$ is the perturbation to the metric due to the gravitational
waves and $\eta_{\mu\nu}$ is the flat Minkowski metric .  We now introduce a
\emph{Local Lorentz Frame} (LLF) for particle $A$.  The LLF of particle $A$ is
a coordinate system $x^\alpha$ in which
\begin{equation}
g_{\mu\nu}(A) = \eta_{\mu\nu}
\end{equation}
and
\begin{equation}
g_{\mu\nu,\alpha}(A) = 0,
\end{equation}
where $g_{\mu\nu}(A)$ is the value of the metric at point $A$. This LLF
is equivalent to a Cartesian coordinate system defined by three orthogonally
pointing gyroscopes carried by particle $A$. The curvature of spacetime means
that the coordinate system is not exactly Cartesian, but it can be shown that
this deviation is second order in the spatial distance from the
particle\cite{MTW73}. This means that along the worldline of particle $A$ the
metric is
\begin{equation}
g_{\mu\nu} = \eta_{\mu\nu} + \frac{\mathcal{O}\left(|\vec{x}|^2\right)}{R^2}
\end{equation}
where $\vec{x}$ is the distance from the particle and $R \sim
|R_{\alpha\beta\gamma\delta}|$. We can write the Cartesian coordinates of the
LLF of $A$ as $x^\mu = (x^0,x^i)$, where $x^0 = t$ is the timelike coordinate
and $x^i$ are the three Cartesian coordinates. Then in
the LLF of particle $A$ the equation of geodesic deviation becomes
\begin{equation}
\frac{\partial^2 \xi^{j}}{\partial t^2} = -
\tensor{R}{^j_{\alpha\beta\gamma}}u^\alpha\xi^\beta u^\gamma =
-\tensor{R}{^j_{0k0}} \xi^k,
\end{equation}
since $u = (1,0,0,0)$.  The presence of the gravitational waves are encoded in
the curvature $R_{\alpha\beta\gamma\delta}$ which satisfies the wave equation
\begin{equation}
\eta^{\mu\nu}R_{\alpha\beta\gamma\delta,\mu\nu} = 0.
\end{equation}
In the Local Lorentz frame, the components of $\vec{\xi}$ are just the
coordinates of $B$.  
In the LLF of $A$ we may write
\begin{equation}
\label{eq:bcoords}
\xi^j = \xi_{(0)}^j + \delta \xi^j,
\end{equation}
where $\xi_{(0)}^j$ is the unperturbed location of particle $B$ and $\delta
\xi^j$ is the change in the position of $B$ caused by the gravitational wave.
Substituting equation (\ref{eq:bcoords}) into the equation of geodesic
deviation, we obtain
\begin{equation}
\label{eq:particledev}
\frac{\partial^2 \delta \xi^j}{\partial t^2} \approx - \tensor{R}{^j_{0k0}} \xi_{(0)}^k =
-R_{j0k0} \xi_{(0)}^k,
\end{equation}
where we have used $\eta_{\mu\nu}$ to lower the spatial index $j$ of the
Riemann tensor. For a weak gravitational wave, all the components of
$R_{\alpha\beta\gamma\delta}$ are completely determined by $R_{j0k0}$.
Furthermore, it can be shown that the $3\times3$ symmetric matrix $R_{j0k0}$,
which we would expect to have $6$ independent components, has only $2$
independent components due to the Einstein equations and the Biancci identity.
We define the (transverse traceless) gravitational wave field,
$h_{jk}^\mathrm{TT}$, by
\begin{equation}
-\frac{1}{2} \frac{\partial^2 h_{jk}^\mathrm{TT}}{\partial t^2} \equiv
R_{j0k0}^\mathrm{TT}.
\label{eq:hjkdef}
\end{equation}
Using this definition in equation (\ref{eq:particledev}), we obtain
\begin{equation}
\delta \xi^j = \frac{1}{2} h_{jk}^\mathrm{TT} \xi_{(0)}^k.
\label{eq:gwxeffect}
\end{equation}
If we orient our coordinates so the gravitational waves propagate in the
$z$-direction, so $h_{jk}^\mathrm{TT}(t-z)$, then the only non-zero components
of $h_{jk}^\mathrm{TT}$ are $h_{xx}^\mathrm{TT}$, $h_{yy}^\mathrm{TT}$,
$h_{xy}^\mathrm{TT}$ and $h_{yx}^\mathrm{TT}$. Since $h_{jk}^\mathrm{TT}$ is
symmetric and traceless, these components satisfy
\begin{align}
h_{xx}^\mathrm{TT} &= - h_{yy}^\mathrm{TT}, \\
h_{xy}^\mathrm{TT} &= h_{yx}^\mathrm{TT}.
\end{align}
For two more particles $C$ and $D$ separated by
\begin{equation}
\zeta^j = \zeta^j_{(0)} + \delta \zeta^j,
\end{equation}
as shown in figure \ref{f:particles} (b), the effect of the gravitational wave
is then given by 
\begin{equation}
\delta \zeta^j = \frac{1}{2} h_{jk}^\mathrm{TT} \zeta_{(0)}^k.
\label{eq:gwyeffect}
\end{equation}
Taking the two particles $A$ and $B$ to lie on the $x$-axis of the LLF of
particle $A$ with separation $x_{(0)}$, without loss of generality, we may
write
\begin{equation}
\xi = (x_{(0)} + \delta x,0,0),
\label{eq:gwxcoord}
\end{equation}
where $\delta x$ is the displacement of particle $B$ caused by the
gravitational wave. Similarly, if particles $C$ and $D$ lie on the $y$-axis of
the LLF of particle $C$ with separation $y_{(0)}$, we may write
\begin{equation}
\zeta = (0,y_{(0)} + \delta y,0)
\label{eq:gwycoord}
\end{equation}
where $\delta y$ is the displacement of particle $D$ caused by the
gravitational wave. 

\begin{figure}[p]
\begin{center}
\includegraphics[width=\linewidth]{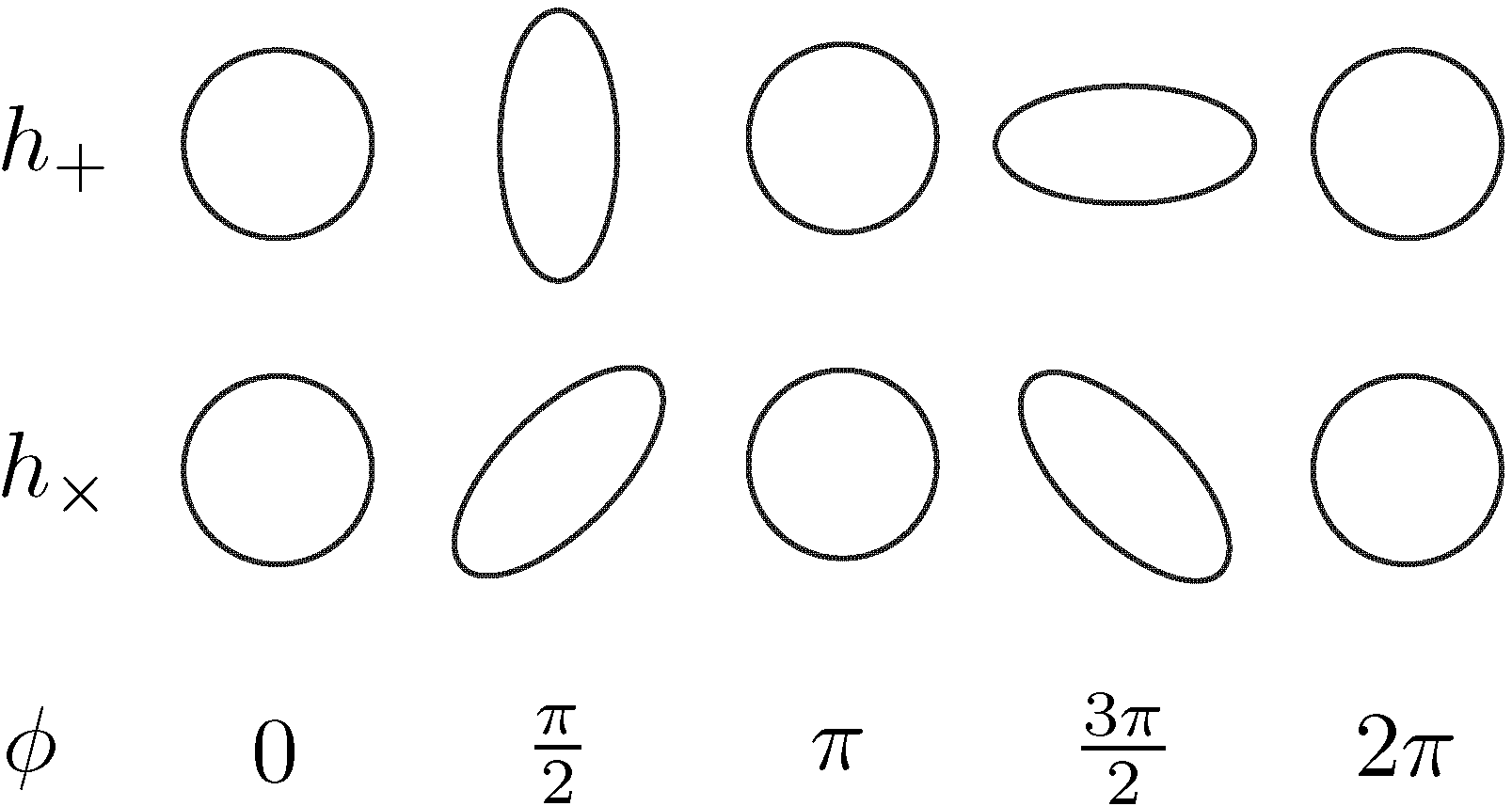}
\end{center}
\caption[Effect of a Gravitational Wave Polarizations on a Ring of Particles]{%
\label{f:rings}
The effect of the two polarizations $h_+$ and $h_\times$ of a sinusoidal
gravitational wave propagating through the page on a ring of test particles.
As the phase $\phi$ of the gravitational wave changes through a complete
cycle, the rings are distorted.
}
\end{figure}

We define the two independent components of the gravitational wave to be
\begin{align}
h_{+} &= h_{xx}^\mathrm{TT} = - h_{yy}^\mathrm{TT}, \\
h_{\times} &= h_{xy}^\mathrm{TT} = h_{yx}^\mathrm{TT}
\end{align}
which we call the \emph{plus} and \emph{cross} polarizations of the
gravitational wave respectively.  The influence of a linearly $+$ polarized
gravitational wave propagating in the $z$-direction on the particles $A,B,C,D$
is then given by substituting equations (\ref{eq:gwxcoord}) and
(\ref{eq:gwycoord}) into (\ref{eq:gwxeffect}) and (\ref{eq:gwyeffect})
respectively to obtain
\begin{align}
\delta x(t-z) &= \frac{1}{2} h_{xx}^\mathrm{TT}(t-z) x_{(0)},\\
\delta y(t-z) &= -\frac{1}{2} h_{yy}^\mathrm{TT}(t-z) y_{(0)}
\end{align}
Similarly, for a linearly $\times$ polarized gravitational wave propagating in the
$z$-direction the effect on the particles is
\begin{align}
\delta x(t-z) &= \frac{1}{2} h_{xy}^\mathrm{TT}(t-z) y_{(0)},\\
\delta y(t-z) &= \frac{1}{2} h_{yx}^\mathrm{TT}(t-z) x_{(0)}.
\end{align}
Figure \ref{f:rings} shows the effect of $h_{+}$ and $h_{\times}$ on a ring
of particles that lie in the $xy$ plane. We can see for the plus polarization
that the effect of a gravitational wave is to stretch the ring
in the $x$ direction, while squeezing it in the $y$ direction for the first
half of a cycle and then squeeze in the $x$ direction and stretch in the $y$
direction for latter half of the cycle.  There is therefore a relative change
in length between the two particles $AB$ and $CD$ as a gravitational wave
passes.  The overall effect of a gravitational wave containing both polarizations
propagating in the $z$ direction is
\begin{align}
\label{eq:deltax}
\delta x(t-z) &= \frac{1}{2}\left[h_{+}(t-z) x_{(0)} + h_{\times}(t-z) y_{(0)}\right],\\
\delta y(t-z) &= \frac{1}{2}\left[-h_{+}(t-z) y_{(0)} + h_{\times}(t-z) x_{(0)}\right].
\label{eq:deltay}
\end{align}
It is the change in the distance between a pair of particles that we attempt
to measure with gravitational wave detectors. We can see from equation
(\ref{eq:deltay}) that the change in length is proportional to the original
distance between the test masses. For a pair of test masses separated by a
length $L$, we define the \emph{gravitational wave strain} $h$ to be the
fractional change in length between the masses
\begin{equation}
h \equiv \frac{1}{2} \frac{\Delta L}{L}.
\end{equation}
The reason to include a factor of $1/2$ in this definition will become
apparent when we discuss measuring gravitational wave strain with an
interferometer.  

\section{The LIGO Gravitational Wave Detectors}
\label{s:ifos}

\begin{figure}[p]
\begin{center}
\includegraphics[width=\linewidth]{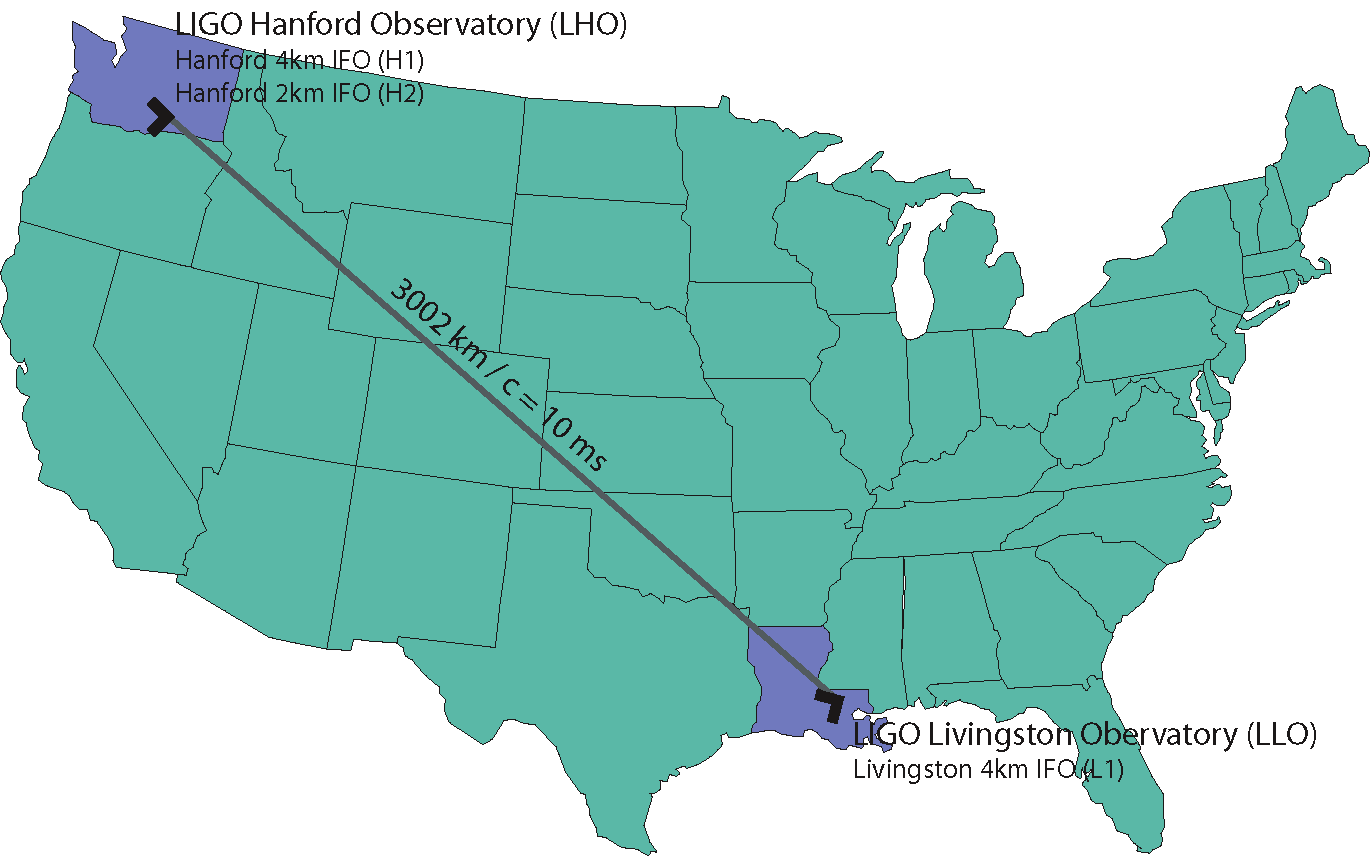}
\end{center}
\caption[Location of LIGO Interferometers]{%
\label{f:usmap}
The location of the three LIGO interferometers. There are two interferometers
at the LIGO Hanford Observatory (LHO) in Washington and one interferometer at
the LIGO Livingston Observatory in Louisiana.
}
\end{figure}

Several major efforts are
underway\cite{Barish:1999,Acernese:2002,Luck:1997hv} to measure the strain
produced by a gravitational wave using \emph{laser interferometry}. The
results in this thesis are based on data from the Laser Interferometer
Gravitational wave Observatory (LIGO). LIGO operates three
\emph{power-recycled-Fabry-Perot-Michelson} interferometers in the United
States. Two of these are co-located at the LIGO Hanford Observatory, WA (LHO)
and one at the LIGO Livingston Observatory (LLO). The interferometers at LHO
are 4~km and 2~km in arm length and are referred to as H1 and H2,
respectively. The interferometer at LLO is a 4~km long interferometer referred
to as L1. The locations and names of the detectors are shown in figure
\ref{f:usmap}.  As we saw in chapter \ref{ch:introduction}, to detect the
gravitational wave strain produced by typical astrophysical sources we need to
measure $h \sim 10^{-22}$. If we separate our test masses by a distance of
$4$~km (a practical distance for earthbound observatories) the challenge
faced by gravitational wave astronomers is to measure changes of length of
order
\begin{equation}
\Delta L \sim 10^{-22} \times 10^4\,\mathrm{m} \sim 10^{-18}\,\mathrm{m}.
\end{equation}

\subsection{The Design of the LIGO Interferometers}
\label{ss:ligoifos}

In an interferometric gravitational wave detector the freely falling masses
described in the previous section are the mirrors that form the arms of the
interferometer\footnote{The mirrors in an Earth bound gravitational wave
observatory are not truly freely falling as they are accelerated by the
gravitational field of the Earth. It can be shown that the horizontal motion
of suspended mirrors is the same as that of freely falling test masses.} and
laser light is used to measure the change in length between the mirrors.  The
challenge facing experimenters constructing a gravitational wave
interferometer is to measure changes of length of order $\sim
10^{-18}\,\mathrm{m}$ using laser light has a wavelength of $\lambda_l \sim
10^{-6}$~m.  It should be noted that measuring a phase shift
\begin{equation}
\Delta \Phi \sim \frac{\Delta L}{\lambda_l} \sim 10^{-12}
\end{equation}
is a factor $10^{12}$ more sensitive than the interferometers used by
Michelson and Morely to disprove the existence of the ether.

\begin{figure}[p]
\begin{center}
\includegraphics[width=\linewidth]{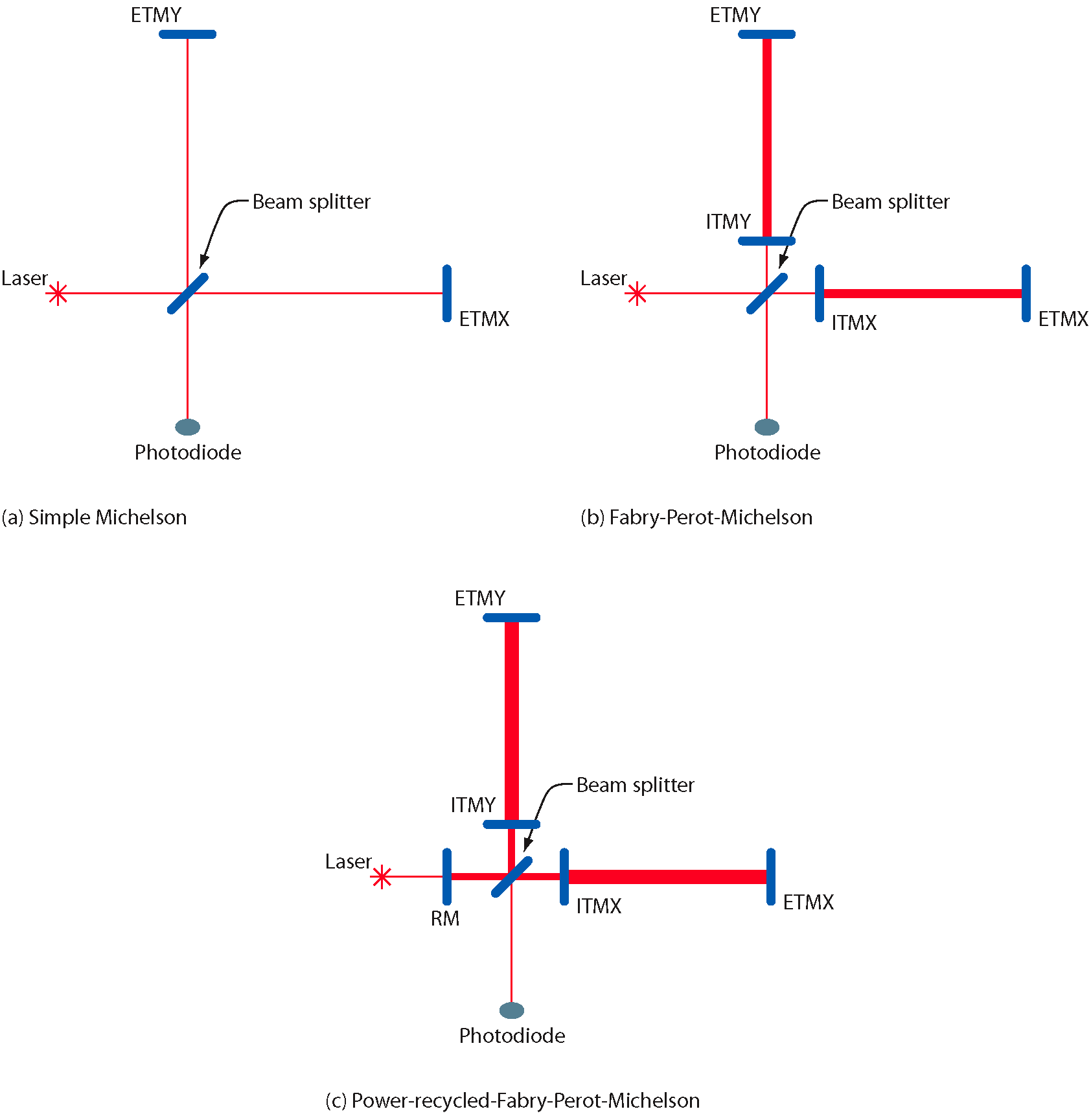}
\end{center}
\caption[Optical Configuration of LIGO]{%
\label{f:ifodesign}
The possible optical configurations of first generation laser interferometers.
The inner $x$ and $y$ test masses (mirrors) are denoted ITMX and ITMY
respectively, and the end $x$ and $y$ test masses (mirrors) are denoted ETMX
and ETMY respectively. The recycling mirror is denoted by RM.  Initial LIGO is
a power-recycled-Fabry-Perot interferometer, type (c) in this figure.
}
\end{figure}

A schematic of a \emph{simple Michelson} interferometer is illustrated in
figure~\ref{f:ifodesign}~(a).  Laser light is shone on a \emph{beam
splitter} which reflects half the light into the \emph{$x$-arm} and transmits
half the light into the \emph{$y$-arm} of the interferometer. The light
travels a distance $L$ in each arm and then is reflected back towards the beam
splitter by the \emph{end test masses}. These masses are equivalent to the
test masses $B$ and $D$ in section~\ref{s:effect}.  Consider the light in the
$x$-arm. For the laser light, the spacetime interval between the beam splitter
and the end test mass is given by
\begin{equation}
ds^2 = g_{\mu\nu}\, dx^\mu\, dx^\nu = 0.
\label{eq:dist}
\end{equation}
In the presence of a plus polarized, sinusoidal, gravitational wave traveling
in the $z$-direction, equation (\ref{eq:dist}) becomes
\begin{equation}
c^2 dt^2 = \left[1 + h_{+}(t-z)\right] dx^2 + \left[1 - h_{+}(t-z)\right] dy^2 + dz^2.
\end{equation}
We can measure the response of the interferometer to a gravitational wave by
considering the phase shift of light in the arms. The phase that the light
acquires propagating from the beam splitter to the $x$-end test mass and
back is given by\cite{Saulson:1994}
\begin{equation}
\begin{split}
\Phi_x &= \int_0^{\tau_\mathrm{RT}} 2\pi f_l\, dt \\
&= \frac{1}{c} \int_0^L 2\pi f_l \sqrt{1 + h_{+}}\,dx -
\frac{1}{c} \int_L^0 2\pi f_l \sqrt{1 + h_{+}}\,dx \\
&\approx \frac{4\pi f_l L}{c} \left(1 + \frac{h_{+}}{2}\right),
\end{split}
\end{equation}
where $\tau_\mathrm{RT}$ is the round trip time of the light and $f_l$ is
its frequency. We have discarded higher order terms in $h_+$ as their effect
is negligible.  We can see that the phase shift acquired in the $x$-arm due to the
gravitational wave is
\begin{equation}
\delta \Phi_x = \frac{2\pi}{\lambda_l} h_{+} L.
\end{equation}
A similar calculation shows that the phase shift acquired in the $y$-arm is
\begin{equation}
\delta \Phi_y = - \frac{2\pi}{\lambda_l} h_{+} L
\end{equation}
and so the difference in phase shift between the arms is
\begin{equation}
\Delta \Phi = \frac{4\pi}{\lambda_l} h_{+} L.
\end{equation}
A typical astrophysical source of gravitational radiation of interest to LIGO,
has a frequency $f_\mathrm{GW} \sim 100$~Hz. Therefore the wavelength of the
gravitational wave is $\lambda_\mathrm{GW} \sim 3000$~km. If
$\tau_\mathrm{RT} = 1 / f_\mathrm{GW}$ there will be no phase shift of the
light at leading order in $h_+$. The light spends exactly one gravitational
wave period in the arm and so the phase shift acquired by positive values of
$h_+(t-z)$ is canceled out by the phase shift due to negative values of
$h_+(t-z)$. The interferometer achieves maximum sensitivity when the light
spends half a gravitational wave period in the arms, that is
\begin{equation}
L = \frac{\lambda_\mathrm{GW}}{2} \sim 1000\,\mathrm{km}
\end{equation}
which is a hopelessly impractical length for a earthbound detector. Instead,
a simple Michelson interferometer is enhanced by placing two additional
mirrors in the arms of the interferometer near the beam splitter, as shown in
figure~\ref{f:ifodesign}~(b). These inner $x$ and $y$ test masses (referred to
as ITMX and ITMY) are designed in LIGO to store the light in the arms for
approximately one half of a gravitational wave period.  The mirrors create a
\emph{Fabry-Perot cavity} in each arm that stores the light for $B \sim 200$
bounces, giving a phase shift of
\begin{equation}
\Delta \Phi = 4\pi \frac{L}{\lambda_l} B h_{+} \sim 
10 \times \frac{4 \times 10^3\,\mathrm{m}}{10^{-6}\,\mathrm{m}} 
\times 200 \times h_+.
\end{equation}
For a gravitational wave strain of $h \sim 10^{-22}$, this increases
$\Delta\Phi$ by 3 orders of magnitude to a phase shift $\Delta \Phi \sim
10^{-9}$.  Further increasing $B$ does not gain additional sensitivity,
however, as storing the light for longer than half a gravitational wave period
causes it to lose phase shift as the sign of the gravitational wave strain
changes.

Is it possible to measure a phase shift of $10^{-9}$ using a
Fabry-Perot-Michelson interferometer? We measure the phase shift by averaging
the light at the photodiode over some period, $\tau$. Let $N$ be the number of
photons from the laser arriving at the photodiode in the time $\tau$. The
measured number of photons in the averaging interval is a Poisson process,
with probability distribution function for $N$ given by
\begin{equation}
p(N) = \frac{ \bar{N} ^{N} \exp \left(-\bar{N}\right) } {N!},
\end{equation}
where $\bar{N}$ is the mean number of photons per interval $\tau$.
The $1\sigma$ uncertainty in the number of photons arriving in the averaging
time is therefore
\begin{equation}
\Delta N = \sqrt{\bar{N}}.
\end{equation}
The accuracy to which we can measure the phase shift for a given input
laser power is constrained by the uncertainty principle,
\begin{equation}
\Delta t \, \Delta E \ge \frac{\hbar}{2}
\label{eq:uncertainty}
\end{equation}
as follows. The energy of the light arriving at the photodiode in time $\tau$ is
\begin{equation}
E = \hbar \frac{2\pi c}{\lambda_l} N,
\end{equation}
which, due to the counting of photons, has uncertainty 
\begin{equation}
\Delta E = \hbar \frac{2\pi c}{\lambda_l} \sqrt{\bar{N}}.
\label{eq:uncertdeltae}
\end{equation}
The uncertainty in the measured the phase is related to the uncertainty
in the time that a wavefront reaches the beam splitter , i.e.
\begin{equation}
\Delta\Phi = 2\pi c \frac{\Delta t }{ \lambda_l }
\label{eq:uncertdeltat}
\end{equation}
Substituting equation (\ref{eq:uncertdeltae}) and (\ref{eq:uncertdeltat})
into equation (\ref{eq:uncertainty}), we obtain
\begin{equation}
\Delta t \, \Delta E = \frac{\Delta \Phi \lambda_l}{2\pi c} \hbar \frac{2\pi
c}{\lambda_l} \sqrt{\bar{N}} \ge \frac{\hbar}{2}.
\end{equation}
The accuracy which which we can measure the phase is therefore no better than
\begin{equation}
\Delta \Phi \ge \frac{1}{\sqrt{\bar{N}}}.
\end{equation}
Hence photon counting statistics limits the accuracy with which the phase
shift can be measured by this method, and this equation tells us how many
photons we need in an averaging period to measure a given phase shift. We need
at least
\begin{equation}
N \ge \frac{1}{2 \left(\Delta\Phi\right)^2}
\end{equation}
photons to measure the phase shift. The optimal averaging time for a
gravitational wave with frequency $f_\mathrm{GW}$ is half a period so that the
light acquires the maximum phase shift, that is
\begin{equation}
\tau \approx \frac{1}{2 f_\mathrm{GW}}.
\end{equation}
The intensity of laser light required to measure a phase shift of $10^{-9}$
for a gravitational wave of $f_\mathrm{GW} \sim 100$~Hz is then
\begin{equation}
\begin{split}
I &= N \left(\frac{2 \pi \hbar c}{\lambda_l}\right) \left( \frac{1}{2 f_\mathrm{GW}}
\right)^{-1} \\
&= \left(\frac{1}{\Delta\Phi}\right)^2 \left(\frac{2 \pi \hbar c}{\lambda_l}\right) 2 f_\mathrm{GW} \\
&\sim \left(\frac{1}{10^{-9}}\right)^2 \left(\frac{10^2 \times 10^{-34} \times
10^{8}}{10^{-6}}\right) 10^2 \sim 10^2\,
\mathrm{W},
\end{split}
\end{equation}
however, lasers used in the first generation of interferometers have a typical
output power of $\sim 5$~W. To increase the power in the interferometer, the
final enhancement to the basic design of our interferometer is the addition of a
\emph{power recycling mirror} (RM) between the beam splitter and the laser, as
shown in figure~\ref{f:ifodesign}~(c). This mirror reflects some of the
(otherwise wasted) laser light back into the interferometer and increases the
power incident on the beam splitter so that the phase shift due to a
gravitational wave of order $h \sim 10^{-23}$ can be measured.

The laser light must be resonant in the power recycling and Fabry-Perot
cavities, to achieve the required power build up in the interferometer. This
requires a complicated \emph{length sensing and control
system}\cite{Fritschel:2001} which continuously monitors the positions of the
mirrors in the interferometer and applies feedback motions via electromagnetic
actuators. The interferometer is said to be \emph{locked} when the control
system achieves a stable resonance. The optics and the servo loop that
controls their positions form the core systems of the interferometer; however
the many subtleties involved in the design and operation of these detectors
are outside the scope of this thesis.

\subsection{Noise sources in an Interferometer}
\label{ss:noise}

\begin{figure}[p]
\begin{center}
\includegraphics[width=\linewidth]{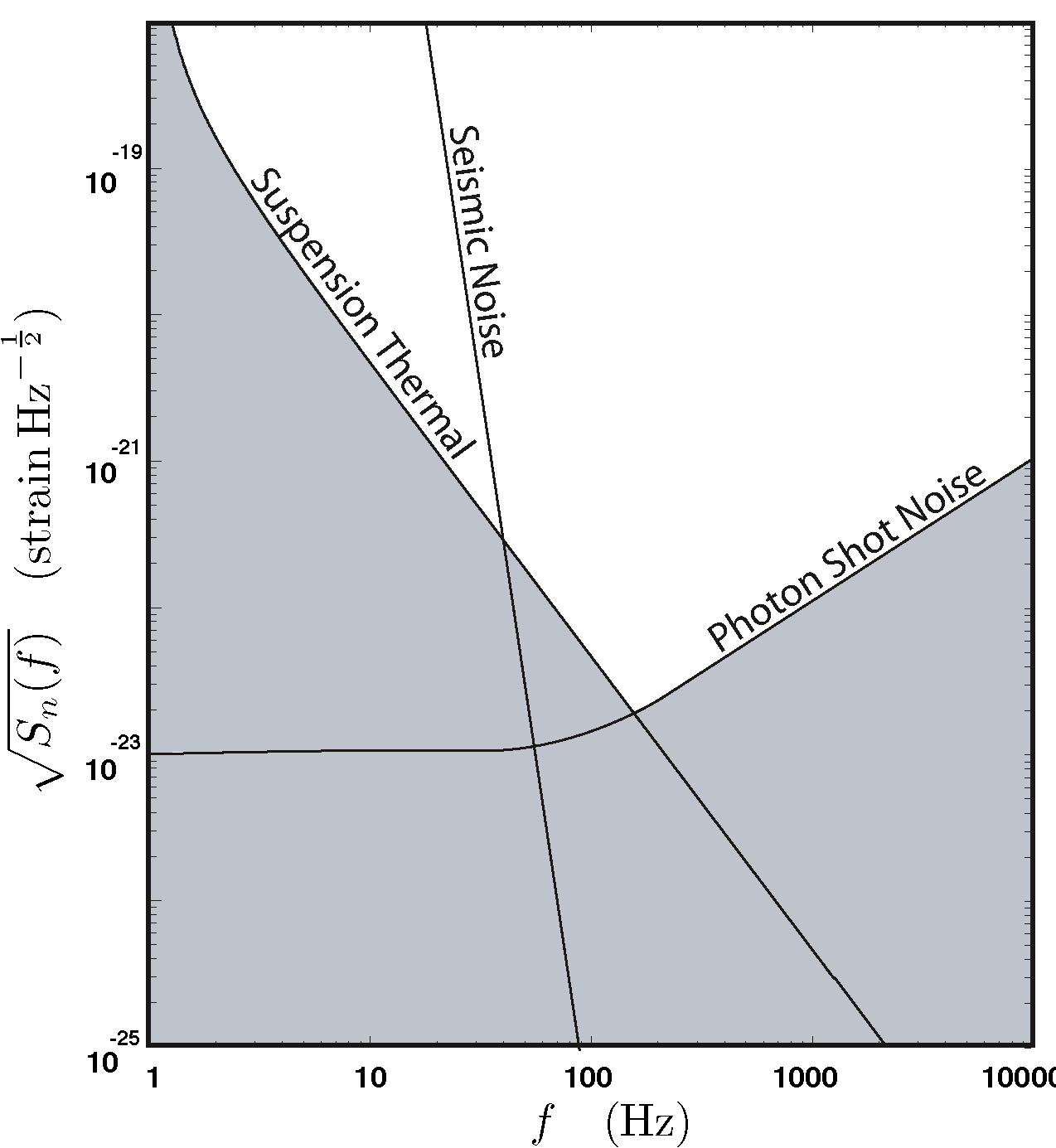}
\end{center}
\caption[Fundamental Noise Sources of Initial LIGO]{%
\label{f:design_noisecurve}
The fundamental noise sources of LIGO.
}
\end{figure}

In reality, there are many sources of noise which can result in an apparent
phase shift of the laser light.  We define the interferometer strain signal,
$s$, to be the relative change in the lengths of the two arms of the
interferometer
\begin{equation}
s(t) = \frac{\Delta L_x - \Delta L_y}{L}.
\end{equation}
This signal has two major additive components: (i) a gravitational wave signal
$h(t)$ and (ii) all other noise sources $n(t)$.  The task of gravitational
wave data analysts is to search for astrophysical signals hidden in this data.
The primary goal of the experimenters engaged in commissioning the LIGO
detectors is the reduction of the noise appearing in $s(t)$.  The noise in
interferometers is measured as the \emph{amplitude spectral density}
$\tilde{h}(f)$. This is the square root of the power spectral density of the
interferometer strain in the absence of gravitational wave signals.
Figure~\ref{f:design_noisecurve} shows the target noise spectral density of
the initial LIGO detectors. There are three fundamental noise sources that
limit the sensitivity of these detectors: 
\begin{enumerate}
\item \emph{Seismic noise.} This is the dominant noise at low frequencies, $f
\lesssim 40$~Hz.  Seismic motion of the earth couples through the suspensions
of the mirrors and causes them to move. To mitigate this, a system of coupled
oscillators is used to isolate the mirror from the ground motion. 

\item \emph{Suspension thermal noise.} This noise source limits the
sensitivity of the interferometer in the range $40$~Hz~$\lesssim f \lesssim
200$~Hz. The steel wire suspending the mirror is at room temperature and
thermal motion of the particles in the wire produce motion of the mirror and
change the arm length.

\item \emph{Photon shot noise.} At high frequencies, $f \gtrsim 200$~Hz, the
noise is dominated by the shot noise due to the photon counting statistics
discussed in the previous section.
\end{enumerate}
For a detailed review of the noise sources present in LIGO's kilometer scale
interferometers, we refer the reader to \cite{Adhikari:thesis}. 

\subsection{Calibration of the Data}
\label{ss:calibration}

\begin{figure}[p]
\begin{center}
\includegraphics[width=\linewidth]{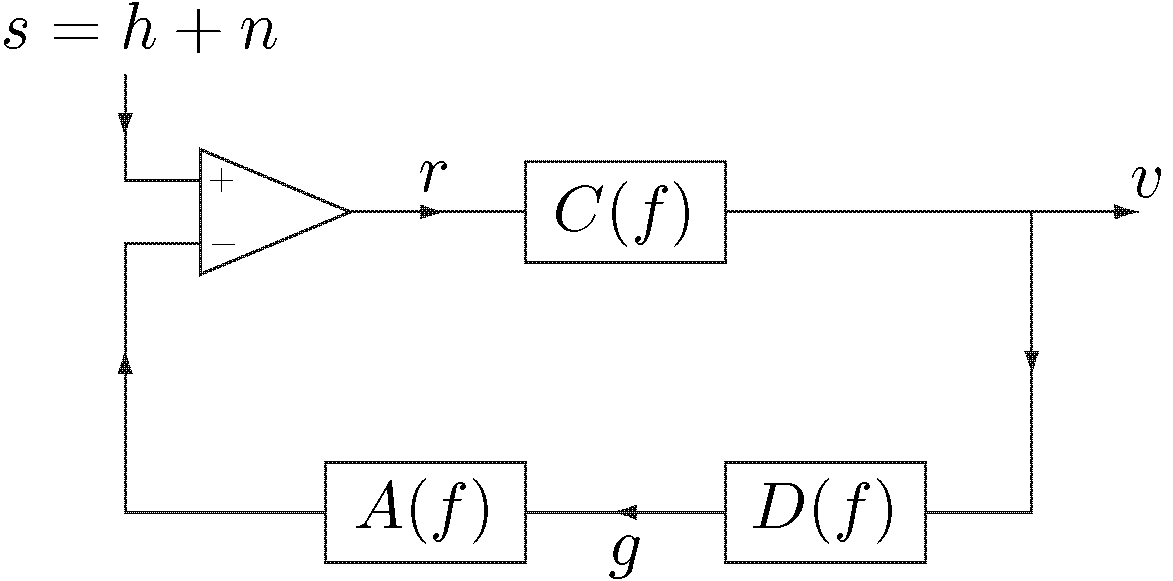}
\end{center}
\caption[Differential mode control servo loop]{%
\label{f:darmloop}
The differential more servo control loop. The figure shows the positions of
three filters: the sensing function $C(f)$, the digital filter $D(f)$ and
the actuation function $A(f)$. The input signal is $s$ and the measured signal
is the error signal $v$. $r$ is the residual length of the cavity and $g$ is
the control signal.
}
\end{figure}

We do not directly record the interferometer strain $s(t)$ but rather the
error signal $v(t)$ of the feedback loop used to control the differential
lengths of the arms. This signal, designated {LSC-AS\_Q} in LIGO,
contains the gravitational wave signal along with other noise. The
interferometer strain is reconstructed from the error signal in the
frequency domain by \emph{calibrating} $v(t)$ using the \emph{response
function} $R(f)$ of the instrument:
\begin{equation}
\tilde{s}(f) = R(f) \tilde{v}(f).
\end{equation}
The response function depends on three elements of the feedback control loop
shown in figure~\ref{f:darmloop}: the sensing function $C(f)$; the actuation
function $A(f)$; the digital feedback filter $D(f)$\cite{Gonzalez:2002}. 

The sensing function $C(f)$ measures the response of the arm cavities to
gravitational waves. It depends on the light power in the arms, which changes
over time as the alignment of the mirrors change.  The actuation function
$A(f)$ encodes the distance the mirrors move for the applied voltage at the
electromagnets. The dominant contribution to this is the pendulum response of
the suspended mirrors.  The digital filter $D(f)$ converts the error signal
$v(t)$ into a control signal that is sent as actuation to the mirrors to keep
the cavities resonant.

If $\tilde{g}(f)$ is the Fourier transform of the control signal applied to
the mirrors, then the residual motion of the mirrors is given by 
\begin{equation}
\tilde{r}(f) = \tilde{s}(f) - A(f)\tilde{g}(f)
\label{eq:reslength}
\end{equation}
as seen in figure~\ref{f:darmloop}. The corresponding error signal is
\begin{equation}
\tilde{v}(f) = C(f) \tilde{r}(f)
\end{equation}
and following around the servo control loop we obtain
\begin{equation}
\tilde{g}(f) = D(f) \tilde{v}(f) = D(f) C(f) \tilde{r}(f).
\label{eq:ctrlsig}
\end{equation}
Substituting equation (\ref{eq:ctrlsig}) into equation (\ref{eq:reslength})
and solving for $\tilde{r}(f)$, we obtain
\begin{equation}
\tilde{r}(f) = \frac{\tilde{s}(f)}{1 + A(f)D(f)C(f)} = \frac{\tilde{s}(f)}{1 + G(f)},
\end{equation}
where $G(f)$ is the \emph{open loop gain} of the interferometer, defined by
$G(f) = A(f)D(f)C(f)$. The error signal is then
\begin{equation}
\tilde{v}(f) = C(f) \tilde{r}(f) = \tilde{s}(f) \frac{C(f)}{1 + G(f)}
\end{equation}
and hence
\begin{equation}
R(f) = \frac{1 + G(f)}{C(f)}.
\end{equation}

The value of the digital filter $D(f)$ is a known at all times.  The
actuation function can be measured by configuring the interferometer as a
simple Michelson, driving a mirror and counting the number of fringes that
appear at the photodiode for a given applied signal. This provides a measure
of the displacement of the mirror for a given control signal.  Since $A(f)$ is
due to the pendulum response of the mirror and known filters used in the
electronics that drive the motion of the mirror, it does not change and its
value can be established before data taking. A sinusoidal
signal of known amplitude that sweeps up in frequency is added to the control
signal after the interferometer is brought into resonance. By comparing the
amplitude of this \emph{calibration sweep} in the output of the detector to
the known input, the value of the open loop
gain and (and hence the sensing function) can be determined as a function of
frequency. The values of the
sensing function and open loop gain at the time of calibration are denoted
$C_0(f)$ and $G_0(f)$.

Although the LIGO detectors have an alignment control system that tries to
keep the power in the arms constant, the power in the cavity can still change
significantly over the course of data taking. These fluctuations in power mean
that the sensing function can change on time scales of order minutes or hours.
To measure $C(f)$ during data taking sinusoidal signals of known amplitude and
frequency $f_\mathrm{cal}$ added to the control signals that drive the
mirrors. These calibration signals show up as peaks in the spectrum and are
called \emph{calibration lines}. By measuring the amplitude of a calibration
line  over the course of the run compared to the time at which the calibration
sweep was taken, we may measure the change in the sensing function
\begin{equation}
C(f;t) = \alpha(t) C_0(f),
\end{equation}
where $\alpha(t)$ is the ratio of the calibration line amplitude at time $t$
to the reference time. We also allow the digital gain of the feedback loop to
vary by a known factor $\beta(t)$ so 
\begin{equation}
D(f;t) = \beta(t) D_0(f).
\end{equation}
The response function at any given time, $t$, becomes
\begin{equation}
R(f;t) = \frac{1 + \alpha(t)\beta(t)G_0(f)}{\alpha(t)C_0(f)}.
\label{eq:calibration}
\end{equation}
To analyze the interferometer data we therefore need the error signal, $v(t)$,
which contains the gravitational wave signal, the functions $C_0(f)$ and
$G_0(f)$, which contain the reference calibration, and the values of
$\alpha(t)$ and $\beta(t)$, which allow us to properly calibrate the data.

\section{Gravitational Waves from Binary Inspiral}
\label{s:inspiralgw}

\begin{figure}[p]
\begin{center}
\includegraphics[width=\linewidth]{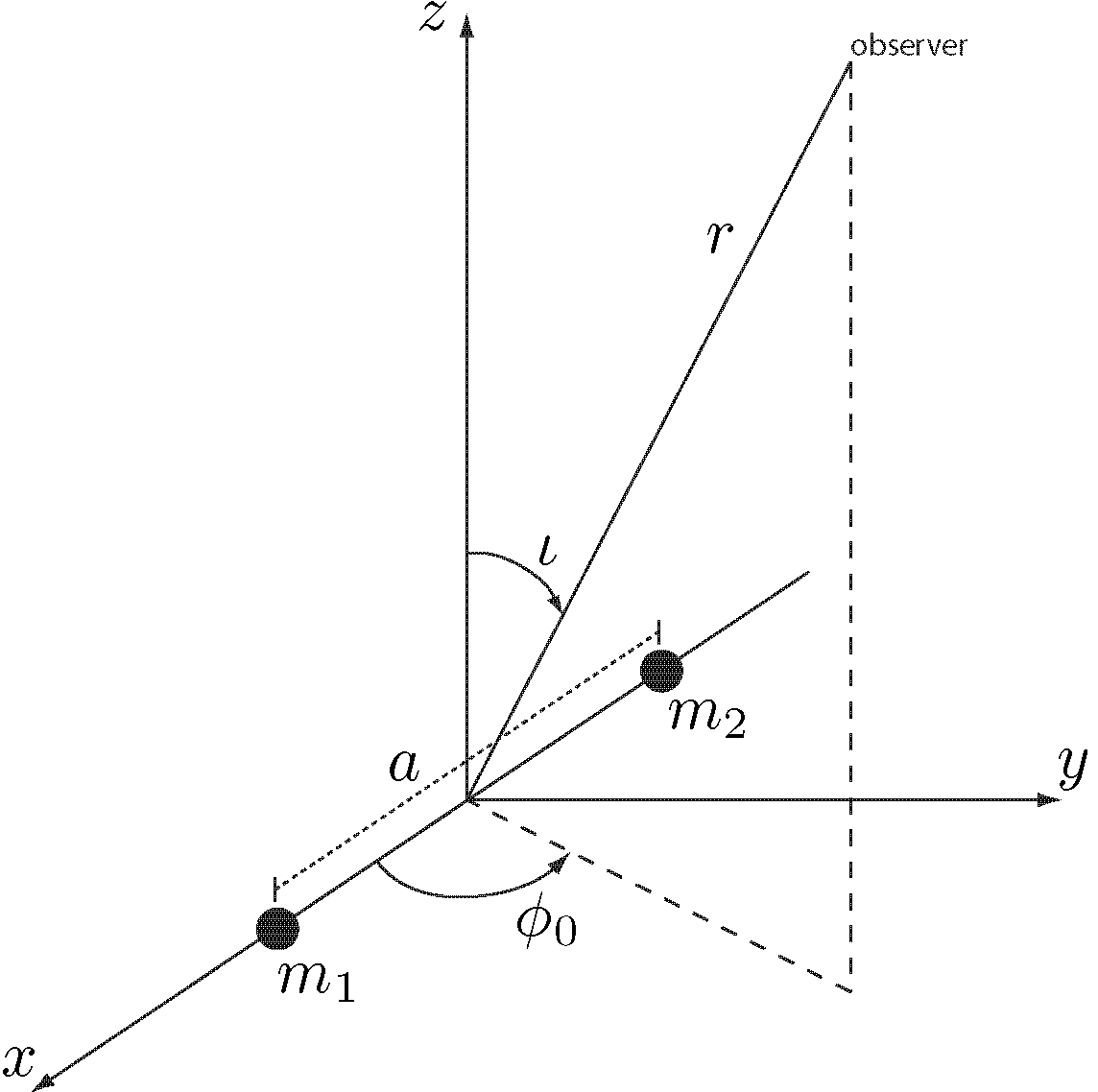}
\end{center}
\caption[Coordinates Used to Describe Gravitational Radiation from a Binary]{%
\label{f:binary}
The parameters of a binary system with rotational axis aligned along the
$z$-axis.
}
\end{figure}

Consider a circular binary system comprised of two black holes $m_1, m_2 \sim
M_\odot$, separated by a distance $a$. If $a \gg 2GM/c^2$, where $M = m_1 +
m_2$, then Newtonian gravity will provide a reasonably accurate description of
the binary dynamics.  If we neglect higher order multipoles, the gravitational
wave field is determined by the \emph{quadrupole formula}\cite{MTW73}
\begin{equation}
h_{jk}^\mathrm{TT} = 
\frac{2G}{c^4r} \frac{d^2 \mathcal{I}_{jk}^\mathrm{TT}(t - r)}{dt^2},
\label{eq:quadrupole}
\end{equation}
where $\mathcal{I}_{jk}$ is the quadrupole moment of the binary, defined by
\begin{equation}
\mathcal{I}_{jk} =
\int \rho(\boldsymbol{x})x_j x_k\,d^3 x
\label{eq:massquad}
\end{equation}
and $\mathcal{I}_{jk}^\mathrm{TT}$ is the transverse traceless part of of
$\mathcal{I}_{jk}$. Since the binary can be described by Newtonian theory,
Kepler's laws are satisfied and the orbital angular velocity is
\begin{equation}
\Omega = \sqrt{\frac{GM}{a^3}}.
\end{equation}
In a Cartesian coordinate system $(x,y,z)$ with origin at
the center of mass of the binary as shown in figure~\ref{f:binary},
the mass distribution of the binary is given, in the point mass approximation,
by
\begin{equation}
\begin{split}
\rho(\boldsymbol{x}) &=
m_1\left[\delta(x - r_1 \cos\Omega t) \delta(y - r_1 \sin\Omega
t)\delta(z)\right] \\
&\quad +
m_2\left[\delta(x + r_2 \cos\Omega t) \delta(y + r_2 \sin\Omega
t)\delta(z)\right],
\label{eq:binarymassdist}
\end{split}
\end{equation}
where
\begin{align}
r_1 &= \frac{m_2}{m_1 + m_2} a, \\
r_2 &= \frac{m_1}{m_1 + m_2} a.
\end{align}
Introduce a second, spherical polar coordinate system labeled
$(r,\iota,\phi_0)$ related to the Cartesian coordinate system by
\begin{align}
\boldsymbol{e}_{\hat{\iota}} &= 
\cos \iota \cos \phi_0 \boldsymbol{e}_{\hat{x}} 
+ \cos \iota \sin \phi_0 \boldsymbol{e}_{\hat{y}} 
- \sin \iota \boldsymbol{e}_{\hat{z}}, \\
\boldsymbol{e}_{\hat{\phi}_c} &= 
- \sin \phi_0 \boldsymbol{e}_{\hat{x}} 
+ \cos \phi_0 \boldsymbol{e}_{\hat{y}}.
\end{align}
To calculate the gravitational radiation $h_{jk}^\mathrm{TT}$ seen by an
observer at a position $(r,\iota,\phi_0)$ relative to the center of mass of
the binary we first calculate the quadrupole moment of the mass distribution
in the frame of the binary. 
The non-zero components of $\mathcal{I}_{jk}$ are $\mathcal{I}_{xx}$,
$\mathcal{I}_{yy}$ and $\mathcal{I}_{xy} = \mathcal{I}_{yx}$. The detailed
derivation of  $\mathcal{I}_{xx}$ following from equation (\ref{eq:massquad})
gives
\begin{equation}
\begin{split}
\mathcal{I}_{xx} &= \int 
m_1\left[\delta(x - r_1 \cos\Omega t) \delta(y - r_1 \sin\Omega
t)\delta(z)\right] \\
&\quad +
m_2\left[\delta(x + r_2 \cos\Omega t) \delta(y + r_2 \sin\Omega
t)\delta(z)\right] x^2\, d^3 x \\
&= (m_1 r_1^2 + m_2 r_2^2) \cos^2 \Omega t \\
&= \left[ m_1 \left(\frac{m_2}{m_1 + m_2}\right)^2 + 
m_2 \left(\frac{m_1}{m_1 + m_2}\right)^2 \right] a^2\cos^2\Omega t \\
&= \left(  \frac{m_1 m_2^2 + m_2 m_1^2}{(m_1 + m_2)^2}\right) a^2\cos^2\Omega t \\
&= \left( \frac{m_1 m_2 (m_1 + m_2)}{(m_1 + m_2)^2} \right) a^2\cos^2\Omega t \\
&= \mu a^2\cos^2\Omega t \\
&= \frac{1}{2} \mu a^2 \left(1 + \cos 2 \Omega t \right).
\end{split}
\end{equation}
Here
\begin{equation}
\mu = \frac{m_1 m_2}{m_1 + m_2} 
\end{equation}
is the \emph{reduced mass} of the binary and we have used
\begin{equation}
\begin{split}
\int \delta(x-x_0) f(x)\, d^3x &= f(x_0), \\
\int \delta(x)\, d^3x = 1.
\end{split}
\end{equation}
The other components are derived in a similar way to give
\begin{equation}
\mathcal{I}_{yy} = \frac{1}{2} \mu a^2 \left(1 - \cos 2 \Omega t \right)
\end{equation}
and 
\begin{equation}
\mathcal{I}_{xy} =
\mathcal{I}_{yx} =
\frac{1}{2} \mu a^2 \sin 2 \Omega t.
\end{equation}
The second time derivative of the quadrupole moment is then
\begin{align}
\ddot{\mathcal{I}}_{xx} &= - 2\mu a^2 \Omega^2 \cos 2\Omega t \\
\ddot{\mathcal{I}}_{yy} &= 2\mu a^2 \Omega^2 \cos 2\Omega t \\
\ddot{\mathcal{I}}_{xy} &=
\ddot{\mathcal{I}}_{yx} = - 2\mu a^2 \Omega^2 \sin 2\Omega t
\end{align}
in the frame of the binary. We transform these to the frame of the observer
using the standard relations
\begin{align}
A'_{ij} &= 
\frac{\partial x_k}{\partial x_i'}
\frac{\partial x_l}{\partial x_j'} A_{kl} \\
\hat{\boldsymbol{e}}_{i}' &= 
\frac{\partial x_j}{\partial x_i'}\hat{\boldsymbol{e}}_{j}
\end{align}
to obtain
\begin{equation}
\begin{split}
\ddot{\mathcal{I}}_{\iota\iota} &= 
\ddot{\mathcal{I}}_{xx} \cos^2 \iota \cos^2 \phi_0 
+  \ddot{\mathcal{I}}_{yy} \cos^2 \iota \sin^2 \phi_0 
+ 2  \ddot{\mathcal{I}}_{xy} \cos^2\iota\sin\phi_0\cos\phi_0 \\
&= - 2\mu a^2 \Omega^2 \cos(2\Omega t) \cos^2\iota \cos 2\phi_0
-2 \mu a^2 \Omega^2 \sin(2\Omega t) \cos^2\iota \sin 2\phi_0 \\
&= -2 \mu a^2 \Omega^2 \cos^2\iota \cos\left(2\Omega t - 2 \phi_0\right).
\end{split}
\end{equation}
Similar transformations give the other components of $\ddot{\mathcal{I}}_{jk}$
\begin{align}
\ddot{\mathcal{I}}_{\phi_0\phi_0} &=
2\mu a^2 \Omega^2 \cos(2\Omega t) \cos 2\phi_0
+ \frac{1}{2} \mu a^2 \Omega^2 \sin(2\Omega t) \sin 2\phi_0 \nonumber\\
&= 2\mu a^2\Omega^2\cos\left(2\Omega t - 2 \phi_0\right), \\
\ddot{\mathcal{I}}_{\iota\phi_0} = 
\ddot{\mathcal{I}}_{\phi_0\iota} &= 
- 2 \mu a^2 \Omega^2 \cos\iota \sin\left(2\Omega t - 2\phi_0\right).
\end{align}
Since these are the transverse components of $\ddot{\mathcal{I}}_{ij}$, we can
simply remove their trace to obtain $\ddot{\mathcal{I}}_{ij}^\mathrm{TT}$:
\begin{equation}
\begin{split}
\ddot{\mathcal{I}}_{\iota\iota}^\mathrm{TT} = - \ddot{\mathcal{I}}_{\phi_0\phi_0}^\mathrm{TT} &= 
\ddot{\mathcal{I}}_{\iota\iota} - \frac{1}{2}
\left( \ddot{\mathcal{I}}_{\iota\iota} + 
\ddot{\mathcal{I}}_{\phi_0\phi_0} \right) = 
\frac{1}{2} \left(  \ddot{\mathcal{I}}_{\iota\iota} - 
\ddot{\mathcal{I}}_{\phi_0\phi_0}
\right) \\
&= - \mu a^2\Omega^2 \left(1 + \cos^2\iota\right) \cos\left(2\Omega t - 2\phi_0\right) ,\\
\ddot{\mathcal{I}}_{\iota\phi_0}^\mathrm{TT} = \ddot{\mathcal{I}}_{\phi_0\iota}^\mathrm{TT}  &=
2\mu a^2 \Omega^2 \cos\iota \sin(2\Omega t - 2\phi_0)
\end{split}
\end{equation}
It is now a simple matter to compute the form of the gravitational
radiation using the quadrupole formula in equation (\ref{eq:quadrupole})
\begin{align}
h_{\iota\iota}^\mathrm{TT} &= 
\frac{2G}{c^4\, r} \ddot{\mathcal{I}}_{\iota\iota}^\mathrm{TT} =
-\frac{2G\mu a^2\Omega^2}{c^4\, r} (1 + \cos^2\iota) \cos\left(2\Omega t - 2\phi_0\right) \\
h_{\iota\phi_0}^\mathrm{TT} &= 
\frac{2G}{c^4\, r} \ddot{\mathcal{I}}_{\iota\iota}^\mathrm{TT} =
-\frac{4G \mu a^2 \Omega^2}{c^4\, r} \cos\iota \sin\left(2\Omega t - 2\phi_0\right).
\end{align}
We can further simplify these equations by using Kepler's third law
\begin{equation}
a = \left(\frac{GM}{\Omega^2}\right)^\frac{1}{3}
\label{eq:kepleragain}
\end{equation}
and defining the \emph{gravitational wave frequency} $f$ which is twice the
orbital frequency 
\begin{equation}
f  = 2\left(\frac{\Omega}{2\pi}\right).
\end{equation}
If we use the basis vectors $\boldsymbol{e}_{\iota}$ and
$\boldsymbol{e}_{\phi_0}$ as the polarization axes of the gravitational wave
we obtain
\begin{align}
h_{+}(t) &= h_{\iota\iota}^\mathrm{TT} = - \frac{2G}{c^4 r} \mu (\pi G M
f)^\frac{2}{3} (1 + \cos^2 \iota ) \cos(2\pi f t - 2 \phi_0) \\
h_{\times}(t) &= h_{\iota\phi_0}^\mathrm{TT} = - \frac{4G}{c^4 r} \mu (\pi G M
f)^\frac{2}{3} \cos\iota \sin(2\pi f t - 2 \phi_0)
\end{align}
If the vector $\boldsymbol{e}_r$ points from the binary to our gravitational
wave detector, the angles $\iota$ and $\phi_0$ are known as the
\emph{inclination angle} and the \emph{orbital phase} and $r$ is the
luminosity distance from the detector to the binary.

\begin{figure}[p]
\begin{center}
\includegraphics[width=\linewidth]{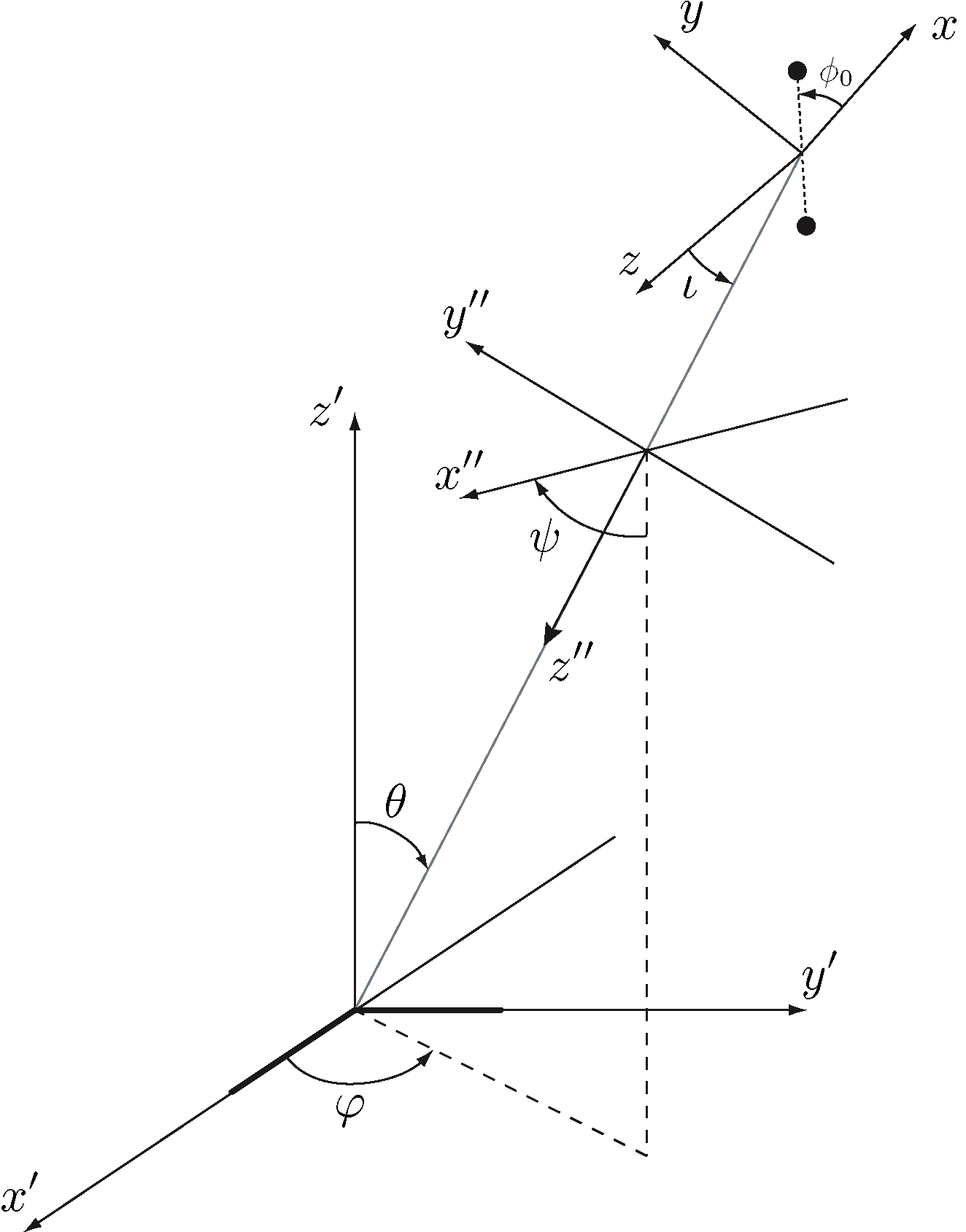}
\end{center}
\caption[Coordinates Used to Describe the Response of a Gravitational Wave Interferometer]{%
\label{f:euler}
Euler angles of a binary system relative to the detector frame $x',y',z'$. The
frame of radiation basis is shown as $x''$ and $y''$.
}
\end{figure}

\begin{figure}[p]
\begin{center}
\includegraphics[width=\linewidth]{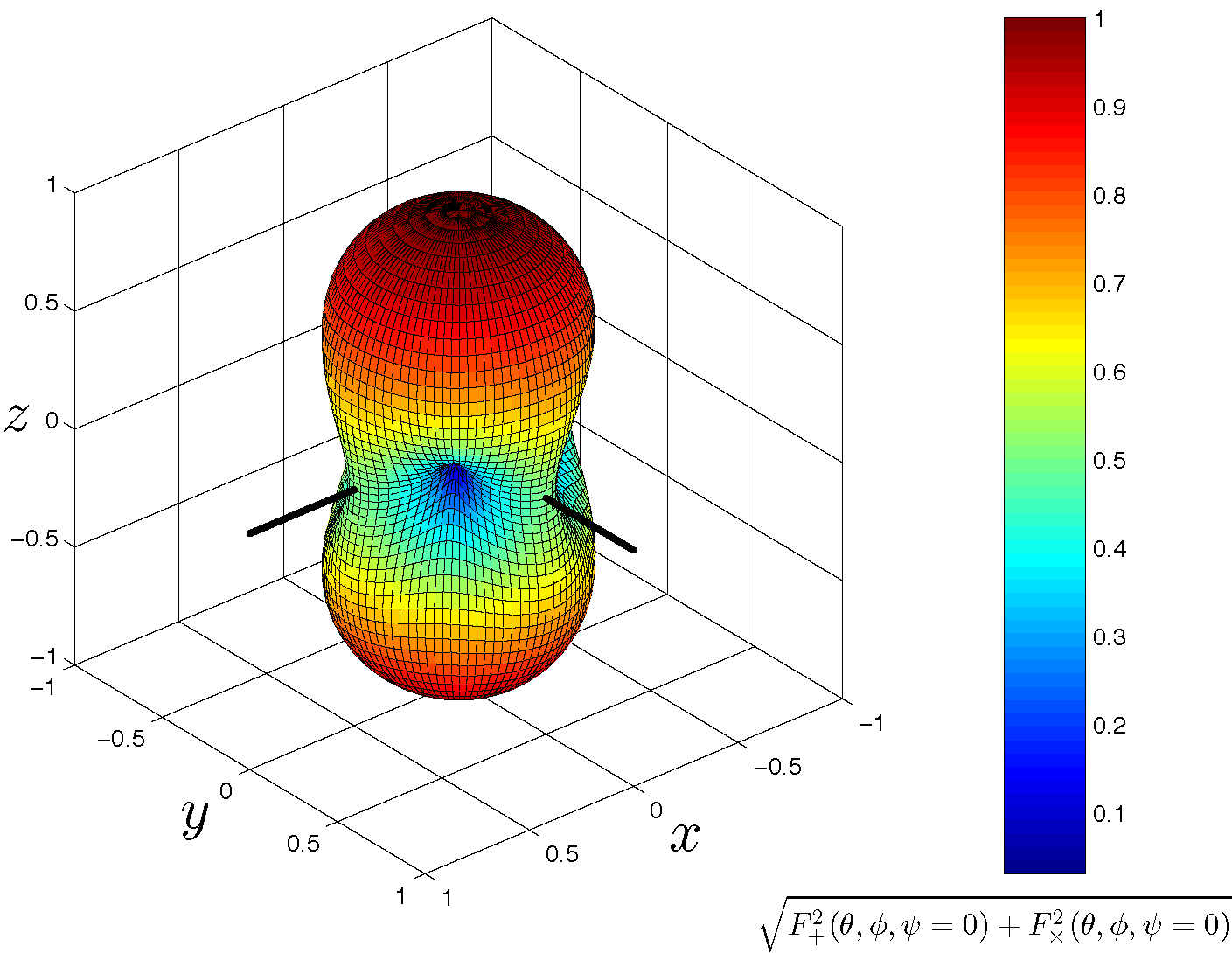}
\end{center}
\caption[Antenna Response of an Interferometer]{%
\label{f:beampattern}
Level surface of the detector response function. The directions of the
interferometer arms are shown.
}
\end{figure}

We assume that the binary evolves through a sequence of quasi-stationary
circular orbits. The orbital energy for a binary with given separation, $a$,
is given by the standard Newtonian formula
\begin{equation}
E = -\frac{1}{2} \frac{G \mu M}{a}.
\end{equation}
The loss energy due to quadrupolar gravitational radiation is\cite{MTW73}
\begin{equation}
\frac{dE}{dt} = - \frac{G}{5c^5} \left\langle 
\frac{d^3 \mathcal{I}^\mathrm{TT}_{jk}}{dt^3}
\frac{d^3 \mathcal{I}^\mathrm{TT}_{jk}}{dt^3}
\right\rangle = \frac{32G^4}{5c^5} \frac{M^3 \mu^2}{a^5}
\end{equation}
and so the inspiral rate for circular orbits is given by
\begin{equation}
\frac{da}{dt} = \frac{dE}{dt}\frac{da}{dE} = - \frac{64G^3}{5c^5}\frac{\mu
M^2}{a^3}.
\end{equation}
The evolution of $a$ as a function of time can therefore be obtained by
integrating
\begin{align}
a^3 \, da &= - \frac{64G^3}{5c^5} \mu M^2 \, dt \\
\frac{a^4}{4} &= \frac{64G^3}{5c^5} \mu M^2 (t_c - t)
\label{eq:afourth}
\end{align}
and so
\begin{equation}
a(t) = \left(\frac{256G^3}{5c^5} \mu M^2 \right)^{\frac{1}{4}}
       \left(t_c - t\right)^{\frac{1}{4}}
\label{eq:aoft}
\end{equation}
which tells us that the orbit shrinks as orbital energy is lost in the form of
gravitational waves. As the orbit shrinks, the orbital frequency increases and
hence the gravitational wave frequency and amplitude increase. We call this
type of evolution a \emph{chirp} waveform.  The evolution of the gravitational
wave frequency $f(t)$ can be obtained by substituting Kepler's third law,
equation (\ref{eq:kepleragain}), into equation (\ref{eq:afourth}) to obtain
\begin{equation}
\begin{split}
\left(GM\right)^{\frac{4}{3}} \Omega^{-\frac{8}{3}} 
&= \frac{256G^3}{5c^5} \mu M^2 (t_c - t) \\
&= \frac{256G^3}{5c^5}\eta M^4 \frac{(t_c - t)}{M},
\end{split}
\end{equation}
where we have defined $\eta = \mu / M$. From this, we may obtain
\begin{equation}
\Omega^{-\frac{8}{3}} = 
\frac{256}{c^8} \left(GM\right)^\frac{8}{3} \frac{c^3 \eta}{5GM}(t_c - t).
\end{equation}
If we define $\Theta(t)$ as the dimensionless time variable
\begin{equation}
\Theta(t) = \frac{c^3\eta}{5GM}(t_c - t),
\label{eq:thetadef}
\end{equation}
then we obtain
\begin{equation}
\Omega^{-\frac{8}{3}} = \left(\frac{8GM}{c^3}\right)^\frac{8}{3}
\Theta(t)
\end{equation}
which written in terms of the gravitational wave frequency $f = \Omega / \pi$
is
\begin{equation}
f(t) = \frac{c^3}{8\pi GM} \left[\Theta(t)\right]^{-\frac{3}{8}}.
\label{eq:foftgw}
\end{equation}
We define the \emph{cosine chirp} $h_c$  and the \emph{sine chirp} $h_s$ as 
\begin{align}
\label{eq:coschirp}
h_c(t) & = \frac{2}{c^2}\left(\frac{\mu}{M_\odot}\right) 
\left[\pi G M f(t)\right]^{\frac{2}{3}} 
\cos\left[2 \phi(t)  - 2\phi_0\right], \\
h_s(t) & = \frac{2}{c^2}\left(\frac{\mu}{M_\odot}\right) 
\left[\pi G M f(t)\right]^{\frac{2}{3}} 
\sin\left[2\phi(t) - 2\phi_0\right],
\label{eq:sinechirp}
\end{align}
where the orbital phase $\phi(t)$ is
\begin{equation}
\phi(t) = 2\pi \int f(t) \, dt
\end{equation}
and $f(t)$ is given by equation (\ref{eq:foftgw}). 
The $+$ and $\times$ waveforms are
\begin{align}
\label{eq:hpluswave}
h_+(t) &= - \frac{GM_\odot}{c^2 r} (1 + \cos^2\iota) h_c(t), \\
h_\times(t) &= - \frac{2GM_\odot}{c^2 r} \cos\iota h_s(t).
\end{align}
If the arms of the interferometer form a second Cartesian axis, $(x',y',z')$,
then we may define the position of the binary relative to the detector by the
spherical polar coordinates $(r,\theta,\varphi)$. It can be shown that the
gravitational waves from the binary will produce a
strain\cite{1987MNRAS.224..131S}
\begin{equation}
h(t) = F_+ h_+(t) + F_\times h_\times(t)
\end{equation}
at the detector, where the \emph{antennae pattern} functions $F_+$ and
$F_\times$ of the detector are given by
\begin{align}
F_+ &= -\frac{1}{2}(1 + \cos^2\theta) \cos 2\varphi \cos 2 \psi - 
\cos\theta \sin 2\varphi \sin 2\psi, \\
F_\times &= +\frac{1}{2}( 1 + \cos^2 \theta) \cos 2\varphi \sin 2\psi -
\cos\theta \sin 2\varphi \cos 2 \psi.
\label{eq:ftimesfunc}
\end{align}
The angle $\psi$ is the third Euler angle that translates from the detectors
frame to the radiation frame. The radiation frame is related to the frame of
the binary by the angles $\iota$ and $\phi_0$, as shown in
figure~\ref{f:euler}. Figure~\ref{f:beampattern} shows the magnitude of the 
strain produced in an interferometer by binary with $\iota = \psi = 0$ at
various positions on the sky. It can be seen that the response of the
detector is essentially omnidirectional, with the maximum sensitivity
occurring when the source lies on the $z$-axis of the detector. Notice that
there are four dead spots in the beam patters where the response of the
interferometer is zero. These correspond to the locations where the binary is
in the plane of the interferometer positioned half way between the $x$ and
$y$-axes. We will often refer to an \emph{optimally oriented} binary system.
This is a binary located at sky position $\theta = 0$ or $\pi/2$, (i.e. above
or below the zenith of the detector) with an inclination angle of $\iota = 0$.
It is so called as this is the position in which the response of the detector
to the binary is a maximum.

\subsection{Higher Order Corrections to the Quadrupole Waveform}

In the previous section we only considered the lowest order multipole
radiation from a binary. The goal is to write down a waveform that is
sufficiently accurate to use matched filtering to search for signals in
detector noise. This requires accurate knowledge of the phase throughout the
LIGO frequency band. In addition to higher order multipoles that contribute to the
energy loss, there are relativistic corrections to the quadrupole formula and
effects such as frame dragging and scattering of the gravitational wave by the
gravitational field of the binary that change the phase evolution. Matched
filtering is less sensitive to the amplitude evolution, however, so we may use
the \emph{restricted post-Newtonian waveform} as the matched filter template.
The restricted post-Newtonian waveform models the amplitude evolution using
the quadrupole formula, but includes higher-order $v/c$ corrections to the phase
evolution. The formula for the orbital phase used in searches for binaries of
component mass $\le 3\,M_\odot$ is given by equation (7) of
\cite{Blanchet:1996pi}
\begin{equation}
\begin{split}
\phi(t) &= \phi_0 - \frac{1}{\eta} \left[ 
\Theta^\frac{5}{8} + \left(\frac{3715}{8064} + \frac{55}{96}\eta\right)
\Theta^\frac{3}{8} - \frac{3\pi}{4} \Theta^\frac{1}{4} \right. \\
&\quad 
+\left. \left(\frac{9\,275\,495}{14\,450\,688} + \frac{284\,875}{258\,048}\eta +
\frac{1855}{2048} \eta^2\right) \Theta^\frac{1}{8} \right],
\label{eq:biwwphase}
\end{split}
\end{equation}
where 
$\phi_0$ and $t_c$ are the orbital phase and time at which the binary coalescences and
$\Theta$ is defined in equation (\ref{eq:aoft}).

\subsection{The Stationary Phase Approximation}
\label{ss:stationaryphase}

We will see in chapter \ref{ch:findchirp} that we require the Fourier
transforms, $\tilde{h}_c(f)$ and $\tilde{h}_s(f)$, of the inspiral waveforms
rather than the time domain waveforms given above. In the search code, we
could compute $\tilde{h}_c(f)$ using the Fourier transform of $h_c(t)$.
This is computationally expensive, however, as it requires an additional
Fourier Transform for each mass pair to be filtered. An alternative 
method is to use the stationary phase approximation\cite{Mathews:1992} to
express the chirp waveforms directly in the frequency
domain\cite{WillWiseman:1996,Cutler:1994}. Given a function
\begin{equation}
B(t) = A(t) \cos 2 \phi(t)
\end{equation}
where
\begin{equation}
\frac{d\ln A}{dt} \ll \frac{d\phi}{dt}
\end{equation}
and
\begin{equation}
\frac{d^2\ln A}{dt^2} \ll \left(\frac{d\phi}{dt}\right)^2
\end{equation}
then the stationary phase approximation to the Fourier transform of $B(t)$ is
given by
\begin{equation}
\begin{split}
\tilde{B}(f) &= \frac{1}{2} A(t) \left(\frac{df}{dt}\right)^{-\frac{1}{2}}
\exp\left[ -i \left(2\pi f t' - 2 \phi(f) - \frac{\pi}{4} \right)\right] \\
&= \frac{1}{2} A(t) \left(\frac{df}{dt}\right)^{-\frac{1}{2}}
\exp\left[ -i \Psi(f) \right], \\
\end{split}
\label{eq:spexpression}
\end{equation}
where $t'$ is the time at which
\begin{equation}
\frac{d\phi(t)}{dt} = \pi f.
\end{equation}
Now it is simple to calculate
\begin{align}
\label{eq:spdfdt}
\frac{df}{dt} 
     &= \frac{d}{dt} \left(\frac{\Omega}{\pi}\right) 
      = \frac{d}{dt} \left(\frac{M^\frac{1}{2}}{\pi} a^{-\frac{3}{2}}\right) \\
     &= \frac{M^\frac{1}{2}}{\pi} \frac{dr}{dt} \left(-\frac{3}{2} a^{-\frac{5}{2}}\right) \\
     &= \frac{M^\frac{1}{2}}{\pi} \left(-\frac{64}{5} \frac{\mu M^2}{r^3}\right)
        \left(-\frac{3}{2} a^{-\frac{5}{2}}\right) \\
     &= \frac{96}{5} \frac{M^\frac{5}{2} \mu}{\pi} a^{-\frac{11}{2}} \\
&= \frac{96}{5} \pi^\frac{8}{3} \mu M^\frac{2}{3} f^\frac{11}{3} \\
&= \frac{96}{5} \pi^\frac{8}{3} \mathcal{M}^\frac{5}{3} f^\frac{11}{3}
\end{align}
where we have defined the \emph{chirp mass} by
\begin{equation}
\mathcal{M} = \mu^\frac{3}{5} M^\frac{2}{5}.
\end{equation}
To obtain the phase function $\Psi(f)$ we note that
\begin{equation}
f = \frac{1}{\pi} \frac{d\phi}{dt}
\end{equation}
and we can invert the series in equation (\ref{eq:biwwphase}) to write
$\Theta$ as a function of $f$. Substituting this result and the result for
$df/dt$ into the equation for the stationary phase approximation, equation
(\ref{eq:spexpression}), we obtain the form of the inspiral chirps that we
will use in matched filtering
\begin{equation}
\label{eq:spcosderiv}
\tilde{h}_c(f)=\frac{2GM_\odot}{c^2 r}
\left(\frac{5\mu}{96M_\odot}\right)^\frac{1}{2}
\left(\frac{M}{\pi^2M_\odot}\right)^\frac{1}{3}
f^{-\frac{7}{6}}\, \left( \frac{GM_\odot}{c^3} \right)^{-\frac{1}{6}}\,
e^{i\Psi(f;M,\eta)}.
\end{equation}
with the phase evolution given by
\begin{equation}
\begin{split}
\Psi(f;M,\eta) &= 2\pi ft_c-2\phi_0-\pi/4+\frac{3}{128\eta}\biggl[x^{-5}+
\left(\frac{3715}{756}+\frac{55}{9}\eta\right)x^{-3}
-16\pi x^{-2} \\
&\quad +\left(\frac{15\,293\,365}{508\,032}+\frac{27\,145}{504}\eta
+\frac{3085}{72}\eta^2\right)x^{-1}\biggr],
\label{eq:spphasederv}
\end{split}
\end{equation}
where $x=(\pi M f G/c^3)^{1/3}$. Notice that in the definition of
$\tilde{h}_c(f)$ we have absorbed the amplitude term $2GM_\odot/c^2 r$ from
$\tilde{h}_{+}$. This allows us to place $\tilde{h}_c$ at a cannonical
distance of $r = 1$~Mpc, as discussed later. Physically the chirp waveform
should be terminate when the orbital inspiral turns into a headlong plunge,
however the frequency at which this happens is not known for a pair of
comparably massive objects. We therefore terminate the waveform at the
gravitational wave frequency of a test particle in the innermost stable
circular orbit of Schwarzschild (ISCO)\cite{Wald:1984}
\begin{equation}
f_\mathrm{isco} = \frac{c^3}{6\sqrt{6}\pi GM},
\end{equation}
which is a reasonable approximation of the termination
frequency\cite{Droz:1999qx}. Since the sine chirp is simply the orthogonal
waveform to the cosine chirp, we have
\begin{equation}
\tilde{h}_s(f)=i\tilde{h}_c(f).
\label{eq:spsinderiv}
\end{equation}
Together equations (\ref{eq:spcosderiv}), (\ref{eq:spphasederv}) and
(\ref{eq:spsinderiv}) give the form of the chirps that we will use for the
application of matched filtering discussed in chapter \ref{ch:findchirp}.

\newpage

\Chapter{Binary Black Hole MACHOs}
\label{ch:macho}

One of the most interesting current problems in astrophysics is that of
\emph{dark matter}. Dark matter is so called because it has eluded detection
through its emission or absorption of electromagnetic radiation. Our knowledge
of its existence comes from its gravitational interaction with luminous matter
in the universe. There have been several ideas proposed to explain the nature
of dark matter; chief among these are \emph{weakly interacting massive
particles} (WIMPs) and \emph{massive astrophysical compact halo objects}
(MACHOs)\cite{Griest:1990vu}.  WIMPs, supersymmetric particles produced as a
relic of the big bang, are outside the scope of this thesis\footnote{We refer
to \cite{Griest:1995gs} for a review of the nature of dark matter.}. No
compelling reason exists to think that WIMPs will produce significant
gravitational waves. In this chapter, we review the evidence for dark matter
in the form of MACHOs in the Galactic halo. The nature of MACHOs is unknown;
we review a proposal that suggests that if MACHOs are primordial black holes
(PBHs) formed in the early universe, then some of the PBH MACHOs may be in
binary systems\cite{Nakamura:1997sm}. Searching for gravitational waves from
the inspiral and coalescence of these binary black hole MACHOs (BBHMACHOs) is
the motivation for this thesis.

\section{Dark Matter In The Galactic Halo}
\label{s:darkmatter}

\begin{figure}[p]
\begin{center}
\includegraphics[width=\linewidth]{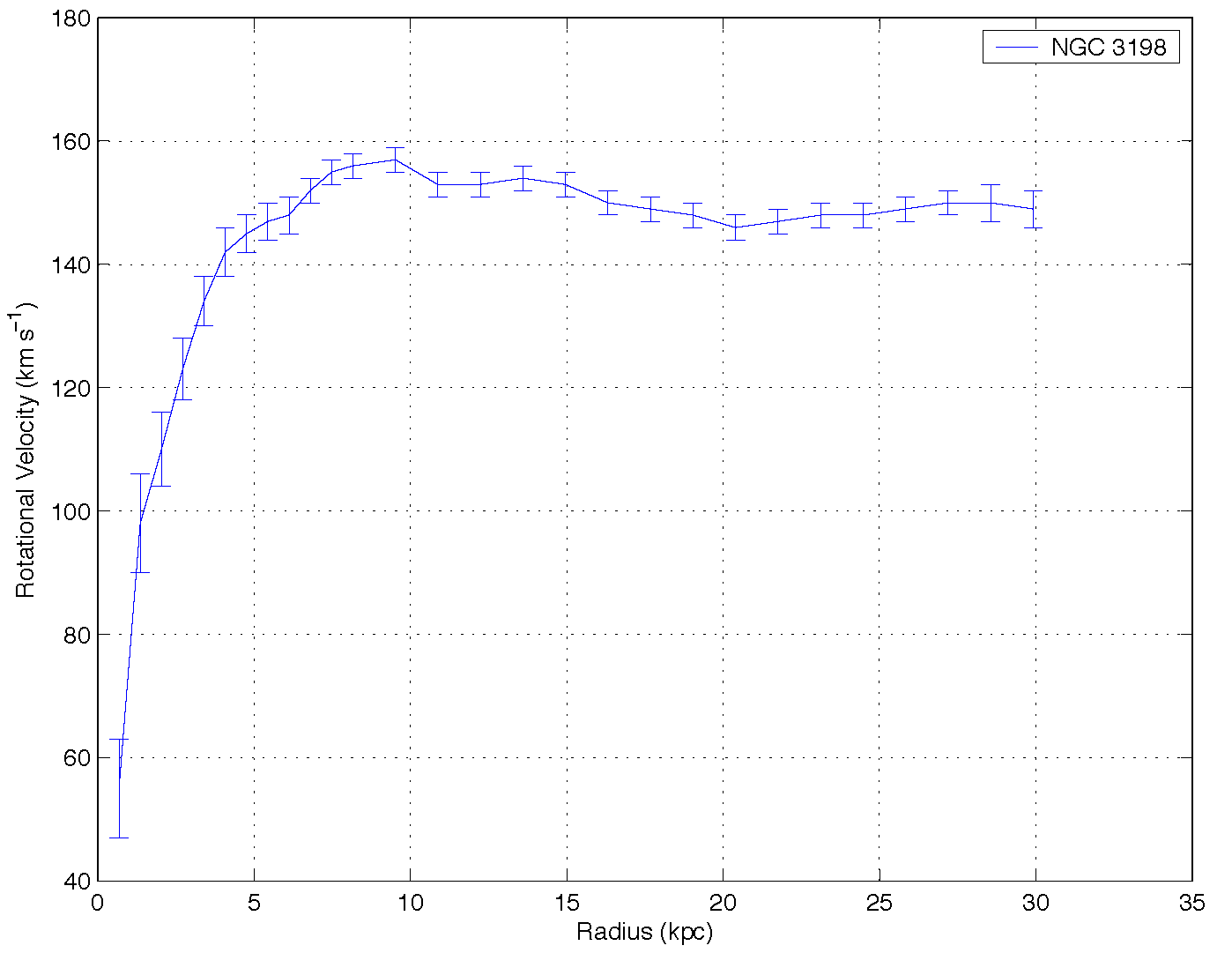}
\end{center}
\caption[H1 Rotation Curve of NGC 3198]{%
\label{f:rotcurves}
The results of observations of the hydrogen $21$~cm line of the spiral galaxy
NGC 3198 show that the rotation curve is flat out to the last measured point
at $30$~kpc\cite{1989A&A...223...47B}. This implies a large discrepancy
between the observed rotation curve and that predicted from light
observations.
}
\end{figure}

Dark matter is detected by its gravitational interaction with luminous matter.
Strong evidence for the presence of dark matter in the universe comes from the
study of galactic rotation curves: measurements of the velocities of luminous
matter in the disks of spiral galaxies as a function of galactic radius.  
Consider a simple rotational model for the disk of a spiral galaxy.
Consider a star with mass $m_s$ orbiting at radius $r$ outside the disk of
the galaxy. Newtonian dynamics tells us that if the mass inside radius
$r$ is $m_g$ then
\begin{equation}
\frac{Gm_g m_s}{r^2} = \frac{m_s v_s^2}{r}
\label{eq:newtongalaxy}
\end{equation}
where $v_s$ is the velocity of the star and $G$ is the gravitational constant. 
Let us suppose that as we increase $r$, the change in the $m_g$ is negligible,
which is a reasonable assumption towards the edge of the disk of a typical
spiral galaxy.  We can see from equation (\ref{eq:newtongalaxy}) that we would
expect the velocity of stars at the edge of the galactic disk to fall off as 
\begin{equation}
v_s \propto \frac{1}{\sqrt{r}},
\end{equation}
when $r \gg R$, where $R$ is the radius containing most of the disk matter.
Galactic rotation curves, determined using the Doppler shift of the $21$~cm
hydrogen line, have been measured for several galaxies\cite{Sancisi:1987}. It
is found that the rotation curves do not fall off as expected. Instead the
rotational velocities of galactic matter are measured to be constant out to
the edge of the visible matter in the disk, as shown in
figure~\ref{f:rotcurves}.  This surprising result suggests that 
80\%--90\% of the matter in spiral galaxies is in the form of dark matter
stretching out at least as far as the visible light.

A typical argument to understand the formation of galactic disks from baryonic
matter considers an initially spherical distribution of baryonic matter
rotating with some angular momentum, $L$. Over time the matter will lose
energy through inelastic collisions. Since the angular momentum of the system
is conserved, the initial distribution must collapse to a rotating disk.  On
the other hand, if the initially spherical distribution is composed of dark
matter instead of baryons, the collisions will be elastic because the dark
matter is weakly interacting. As a result of this, dark matter initially
distributed in an isotropic sphere will maintain this distribution over time.
Since we do not expect a spherical dark matter halo to collapse to a disk, the
simplest possible assumption is that the dark halo is a spherical, isothermal
distribution of dark matter. This suggests that dark matter will be
distributed in an extended halo encompassing the luminous matter of a galaxy.
If we assume that the density of the dark matter
is $\rho(r)$ then the mass within a thin shell of a spherical halo is
\begin{equation}
dM(r) = 4\pi r^2 \rho(r)\, dr,
\label{eq:simplehalodensity}
\end{equation}
where $dr$ is the thickness of the shell.
Using Newtonian dynamics, the velocity $v$ of a particle of mass $m$ at
radius $r$ is
\begin{equation}
\begin{split}
\frac{GM(r)m}{r^2} &= \frac{mv^2}{r} \\
v^2 &= \frac{GM(r)}{r}.
\end{split}
\end{equation}
The galactic rotation curves tell us that the velocity is independent of the
radius, so
\begin{equation}
M(r) = \frac{v^2r}{G}.
\end{equation}
Differentiating this with respect to $r$ and substituting the result into
equation (\ref{eq:simplehalodensity}), we obtain
\begin{equation}
\frac{dM(r)}{dr} = \frac{v^2}{G} = 4\pi r^2\rho(r)
\end{equation}
which gives
\begin{equation}
\rho(r) = \frac{v^2}{4\pi r^2 G}.
\label{eq:simplehalodensity2}
\end{equation}
If we assume that the dark and visible matter are in thermal equilibrium, we
may use the measured rotational velocity of local stars about the galactic
center as the velocity of the dark matter. 

We can easily estimate the density of dark matter in the neighborhood of the
Earth $\rho(r_E)$ as follows. The earth is approximately $8$~kpc
from the galactic center and the rotational velocity of objects at this radius
is $v\sim 200$~$\mathrm{km\,s}^{-1}$. Using these values in equation
(\ref{eq:simplehalodensity2}), we find
\begin{equation}
\rho(r_E) = 7.6 \times 10^{-25}\, \mathrm{g}\,\mathrm{cm}^{-3}.
\end{equation}
More sophisticated modeling of the
Galaxy\cite{1995ApJ...449L.123G}, suggests that the local halo density is
\begin{equation}
\rho(r_E) = 9.2_{-3.1}^{+3.8} \times 10^{-25}\, \mathrm{g}\,\mathrm{cm}^{-3}
\end{equation}
or approximately $0.01\,M_\odot\,\mathrm{pc}^{-3}$.

Equation (\ref{eq:simplehalodensity2}) applies only at intermediate radial
distances. The data at small $r$ is consistent with the dark matter having a
constant \emph{core density} $\rho_c$ within a \emph{core radius}
$a$\cite{Rix:1996}. The halo density then becomes
\begin{equation}
\rho(r) = \frac{\rho_c}{1 + \left(\frac{r}{a}\right)^2}.
\label{eq:simplehalodensity3}
\end{equation}
The values of $\rho_c$ and $a$ are obtained by fitting measured galactic
rotation curves to equation (\ref{eq:simplehalodensity3}) using data near the
galactic center.  There is, in fact, no evidence to suggest that halos are
exactly spherical. In fact the halo density may be flattened\cite{Rix:1996}. For a
flattened halo a model of the dark matter density becomes
\begin{equation}
\rho(R,z) = \frac{\rho_c r^2_c}{a^2 + R^2 + z^2/q^2}
\label{eq:simplehalodensity4}
\end{equation}
where $R$ and $z$ are galactocentric cylindrical coordinates and $q$ is a
parameter that describes the flattening of the halo. At present there is no
measurement of the extent of galactic halos beyond the luminous matter. For
the Milky Way it is thought that the halo extends out to a radius of $\sim
50$kpc, although it is possible that it extends all the way out to the Andromeda
galaxy at $\sim 700$~kpc.

\section{MACHOs in the Galactic Halo}
\label{s:machos}

Galactic rotation curves provide strong evidence that spiral galaxies such
as the Milky Way are surrounded by a large quantity of dark matter, but 
tell us nothing about the nature of this dark matter.  A variety of
candidates have been proposed to explain the nature of dark matter. These
generally fall into two classes. The first class consists of elementary
particles such as axions\cite{Weinberg:1977ma} or weakly interacting massive
particles (WIMPs)\cite{Goodman:1984dc}. 
Such dark matter candidates
are outside the scope of this thesis. Active searches for WIMPs and axions are
underway and we refer to \cite{Griest:1995gs} for a review of the particle
physics dark matter candidates.  The second class of dark matter candidates
are known as \emph{massive astrophysical compact halo objects} or MACHOs.
MACHOs are objects such as brown dwarfs (stars with insufficient mass to burn
hydrogen), red dwarfs (stars with just enough mass to induce nuclear fusion),
white dwarfs (remnants of $1$--$8\,M_\odot$ stars) or black holes located in
the halos of galaxies.  Optical and infrared observations in the early 1990's
were not sensitive enough to constrain the fraction of the halo in
MACHOs\cite{1994MNRAS.266..775K} and the method of \emph{gravitational
lensing} was suggested as a method for detecting halo dark matter in the form
of MACHOs\cite{Paczynski:1985jf}.

\subsection{Gravitational Lensing of Light}
\label{ss:microlensing}

\begin{figure}[p]
\begin{center}
\includegraphics[width=\linewidth]{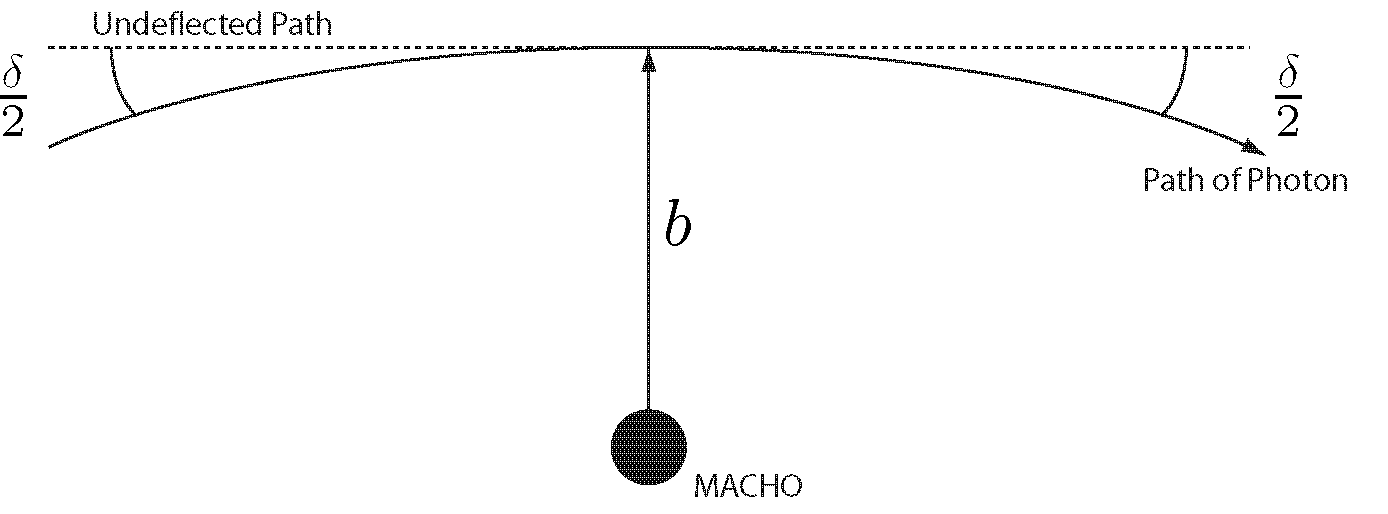}
\end{center}
\caption[Scattering of Light in Schwarzschild Spacetime]{%
\label{f:scattering}
A photon will be scattered by the curvature of spacetime caused to the
gravitational field of a MACHO. If the closest approach of the photon is 
$b$, it will be deflected by an angle $\delta = 4GM / c^2 b$.
}
\end{figure}

\begin{figure}[p]
\begin{center}
\includegraphics[width=\linewidth]{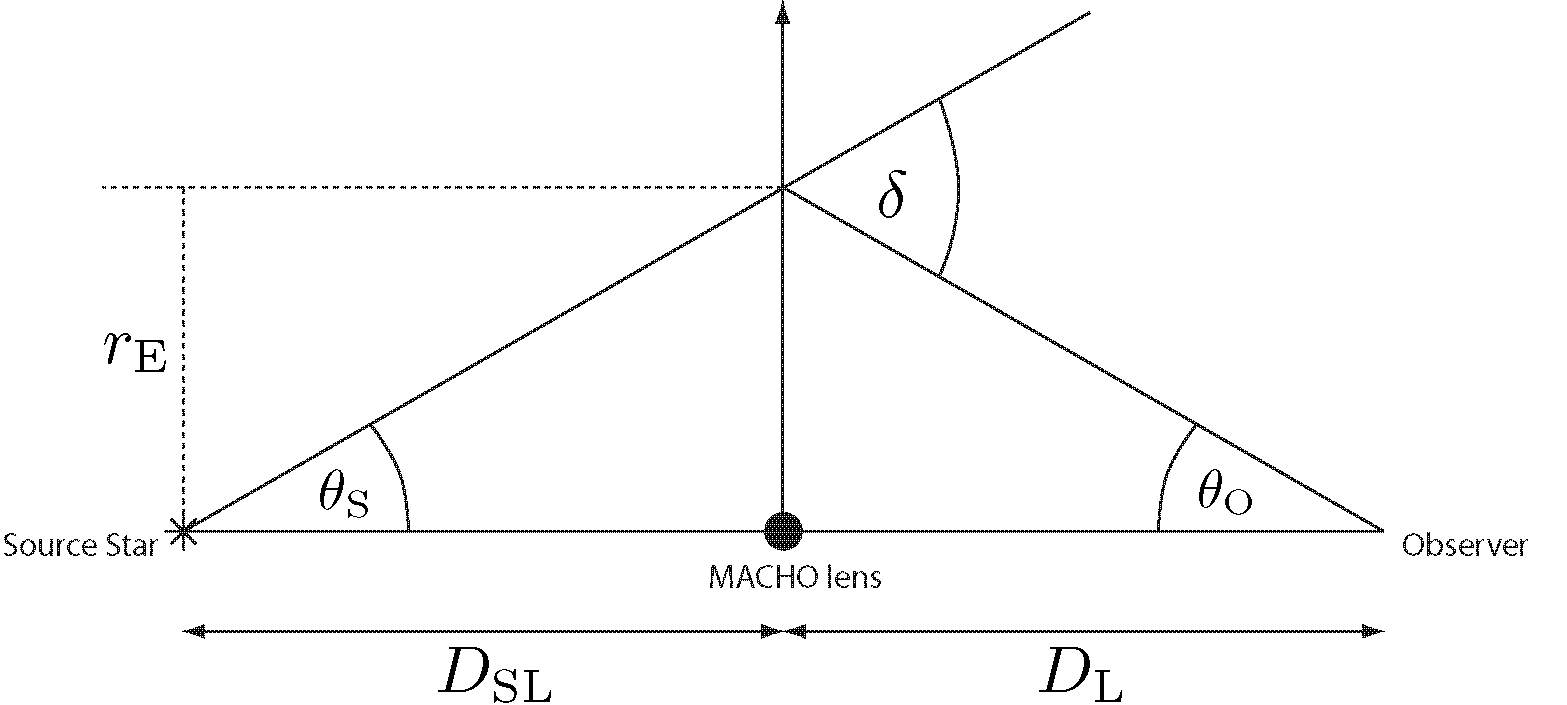}
\end{center}
\caption[Gravitational Lensing of Light By a MACHO]{%
\label{f:macholens}
The geometry of microlensing of light from a source star by a MACHO showing
the definition of the Einstein radius, $r_\mathrm{E}$.
}
\end{figure}

Gravitational lensing is caused by the bending of light around a massive
object. Assume that a MACHO produces a spherically symmetric gravitational
field; the geometry of spacetime around the MACHO satisfies the Schwarzschild
solution. Consider the scattering of light by a MACHO shown in figure
\ref{f:scattering}, where $b$ the \emph{impact parameter} of the light, 
the minimum distance of the photon to the MACHO.   Recall that the
lightlike orbits of Schwarzschild spacetime satisfy\cite{Wald:1984}
\begin{equation}
\frac{d^2 u}{d\phi^2} + u = \frac{3GM}{c^2 b}u^2
\label{eq:schphotonorbit}
\end{equation}
where $u = b/r$ and $M$ is the mass
of the MACHO. If $R$ is the size of the
MACHO, then
\begin{equation}
\frac{3GMu^2}{c^2ub} = \frac{3GM}{c^2R} \frac{R}{b} \ll 1
\end{equation}
if $b \gg R$. we can solve equation (\ref{eq:schphotonorbit}) perturbatively
in the small parameter $\epsilon = R / b$ as follows. Write
\begin{equation}
u = u_0 + \epsilon u_1 + \cdots
\end{equation}
and substitute into equation (\ref{eq:schphotonorbit}) to get
\begin{equation}
u_0'' + \epsilon u_1'' + \cdots + u_0 + \epsilon u_1 + \cdots 
= \frac{3GM}{c^2R}\left(u_0 + \epsilon u_1 + \cdots\right)^2 
\end{equation}
where prime denotes differentiation with respect to $\phi$. At leading order,
\begin{equation}
u_0'' + u_0 = 0
\end{equation}
has solution
\begin{equation}
u_0 = A \sin \left(\phi + \phi_0\right)
\end{equation}
where $A$ and $\phi_0$ are constants. We are free to choose any value for
$\phi_0$ as it simply chooses an orientation for the axes in figure
\ref{f:scattering}, so let $\phi_0 = 0$. Since $\phi = \pi / 2$
gives the distance of closed approach, we find $A = b / r_\mathrm{min} = 1$.
Now $u_1$ satisfies
\begin{equation}
\begin{split}
u_1'' + u_1 &= \frac{3GM}{c^2 R} \sin^2 \phi \\
&= \frac{3GM}{2c^2 R}\left(1 - \cos 2\phi\right).
\label{eq:u1eq}
\end{split}
\end{equation}
Inspection suggests a solution of the form
\begin{equation}
u_1 = \frac{3GM}{2c^2 R} + \alpha \cos 2\phi.
\end{equation}
Substituting this into equation (\ref{eq:u1eq}) we find that
\begin{equation}
-4\alpha \cos 2\phi + \frac{3GM}{2c^2 R} + \alpha\cos 2\phi =
\frac{3GM}{2c^2 R} - \frac{3GM}{2c^2 R} \cos 2\phi
\end{equation}
and so $\alpha = GM/2c^2 R$. The solution for $u$, up to first order, is
therefore
\begin{equation}
u = \sin \phi + \frac{GM\epsilon}{2c^2 R}\left(3 + \cos 2\phi\right).
\end{equation}
As $r \rightarrow \infty$, $u \rightarrow 0$ and $\phi \rightarrow - \delta/2$,
so
\begin{equation}
0 = \sin \left(-\frac{\delta}{2}\right) + 
\frac{GM}{2c^2 b} \left(3 + \cos(-\delta)\right)
\end{equation}
For small $\delta$, $\sin(-\delta/2) \approx -\delta/2$ and $\cos(-\delta)
\approx 1$, so
\begin{align}
0 &\approx - \frac{\delta}{2} + \frac{2GM}{c^2 b} \\
\delta &\approx \frac{4GM}{c^2b}
\end{align}
The total deflection of the light is therefore
\begin{equation}
\delta = \frac{4GM}{c^2 b}.
\end{equation}
Suppose a MACHO lens is at a distance $D_\mathrm{SL}$ from a source star and
an observer is at a distance $D_\mathrm{L}$ from the MACHO as shown in figure
\ref{f:macholens}. Then the ray of light from the source that encounters
the MACHO with critical impact parameter $r_\mathrm{E}$ will reach the
observer. Simple geometry, using the small angle approximations, shows that
\begin{equation}
\delta = \theta_\mathrm{S} + \theta_\mathrm{O} =
\frac{r_\mathrm{E}}{D_\mathrm{L}} + \frac{r_\mathrm{E}}{D_\mathrm{SL}} =
\frac{4GM}{c^2r_\mathrm{E}}
\end{equation}
therefore the observer sees the lens light when the ray is at the
\emph{Einstein radius}, $r_\mathrm{E}$, given by
\begin{equation}
r_\mathrm{E} = \sqrt{\frac{4GM}{c^2} \frac{D_\mathrm{SL} D_\mathrm{L}}
{D_\mathrm{SL} + D_\mathrm{L}}}.
\end{equation}
If the source, MACHO and observer are collinear, as shown in figure
\ref{f:macholens} the observer sees a bright ring of radius
$r_\mathrm{E}$ around the MACHO. The angular radius of this ring is the
\emph{Einstein angle},
\begin{equation}
\theta_\mathrm{E} = \sqrt{\frac{4GM}{c^2} \frac{D_\mathrm{SL}}
{D_\mathrm{L}\left(D_\mathrm{SL} + D_\mathrm{L}\right)}}.
\end{equation}
In the realistic case of slight misalignment, then the lensed star will appear
as two small arcs. Consider a MACHO of mass $0.5\,M_\odot$ at a distance of $D
= 25$~kpc lensing a star in the Large Magellanic Cloud (LMC) at a distance of
$50$~kpc. Then
\begin{equation}
\theta_\mathrm{E} = \sqrt{\frac{2GM}{Dc^2}} \approx 10^{-10} \approx 2" \times
10^{-5},
\end{equation}
too small to be resolved by optical telescopes. Fortunately the
lensing produces an apparent amplification of the source star by a factor
\cite{1964MNRAS.128..295R}
\begin{equation}
A = \frac{v^2 + 2}{v\sqrt{v^2 + 4}},
\label{eq:lightcurve}
\end{equation}
where $v = \beta / \theta_\mathrm{E}$, and $\beta$ is the angle between the
observer-lens and observer-star lines. Since objects in the halo are in
motion,
\begin{equation}
\beta(t) = \sqrt{ (v_\perp t)^2 + \beta_\mathrm{min}^2 },
\end{equation}
where $v_\perp$ is the transverse velocity of the lens relative to the 
line of sight, $\beta_\mathrm{min}$ is the closest approach of the lens to
the source, and $t$ is the time to the point of closest
approach\cite{Paczynski:1985jf,Griest:1990vu}. 
Searches for the amplification of stars caused by gravitational lensing of
$\theta_\mathrm{E} \sim $ micro arc seconds are referred to as
\emph{gravitational microlensing surveys.} Such surveys measure
magnification of the star and the duration of the microlensing event. 
Unfortunately it is not possible to determine the \emph{size} of the
the lens from these measurements.

\subsection{Gravitational Microlensing Surveys}

Several research groups are engaged in the search for microlensing events from
dark matter\cite{Alcock:2000ph,Afonso:2002xq}. By monitoring a large
population of well resolved background stars such as the LMC, constraints can
be placed on the MACHO content of the halo. The MACHO project has conducted a
5.7 year survey monitoring 11.9 million stars in the LMC to search for
microlensing events\cite{Alcock:2000ph} using an automated search algorithm to
monitor the light curves of LMC stars. Optimal filtering is used to search for
light curves with the characteristic shape given by equation
(\ref{eq:lightcurve}).

Since the effect of microlensing is achromatic, light curves are monitored in
two different frequency bands to reduce the potential background sources
which may falsely contribute to the microlensing rate. Background events
include variable stars in the LMC (known as bumpers\cite{1996astro.ph..6165A}), 
which can usually be rejected as the fit of the light curves to true
microlensing curves is poor. Supernovae occurring behind the LMC are the most
difficult to cut from the analysis. The MACHO project reported 28 candidate
microlensing events in the 5.7 year survey of which 10 were thought to be
supernovae behind the LMC and 2--4 were expected from lensing by known stellar
populations.  They report an excess of 13--17 microlensing events, depending
on the selection criteria used.

The \emph{optical depth}, $\tau$, is the probability that a given source star
is amplified by a factor $A > 1.34$\cite{Paczynski:1985jf}. This is just the
probability that the source lies on the sky within a disk of radius
$\theta_\mathrm{E}$ around a microlensing object and is given
by\cite{Alcock:1995zx}
\begin{equation}
\tau = \frac{4\pi G}{c^2} \int_0^{L} \rho(l) \frac{l(L - l)}{L}\,dl,
\end{equation}
where $L = D_\mathrm{SL} + D_\mathrm{L}$ is the observer-star distance and $l
= D_\mathrm{L}$ is the observer-lens distance. For the spherical halo given in
equation (\ref{eq:simplehalodensity3}) with density
\begin{equation}
\rho(r) = 0.0079 \frac{R_0^2 + a^2}{r^2 + a^2} \,M_\odot\, \mathrm{pc}^{-3},
\end{equation}
where $R_0 = 8.5$~kpc is the Galactocentric radius of the Sun and a galactic
core radius of $a = 5$~kpc, the predicted optical depth towards the LMC
(assumed to be at $50$~kpc) is\cite{Alcock:1995zx}
\begin{equation}
\tau_\mathrm{LMC} = 4.7 \times 10^{-7}.
\end{equation}
The optical depth towards the LMC measured by the MACHO project microlensing
surveys is
\begin{equation}
\tau_\mathrm{LMC} = 1.2_{-0.3}^{+0.4} \times 10^{-7}.
\end{equation}
This suggests that the fraction of the halo in MACHOs is less that $100\%$,
but does not exclude a MACHO halo.

The number of observed MACHO events and the time scales of the light curves
can be compared with various halo models. The MACHO project has performed a
maximum-likelihood analysis in which the halo MACHO fraction $f$ and MACHO
mass $m$ are free parameters. For the standard spherical halo, they find the
most likely values are $f = 20\%$ and $m = 0.45\,M_\odot$. The $95\%$
confidence interval of on the MACHO halo faction is $f = 8\%$--$50\%$ and the
$95\%$ confidence interval of on the MACHO mass is $0.15$--$0.9\,M_\odot$. The
total mass in MACHOs out to $50$~kpc is found to be $9_{-3}^{+4} \times
10^{10}\,M_\odot$, independent of the halo model\cite{Alcock:2000ph}.  The
EROS collaboration has recently published results of a search for microlensing
events towards the Small Magellanic Cloud (SMC)\cite{Afonso:2002xq}. The EROS
result further constrains the MACHO fraction of a standard halo in the mass
range of interest to less than $25\%$; they do not exclude a MACHO component
of the halo, however.

\section{Gravitational Waves from Binary Black Hole MACHOs}
\label{s:bbhmacho}

Since the microlensing surveys have shown that $\sim 20\%$ of the halo dark
matter may be in the form of $\sim 0.5\,M_\odot$ MACHOs, it is natural to ask
what the MACHOs may be. As we mentioned above, it has been proposed that
MACHOs could be baryonic matter in the form of brown dwarfs, objects lighter
than $\sim 0.1\,M_\odot$ that do not have sufficient mass to sustain fusion,
however, this is inconsistent with the observed masses of MACHOs. The fraction
of the halo in red dwarfs, the faintest hydrogen burning stars with masses
greater than $\sim 0.1\,M_\odot$, can be constrained using the Hubble Space
Telescope. Hubble observations may also be used to constrain the fraction of
the halo in brown dwarfs. The results indicate that brown dwarfs make up less
than $\sim 3\%$ and red dwarfs less than $\sim 1\%$ of the
halo\cite{Graff:1995ru,Graff:1996rz}.  A third possible candidate for baryonic
MACHOs is a population of ancient white dwarfs in the halo. White dwarfs are
the remnants of stars of mass $1$--$8\,M_\odot$ and have masses of $\sim
0.6\,M_\odot$. Although they seem to be natural candidates for MACHOs,
proper motion searches for halo white dwarfs have been conducted and no candidates have
been found\cite{2002A&A...389L..69G,2002ApJ...573..644N,Creze:2004gs}.
Creeze~\emph{et~al.} combined the results of previous surveys to find that
$4\%$ ($95\%$ confidence level) of the halo is in the form of white
dwarfs\cite{Creze:2004gs}. 

It is possible that there is an over dense clump of MACHOs in the direction of
the LMC\cite{1996ApJ...473L..99N}, the lenses are located in the LMC
itself\cite{Salati:1999gd} or the lenses are in the disk of the
galaxy\cite{Evans:1997hq}.  If the MACHOs detected by microlensing are truly
in the halo, however, it is possible that MACHOs are non-baryonic matter such
as black holes \cite{Finn:1996dd,Nakamura:1997sm}. Black holes of mass $\sim
0.5\,M_\odot$ could not have formed as a product of stellar evolution and so
they must have been formed in the early
universe\cite{1967SvA....10..602Z,1974MNRAS.168..399C}.  Several mechanisms
have been proposed to form primordial black holes with the masses consistent
with the MACHO observations. These include multiple scalar fields during
inflation\cite{Yokoyama:1995ex}, chaotic inflation\cite{Yokoyama:1999xi} or
reduction of the speed of sound during the QCD phase
transition\cite{Jedamzik:1996mr}. We do not consider these formation
mechanisms in detail here; it is sufficient for our purposes that PBHs with masses
consistent with microlensing observations can form.  If the MACHOs are
primordial black holes then there must be a large number of them in the halo.
The total mass in MACHOs out to $50$~kpc is $9\times 10^{10}\,M_\odot$, as
measured by microlensing surveys. If these are $0.5\,M_\odot$ PBHs then there
will be at least $\sim 1.8 \times 10^{11}$ PBHs in the halo. With such a large
number of PBHs in the halo it is natural to assume that some of these may be
in binary systems.

Nakamura~\emph{et~al.}\cite{Nakamura:1997sm} considered PBHs formed when the
scale factor of the universe $R$, normalized to unity at the time of
matter-radiation equality, is
\begin{equation}
R_f = \sqrt{GM_\mathrm{BH}}{c^2L_\mathrm{eq}} =
1.2\times 10^{-8} \left(\frac{M_\mathrm{BH}}{M_\odot}\right)^\frac{1}{2}
(\Omega h^2),
\end{equation}
where $L_\mathrm{eq}$ is the Hubble horizon scale at the time of
matter-radiation equality, $\Omega$ is the fraction of the closure density in
PBHs and $h$ is the Hubble parameter in units of $100$~km~s$^{-1}$. The age
and temperature of the universe at this epoch are $\sim 10^{-5}$~seconds and
$\sim 1$~GeV, respectively. By considering a pair of black holes that have
decoupled from the expansion of the universe to form a bound system
interacting with a third black hole, which gives the pair angular momentum to
form a binary, they showed that the distribution of the semi-major axis, $a$,
and eccentricity, $e$ of a population of binary black hole MACHOs is
\begin{equation}
f(a,e)\, da\, de = 
\frac{ 3ea^{\frac{1}{2}} }
{ 2\bar{x}^{\frac{3}{2}} (1-e^2)^{\frac{3}{2}}  } \, da\, de 
\label{eq:semieccdist}
\end{equation}
where $\bar{x}$ is the mean separation of the black hole MACHOs at the time of
matter-radiation equality, given by
\begin{equation}
\bar{x} = 1.1 \times 10^{16} \left(\frac{M}{M_\odot}\right)^{\frac{1}{3}}
\left(\Omega h^2\right)^{-\frac{4}{3}} \,\mathrm{cm},
\end{equation}
The coalescence time of a binary by the emission of gravitational waves is
approximately given by
\cite{Peters:1964}
\begin{equation}
t = t_0 \left(\frac{a}{a_0}\right)^4 \left(1 - e^2\right)^{\frac{7}{2}},
\label{eq:peters}
\end{equation}
where $t_0 = 10^{10}$~years and
\begin{equation}
a_0 = 2 \times 10^{11}
\left(\frac{M}{M_\odot}\right)^{\frac{3}{4}}\,\mathrm{cm}
\end{equation}
is the semimajor axis of a binary with circular orbit which coalesces in time
$t_0$. Integrating equation (\ref{eq:semieccdist}) for fixed $t$ using equation
(\ref{eq:peters}), Nakamura~\emph{et~al.}\cite{Nakamura:1997sm} found the
probability distribution $f_t(t)$ for the coalescence time is
\begin{equation}
f_t(t)\,dt = \frac{3}{29}\left[
\left(\frac{t}{t_\mathrm{max}}\right)^{\frac{3}{37}} -
\left(\frac{t}{t_\mathrm{max}}\right)^{\frac{3}{8}}\right] \frac{dt}{t},
\end{equation}
where $t_\mathrm{max} = t_0(\bar{x}/a_0)^4$. The number of coalescing binaries
with $t \sim t_0$ is then $\sim 5 \times 10^{8}$ for $\Omega h^2 = 0.1$, so
the event rate of coalescing binaries is $\sim 5 \times 10^{-2}$ events per
year per galaxy. Ioka~\emph{et~al.}\cite{Ioka:1998nz} performed more detailed
studies of binary black hole MACHO formation in the early universe and found
that, within a $50\%$ error, the distribution function and the rate of
coalescence given in \cite{Nakamura:1997sm} agree with numerical simulations.
The event rate of coalescing binary black hole MACHOs is therefore
\begin{equation}
R_\mathrm{BBHMACHO} = 5\times 10^{-2}\times 2^{\pm
1}\,\mathrm{yr}^{-1}\,\mathrm{galaxy}^{-1}.
\end{equation}
This rate is significantly higher than the coalescence rate of
binary neutron stars, which is\cite{Kalogera:2004tn}
\begin{equation}
R_\mathrm{BNS} = 8.3\times 10^{-5}\times 2^{\pm
1}\,\mathrm{yr}^{-1}\,\mathrm{galaxy}^{-1}.
\end{equation}
It must be emphasized that several neutron star binaries have been
observed, but there are no observations of black hole MACHO binaries.

\begin{figure}[p]
\begin{center}
\includegraphics[width=\linewidth]{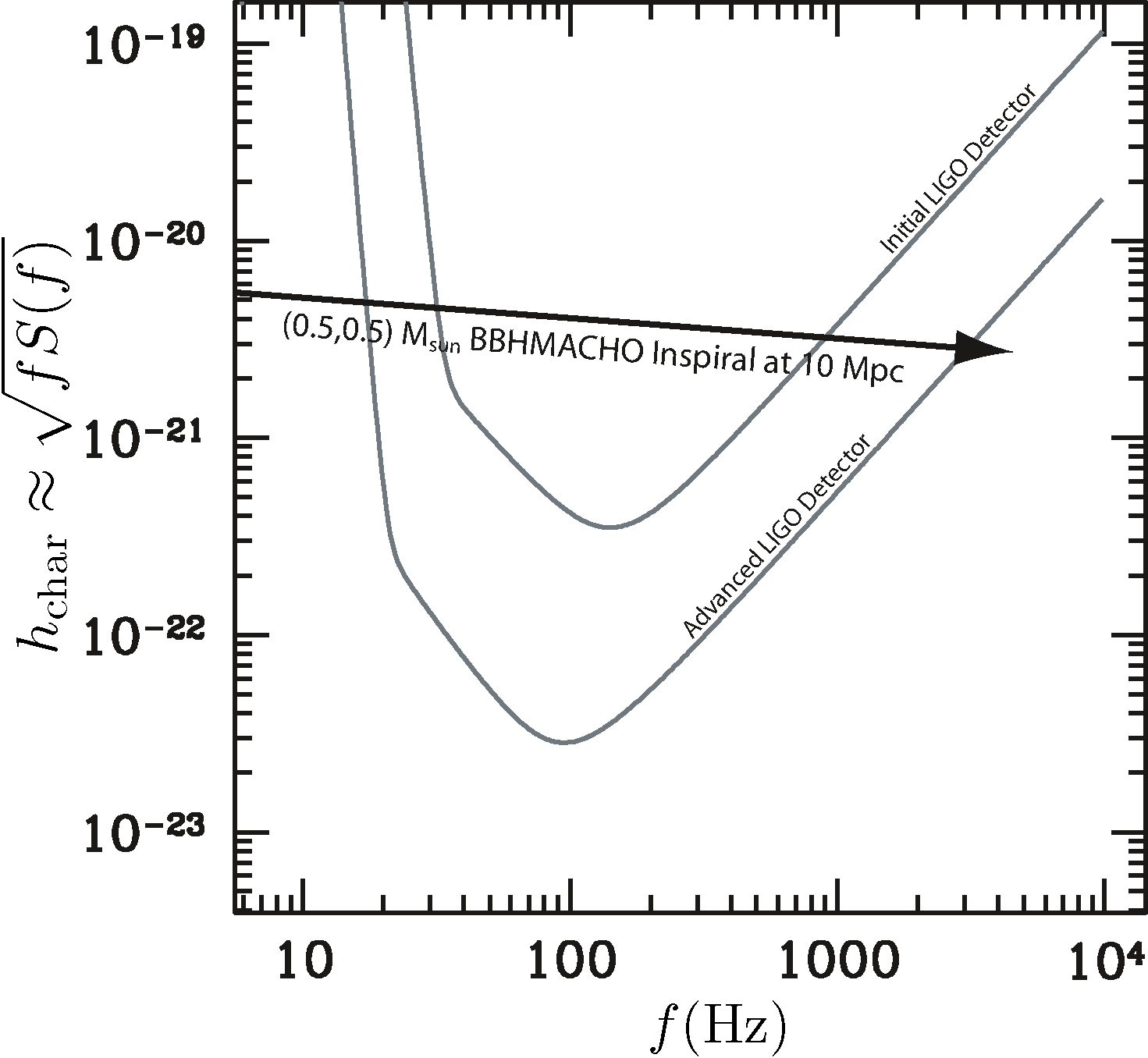}
\end{center}
\caption[Sensitivity of LIGO to a Binary Black Hole MACHO Inspiral]{%
\label{f:machosensitivity}
The sensitivity of LIGO to a binary black hole MACHO inspiral can be
considered in terms of comparison between the characteristic strain
$h_\mathrm{char}$ of the inspiral with the RMS noise curve of the detector. It
can be seen that if binary black hole MACHOs exist, they could be an excellent
source for LIGO.
}
\end{figure}

The distance to which we can detect a binary inspiral is usually expressed in
terms of the \emph{characteristic strain}, $h_\mathrm{char}$ of the binary. This
represents the intrinsic amplitude of the signal at some frequency times the
square-root of the number of cycles over which the signal is observed at that
frequency
\begin{equation}
h_\mathrm{char}(f) = |f \tilde{h}(f)| \approx h \sqrt{n}
\end{equation}
 For an inspiral signal this is given by\cite{Thorne:1982cv}
\begin{equation}
h_\mathrm{char}(f)  =  4 \times 10^{-21} \left(\frac{\mathcal{M}}{{M_\odot}}\right)^\frac{5}{6}
\left(\frac{f}{100\,\mathrm{Hz}}\right)^{-\frac{1}{6}} \left(\frac{r}{20\,
\mathrm{Mpc}}\right)^{-1},
\end{equation}
where $r$ is the distance to the binary and $\mathcal{M}$ is the chirp mass.
For comparison with signal strength, the detector sensitivity is better
expressed in terms of the root mean square (RMS) dimensionless strain per
logarithmic frequency interval
\begin{equation}
h_\mathrm{rms} = \sqrt{f S_n(f)},
\end{equation}
where $s$ is the detector strain output in the absence of a gravitational wave
signal and $S_n(f)$ is the power spectral density of $s$. If the value of
$h_\mathrm{char}
> \left(\textrm{a few}\right) \times h_\mathrm{rms}$, then the binary will be
detectable. Figure~\ref{f:machosensitivity} shows the characteristic strain of
a binary consisting of two $0.5\,M_\odot$ black holes at $10$~Mpc compared to
the RMS noise for initial LIGO. It can be seen that the inspiral signal lies
significantly above the noise, so these MACHO binaries could be excellent
source for initial LIGO. Nakamura~\emph{et~al.}\cite{Nakamura:1997sm} showed
that the rate of MACHO binaries could be as high as $3$~yr$^{-1}$ at a
distance of $15$~Mpc, under their model assumptions.

\section{Binary Black Hole MACHO Population Model}
\label{s:bbhmachopopulation}

The goal of this thesis is to search for gravitational waves from binary black
hole MACHOs described in the previous section. In the absence of a detection,
however, we wish to place an \emph{upper limit} on the rate of binary black
hole MACHO inspirals in the galaxy. We can then compare the predicted rate
with that determined by experiment. We will see later that in order to
determine an upper limit on the rate, we need to measure the \emph{efficiency}
$\varepsilon$ of our search to binary black hole MACHOs in the galactic halo.
We do this using a Monte Carlo simulation which generates a population of
binary black hole MACHOs according to a given probability density function
(PDF) of the binary black hole MACHO parameters. Using the set of parameters
generated by sampling the PDF, we can simulate the corresponding inspiral
waveforms on a computer. We then digitally add the simulated waveforms to the
output of the gravitational wave detector. By analyzing the interferometer
data containing the simulated signals, we can determine how many events from
the known source distribution we find. The efficiency of the
search is then simply
\begin{equation}
\varepsilon = 
\frac{\textrm{number of signals found}}{\textrm{number of signals injected}}.
\end{equation}
Recall that an inspiral waveform is described by the following nine 
parameters:
\begin{equation*}
\begin{split}
t_c &\quad \textrm{the end time of the inspiral}, \\
m_1 &\quad \textrm{the mass of the first binary component}, \\
m_2 &\quad \textrm{the mass of the second binary component}, \\
\iota &\quad \textrm{the inclination angle of the binary}, \\
\phi_0 &\quad \textrm{the oribital phase of the binary}, \\
\psi &\quad \textrm{the polarization angle of the binary}, \\
(\theta,\varphi) &\quad \textrm{the sky coordinates of the binary},\\
r  &\quad \textrm{the distance to the binary}.
\end{split}
\end{equation*}
To simulate a population of BBHMACHOs in the halo we need to generate a list
of these parameters that correctly samples their distributions.

We first address the generation of inspiral end time, $t_c$. The nature of the
noise in the interferometers changes with time, as does the orientation of the
detectors with respect to the galaxy as the earth rotates about its axis over
a sidereal day. To sample the changing nature of the detector output, the
Monte Carlo population that we generate contains many inspiral signals with
end times distributed over the course of the science run. We generate values
of $t_c$ at fixed intervals starting from a specified time $t_0$. The fixed
interval is chosen to be $2048+\pi \approx 2051.141592653$~sec. This
allows us to inject a significant number of signals over the course of the two
month run with the signals far enough apart that they do not dominate the
detector output. The interval is chosen to be non-integer to avoid any
possible periodic behavior associated with data segmentation.
The start time for the Monte Carlo, $t_0$, is chosen from a uniform random
distribution in the range $t_\mathrm{start} - 2630/\pi \le t_0 \le
t_\mathrm{start}$, where $t_\mathrm{start}$ is the time at which the science
run begins. We stop generating inspiral parameters when $t_c > t_\mathrm{end}$
the time at which the second science run ends. For each generated end time,
$t_c$ we generate the other inspiral waveform parameters.

We obtain the distribution of the mass parameters $(m_1,m_2)$ from the
microlensing observations of MACHOs in the galactic halo, described in section
\ref{ss:microlensing}, which suggest that the most likely MACHO mass is between
$0.15$ and $0.9\,M_\odot$. In the absence of further information on the mass
distribution we simply draw each component mass from a uniform
distribution in this range. We increase the range slightly to better measure
the performance of our search at the edge of the parameter space. We also note
that the search for binary neutron stars covers the mass range $1$ to
$3\,M_\odot$, so we continue the BBHMACHO search up to $1\,M_\odot$ rather
than terminating it at $0.9\,M_\odot$. We therefore generate each BBHMACHO
mass parameter, $m_1$ or $m_2$, from a uniform distribution of masses between
$0.1$ and $1.0\,M_\odot$.

The angles $\iota$ and $\phi_0$ are generated randomly to reflect a uniform
distribution in solid angle; $\cos \iota$ is uniform between $-1$ and $1$ and
$\phi_0$ is uniform between $0$ and $2\pi$. The polarization angle $\psi$ is
also generated from a uniform distribution between $0$ and $2\pi$.

To generate the spatial distribution of BBHMACHOs, we assume that the
distribution in galactocentric cylindrical coordinates, $(R,\theta,z)$,
follows the halo density given by equation (\ref{eq:simplehalodensity4}),
that is,
\begin{equation}
\rho(r) \propto \frac{1}{a^2 + R^2 + z^2/q^2}
\end{equation}
where $a$ is the halo core radius and $q$ is the halo flattening parameter. We
can see that this distribution is independent of the angle $\theta$, so we
generate $\theta$ from a uniform distribution between $0$ and $2\pi$.  If
we make the coordinate change $z/q \rightarrow z$, we may obtain a
probability density function (PDF) for the spatial distribution of
the MACHOs given by
\begin{equation}
f(R,z)\,R\, dR\, dz \propto \frac{1}{a^2 + R^2 + z^2}\,R\, dR\, dz.
\label{eq:spacepdf}
\end{equation}
We wish to randomly sample this PDF to obtain the spatial distribution of
the BBHMACHOs. Once we have obtained a value of the new coordinate $z$, we
simply scale by $q$ to obtain the original value of $z$. Recall that for a
probability density function $f(x)$ the cumulative distribution $F(X)$
given by
\begin{equation}
F(X) = \int_{-\infty}^X f(x)\, dx
\end{equation}
with $F(X) \in [0,1]$ for all $f(x)$. If we generate a value of $u$ from
a uniform distribution between $[0,1]$ and solve
\begin{equation}
\int_{-\infty}^x f(x')\, dx' = u
\label{eq:goldenrule}
\end{equation}
for $x$ then we will uniformly sample the probability distribution given by
$f(x)$.  Notice, however, that PDF in equation (\ref{eq:spacepdf}) is a
function of the two random variables $R$ and $z$, rather than a single
variable $x$ as in equation (\ref{eq:goldenrule}).  Let us make the coordinate
change
\begin{equation}
\begin{split}
R &= r \cos \varphi, \\
z &= r \sin \varphi \\
\label{eq:probcoordtrans}
\end{split}
\end{equation}
and so
\begin{equation}
\begin{split}
\tan \varphi &= \frac{z}{R}, \\
r^2 &= R^2 + z^2. \\
\end{split}
\end{equation}
Equation (\ref{eq:spacepdf}) becomes
\begin{equation}
\begin{split}
\mathcal{K} \int\int \frac{1}{a^2 + R^2 + z^2}\,R\, dR\, dz 
&= \mathcal{K} \int\int \frac{r\cos\varphi}{a^2+r^2}\,r\, dr \, d\varphi \\
&= \mathcal{K} \int_{-1}^{1} d\sin\varphi \int_0^{r_{\mathrm{max}}} \frac{r^2}{a^2+r^2}\, dr \\
&= \mathcal{K} \left[\sin\varphi\right]_{-1}^1 \left[r -
a\arctan\left(\frac{r}{a}\right)\right]_0^{r_{\mathrm{max}}},
\end{split}
\label{eq:bothpdfs}
\end{equation}
where $r_\mathrm{max} = 50$~kpc is the extent of the halo and $\mathcal{K}$ is
a constant that normalizes the PDF to unity. We can see immediately from
equation (\ref{eq:bothpdfs}) that $\sin\varphi$ is uniformly
distributed between $-1$ and $1$. Now consider the PDF for $r$ given by
\begin{equation}
f(r) = \mathcal{K} \left[r - a\arctan\left(\frac{r}{a}\right)\right]_0^{r_{\mathrm{max}}}
\end{equation}
with normalization constant
\begin{equation}
\mathcal{K} = 
\left[R_\mathrm{max} -
a\arctan\left(\frac{R_\mathrm{max}}{a}\right)\right]^{-1}
\end{equation}
To sample the distribution for $r$, generate a random variable $u$
uniform between $0$ and $1$ and find the root of
\begin{equation}
r - a\arctan\left(\frac{r}{a}\right) - u \left[R_\mathrm{max} -
a\arctan\left(\frac{R_\mathrm{max}}{a}\right)\right] = 0.
\label{eq:rroot}
\end{equation}
We can see that the value of $r$ that solves equation (\ref{eq:rroot}) must
lie between $0$ and $R_\mathrm{max}$ and that the left hand side is a
monotonically increasing function of $r$. We may therefore use a simple 
bisection to solve for the value of $r$. The values of $r$
and $\varphi$ are easily inverted for $R$ and $z$ using equation
(\ref{eq:probcoordtrans}). 

\begin{figure}[p]
\begin{center}
\includegraphics[width=\linewidth]{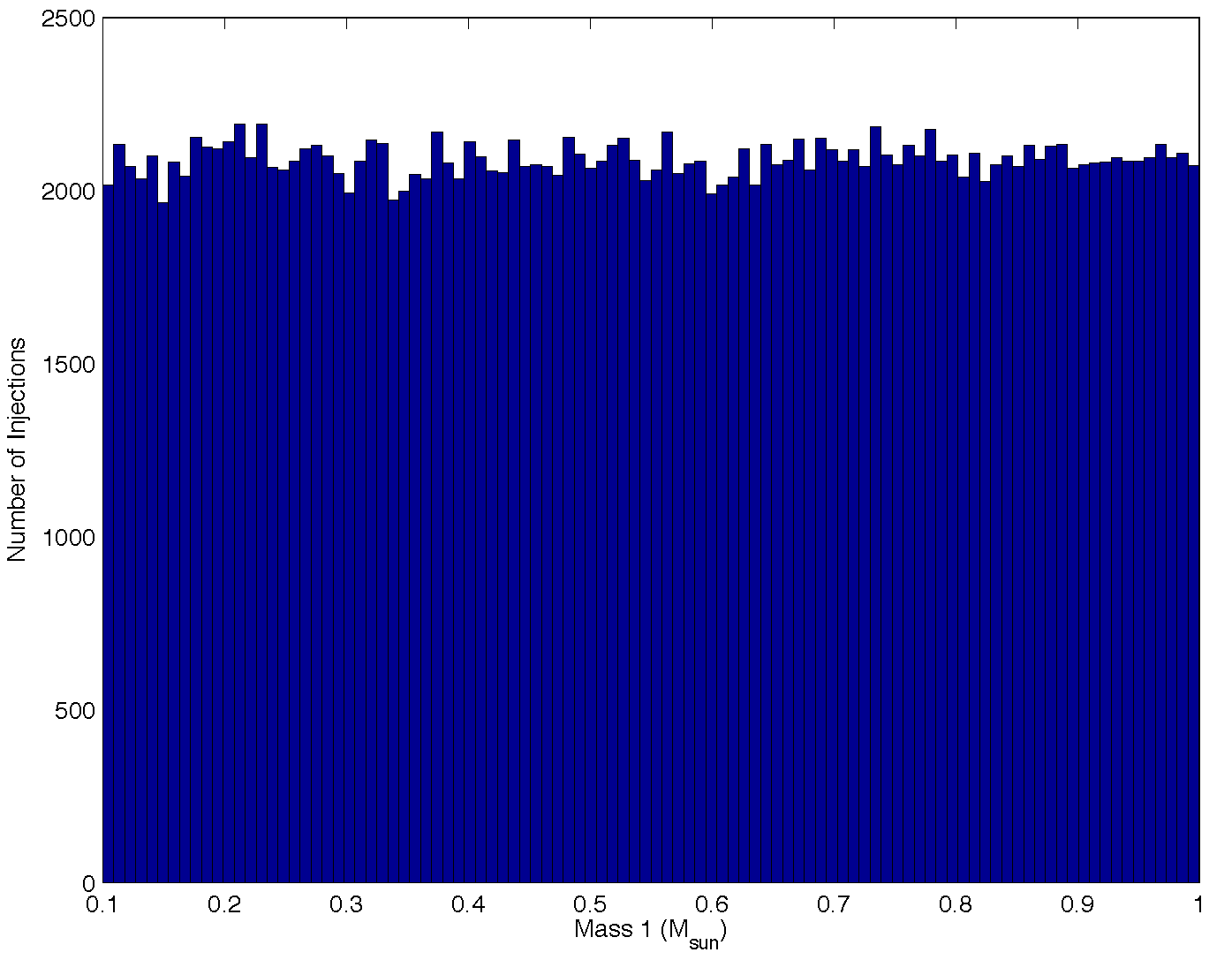}
\end{center}
\caption[Histogram of BBHMACHO Monte Carlo Mass Distribution]{
\label{f:m1_hist}
The BBHMACHO population Monte Carlo code is used to simulate a distribution of
$209048$ coalescing binaries and a histogram is made of the first component
mass, $m_1$ to confirm that it is uniformly distributed over the expected
range.  Similar tests are performed for the second mass parameter, $m_2$, the
galactocentric longitude, $\theta$, the inclination angle, $\iota$, the
polarization angle, $\psi$, and the coalescence phase, $\phi_c$.
}
\end{figure}

\begin{figure}[p]
\begin{center}
\includegraphics[width=\linewidth]{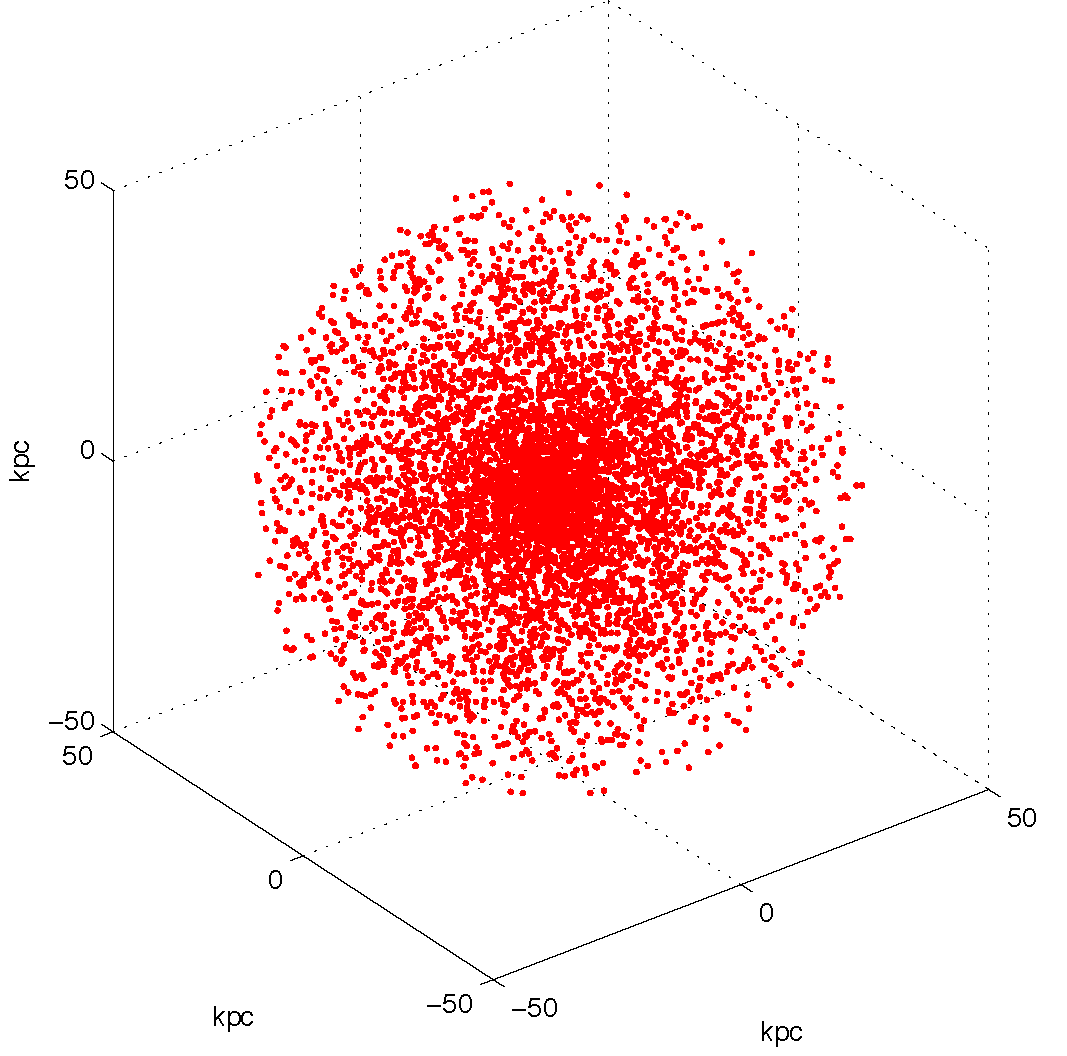}
\end{center}
\caption[Spatial Distribution of Binary Black Hole MACHOs in Galactocentric
Coordinates]{%
\label{f:spherical_cartesian}
The spatial distribution of $5000$ simulated BBHMACHO binaries in a spherical
$q=1$ Galactic halo of size $R_\mathrm{max} = 50\,\mathrm{kpc}$ with a core
radius $a = 8.5\,\mathrm{kpc}$ shown in galactocentric coordinates. Each point
in the figure corresponds to a simulated binary black hole MACHO injection.
}
\end{figure}

\begin{figure}[p]
\begin{center}
\includegraphics[width=\linewidth]{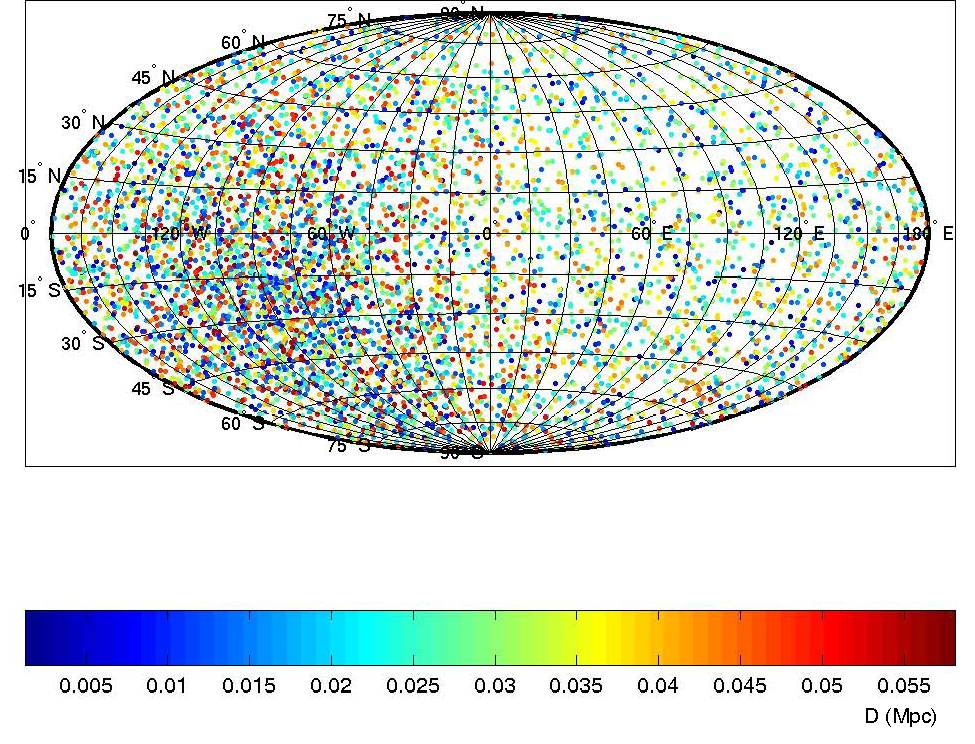}
\end{center}
\caption[Spatial Distribution of Binary Black Hole MACHOs Seen From Earth]{
\label{f:spherical_equatorial}
The spatial distribution of $5000$ simulated BBHMACHO binaries in a spherical,
$q=0$, Galactic halo of size $R_\mathrm{max} = 50\,\mathrm{kpc}$ with a core
radius $a = 8.5\,\mathrm{kpc}$ shown in equatorial coordinates. Each point
in the figure corresponds to a simulated binary black hole MACHO injection.
The color of the point shows the distance from the center of the earth to the
binary. Note the dense clump of binaries in the southern hemisphere, towards
the center of the Galaxy.
}
\end{figure}

This method was implemented in lalapps\_minj and figure~\ref{f:m1_hist} shows
a histogram of the first mass parameter generated the by the Monte Carlo code.
It can be seen that this is uniform between $0.1$ and $1.0\,M_\odot$, as
expected. Figure \ref{f:spherical_cartesian} shows the spatial distribution
of BBHMACHO binaries for a spherical, $q=1$, halo that extends to
$R_\mathrm{max} = 50\,\mathrm{kpc}$ with a core radius of $a = 5$~kpc.  Since
the software that simulates inspiral waveforms expects the position of the
inspiral to be specified in equatorial coordinates, the population Monte Carlo
code also generates the coordinates of the inspiral as longitude, latitude and
distance from the center of the earth, as shown in figure
\ref{f:spherical_equatorial}.  We will return to the use of population Monte
Carlos in chapter \ref{ch:result}.

\newpage

\Chapter{Binary Inspiral Search Algorithms}
\label{ch:findchirp}

Using equation (\ref{eq:hpluswave})--(\ref{eq:ftimesfunc}), we may write the
gravitational wave strain induced in the interferometer as
\begin{equation}
h(t) = \frac{A(t)}{\mathcal{D}} \cos\left( 2 \phi(t) - \theta \right),
\label{eq:rootwaveform}
\end{equation}
where
\begin{equation}
A(t) = - \frac{2G\mu}{c^4} \left[ \pi GM f(t) \right]^\frac{2}{3}
\end{equation}
and $\mathcal{D}$ is the \emph{effective distance}, given by
\begin{equation}
\mathcal{D} = \frac{r}{\sqrt{F_+^2 (1 + \cos^2 \iota)^2/4 + F_\times^2 \cos^2 \iota}}.
\end{equation}
The phase angle $\theta$ is
\begin{equation}
\tan \theta = \frac{F_\times 2\cos \iota}{F_+(1 + \cos^2 \iota)}
\end{equation}
and $\phi(t)$ is given by equation (\ref{eq:biwwphase}).
In this chapter we address the problem of finding such a signal hidden in
detector noise. The detection of signals of known form in noise is a classic
problem of signal processing\cite{wainstein:1962} and has been studied in the
context of binary inspiral in \cite{Finn:1992wt,Finn:1992xs}. This material is
reviewed in section~\ref{s:detectiontheory}. The particular implementation
used to extract inspiral signals from interferometer data in a computationally
efficient manner is presented in section~\ref{s:matchedfilter}.

\section{Detection of Gravitational Waves in Interferometer Noise}
\label{s:detectiontheory}

Our goal is to determine if the (calibrated) output of the interferometer
$s(t)$ contains a gravitational wave in the presence of the detector noise
described in section~\ref{ss:noise}. When the interferometer is operating
properly
\begin{equation}
s(t) = \begin{cases}
n(t) + h(t) & \text{signal present},\\
n(t) & \text{signal absent}.
\end{cases}
\end{equation}
The instrumental noise $n(t)$ arises from naturally occurring random processes
described mathematically by a probability distribution function. The
\emph{optimum receiver} for the signal $h(t)$ takes as input the
interferometer data and returns as its output the conditional probability
$P(h|s)$ that the signal $h(t)$ is present given the data $s(t)$. The
conditional probability that the signal is not present, given the data is then
$P(0|s) = 1 - P(h|s)$. The probabilities $P(h|s)$ and $P(0|s)$ are \emph{a
posteriori} probabilities. They are the result of an experiment to
search for the signal $h(t)$. The probability that the
signal is present before we conduct the experiment is the \emph{a priori}
probability $P(h)$. Similarly, $P(0) = 1 - P(h)$ is the \emph{a priori}
probability that the signal is absent.

The construction of the optimal receiver depends on the following elementary
probability theory. The probability that two events $A$ and $B$ occur
is given by
\begin{equation}
P(A,B) = P(A) P(B|A) = P(B) P(A|B),
\label{eq:probproduct}
\end{equation}
allowing us to relate the two conditional probabilities by
\begin{equation}
P(A|B) = \frac{P(A,B)}{P(B)} = \frac{P(A)P(B|A)}{P(B)}.
\label{eq:protobayes}
\end{equation}
If instead of a single event, $A$, suppose we have a complete set of mutually
exclusive events $A_1, A_2, \ldots, A_K$. By mutually exclusive we mean that
two or more of these events cannot occur simultaneously and by complete we
mean that one of them must occur. Now suppose $B$ is an event that
can occur only if one of the $A_k$ occurs. Then the probability that $B$
occurs is given by
\begin{equation}
P(B) = \sum_{k=1}^K P(A_k)P(B|A_k).
\label{eq:totalprob}
\end{equation}
Equation (\ref{eq:totalprob}) is called the \emph{total probability formula}.
Now let us suppose that $B$ is the result of an experiment and we want to know
the probability that it was event $A_k$ that allowed $B$ to happen. This can
be obtained by substituting equation (\ref{eq:totalprob}) into equation
(\ref{eq:protobayes}) to get
\begin{equation}
P(A_k|B) = \frac{P(A_k)P(B|A_k)}{P(B)} 
= \frac{P(A_k)P(B|A_k)}{\sum_{j=1}^K P(A_j)P(B|A_j)}.
\label{eq:bayeslaw}
\end{equation}
Equation (\ref{eq:bayeslaw}) is \emph{Bayes' theorem}. The probability
$P(A_k)$ is the \emph{a priori} probability of event $A_k$ occurring and
$P(A_k|B)$ is the \emph{a posteriori} probability of $A_k$ occurring given
that the outcome of our experiment $B$ occurred. The conditional probability
$P(B|A_k)$ is called the \emph{likelihood}.

Now suppose that set $\{A_k\}$ contains only to two members: ``the signal is
present'' and ``the signal is absent''. The \emph{a priori} probabilities of
these events are $P(h)$ and $P(0)$, as discussed earlier. We consider $B$ to
be the output of the interferometer for a particular experiment. We can
use Bayes' theorem to compute the \emph{a posteriori} probability that the
signal is present, given the output of the detector:
\begin{equation}
P(h|s) = \frac{P(h)P(s|h)}{P(s)}
\label{eq:pofhgivens1}
\end{equation}
where $P(s)$ is the \emph{a priori} probability of obtaining the detector
output and $P(s|h)$ is the likelihood function. $P(s|h)$ is the probability of
obtaining the detector output given that the signal is present in the data.
The probability of obtaining the detector output is given by
\begin{equation}
P(s) = P(h)P(s|h) + P(0)P(s|0)
\label{eq:pofs}
\end{equation}
since the signal is either present or not present.  Substituting equation
(\ref{eq:pofs}) into (\ref{eq:pofhgivens1}), we write
\begin{equation}
P(h|s) = \frac{P(h)P(s|h)}{P(h)P(s|h) + P(0)P(s|0)}.
\label{eq:pofhgivens2}
\end{equation}
Dividing the numerator and denominator on the right hand side of equation
(\ref{eq:pofhgivens2}) by $P(h)P(s|0)$ we obtain
\begin{equation}
P(h|s) = \frac{P(s|h)/P(s|0)}{[P(s|h)/P(s|0)] + [P(0)/P(h)]}.
\label{eq:pofhgivens3}
\end{equation}
Define the likelihood ratio
\begin{equation}
\Lambda = \frac{P(s|h)}{P(s|0)}
\label{eq:likelihooddef}
\end{equation}
so that equation (\ref{eq:pofhgivens3}) becomes
\begin{equation}
P(h|s) = \frac{\Lambda}{\Lambda + [P(0)/P(h)]}.
\label{eq:pofhgivens4}
\end{equation}
Similarly, we find that the probability that the signal is absent is given by
\begin{equation}
P(0|s) = 1 - P(h|s) = \frac{P(0)/P(h)}{\Lambda + [P(0)/P(h)]}.
\label{eq:pofnothgivens}
\end{equation}
Using equations (\ref{eq:pofhgivens4}) and (\ref{eq:pofnothgivens}), we find
that the ratio of the \emph{a posteriori} probabilities is
\begin{equation}
\frac{P(h|s)}{P(0|s)} = \Lambda\frac{P(h)}{P(0)}.
\label{eq:postratio}
\end{equation}

We now construct a decision rule for present or absence of the signal.  If
$P(h|s)$ is large (close to unity) then it is reasonable to conclude that the
signal is present.  Conversely, if $P(h|s)$ is small (close to zero) then we
may conclude that the signal is absent.  Therefore we may set a threshold
$P_\ast$ on this posterior probability as our decision rule is
\begin{alignat}{2}
P(h|s) &\ge P_* &\quad&\text{decide the signal is present}, \label{eq:ney1} \\
P(h|s) &< P_* &&\text{decide the signal is not present}.\label{eq:ney2} 
\end{alignat}
Given this decision rule there are two erroneous outcomes.  If $P(h|s)
\ge P_*$ and the signal is not present, we call this a \emph{false alarm}; our
decision that the signal is present was incorrect. Conversely, if
$P(h|s) < P_*$ and the signal is present, we have made a \emph{false
dismissal}. Each possible outcome has an associated probability
\begin{align}
F  &&&\text{probability that we have a false alarm} \\
F' &= 1 - D &&\text{probability that we have a false dismissal},
\end{align}
where $D$ is the probability of a correct detection.
To construct the posterior probability, $P(h|s)$ we need
the unknown \emph{a priori} probabilities, $P(h)$ and $P(0)$. We see from
equation (\ref{eq:pofhgivens4}), however, that $P(h|s)$ is a monotonically
increasing function of the likelihood. The ratio of the \emph{a priori}
probabilities, $P(h)/P(0)$, is a constant that does not involve the result of
our experiment. Therefore we can define the output of our optimum receiver to be
the device which, given the input data $s(t)$, returns the likelihood ratio
$\Lambda$.  For the receiver to be optimal in the Neyman-Pearson sense the
detection probability should be maximized for a given false alarm rate, $F$.
Rule (\ref{eq:ney1})--(\ref{eq:ney2}) is optimal in the Neyman-Pearson sense.

We now consider the construction of $\Lambda$ for the interferometer data
$s(t)$ and the gravitational wave signal $h(t)$. Assume that the noise
is stationary and Gaussian with zero mean value
\begin{equation}
\left\langle n(t) \right\rangle = 0
\end{equation}
where angle brackets denote averaging over different ensembles of the
noise. The  (one sided) power spectral density $S_n(|f|)$ of the noise is
defined by
\begin{equation}
\left\langle \tilde{n}(f) \tilde{n}(f') \right\rangle = \frac{1}{2} S_n(|f|)
\delta(f-f')
\label{eq:ospsddef}
\end{equation}
where $\tilde{n}(f)$ is the Fourier transform of $n(t)$. We wish to compute
the quantity
\begin{equation}
\Lambda = \frac{P(s|h)}{P(s|0)},
\end{equation}
however since the probabilities $P(s|h)$ and $P(s|0)$ are usually zero, so in
calculating the likelihood ratio, we must get rid of the indeterminacy by
writing
\begin{equation}
\Lambda = \frac{P(s|h)}{P(s|0)}
= \frac{p(s|h)\,ds}{p(s|0)\,ds}
= \frac{p(s|h)}{p(s|0)}.
\end{equation}
Instead of using the zero probabilities where $P(s|h)$ and $P(s|0)$, we use the
corresponding probability densities $p(s|h)$ and $p(s|0)$. The probability
density of obtaining a particular instantiation of detector noise
is\cite{Finn:1992wt}
\begin{equation}
p(n) = \mathcal{K} \exp\left[-\frac{1}{2} (n|n)\right]
\end{equation}
where $\mathcal{K}$ is a normalization constant and the inner product
$(\cdot|\cdot)$ is given by
\begin{equation}
\label{eq:fullinnerproduct}
  (a\mid b) \equiv \int_{-\infty}^\infty df\,
  \frac{\tilde{a}^\ast(f)\tilde{b}(f)+\tilde{a}(f)\tilde{b}^\ast(f)}
       {S_n(|f|)}.
\end{equation}
The probability density of obtaining the interferometer output, $s(t)$, in the
absence of signal, i.e. $s(t) = n(t)$, is therefore
\begin{equation}
p(s|0) = p(s) = \mathcal{K} \exp\left[-\frac{1}{2} (s|s)\right]
\end{equation}
The probability density of obtaining $s(t)$ in the presence of a signal, i.e.
when $s(t) = n(t) + h(t)$, is given by 
\begin{equation}
p(s|h) = p(s-h) = \mathcal{K} \exp\left[-\frac{1}{2} (s-h|s-h)\right]
\end{equation}
where we have used $n(t) = s(t) - h(t)$. Therefore the likelihood ratio
becomes
\begin{equation}
\begin{split}
\Lambda &= \frac{p(s|h)}{p(s|0)} = \frac{p(s-h)}{p(s)} \\
&= \frac{\exp\left[-\frac{1}{2} (s-h|s-h)\right]}{\exp\left[-\frac{1}{2} (s|s)\right]} \\
&= \exp\left\{-\frac{1}{2}\left[(s|s) - 2(s|h) - (h|h)\right] + \frac{1}{2}(s|s)\right\} \\
&= \exp\left[(s|h) - \frac{1}{2}(h|h)\right]
\label{eq:like1}
\end{split}
\end{equation}
where $(s|h)$ depends on the detector output and $(h|h)$ is constant for a
particular $S_n(|f|)$ and $h$. Since the likelihood ratio is a monotonically
increasing function of $(s|h)$ we can threshold on $(s|h)$ instead of the
posterior probabilities. Our optimal receiver is involves the construction of
$(s|h)$ followed by a test
\begin{equation}
\begin{split}
(s|h) &\ge x_\ast \quad \text{the signal is present}, \\
(s|h) &< x_\ast \quad \text{the signal is not present}.
\end{split}
\end{equation}
For a given $h(t)$, the inner
product in equation (\ref{eq:fullinnerproduct}), is a linear map from the
infinite dimensional vector space of signals to $\mathbb{R}$. Therefore the
optimal receiver is a linear function of the input signal $s(t)$. Both the
output of a gravitational wave interferometer and inspiral signals that we are
searching for are real functions of time, so 
\begin{align}
\tilde{s}^\ast(f) &= \tilde{s}(-f) \\
\tilde{h}^\ast(f) &= \tilde{h}(-f)
\end{align}
and the inner product in equation (\ref{eq:fullinnerproduct}) becomes
\begin{equation}
\left(a\mid b\right) = 2 \int_{-\infty}^{\infty}df\,
\frac{\tilde{a}(f)\tilde{b}^\ast(f)}{S_n\left(\left|f\right|\right)}.
\label{eq:innerproduct}
\end{equation}

If we receive only noise, then the mean of $(s|h)$ over an ensemble of detector
outputs is 
\begin{equation}
\begin{split}
\left\langle (s|h) \right\rangle &= \left\langle (n|h) \right\rangle \\
&= \int_{-\infty}^{\infty} 
   \frac{\langle\tilde{n}(f)\rangle \tilde{h}^\ast(f)}{S_n(|f|)} \\
&= 0
\end{split}
\end{equation}
since $\langle n(t) \rangle = 0$. The variance of $(s|h)$ in the absence of
a signal is
\begin{equation}
\begin{split}
\left\langle(s|h)^2\right\rangle 
&= 4 \left\langle \int_{-\infty}^\infty \int_{-\infty}^\infty \,df\,df'\,
\frac{\tilde{n}(f)\tilde{h}^\ast(f) \tilde{n}^\ast(f')\tilde{h}(f')}
{S_n(|f|)\,S_n(|f'|)} \right\rangle \\
&= 4 \int_{-\infty}^\infty \int_{-\infty}^\infty \,df\,df'\,
\frac{\left\langle \tilde{n}(f)\tilde{n}^\ast(f')\right\rangle\tilde{h}^\ast(f)\tilde{h}(f')}
{S_n(|f|)\,S_n(|f'|)} \\
&= 4 \int_{-\infty}^\infty \int_{-\infty}^\infty \,df\,df'\,
\frac{\frac{1}{2}S_n(|f'|)\delta(f-f') \tilde{h}^\ast(f)\tilde{h}(f')}
{S_n(|f|)\,S_n(|f'|)} \\
&= (h|h)
\end{split}
\label{eq:filtervariance}
\end{equation}
where we have used the definition of the one sided power spectral density from
equation (\ref{eq:ospsddef}).  In the presence of signal and noise, then the
mean of $(s|h)$ is
\begin{equation}
\left\langle (n+h|h) \right\rangle = (\langle n \rangle + h|h) = (h|h).
\end{equation}
We can also show that the variance of $(s|h)$ in the presence of a signal is
\begin{equation}
\left\langle \left[ (s|h) - (h|h) \right]^2 \right\rangle
= \left\langle \left[ (n|h) \right]^2 \right\rangle
= (h|h).
\end{equation}
Therefore the quantity $(h|h)$ is the variance of the output of the optimal
receiver, $(s|h)$, and we denote it by
\begin{equation}
\sigma^2 \equiv (h|h).
\label{eq:sigmasqdef}
\end{equation}

Now suppose that the signal we wish to recover has an unknown amplitude,
$\mathcal{A}$. The above discussion holds with $h(t) \rightarrow
\mathcal{A}h(t)$ and, from equation (\ref{eq:like1}), the likelihood ratio becomes
\begin{equation}
\Lambda = \exp\left[\mathcal{A}(s|h) - \frac{1}{2}\mathcal{A}^2(h|h)\right]
\end{equation}
which is again monotonic in $(s|h)$, and so our previous choice of optimal
statistic and decision rule hold. Now we are ready to consider the case of a
gravitational wave inspiral signal of the form given in equation
(\ref{eq:rootwaveform}).  The likelihood ratio now becomes a function of
$\theta$
\begin{equation}
\Lambda'(\theta) = 
p(\theta) \exp\left\{D^{-1}(s|A(t)\cos\left[2\phi(t) - \theta\right]) -
\frac{1}{2}D^{-2}(h|h)\right\}.
\end{equation}
Now consider the first inner product in the above exponential. Using
$\cos(\phi - \theta) = \cos\theta\cos\phi + \sin\theta\sin\phi$, we may write
this as
\begin{equation}
\begin{split}
\left(s\big|A(t)\cos\left[2\phi(t) - \theta\right]\right)  &= 
\cos\theta \left(s\big|A(t)\cos\left[2\phi(t)\right]\right) + 
\sin\theta \left(s\big|A(t)\sin\left[2\phi(t)\right]\right)  \\
& = x\cos\theta + y\cos\theta \\
& = |z|\cos(\Phi - \theta)
\end{split}
\end{equation}
where
\begin{align}
x &= |z|\cos\Phi = \left(s\big|A(t)\cos(2\phi(t))\right), \\
y &= |z|\sin\Phi = \left(s\big|A(t)\sin(2\phi(t))\right), \\
|z| &= \sqrt{x^2 + y^2}, \\
\tan \Phi &= \frac{y}{x}.
\end{align}
(The notation $|z|$ will become clear later in this chapter.) To calculate the
likelihood ratio, $\Lambda$, we assume that the unknown phase is uniformly
distributed between $0$ and $2\pi$,
\begin{equation}
p(\theta) = \frac{1}{2\pi},
\end{equation}
and integrate $\Lambda'$ over the angle $\theta$ to obtain
\begin{equation}
\begin{split}
\Lambda &= \int_0^{2\pi} \Lambda'(\theta) \,d\theta
= \frac{1}{2\pi}\int_0^{2\pi}\exp\left[D^{-1}|z|\cos(\Phi - \theta) -
\frac{D^{-2}}{2}(h|h)\right] \,d\theta \\
&= I_0(D^{-1}|z|) e^{-D^{-2}\frac{1}{2}(h|h)}
\end{split}
\end{equation}
where $I_0$ is the modified Bessel function of the first kind of order zero.
Once again, we note that the function $I_0(D^{-1}|z|)$ is a monotonically
increasing function of $|z|$ and so we can threshold on $|z|$ instead of
$\Lambda$. Note that $s$ appears in the expression for the likelihood through
$|z|$ only.

Recall from chapter \ref{ch:inspiral} that we denoted the two orthogonal
phases of the binary inspiral waveform by $h_c$ and $h_s$ given by equations
(\ref{eq:coschirp}) and (\ref{eq:sinechirp})
\begin{align}
h_c(t) & = \frac{2}{c^2}\left(\frac{\mu}{M_\odot}\right) 
\left[\pi G M f(t)\right]^{\frac{2}{3}} 
\cos\left[2 \phi(t)  - 2\phi_0\right], \\
h_s(t) & = \frac{2}{c^2}\left(\frac{\mu}{M_\odot}\right) 
\left[\pi G M f(t)\right]^{\frac{2}{3}} 
\sin\left[2\phi(t) - 2\phi_0\right],
\end{align}
and so for inspiral waveforms we can compute $|z|$ by
\begin{equation}
z = \sqrt{(s|h_c)^2 + (s|h_s)^2}.
\end{equation}
The threshold on $|z|$ would be determined to achieve a given false alarm
probability. We note that in the absence of signal $|z|^2$ is the sum of
squares of two independent Gaussian random variables of zero means and
variance $\sigma^2 = (h_c|h_c) = (h_s|h_s)$. $x$ and $y$ are independent
random variables since $(h_c|h_s) = 0$. It is therefore convenient to work
with a normalized signal-to-noise ratio defined by
\begin{equation}
\rho^2 = \frac{|z|^2}{\sigma^2}
\label{eq:snrrootdef}
\end{equation}
which is $\chi^2$ distributed with two degrees of freedom for Gaussian
detector noise.

If a gravitational wave signal is present, then its location in time is
defined by the \emph{end time} parameter $t_e$ of the waveform. In chapter
\ref{ch:inspiral} we defined the end time of the chirp to be the time at which
the frequency of the gravitational wave reached $f_\mathrm{isco}$, taken as the
gravitational wave frequency of a particle in the innermost stable circular
orbit of Schwarzschild spacetime.  In the above discussion of the optimal
receiver, we implicitly knew the location of the signal in the data to have
$t_e = 0$. Now suppose that the inspiral waveform ends at some unknown time
$t_e$. We may write the signal we are searching for as $h(t'-t_e)$. Consider
the Fourier transform of this signal
\begin{equation}
\begin{split}
\int_{-\infty}^\infty e^{-2\pi i f t'} h(t'-t_e) \, dt' &= 
e^{-2\pi ift_e} \int_{-\infty}^\infty e^{-2\pi i f \tau} h(\tau) \, d\tau \\
&= e^{-2\pi ift_e} \tilde{h}(f).
\end{split}
\end{equation}
where we have used $\tau = t' - t_e$, $dt = d\tau$ and $t' = t_e + \tau$.
The value of the inner product $(s|h_c)$ for a waveform that ends at time $t_e$ is
therefore
\begin{equation}
(s|h_c(t_e)) = 2 \int_{-\infty}^\infty\,df e^{2\pi ift_e}
\frac{\tilde{s}(f)\tilde{h}_c^\ast(f)}{S_n(|f|)}
\label{eq:ipift}
\end{equation}
and the signal-to-noise ratio for a chirp that ends at time $t$ is
\begin{equation}
\rho(t) = \frac{1}{\sigma} \sqrt{ (s|h_c(t))^2 + (s|h_s(t))^2}
\end{equation}
where the quantities $(s|h_c(t))$ and $(s|h_s(t))$ can be obtained by inverse
Fourier transforms of the form in equation (\ref{eq:ipift}).

Now the statistic $\rho(t)$ derived from the likelihood is a function of a
time parameter. For Neyman-Pearson optimal detection, we would integrate over
all possible arrival times and threshold on this value. However, as well as
making a statement about the presence or absence of a signal in the data we
also want to measure the time that the signal occurs. To do this, we use the
method of \emph{maximum likelihood}\cite{helstrom:1995}.  The maximum
likelihood estimator states that the most probable value for the location of
the signal is the time at which the likelihood ratio is maximized.  So to find
a single inspiral signal in a segment of interferometer data, we search for
the maximum of $\rho(t)$. If $\max_t\left[\rho(t)\right] > \rho_\ast$ then we
decide that we have detected a signal at the time of the maximum. When there
is more than one inspiral in the data segment the maximization is not over all
times.

We have now completely specified the solution to the problem of finding a
waveform of unknown amplitude and phase at an unknown time in the data;
our optimum receiver is the \emph{matched filter} of equation equation
(\ref{eq:ipift}). Below we develop the formalism to construct a digital
implementation of the matched filter to search for gravitational wave signals
in interferometer data.

\section{Conventions for Discrete Quantities}
\label{s:conventions}

The raw output of the interferometer is the error signal from the length
sensing and control servo, LSC-AS\_Q, as described in chapter
\ref{ch:inspiral}.  Although this signal is a dimensionless quantity, we say
that it has units of ``counts'' and we denote it by $v(t)$.  The calibrated
detector output is related to the raw detector output by the detector response
function according
to \begin{equation}
\tilde{s}(f) = R(f;t) \tilde{v}(f)
\end{equation}
where $R(f;t)$ is the (complex) response function of the detector at time $t$
and has units of strain/count (see section \ref{ss:calibration}).  In
practice, the interferometer output is a discretely sampled quantity with
sampling interval $\Delta t$, that is $v_j \equiv v(t_j)$ where $t_j = j\Delta
t$.  The digital matched filter operates on a single \emph{data segment}
consisting of $N$ consecutive samples of $v(t_j)$. The length of this data
segment is $T = N\Delta t$~seconds.  Henceforth, we let $N$ be a power of
$2$ and follow the convention that the subscript $j$ refers to discretely
sampled time domain quantities and the subscript $k$ to discretely sampled
frequency domain quantities.  The frequency domain quantity $\tilde{v}(f_k)$
denotes the value of the continuous function $\tilde{v}(f)$ at a particular
frequency, labeled $f_k = k/(N\Delta t)$. If the units of $v_j$ are counts,
then $\tilde{v}(f_k)$ has units of counts/Hz. We define the quantity
$\tilde{v}_k$ by $\tilde{v}_k = \tilde{v}(f_k) / \Delta t$, which has units of
counts. If $k$ is negative, this corresponds to negative frequencies.

\subsection{The Discrete Fourier Transform}
\label{ss:dft}

If $v(t_j)$ is sampled at intervals of $\Delta t$, then the sampling
theorem\cite{Press:1992} tells us that $v(t_j)$ is bandwidth limited to the
frequency range $-f_\mathrm{Ny} \le f \le f_\mathrm{Ny}$, where
\begin{equation}
f_\mathrm{Ny} = \frac{1}{2\Delta t}
\end{equation}
is the \emph{Nyquist critical frequency}. Any power in $v(t)$ at frequencies
above $f_\mathrm{Ny}$ will be aliased into the range $-f_\mathrm{Ny} \le f
\le f_\mathrm{Ny}$, corrupting the signal. To prevent this, signals of
frequency higher than $f_\mathrm{Ny}$ in the interferometer output are removed
using analog low-pass filters before the signal is digitized. Therefore
$v(t_j)$ completely determines the signal $v(t)$ in the band of interest. We
may approximate the Fourier transform of this band limited signal $v(t_j)$ by
\begin{equation}
\tilde{v}(f_k) \rightsquigarrow \sum_{j=0}^{N-1} \Delta t\, v(t_j) e^{-2 \pi i f_k t_j}
= \Delta t \sum_{j=0}^{N-1} v_j e^{-2 \pi i j k / N},
\label{eq:fftapprox}
\end{equation}
where $-(N/2 + 1) \le k \le N/2$ and the symbol $\rightsquigarrow$ means
equal to under discretization.  Notice that the approximation to the
Fourier transform is periodic in $k$ with period $N$ and so 
\begin{equation}
\tilde{v}_{-k} = \tilde{v}_{N-k}\quad k = 1, \ldots, N - 1.
\end{equation}
Thus we let $k$ vary from $0$ to $N-1$ where zero frequency (DC) corresponds
to $k=0$, positive frequencies $0 < f < f_\mathrm{Ny}$ to values in the range
$0 < k < N/2$ and negative frequencies $-f_\mathrm{Ny} < f < 0$ correspond
to values in the range $N/2 < k < N$. The value $k = N/2$ approximates the
value of the Fourier transform at both $-f_\mathrm{Ny}$ and $f_\mathrm{Ny}$;
both these values are equal due to the periodicity of the discrete transform 
defined by\cite{Anderson:2001a}
\begin{equation}
\tilde{v}_k = \sum_{j=0}^{N-1} v_j e^{-i 2 \pi j k / N}.
\label{eq:dftdef}
\end{equation}
We may estimate the discrete inverse Fourier transform
in a similar way, using the relation
\begin{equation}
\Delta f = f_{k+1} - f_k = \frac{k+1}{N\Delta t} - \frac{k}{N\Delta t} =
\frac{1}{N\Delta t}
\end{equation}
to obtain
\begin{equation}
v_j = \frac{1}{N} \sum_{k=0}^{N-1} \tilde{v}_k e^{2 \pi i j k / N}.
\end{equation}

\subsection{Power Spectral Densities}
\label{ss:psdconv}

In equation (\ref{eq:ospsddef}), we defined the one sided power spectral
density $S_n(|f|)$ of $n(t)$ to be 
\begin{equation}
\left\langle\tilde{n}(f) \tilde{n}^\ast(f')\right\rangle = 
\frac{1}{2}S_n(|f|)\delta(f-f')
\end{equation}
where angle brackets denote an average over different realizations of the noise.
If $n(t)$ has units of $U$ then $\tilde{n}(f)$ has units of
$(\mathrm{time}) \times U$. The units $\delta(f-f')$ are $(\mathrm{time})$,
since 
\begin{equation}
\int_{-\infty}^\infty \delta(f)\,df = 1
\end{equation}
is a dimensionless quantity and $df$ has units $(\mathrm{time})^{-1}$.
Therefore we see that $S_n(|f|)$ has units of $(\mathrm{time})\times U^2$.
If we replace $\tilde{n}(f_k)$ with the discretely sampled quantities 
$\tilde{n}_k = \tilde{n}(f_k)$, we obtain
\begin{equation}
\left\langle\tilde{n}_k \tilde{n}_{k'}^\ast\right\rangle = 
\frac{N}{2\Delta t}\ospsd\delta_{kk'}
\label{eq:ospsddisc}
\end{equation}
where $\delta_{kk'}$ is the dimensionless Kronecker $\delta$-function,
obtained by discretization of the continuous $\delta$-function:
\begin{equation}
\delta(f-f') \rightsquigarrow N\Delta t\delta_{kk'}
\end{equation}
Equation (\ref{eq:ospsddisc}) defines \ospsd in terms of the discrete
frequency domain quantities.  The definition in equation (\ref{eq:ospsddisc}) is
equivalent to
\begin{equation}
\ospsd =
\begin{cases}
\frac{\Delta t}{N} \left\langle | \tilde{n}_0 |^2 \right\rangle & k = 0, \\
\frac{\Delta t}{N} \left\langle| \tilde{n}_{N/2} |^2\right\rangle & k = \frac{N}{2}, \\
\frac{\Delta t}{N} \left\langle \left( | \tilde{n}_k |^2 + | \tilde{n}_{N-k} |^2 \right)\right\rangle & \text{otherwise}
\end{cases}
\end{equation}
where the normalization is chosen so that the power spectral
density satisfies the discrete form of Parseval's theorem
\begin{equation}
\Delta t \sum_{j=0}^{N-1} |v_j|^2 = \sum_{k=0}^{N/2} S_v(f_k).
\end{equation}
Parseval's theorem states that the total power in a signal is independent of
whether it is calculated in the time domain or the frequency domain.

The value of \ospsd for white Gaussian noise will be useful to us later, so we
compute it here. If the noise $n(t)$ is zero mean, white noise with variance
$\varsigma^2$, then
\begin{equation}
\begin{split}
\left\langle \tilde{n}_k \tilde{n}_{k'}^\ast \right\rangle 
&= \sum_{j=0}^{N-1} \sum_{j'=0}^{N-1} e^{2\pi i \left(jk - j'k'\right) / N}
\left\langle n_j n_{j'} \right\rangle \\
&= \sum_{j=0}^{N-1} \sum_{j'=0}^{N-1} e^{2\pi i \left(jk - j'k'\right) / N}
\varsigma^2 \delta_{jj'} \\
&= \sum_{j=0}^{N-1} e^{2\pi i j \left(k - k'\right) / N} \varsigma^2 \\
&= N \delta_{kk'} \varsigma^2
\end{split}
\end{equation}
Substituting this into equation (\ref{eq:ospsddisc}), we obtain
\begin{equation}
\frac{N}{2\Delta t}\ospsd\delta_{kk'} = N \delta_{kk'} \varsigma^2
\end{equation}
and so the power spectrum of white Gaussian noise is a constant with value
\begin{equation}
\ospsd = 2\Delta t \varsigma^2.
\end{equation}

\section{Digital Matched Filtering}
\label{s:matchedfilter}

The signal-to-noise ratio (\ref{eq:snrrootdef}) requires us to compute
the time series
\begin{equation}
\label{eq:xcts}
x(t) = 2 \int_{-\infty}^{\infty}df\,e^{2\pi i f t} 
\frac{\tilde{s}(f) \tilde{h_c}^\ast(f)}{S_n\left(\left|f\right|\right)}
\end{equation}
and
\begin{equation}
\label{eq:ycts}
y(t) = 2 \int_{-\infty}^{\infty}df\,e^{2\pi i f t} 
\frac{\tilde{s}(f) \tilde{h_s}^\ast(f)}{S_n\left(\left|f\right|\right)}
\end{equation}
and the normalization constant $\sigma$ that measures that ``amount of noise''
in the detector (for a given inspiral waveform). From the definition of the
inner product in equation (\ref{eq:innerproduct}) and the definition of
$\sigma^2$ in equation (\ref{eq:sigmasqdef}), we explicitly write
\begin{equation}
\label{eq:sigmasqcts}
\sigma^2 = 2 \int_{-\infty}^{\infty}df\,
\frac{\tilde{h_c}^\ast(f)\tilde{h_c}(f)}{S_h\left(\left|f\right|\right)} 
= 2 \int_{-\infty}^\infty 
\frac{\tilde{h_s}^\ast(f)\tilde{h_s}(f)}{S_h\left(\left|f\right|\right)}.
\end{equation}
The signal-to-noise ratio is normalized according to the convention
of Cutler and Flanagan \cite{Cutler:1994}, so that in the case when the detector
output is Gaussian noise, the square of the signal-to-noise ratio averaged
over an ensemble of detectors with different realizations of the noise is
\begin{equation}
\left\langle \rho^2 \right\rangle = 
\frac{1}{\sigma^2} \left\langle x^2 + y^2 \right\rangle = 2,
\end{equation}
as seen from equation (\ref{eq:filtervariance}).

\subsection{Construction of the digital filter using stationary phase chirps}
\label{ss:digitalfilter}

In section \ref{ss:stationaryphase} we derived the stationary phase
approximation to the Fourier transform of the restricted post$^2$-Newtonian
binary inspiral waveform to be
\begin{align}
\label{eq:spcos}
\tilde{h}_c(f)&=\frac{2GM_\odot}{(1\,\mathrm{Mpc})c^2}
\left(\frac{5\mu}{96M_\odot}\right)^\frac{1}{2}
\left(\frac{M}{\pi^2M_\odot}\right)^\frac{1}{3}
f^{-\frac{7}{6}}\, \left( \frac{GM_\odot}{c^3} \right)^{-\frac{1}{6}}\,
e^{i\Psi(f;M,\eta)},\\
\tilde{h}_s(f)&=i\tilde{h}_c(f),
\label{eq:hsorthog}
\end{align}
where $f$ is the gravitational wave frequency in Hz, $M = m_1+m_2$ 
is the total mass of the binary measured in solar masses, $\mu = m_1 m_2 / M$
is the reduced mass and $\eta = \mu/M$.  Note that $\tilde{h}_{c,s}(f)$ have
units of 1/Hz and we have chosen the chirp to be at a canonical distance of
$r = 1\,\mathrm{Mpc}$.  The instrument strain per Hz $\tilde{h}(f)$ is 
a linear superposition of $\tilde{h}_{c,s}(f)$ in the same way as
$h(t)$ is obtained from $h_{c,s}(t)$. The phase evolution to
post$^2$-Newtonian order is given by
\begin{equation}
\begin{split}
\Psi(f;M,\eta) &= 2\pi ft_c-2\phi_0-\pi/4+\frac{3}{128\eta}\biggl[x^{-5}+
\left(\frac{3715}{756}+\frac{55}{9}\eta\right)x^{-3}
-16\pi x^{-2} \\
&\quad +\left(\frac{15\,293\,365}{508\,032}+\frac{27\,145}{504}\eta
+\frac{3085}{72}\eta^2\right)x^{-1}\biggr],
\label{eq:spphase}
\end{split}
\end{equation}
where $x=(\pi M f G/c^3)^{1/3}$. The coalescence phase $\phi_0$ is the orbital
phase, determined by the binary ephemeris, and the coalescence time $t_c$ is
the time at which the bodies collide. The overall value coalescence phase
$\phi_0$ is part of the unknown phase of the matched filter and we set
$\phi_0=0$, respectively.  We set the coalescence time $t_c = 0$, since it is
accounted for by the Fourier transform in equations (\ref{eq:xcts}) and
(\ref{eq:ycts}).  The validity of the stationary phase approximation for
inspiral templates is well established\cite{Droz:1999qx}.

Since the two chirp waveforms $\tilde{h_c}$ and $\tilde{h_s}$ are 
orthogonal, the most efficient algorithm for constructing the time series
$\rho(t)$ uses a single complex inverse FFT rather than computing it from
$x(t)$ and $y(t)$ which requires two real inverse FFTs. We may further
increase efficiency when using stationary phase chirps by splitting the filter
into a part that depends on the data and a part that depends only on the
template parameters. In this section we describe the construction of a digital
matched filter which uses these two tricks.  Consider the discrete form of
equation (\ref{eq:xcts})
\begin{equation}
\begin{split}
\label{eq:xdisc}
x_j &= 2 \frac{1}{N\Delta t} \sum_{k=0}^{N-1} e^{2\pi ijk/N} 
\frac{\tilde{s}(f_k) \tilde{h}_{c}^\ast(f_k)}{\ospsd} \\
&=
2 \frac{\Delta t}{N} \sum_{k=0}^{N-1} e^{2\pi ijk/N} 
\frac{\tilde{s}_k \tilde{h}_{ck}^\ast} {\ospsd}
\end{split}
\end{equation}
where  $\tilde{h}_{ck} \equiv {h}_c(f_k) / \Delta t$. From equation
(\ref{eq:ycts}) we obtain
\begin{equation}
\label{eq:ydisc}
y_j = 2 \frac{\Delta t}{N} \sum_{k=0}^{N-1} e^{2\pi ijk/N} 
\frac{\tilde{s}_k \tilde{h}_{sk}^\ast} {\ospsd}.
\end{equation}
Recall that $s(t)$ and $h(t)$ are real signals. We may use the relations
$\tilde{s}(f) = \tilde{s}^\ast(-f)$ and $\tilde{h}(f) = \tilde{h}^\ast(-f)$ to
write the normalization constant, $\sigma^2$, defined in equation
(\ref{eq:sigmasqcts}), as
\begin{equation}
\begin{split}
\label{eq:sigmasqdisc}
\sigma^2 &= 2 \frac{1}{N\Delta t} \sum_{k=0}^{N-1}
\frac{\tilde{h}_{c}(f_k)\tilde{h}_{c}^\ast(f_k)}{\ospsd}  \\
&=
2 \frac{\Delta t}{N} \sum_{k=0}^{N-1}
\frac{\tilde{h}_{ck}\tilde{h}_{ck}^\ast}{\ospsd} \\
&=
2 \frac{\Delta t}{N} \left( 
\frac{\tilde{h}_{c0}\tilde{h}_{c0}^\ast}{\ospsd} 
+
2 \sum_{k=1}^{N/2-1}
\frac{\tilde{h}_{ck}\tilde{h}_{ck}^\ast}{\ospsd}
+
\frac{\tilde{h}_{cN/2}\tilde{h}_{cN/2}^\ast}{\ospsd} 
\right).
\end{split}
\end{equation}
Since earth based gravitational wave detectors have no useful low frequency
response, henceforth we set the DC ($k=0$) term to zero. In addition to this, we
assume that there is no power at the Nyquist frequency, as the low pass filter
that band limits the interferometer data to frequencies below
$f_\mathrm{Ny}$ falls off rapidly as the Nyquist frequency is approached.
Therefore we may also set the $k=N/2$ term to zero. We assume this for all
frequency domain quantities.

Now we may write the cosine phase of the filter given in equation
(\ref{eq:xdisc}) as
\begin{equation}
\begin{split}
x_j &= 
2\frac{\Delta t}{N}
\left[
  \sum_{k=N/2+1}^{N-1} e^{2\pi ijk/N} 
  \frac{\tilde{s}_k \tilde{h}_{ck}^\ast}{\ospsd}
  +
  \sum_{k=1}^{N/2-1} e^{2\pi ijk/N} 
  \frac{\tilde{s}_k \tilde{h}_{ck}^\ast}{\ospsd}
\right] \\
&= 2\frac{\Delta t}{N}
\left[
  \sum_{k=1}^{N/2-1} e^{-2\pi ijk/N} 
  \frac{\tilde{s}_k^\ast \tilde{h}_{ck}}{\ospsd}
  +
  \sum_{k=1}^{N/2-1} e^{2\pi ijk/N} 
  \frac{\tilde{s}_k \tilde{h}_{ck}^\ast}{\ospsd}
\right] \\
&= 2\frac{\Delta t}{N}(Q_j^\ast + Q_j)
\end{split}
\end{equation}
where we have used the fact that $f_k = f_{N-k}$. $Q_j$ is defined to be
\begin{equation}
\label{eq:Qdef}
Q_j = \sum_{k=1}^{N/2-1} e^{2\pi ijk/N} 
  \frac{\tilde{s}_k \tilde{h}_{ck}^\ast}{\ospsd}.
\end{equation}
The sine phase of the filter given in equation (\ref{eq:ydisc}) can similarly
be written as
\begin{equation}
\begin{split}
\label{eq:yinter}
y_j &=
2\frac{\Delta t}{N}
\left[
  \sum_{k=N/2+1}^{N-1} e^{2\pi ijk/N} 
  \frac{\tilde{s}_k \tilde{h}_{sk}^\ast}{\ospsd}
  +
  \sum_{k=1}^{N/2-1} e^{2\pi ijk/N} 
  \frac{\tilde{s}_k \tilde{h}_{sk}^\ast}{\ospsd}
\right] \\
&= 
2\frac{\Delta t}{N}
\left[
  \sum_{k=1}^{N/2-1} e^{-2\pi ijk/N} 
  \frac{\tilde{s}_k^\ast \tilde{h}_{sk}}{\ospsd}
  +
  \sum_{k=1}^{N/2-1} e^{2\pi ijk/N} 
  \frac{\tilde{s}_k \tilde{h}_{sk}^\ast}{\ospsd}
\right].
\end{split}
\end{equation}
Using $\tilde{h}_s = i \tilde{h}_c$, equation (\ref{eq:yinter}) becomes
\begin{equation}
\begin{split}
y_j &= 
2\frac{\Delta t}{N}
\left[
  \sum_{k=1}^{N/2-1} e^{-2\pi ijk/N} 
  \frac{\tilde{s}_k^\ast i\tilde{h}_{ck}}{\ospsd}
  +
  \sum_{k=1}^{N/2-1} e^{2\pi ijk/N} 
  \frac{\tilde{s}_k (-i)\tilde{h}_{ck}^\ast}{\ospsd}
\right] \\
& = 
-2i\frac{\Delta t}{N}
\left[
  - \sum_{k=1}^{N/2-1} e^{-2\pi ijk/N} 
  \frac{\tilde{s}_k^\ast \tilde{h}_{ck}}{\ospsd}
  +
  \sum_{k=1}^{N/2-1} e^{2\pi ijk/N} 
  \frac{\tilde{s}_k \tilde{h}_{ck}^\ast}{\ospsd}
\right] \\
& = 
2\frac{\Delta t}{N}i(Q_j^\ast - Q_j).
\end{split}
\end{equation}
Thus the outputs of the filter for the two phases are
\begin{align}
x_j &= \Re z_j, \\
y_j &= \Im z_j.
\end{align}
The quantity $z_j$ is defined to be
\begin{equation}
\begin{split}
\label{eq:zdef}
z_j &= 4 \frac{\Delta t}{N}\sum_{k=1}^{N/2-1} e^{2\pi ijk/N} 
\frac{\tilde{s}_k \tilde{h}_{ck}^\ast}{\ospsd}  \\
&= \frac{\Delta t}{N}\sum_{k=0}^{N-1} e^{2\pi ijk/N} \tilde{z}_k
\end{split}
\end{equation}
where
\begin{equation}
\label{eq:ztildedef}
\tilde{z}_k = \left\{
\begin{array}{ll}
4 \frac{\tilde{s}_k \tilde{h}_{ck}^\ast}{\ospsd} 
  \quad\quad & 0 < k < \frac{N}{2},\\
\\
0 & \mathrm{otherwise}.
\end{array}
\right.
\end{equation}
We can now compute the square of the signal-to-noise ratio
\begin{equation}
\rho^2(t_j) = \frac{x_j^2 + y_j^2}{\sigma^2} = \frac{1}{\sigma^2}|z_j|^2
\end{equation}
by a single complex inverse Fourier transform and threshold on $\rho^2 \ge
\rho^2_\ast$.  Since we choose the template $\tilde{h}_c(f)$ to be at a
canonical distance of $1$~Mpc, the effective distance $\mathcal{D}$ to a chirp
detected with signal to noise ratio $\rho^2$ can be established as
\begin{equation}
\mathcal{D} = \frac{\sigma}{\rho} \,\mathrm{Mpc}.
\label{eq:effdistdef}
\end{equation}
Recall that $\sigma$ is a measure of the noise in the interferometer output;
it is a measure of the sensitivity of the detector.  Larger values of $\sigma$
correspond to a quieter detector (due to the $1/\ospsd$ term in the expression
for $\sigma$) and smaller values to a noisier detector. 

\subsection{Details of Filter Implementation}
\label{ss:dirtydetails}

The calibrated detector output is related to the raw detector output by the
detector response function according to
\begin{equation}
\tilde{s}(f) = R(f;t) \tilde{v}(f)
\end{equation}
where $R(f;t)$ is the (complex) response function of the detector at a
specific time $t$, as described in chapter~\ref{ch:inspiral}. In practice, we
compute the uncalibrated power spectral density $S_v(|f_k|)$ from the raw data
and then the calibrated power spectral density, $\ospsd$, in the denominator
of (\ref{eq:zdef}) is
\begin{equation}
\ospsd = |R(f;t)|^2 S_v(|f_k|).
\end{equation}
Further details of the computation of $S_v(|f_k|)$ are given in sections
\ref{ss:psd} and \ref{ss:invspec}.

Typical values of the variance of $v(t)$ for initial LIGO data are $10^3$;
however, the response function $R(f)$ has magnitude $\sim 10^{-22}$ at the most
sensitive frequencies of the instrument.  This means that $\ospsd \sim
10^{-44}$ which is beyond the range of 4-byte floating point
numbers\footnote{The smallest non-zero value that can be stored in an IEEE~754
floating point number is $1.17549435\times 10^{-38}$.}. We may store such
values as 8-byte floating point numbers, but this is wasteful of memory since
the extra precision of an 8-byte number is not needed. Therefore when we
implement the digital filter, we multiply the response function $R(f)$ by a
scaling variable $d$ which typically has values of $d = 2^{69}$ for initial
LIGO data. This scales all frequency domain quantities to have approximately
order unity.  Therefore equation (\ref{eq:zdef}) becomes
\begin{equation}
\label{eq:zdefcal}
z_j = 4 \frac{\Delta t}{N} \sum_{k=1}^{N/2-1} e^{2\pi ijk/N} 
  \frac{dR\tilde{v}_k\, d\tilde{h}_{ck}^\ast}
  {d^2|R|^2S_v\left(\left|f_k\right|\right)}
\end{equation}
and (\ref{eq:sigmasqdisc}) becomes
\begin{equation}
\label{eq:sigmasqdisccal}
\sigma^2 = 4 \frac{\Delta t}{N} \sum_{k=1}^{N/2-1}
\frac{d^2 \tilde{h}_{ck}\tilde{h}_{ck}^\ast}
{d^2|R|^2S_v\left(\left|f_k\right|\right)}. 
\end{equation}
Notice that we must also multiply the chirp by $d$ so that all the factors of
$d$ cancel in the signal-to-noise ratio $\rho(t_j)$ and the normalization
constant $\sigma^2$. In fact this is convenient as it brings the value of
the $\tilde{h}_c(f_k)$ to order unity for chirps that would produce a
signal-to-noise ratio of order unity. From equation (\ref{eq:spcos}) we obtain
the dimensionless quantity
\begin{equation}
\begin{split}
\label{eq:hck}
d\,\tilde{h}_{ck} &= \frac{d \tilde{h}_c(f_k)}{\Delta t} \\
&= 
\frac{2dGM_\odot}{(1\,\mathrm{Mpc}) c^2}
\left(\frac{5\mu}{96M_\odot}\right)^\frac{1}{2}
\left(\frac{M}{\pi^2M_\odot}\right)^\frac{1}{3}
\left(\frac{GM_\odot}{c^3\Delta t}\right)^{-\frac{1}{6}}
\left( f\,\Delta t \right)^{-\frac{7}{6}} \\
&\quad\quad\times\exp\,[i\Psi(f_k;M,\eta)] \Theta\left(k-k_\mathrm{isco}\right)\\
&=
\sqrt{\mathcal{T}(M,\mu)}\left(\frac{k}{N}\right)^{-\frac{7}{6}}
e^{i\Psi\left(f_k;M,\eta\right)} \Theta\left(k-k_\mathrm{isco}\right)
\end{split}
\end{equation}
where the term $\Theta\left(k-k_\mathrm{isco}\right)$ ensure that the chirp is
terminated at the frequency of the innermost stable circular orbit of
Schwarzschild. The function $\Psi(f_k;M,\eta)$ is the value of the
post$^2$-Newtonian phase evolution, which is given by (\ref{eq:spphase}), at
the frequency $f_k$. The quantity $\mathcal{T}(M,\mu)$ in equation
(\ref{eq:hck}) is called the \emph{template dependent normalization constant}
and is given by
\begin{equation}
\mathcal{T}(M,\mu) = \left[
\left(\frac{2dGM_\odot}{(1\,\mathrm{Mpc})c^2}\right)
\left(\frac{5\mu}{96M_\odot}\right)^\frac{1}{2}
\left(\frac{M}{\pi^2M_\odot}\right)^\frac{1}{3}
\left(\frac{GM_\odot}{\Delta tc^3}\right)^{-\frac{1}{6}}
\right]^2.
\end{equation}
Note that $\mathcal{T}(M,\eta)$ depends on the masses of the template and as
such must be recomputed once per template. If we substitute equation
(\ref{eq:hck}) into equation (\ref{eq:sigmasqdisccal}) we obtain
\begin{equation}
\label{eq:sigmasqts}
\sigma^2 = 4 \frac{\Delta t}{N} \mathcal{T} 
\sum_{k=1}^{k_\mathrm{isco}} 
\frac{\left(\frac{k}{N}\right)^{-\frac{7}{3}}}
{d^2|R|^2S_v\left(\left|f_k\right|\right)}
= 4 \frac{\Delta t}{N} \mathcal{T} \mathcal{S}
\end{equation}
where $\mathcal{S}$ is defined to be
\begin{equation}
\mathcal{S} = 
\sum_{k=1}^{k_\mathrm{isco}} 
\frac{\left(\frac{k}{N}\right)^{-\frac{7}{3}}}{d^2|R|^2S_v\left(\left|f_k\right|\right)}.
\end{equation}
$\mathcal{S}$ is referred to as the \emph{segment dependent normalization}.
It depends on the binary masses only through $k_\mathrm{isco}$, so we compute
and store the array 
\begin{equation}
\mathcal{S}(k_\mathrm{isco}) \quad \quad 1 \le k_\mathrm{isco} \le \frac{N}{2}
\end{equation}
from the input power spectral density. We then select the correct value of
$\mathcal{S}$ for a given mass pair by computing $k_\mathrm{isco} =
f_\mathrm{isco} / \Delta f$.

The signal-to-noise ratio squared is then
\begin{equation}
\rho^2(t_j) = 
\frac{16}{\sigma^2}\left(\frac{\Delta t}{N}\right)^2 \mathcal{T}
\left| 
  \sum_{k=1}^{N/2-1} e^{2\pi ijk/N} 
  \frac{dR\tilde{v}_k \left(\frac{k}{N}\right)^{-\frac{7}{6}} e^{-i\Psi(f_k;M,\eta)}\Theta(k-k_\mathrm{isco})}
       {d^2|R|^2 S_v\left(\left|f_k\right|\right)}
\right|^2 
\label{eq:signaltonoisesq}
\end{equation}
where $\sigma^2$ is now given by equation (\ref{eq:sigmasqts}).  Let us define
$\tilde{q}_k$ by
\begin{equation}
\label{eq:qtildedef}
\tilde{q}_k = 
\begin{cases}
\frac{d\tilde{v}_k \left(\frac{k}{N}\right)^{-\frac{7}{6}} \exp\left[-i\Psi(f_k;M,\eta)\right]}
     {d^2|R|^2S_v\left(\left|f_k\right|\right)} & 0 < k < k_\mathrm{isco}, \\
0 & \text{otherwise}.
\end{cases}
\end{equation}
and $q_j$ as the discrete complex inverse Fourier transform of $\tilde{q}_k$. Then
the signal-to-noise ratio squared is
\begin{equation}
\rho^2(t_j) = \frac{16}{\sigma^2}\left(\frac{\Delta T}{N}\right)^2 \mathcal{T}
\left|q_j\right|^2
\end{equation}
The computation of $\tilde{q}_k$ can further be split into the template
independent computation of
\begin{equation}
\tilde{F}_k = \frac{d\tilde{v}_k \left(\frac{k}{N}\right)^{-\frac{7}{6}}}
{d^2|R|^2S_v\left(\left|f_k\right|\right)}
\end{equation}
and the computation of
\begin{equation}
\tilde{T}_k = \exp\left[i\Psi(f_k;M,\eta)\right] \Theta\left(k-k_\mathrm{isco}\right)
\end{equation}
where $\tilde{F}_k$ is called the \emph{findchirp data segment} and 
$\tilde{T}_k$ is called the \emph{findchirp template}, so
\begin{equation}
\tilde{q}_k = 
\begin{cases}
\tilde{F}_k \tilde{T}_k^\ast & 0 < k < \frac{N}{2},\\
0 & \text{otherwise}.
\end{cases}
\end{equation}

The goal of this separation is to reduce the computational
cost of producing $\rho(t)$ by computing the template and using it to filter
several data segments. For a given power spectral density $S_v(|f_k|)$ we
only need to compute $\mathcal{S}(k_\mathrm{isco})$ once. The findchirp code
is designed to process several data segments, labeled $i = 1,\ldots,M$, at a
time. We can compute $\tilde{F}_k^i$ once for  each data segment and then for
each template we compute $\mathcal{T}$ and $\tilde{T}_k$. 
This reduces the computational cost of filter generation.
Furthermore, we can threshold against the quantity $|q_j|^2$
\begin{equation}
|q|^2_\ast = \frac{\rho^2_\ast} 
{\frac{16}{\sigma^2}\left(\frac{\Delta t}{N}\right)^2 \mathcal{T}}
\label{eq:qopstat}
\end{equation}
thus saving a multiplication per sample point.
The effective distance of an inspiral signal at time $t_j = j\Delta t$ is
given by equation (\ref{eq:effdistdef}) which becomes
\begin{equation}
\mathcal{D} = \frac{\mathcal{T}\mathcal{S}^2}{|q_j|^2}
\end{equation}
in this notation.

\subsection{Recording Triggers}
\label{ss:record}

We call times when the optimal receiver tells us that a signal is
present \emph{inspiral triggers} and record the time of the trigger,
the mass parameters of the template and the value of $\sigma^2$ for the data
segment. There are several complications that mean that simply thresholding on
equation (\ref{eq:qopstat}) is not what we do in practice, however. In section
\ref{ss:impulsetime} we will show that an impulse in the data segment can
cause the filter output event though no chirp is present, and hence cause a
false alarms. Although such events are rare in Gaussian noise, they are quite
common in real detector output, so we construct an addition test on the
presence of absence of the signal, called the $\chi^2$ veto\cite{Allen:2004},
which is described in section \ref{s:chisqcts}. Furthermore, the inspiral
signals that we are searching for are shorter than the length of
a data segment, so we want to allow the possibility of generating multiple
inspiral triggers in a single data segment. We also do not record all
times for which $|q_j|^2 \ge |q|^2_\ast$, as we would soon be flooded with
triggers in real interferometer data. The algorithm that we use to select the
times for which we generate inspiral triggers based on the output of the
matched filter and the $\chi^2$ veto is described in section
\ref{s:maxoverchirp}.

\section{Testing the filtering code}
\label{s:testing}

\subsection{Normalization}
\label{ss:normalization}


Consider the case when the filter input is Gaussian noise, i.e. $\tilde{s}_k =
\tilde{n}_k$ and set $R(f_k)\equiv 1$ and $d = 1$.  Then the expectation
value of the signal-to-noise ratio squared, $\langle \rho^2
\rangle$ is
\begin{equation}
\begin{split}
\langle \rho^2(t_j) \rangle &=
\frac{16}{\sigma^2}\left(\frac{\Delta t}{N}\right)^2 \mathcal{T}
  \sum_{k=1}^{k_\mathrm{isco}} \sum_{k'=1}^{k_\mathrm{isco}} 
  e^{2\pi ij(k-k')/N} 
  \frac{\left\langle \tilde{n}_k \tilde{n}_{k'}^{\ast} \right\rangle 
        \left(\frac{k}{N}\right)^{-\frac{7}{6}} \left(\frac{k'}{N}\right)^{-\frac{7}{6}}
        e^{-i\Psi(f_k)} e^{i\Psi(f_{k'})}}
       {S_n\left(\left|f_k\right|\right)S_n\left(\left|f_{k'}\right|\right)} \\
&= 
\frac{16}{\sigma^2}\left(\frac{\Delta t}{N}\right)^2 \mathcal{T} \\
&\quad\times
  \sum_{k=1}^{k_\mathrm{isco}} \sum_{k'=1}^{k_\mathrm{isco}} 
  e^{2\pi ij(k-k')/N} \left(\frac{1}{2} \frac{N}{\Delta t}  \delta_{kk'} \right) 
  \frac{ S_n\left(\left|f_k\right|\right)
        \left(\frac{kk'}{N^2}\right)^{-\frac{7}{6}}
        e^{i ( \Psi(f_{k'}) - \Psi(f_k) )}}
       {S_n\left(\left|f_k\right|\right)S_n\left(\left|f_{k'}\right|\right)} \\
&= 
\frac{8}{\sigma^2} \frac{\Delta t}{N} \mathcal{T}
  \sum_{k=0}^{N/2}
  \frac{ \left(\frac{k}{N}\right)^{-7/3} }
       {S_n\left(\left|f_k\right|\right)} \\
&= 
\frac{8N}{4\Delta t\, \mathcal{T}\mathcal{S}} \frac{\Delta t}{N} \mathcal{T} \mathcal{S} \\
&= 2,
\label{eq:filternorm}
\end{split}
\end{equation}
where we have used the definition of $\sigma^2$ from equation
(\ref{eq:sigmasqts}) and the definition of the one-sided power spectral
density from equation (\ref{eq:ospsddisc}).

The first test of the code is to check that the normalization
of the filter agrees with equation (\ref{eq:filternorm}) when the response
function, $R$, and dynamic range scaling, $d$, are both set to unity. In order
to exclude issues related to power spectral estimation at this stage of
testing  we set the power spectral density to be the (constant) theoretical
value for white Gaussian noise given by
\begin{equation}
\ospsd = 2 \varsigma^2 \Delta t,
\label{eq:gaussianpsd}
\end{equation}
where $\varsigma^2$ is the variance of the Gaussian noise.  We generate five
data segments containing white Gaussian noise of mean zero and variance
$\varsigma^2 = 64$ at a sample rate of $16384$~Hz. The length of each segment
is $1920$~seconds, so there are $N = 31\,457\,280$ samples per segment. Table
\ref{t:normresults} shows the value of $\langle \rho^2 \rangle$ after
averaging the output $\rho^2(t_j)$ of the filtering code over all output
samples. The values obtained are in good agreement with the theoretical
expectation.  Similar tests were performed with colored Gaussian noise, where
the power spectrum is no longer a constant, and noise colored by a response
function $R(f)$; the average filter output was
consistent with the expected value.  Large and small values of the variance
for the noise, $\varsigma^2$, were also used to test that the dynamic
range scaling factor $d$ was correctly implemented. In all cases the output of
the filtering code was consistent with equation (\ref{eq:filternorm}).

\begin{figure}[p]
\begin{center}
\includegraphics[width=\linewidth]{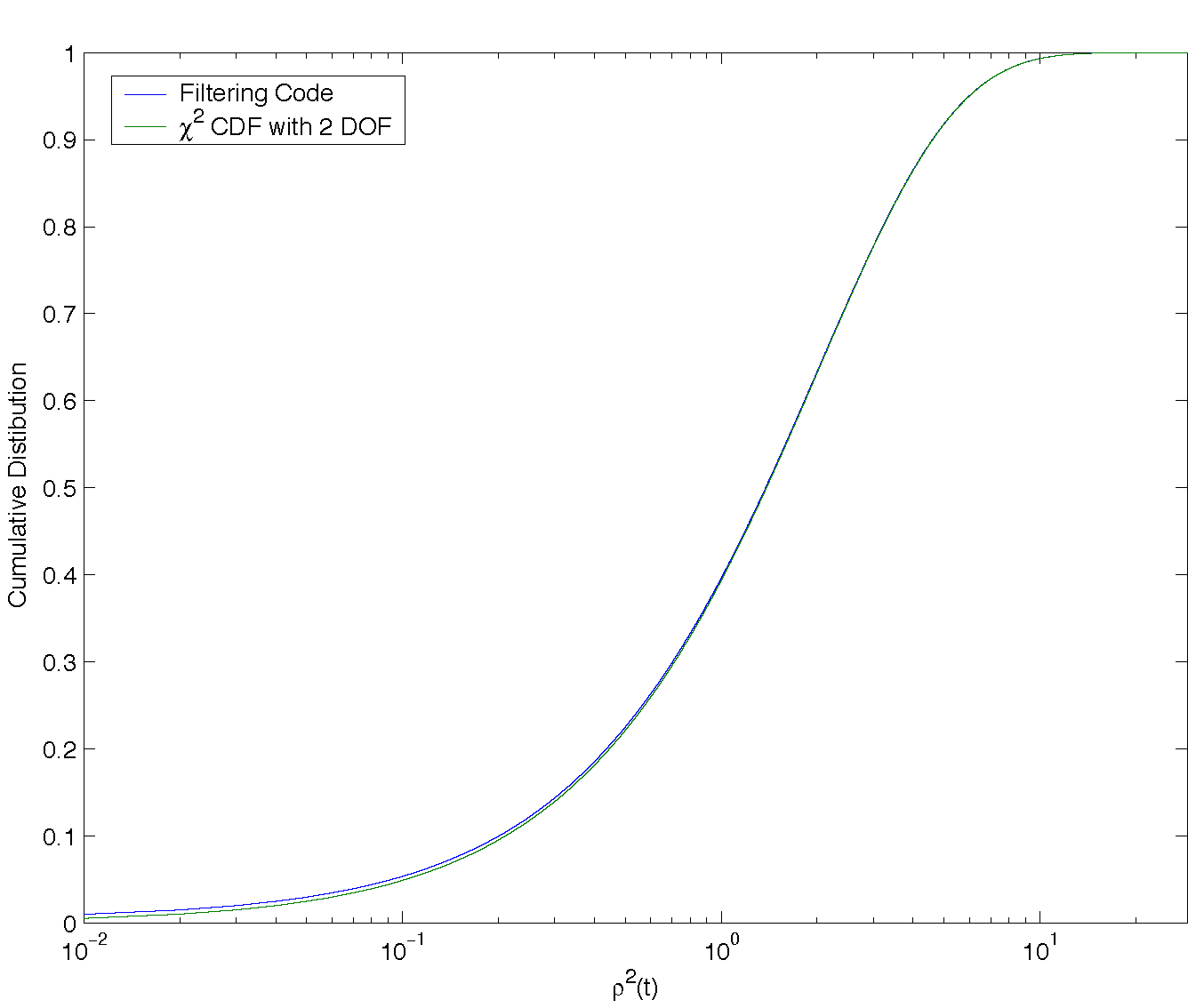}
\end{center}
\caption[Distribution of the Filter Output for Gaussian Noise]{%
\label{f:rhosq_gaussian_cdf}
In the presence of Gaussian noise the expected filter output, $\rho^2(t)$, is
the sum of the squares of two Gaussian distributed quantities and so should be
$\chi^2$ distributed with two degrees of freedom. This figure shows the
cumulative distribution function (CDF) of the filtering code output and the
expected analytic value. The filter input is white Gaussian noise of variance
$\varsigma^2$ and a constant power spectral density of $\ospsd = 2\varsigma^2\delta T$. It
can be seen that there is good agreement between the observed and expected
values.
}
\end{figure}

We may also consider the distribution of the signal-to-noise squared in
the presence of Gaussian noise. It is we can see from the definition of the
filter $x(t)$, given by equation (\ref{eq:xcts}), that it is a linear map from
$s(t)$ to $x(t)$, and similarly for the filter in equation (\ref{eq:ycts}) that
maps $s(t)$ to $y(t)$. If the input signal is $s(t)$ is a Gaussian random
variable, then the filter outputs $x(t)$ and $y(t)$ will be (uncorrelated)
Gaussian random variables. Since the filter output $\rho^2(t)$ is the sum of
the squares of these two Gaussian quantities, it will be $\chi^2$ distributed
with two degrees of freedom. Recall that for a random variable, $X$, the
cumulative density function is defined to be
\begin{equation}
P(x) = \int_{-\infty}^x p(x)\, dx
\end{equation}
where $p(x)$ is the probability density function. For a $\chi^2$ distribution
with $2$ degrees of freedom, this is
\begin{equation}
P(\chi^2) = \int_0^{\chi^2} \frac{e^{-{x}/2}}{2}\,dx.
\label{eq:chisqcdf}
\end{equation}
Figure \ref{f:rhosq_gaussian_cdf} shows the cumulative density function of
$\rho^2(t)$ obtained from one of the data segments in Table
\ref{t:normresults} plotted against the theoretical value given in equation
(\ref{eq:chisqcdf}). Clearly the measured and theoretical values agree very
well.

\subsection{Impulse Time}
\label{ss:impulsetime}

\begin{figure}[p]
\begin{center}
\includegraphics[width=\linewidth]{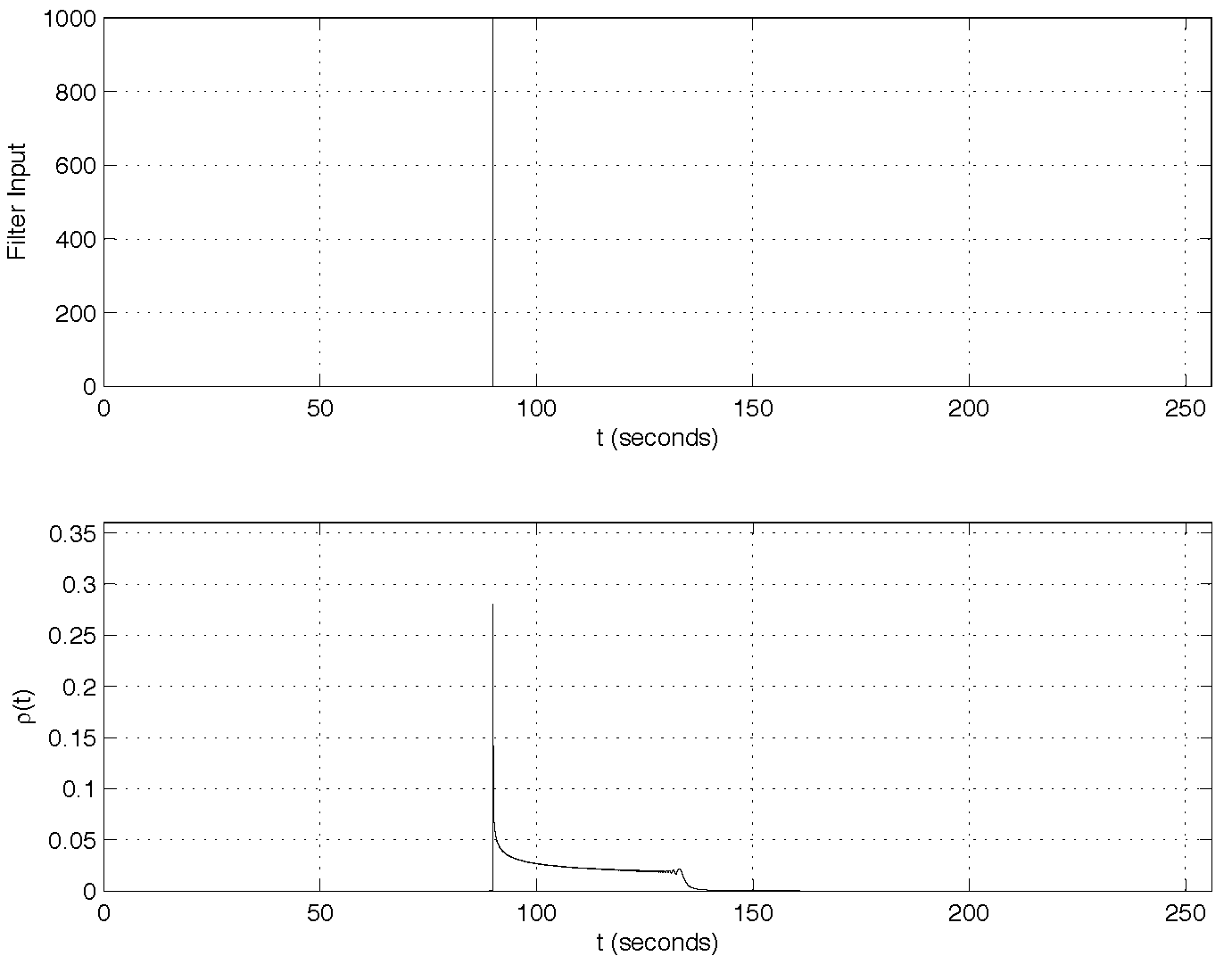}
\end{center}
\caption[Impulse Response of Matched Filter With a Constant Power Spectrum]{
\label{f:impulse_snr}
The top panel shows the filter input which consists of an impulse at $t_0 = 90$.
The power spectrum is set to that of white uncorrelated noise. The bottom panel
shows the output of the filter. The filter output is the sum of the squares of
the time reverse chirps and the maximum of the filter output occurs at the
time of the impulse.
}
\end{figure}

The second test is to examine the output of the filter in the presence of a
delta function and a constant (white) power spectrum. The input to the matched
filter is
\begin{equation}
\tilde{s}_k = \sum_{k=0}^{N-1} \delta_{jl} e^{-2\pi ijk/N} = e^{-2\pi ilk/N}.
\label{eq:deltafft}
\end{equation}
Substituting equation (\ref{eq:deltafft}) into equation
(\ref{eq:signaltonoisesq}), we obtain
\begin{equation}
\rho^2(t_j) = h_c^2(t_e - t_j) + h_s^2(t_e - t_j)
\label{eq:impulse_snr}
\end{equation}
which is the sum of the squares of the time reversed chirps.  Figure
\ref{f:impulse_snr} shows the output of the matched filter with a delta
function input at $t=90$ seconds. The length of the data segment is $256$
seconds, the template has $m_1 = m_2 = 1 M_\odot$ and the low frequency cut
off of the template is $40$~Hz. The length of this template is $43.7$ seconds.
It can be seen that the filter output does indeed follow the form of equation
(\ref{eq:impulse_snr}).  The \emph{impulse time} is the time at which an
impulse in the data would cause the filter output to peak at
$t_e$. We can see from equation (\ref{eq:impulse_snr}) and figure
\ref{f:impulse_snr} that for the filter we have implemented, the impulse time
will be at $t = t_0$, since this is when the maximum of the filter occurs in
the presence of an impulse.

\section{Wrap-around of the Fast Fourier Transform}
\label{s:wraparound}

\begin{figure}[p]
\begin{center}
\includegraphics[width=\linewidth]{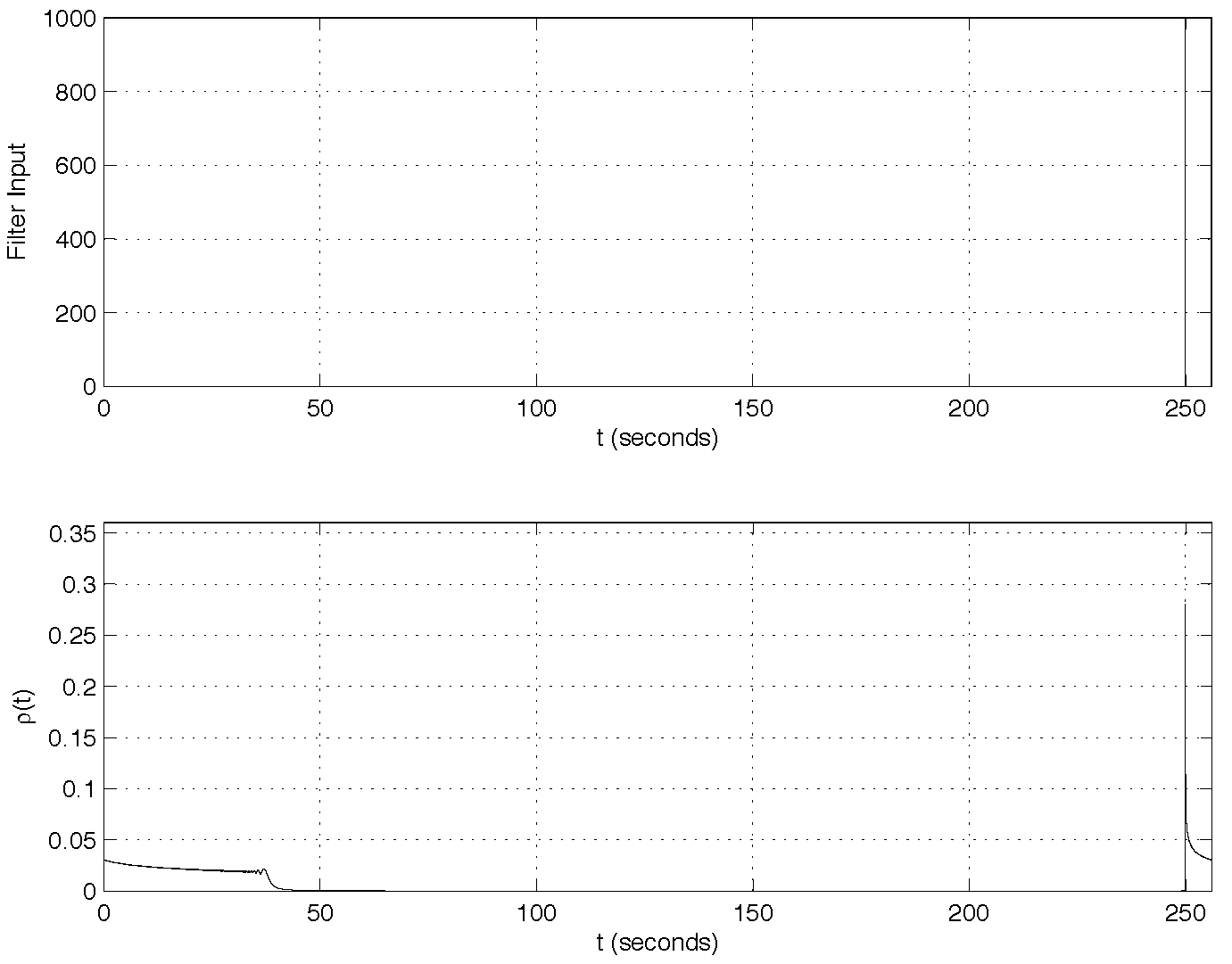}
\end{center}
\caption[Wrap-around of the Matched Filter]{%
\label{f:impulse_wraparound}
The top panel shows the filter input which consists of an impulse at $t_0 = 250$.
The power spectrum is set to that of white uncorrelated noise. The bottom panel
shows the output of the filter. The length of the chirp template is $43.7$
seconds. Notice that the filter output is non-zero for the first $37.7$
seconds of the output due to the wrap-around of the FFT.
}
\end{figure}

A simple experiment serves to demonstrate the effect of periodicity of the
Fast Fourier Transform (FFT) in matched filtering. As with
the example depicted in figure \ref{f:impulse_snr}, we generate an input data
segment of length  $256$ seconds. Now we place the impulse at $t = 250$
seconds, however. Figure \ref{f:impulse_wraparound} shows the input and output
of the filter for such a data segment. Notice that the output of the filter
\emph{wraps around}, so that the first $43.7-6 = 37.7$ seconds of the filter
output is non-zero. This is due to the Fast Fourier Transform  treating the
data as periodic: it identifies $t=0$ and $t=256$. If the impulse was
placed at $t=256$, just before the end of the segment, then the first $t_c$
seconds of $\rho(t)$ would be corrupted, where $t_c$ is the length of the chirp
template. This demonstrates that data at the \emph{start} of the segment is
being correlated with data at the \emph{end} of the segment due to the
wrap-around of the FFT. This is obviously unphysical, so we consider the first
$t_c$ seconds of the signal-to-noise ratio corrupted and ensure that we do not
consider this data when searching for inspiral triggers. We will return to
this problem in section \ref{ss:invspec} when we consider the construction of
the inverse power spectrum $1/\ospsd$ used in the filter.

\section{Power Spectral Estimation}
\label{ss:psd}

Interferometer data is not stationary over long periods of time, so we cannot
simply compute a single value of $S_n(|f|)$ to be used in the matched filter
for all time. We must use a power spectrum that gives the noise level at the
time of the data segment that we are filtering. To do this we use Welch's
method\cite{Welch:1967} to estimate the average power spectral density using
data close in time to the segment we are filtering.

A Welch power spectral density estimate is defined by an FFT length, overlap
length and choice of window function.  We require that the frequency
resolution and length of the power spectrum are the same as those of the data
$\tilde{v}_k$ and template $\tilde{h}_{ck}$. If the data segment is of length
$N$ points with a sampling interval of $\Delta t$, then the power spectrum
must be of length $N/2 + 1$ points with a frequency resolution of $\Delta f =
1/(N\Delta t)$.  (It is possible to generate the average power spectral density
at a different frequency resolution and then interpolate or decimate it to the
correct frequency resolution, however.) For simplicity of implementation,
the length of the data used to compute the power spectrum is the
same as that used in the filter data segment.  To construct the average power
spectrum we take $N_\mathrm{seg}$ data segments of length $N$ from near in
time to the segment being filtered. Each segment overlaps its neighbors by
$N_\mathrm{overlap}$ sample points, so we need
\begin{equation}
N_\mathrm{chunk} = 
N \times N_\mathrm{seg} - ( N_\mathrm{seg} - 1 ) \times N_\mathrm{overlap}
\end{equation}
input data points to compute the average power spectrum. The
$N_\mathrm{chunk}$ input data points are called an \emph{analysis chunk}.
In section \ref{ss:digitalfilter}, we discussed filtering several data
segments through each template in the filtering code; we will see later that
the data segments used in the filtering code have the same length and 
overlap as those used to estimate the power spectrum. 

Recall that since we are computing a discrete Fourier transform of the input
data, any power that is not at a sampled frequency in the power spectrum will
bleed into adjacent bins. This is a particular problem for LIGO data where
there are a lot of spectral line features, caused by power line harmonics or
mirror suspension wire resonances. These features contain a lot of power and,
in general, their frequencies do not lie exactly at sampled frequencies.
To prevent this power bleeding into adjacent bins, we apply a Hann window to
the data before taking the Fourier transform. This is a standard technique and
for further details we refer to the discussion in \cite{Press:1992}. 

To construct an average power spectrum from the $N_\mathrm{seg}$ individual
spectra that are computed, we average the $N_\mathrm{seg}$ values for each
frequency bin. That is, the value of $\ospsd$ at a frequency $f_k$ is the
average of the $N_\mathrm{seg}$ values of the power spectra at $f_k$. In the
standard Welch computation of the power spectral estimate, the mean is used to
average the values in each frequency bin.  Consider using the mean to compute
the average in the presence of a loud signal. If the data that contains the
loud signal is used in the computation of the average spectrum, then $\ospsd$
will contain power due to the signal. This will suppress the
correlation of the signal and the template at those frequencies and cause the
value of the signal-to-noise ratio to be lower than one would obtain if the
average power spectrum is computed from noise alone.  To avoid this problem we
use the median to estimate the average power spectrum. This has two
advantages: \emph{(i)} computational simplicity, as we only need to compute
one PSD and can use it for several data segments and \emph{(ii)} insensitivity
to outliers in the spectra, which means that excess power in one segment does
not corrupt the spectra for neighboring segments. This is useful since the
LIGO data is not truly stationary.

\begin{figure}[p]
\begin{center}
\includegraphics[width=\linewidth]{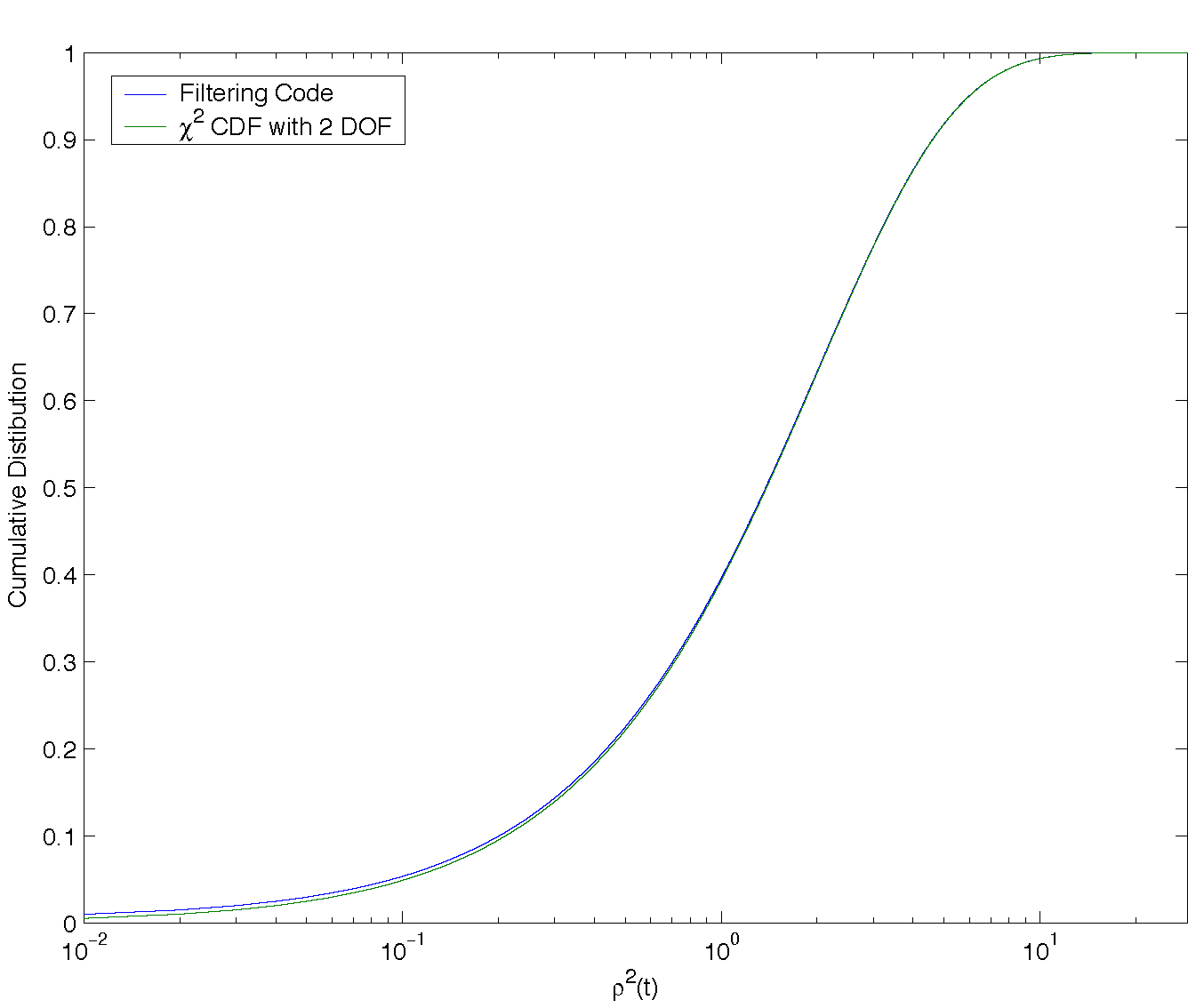}
\end{center}
\caption[Distribution of Filter Output Using Median Power Spectral Estimator]{%
\label{f:rhosq_median_cdf}
In the presence of Gaussian noise the expected filter output, $\rho^2(t)$, is
the sum of the squares of two Gaussian distributed quantities and so should be
$\chi^2$ distributed with two degrees of freedom. This figure shows the
cumulative distribution function (CDF) of the filtering code output and the
expected analytic value. The filter input is white Gaussian noise of length
$256$ seconds and the power spectrum $\ospsd$ is computed from $15$ segments
of white Gaussian noise length $256$ seconds, overlapped by $128$ seconds
using Hann windowing and the median estimator.
}
\end{figure}

For a  Gaussian random variable, the median is a
factor of $\log 2$ larger than the mean. We must therefore divide the median
power spectrum by $\log 2$ to ensure that it has the same normalization as the
mean power spectrum for Gaussian noise. This scaling has the unwanted effect
of suppressing constant features in the spectrum, such as power lines and wire
resonances, by a factor of $\log 2$ compared with the mean spectrum. In
practice we find that this does not have a significant effect on the output of
the filtering code.  For a low number of data segments,
$N_\mathrm{seg}$, the $\log 2$ correction factor is incorrect; the true value
is between $\log 2$ and $1$ and we do not correct for this bias. Figure
\ref{f:rhosq_median_cdf} shows the cumulative distribution of the filter
output in the presence of Gaussian noise, where the average power spectrum is
computed using the median method.  The bias introduced by the
for low $N_\mathrm{seg}$ does not have a significant
effect on the filter output.

\section{Computation of the inverse power spectrum}
\label{ss:invspec}

We observed in section \ref{s:wraparound} that the FFT we use to compute the
match filter treats the data as being periodic and that we had to ignore part
of the filter output that was corrupted due to wraparound of the data.

\begin{figure}[p]
\begin{center}
\includegraphics[width=\linewidth]{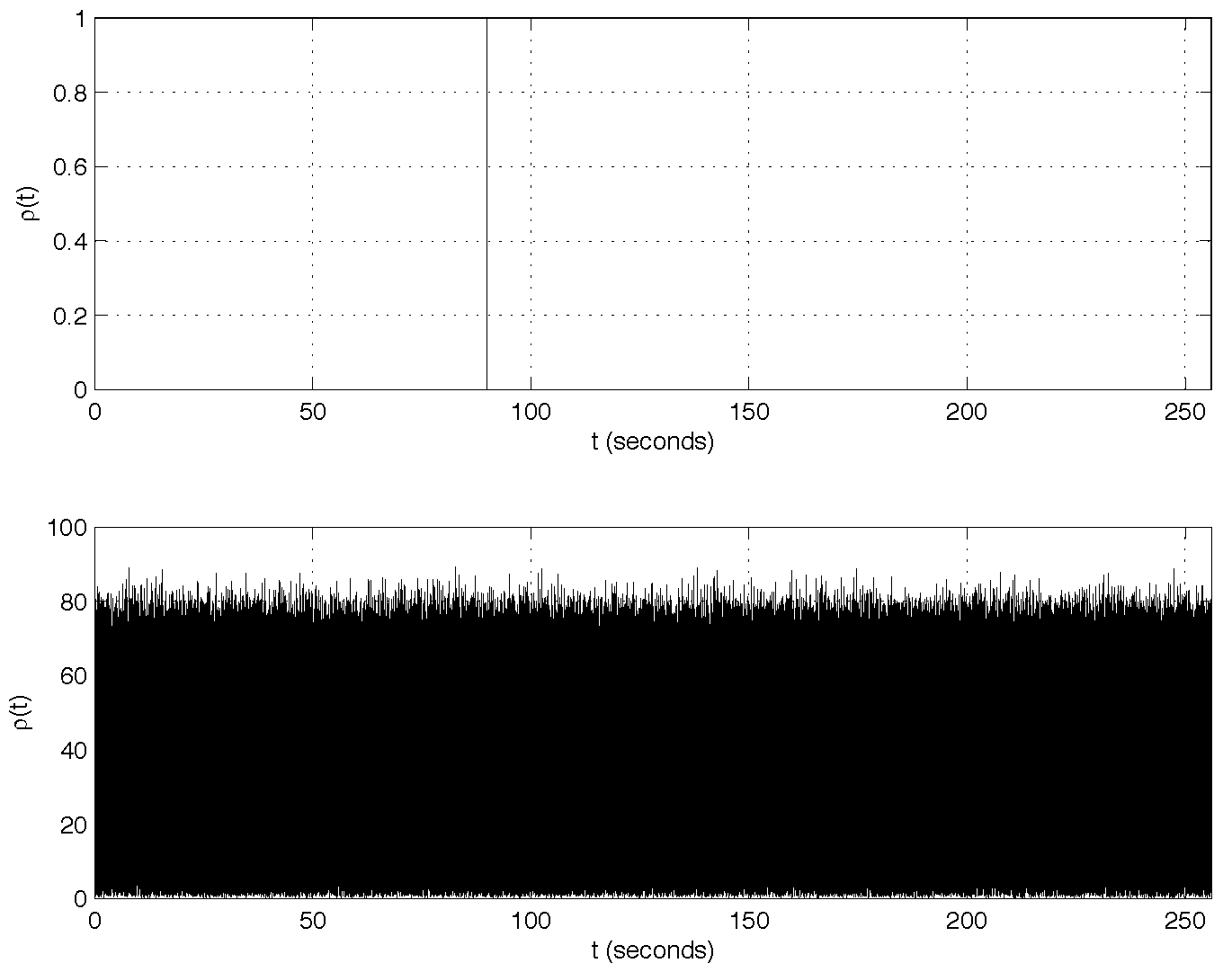}
\end{center}
\caption[Impulse Response of Matched Filter With a Real Power Spectrum]{
\label{f:impulse_spec}
The top panel shows the filter input which consists of an impulse at $t_0 =
90$.  The power spectrum is computed from Gaussian noise of the same length of
the input data using Welch's method. The bottom panel shows the output of the
filter. Due to the fact that the duration of the inverse power spectrum
$1/\ospsd$ in the time domain is the same length as the data segment, the
entire filter output is corrupted due to the wrap around of the FFT.
}
\end{figure}

If we look at the correlation in equation (\ref{eq:signaltonoisesq}), we can
see that we are filtering the data against the inverse power
spectrum as well as the chirp, that is our filter is
\begin{equation}
\frac{\tilde{h}_c^\ast}{\ospsd}.
\end{equation}
Recall that the chirp has a duration that is typically much less than the
length of the data segment, so the effect of wrap-around only corrupts a
region that is the length of the chirp at the start of the data segment.
Unfortunately, the length of the inverse power spectrum, as a time domain
filter, is the same length as the data segment. Figure \ref{f:impulse_spec}
shows the filter output when the input data is an impulse at $t=90$ seconds
and the power spectrum is computed from Gaussian noise using the median
method. Notice that the filter output is non-zero at all times. No matter
where the impulse is placed, the entire filter output would be corrupted
by the inverse power spectrum.  To prevent this, we truncate the square
root of the inverse power spectrum to a length $t_\mathrm{invspectrunc}$
seconds in the time domain. This means that the inverse power spectrum will
have support (i.e non-zero values) for $2t_\mathrm{invspectrunc}$ seconds in
the time domain.  Truncation of the inverse spectrum has the effect of
smoothing out the high $Q$ features (narrow line features, such as power line
harmonics 
or resonances of the mirror suspension wires) and restricting the length
of time that the filter is corrupted. The corrupted regions can then be
ignored when searching for chirps in the filter output.

\begin{figure}[p]
\begin{center}
\includegraphics[width=\linewidth]{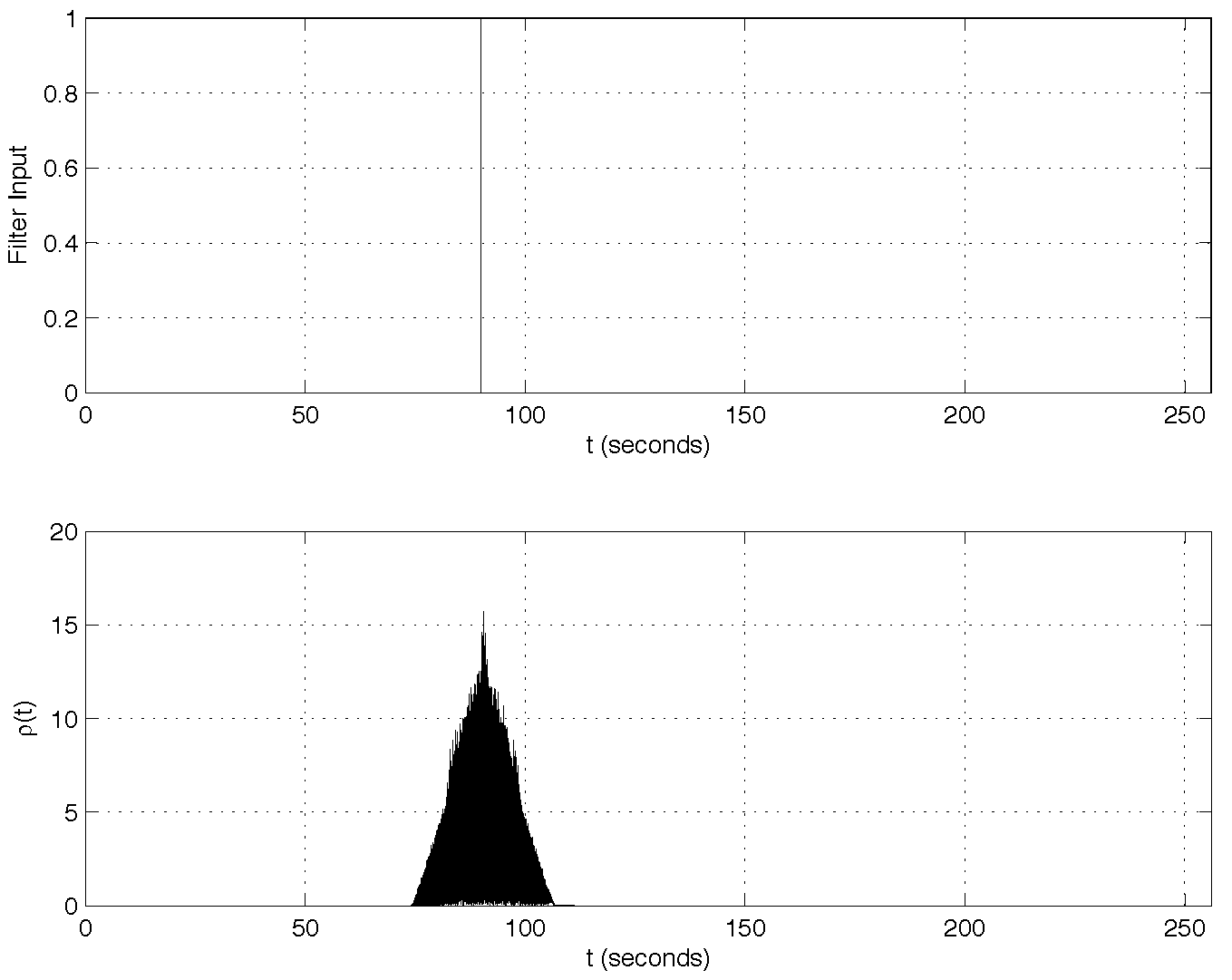}
\end{center}
\caption[Impulse Response of Matched Filter With a Truncated Power Spectrum]{%
\label{f:impulse_inv_spec}
The top panel shows the input to the filtering code which is an impulse at $t
= 90$ seconds. The average power spectrum is computed from typical LIGO noise
and then truncated to $16$ seconds in the time domain. The duration of non-zero
filter output is also $16$ seconds.
}
\end{figure}

The algorithm used to truncate the power spectrum is as follows:
\begin{enumerate}
\item Compute the average power spectrum of the uncalibrated input data
$v(t_j)$ using Welch's method as described in the previous section.
\item Compute the square root of the inverse power spectrum,
\begin{equation}
\sqrt{S^{-1}_v(|f_k|)}.
\end{equation}
\item Set the Nyquist, $(k = N/2)$ and DC $(k = 0)$ components of this to zero.
\item Compute the inverse Fourier transform of $\sqrt{S^{-1}_v(|f_k|)}$ to
obtain the time domain inverse PSD of length $T = N\Delta t$ seconds.  \item
Zero the square root of the inverse spectrum between the time
$t_\mathrm{invspectrunc}/2$ and $(T-t_\mathrm{invspectrunc})/2$
seconds. This sets the length of the square root of the inverse spectrum in
the time domain to be $t_\mathrm{invspectrunc}$ seconds.
\item Fourier transform the time domain quantity back to the frequency domain.
\item Divide by the number of points $N$ to ensure that the inverse power
spectrum is correctly normalized.
\item Square this quantity to recover $\bar{S}^{-1}_v(|f_k|)$.
\item Set the Nyquist and DC frequencies to zero.
\item The (scaled) strain inverse power spectral density is then computed by
\begin{equation}
\frac{1}{d^2\ospsd} = \frac{1}{\left|d \times R(f_k)\right|^2} \bar{S}^{-1}_v(|f_k|).
\label{eq:truncedspec}
\end{equation}
\end{enumerate}
The factor of $1/R(f)$ in equation (\ref{eq:truncedspec}) may add some
additional length to $\bar{S}_n(|f|)$ in the time domain (since $R(f)$ is not
white), but because $R(f)$ is smooth with no sharp line features, this is
insignificant.
The length of the inverse power spectrum is a parameter that we may tune based
on the nature of the data that we are filtering. In the analysis described
in this thesis, we set the length of the inverse power spectrum in the time
domain to 32 seconds.  Figure \ref{f:impulse_inv_spec} shows the filter output
in the presence of an impulse for a truncated power spectrum. There is
non-zero data \emph{before} as well as after the impulse, so we must ignore
data at the end of a segment as well as before.

\section{The $\chi^2$ veto}
\label{s:chisqcts}

Although the matched filter is very good at finding signals in the noise,
transient events in the data will also cause high values of the
signal-to-noise ratio for data containing impulses, as we saw in section
\ref{ss:impulsetime}. To distinguish a high signal-to-noise event due to a
signal from one due to a transient, we use a time-frequency veto known as the
$\chi^2$ veto. This was first proposed in \cite{grasp} and is described in
more detail in \cite{Allen:2004}.  In this section, we review the construction
of the $\chi^2$ veto and in the next sections describe the implementation used
in the filtering code.

Let $u$ and $v$ be two orthonormal time series representing the two phases of a
binary inspiral signal, $h_c(t_j)$ and $h_s(t_j)$.  We divide these waveforms
into $p$ frequency sub-intervals $\{u_l\}$ and $\{v_l\}$, $l=1\ldots p$ with
\begin{eqnarray}
  (u_l|u_m) &=& \frac{1}{p}\delta_{lm} \\
  (v_l|v_m) &=& \frac{1}{p}\delta_{lm} \\
  (u_l|v_m) &=& 0
\end{eqnarray}
and $u=\sum_{l=1}^p u_l$ and $v=\sum_{l=1}^p v_l$.

We then obtain the $2p$ time series
\begin{align}
\{x_l\} & =(s|u_l),\\ 
\{y_l\} & =(s|v_l),
\end{align}
where $s$ is the detector output. Notice that 
\begin{align}
x &= \sum_{l=1}^p x_l=(h|u) \\
y &=\sum_{l=1}^p y_l=(h|v)
\end{align}
so that $(x^2+y^2)/ \sigma^2$ is the signal to noise ratio squared $\rho^2$.
Now, let 
\begin{align}
\Delta x_l &= x_l - \frac{x}{p}, \\
\Delta y_l &= y_l - \frac{y}{p}
\end{align}
and define
\begin{equation}
\chi^2 = \frac{p}{\sigma^2} \sum_{l=1}^p \left[ (\Delta x_l)^2 + (\Delta y_l)^2 \right]
\end{equation}
In the presence of Gaussian noise $s=n$ this statistic is $\chi^2$ distributed
with $\nu=2p-2$ degrees of freedom.  Furthermore, if a signal $h=Au+Bv$ (with signal
to noise squared of $\rho^2_{\mathrm{signal}}=A^2+B^2$) is present along with
Gaussian noise $s=h+n$, then $\chi^2=pr^2$ is still $\chi^2$ distributed
with $\nu=2p-2$ degrees of freedom. Small values of the $\chi^2$ veto mean
that the signal-to-noise ratio has been accumulated in a manner consistent
with an inspiral signal. We apply an additional threshold on $\chi^2$
for triggers that have a high signal-to-noise ratio.

\subsection{Implementation of the Digital $\chi^2$ Veto}
\label{ss:chisqdisc}

Recall that the templates $\tilde{h}_c$ and $\tilde{h}_s$ are normalized such
that
\begin{equation}
\sigma^2 = 4 \frac{\Delta t}{N} \sum_{k=0}^{N/2}
\frac{\tilde{h}_{ck} \tilde{h}_{ck}^\ast } {\ospsd}.
\end{equation}
We construct the $p$ templates $\left\{ \tilde{h}_{c(l)} \right\}$ and 
$\left\{ \tilde{h}_{s(l)} \right\}$, where $l = 1,\ldots,p$, with
\begin{align}
\frac{4\Delta t}{N} \sum_{k=0}^{N/2} 
\frac{\tilde{h}_{ck(l)} \tilde{h}_{ck(m)}^\ast}{\ospsd} &= \frac{1}{p}\delta_{lm}\sigma^2 \\
\frac{4\Delta t}{N} \sum_{k=0}^{N/2} 
\frac{\tilde{h}_{sk(l)} \tilde{h}_{sk(m)}^\ast}{\ospsd} &= \frac{1}{p}\delta_{lm}\sigma^2 \\
\frac{4\Delta t}{N} \sum_{k=0}^{N/2} 
\frac{\tilde{h}_{ck(l)} \tilde{h}_{sk(m)}^\ast}{\ospsd} &= 0
\end{align}
and
\begin{align}
\tilde{h}_{c} &= \sum_{l=1}^{p} \tilde{h}_{c(l)}, \\
\tilde{h}_{s} &= \sum_{l=1}^{p} \tilde{h}_{s(l)}.
\end{align}
We construct the $2p$ filter outputs
\begin{equation}
x_{j(l)} = 4 \frac{\Delta t}{N} 
   \sum_{k=0}^{N/2} e^{2\pi ijk/N} 
   \frac{\tilde{s}_k \tilde{h}_{ck(l)}^\ast}
        {\ospsd}
\end{equation}
and
\begin{equation}
y_{j(l)} = 4 \frac{\Delta t}{N} 
   \sum_{k=0}^{N/2} e^{2\pi ijk/N} 
   \frac{\tilde{s}_k \tilde{h}_{sk(l)}^\ast}
        {\ospsd}
\end{equation}
from which we can recover equations (\ref{eq:xdisc}) and (\ref{eq:ydisc}) by
\begin{equation}
x_j = \sum_{l = 1}^{p} x_{j(l)}
\end{equation}
and
\begin{equation}
\quad y_j = \sum_{l = 1}^{p} y_{j(l)}.
\end{equation}
Consequently, the signal-to-noise ratio can be written as
\begin{equation}
\rho^2(t_j) = \frac{1}{\sigma^2} \left[ \left( \sum_{l = 1}^{p} x_{j(l)} \right)^2 + \left( \sum_{l = 1}^{p} y_{j(l)} \right )^2 \right].
\end{equation}
Let
\begin{equation}
\Delta x_{j(l)} = x_{j(l)} - \frac{x_j}{p}
\end{equation}
and
\begin{equation}
\Delta y_{j(l)} = y_{j(l)} - \frac{y_j}{p}
\end{equation}
and define the quantity
\begin{equation}
\chi^2(t_j) = \frac{p}{\sigma^2} \sum_{l = 1}^p \left[ \left(\Delta x_{j(l)}\right)^2 + \left(\Delta y_{j(l)}\right)^2 \right].
\label{eq:chisqdefn}
\end{equation}
If at any time $t_j$ the signal-to-noise ratio
exceeds the threshold $\rho(t_j) \ge \rho_\ast$ then we compute $\chi^2(t_j)$
 for the data segment. We can then threshold on $\chi^2 < \chi^2_\ast$ to
decide if the signal-to-noise event is consistent with a true inspiral signal.
In section \ref{ss:mismatchedchisq} we discuss a modification to this
threshold, if the template and signal are not exactly matched.

\subsection{Mismatched signal}
\label{ss:mismatchedchisq}

The waveform of an inspiral depends on the masses parameters $M,\eta$ of the two
objects in the binary. If the template being used in the
matched filter does not exactly match the true signal
$h'(M',\eta')$ then the output of the matched filter will be smaller than if
the template was correct. The mismatch can arise for any number of reasons;
for example errors in the theoretical
template mean that the post$^2$-Newtonian waveform does not match the true
inspiral signal (which becomes important at high masses, $M > 3 M_\odot$) or
errors in the calibration function $R(f)$ may change the
amplitude and/or phase of the signal in the data relative to the corresponding
template.  

The loss of signal-to-noise ratio due to mismatch is accompanied by an
increase in $\chi^2$ which requires a modification of the threshold.
Suppose a signal, $Aw$, that is not exactly matched by $u$ or $v$
is present in the data: $s=Aw$ where $w$ is the (normalized) and $A$ is an
amplitude.  (Here we assume no noise.)  With no loss of generality, orient $u$
and $v$ such that $(w|v)=0$.  If $w$ is nearly parallel to $u$, separated by
some parameter difference $\delta x^\alpha$ (which is small) then
\begin{equation}
  w \simeq u + \frac{\partial u}{\partial x^\alpha}\delta x^\alpha 
  + \frac{1}{2}\frac{\partial^2 u}{\partial x^\alpha \partial x^\beta}\delta x^\alpha\delta x^\beta
\end{equation}
so
\begin{equation}
\begin{split}
  (s|u) &\simeq A\left(u + \frac{\partial u}{\partial x^\alpha}\delta x^\alpha 
  + \frac{1}{2} \frac{\partial^2u}{\partial x^\alpha\partial x^\beta}\delta x^\alpha\delta x^\beta\bigl|u\right)
  \\
  &= A\bigl\{ (u|u) + \left(\frac{\partial u}{\partial x^\alpha}\big|u\right)\delta x^\alpha
  + \frac{1}{2} \left(\frac{\partial^2 u}{\partial x^\alpha \partial x^\beta}\bigl|u\right)\delta x^\alpha\delta x^\beta
  \bigr\}\\
  &= A\bigl\{ 1 +
  \frac{1}{2} \left(\frac{\partial^2 u}{\partial x^\alpha \partial x^\beta}\bigl|u\right)\delta x^\alpha\delta x^\beta
  \bigr\}
\end{split}
\end{equation}
since $(w|u)$ is a local maximum for $w=u$; thus $(\frac{\partial u}{\partial x^\alpha}|u)=0$.
Let us define the ``mismatch'' between $w$ and $u$ as
\begin{equation}
ds^2 = 1 - (w|u)
\end{equation}
and then
\begin{equation}
  ds^2 = g_{\alpha\beta}\delta x^\alpha\delta x^\beta =
  - \frac{1}{2} \left(\frac{\partial^ 2u}{\partial x^\alpha \partial x^\beta}\bigl|u\right)\delta x^\alpha\delta x^\beta.
\end{equation}
so
\begin{equation}
g_{\alpha\beta}=-{\frac{1}{2}}\left(\frac{\partial^2 u}{\partial x^\alpha\partial x^\beta}\bigl|u\right).
\end{equation}
Now we compute $\chi^2$.  We can ignore the $v$ terms.  Thus
\begin{equation}
\begin{split}
\chi^2 &= \frac{p}{\sigma^2} \sum_{l=1}^p \left[ (h|u_l) - (h|u)/p \right]^2  \\
&= \frac{p}{\sigma^2} \sum_{i=l}^p \left[ (h|u_l)^2 - 2(h|u)(h|u_l)/p + (h|u)^2/p^2 \right] \\
&= \frac{p}{\sigma^2}\sum_{i=l}^p (h|u_l)^2 - (h|u)^2.
\end{split}
\end{equation}
Using the Schwartz inequality $(h|u_l)^2\le(h|h)(u_l|u_l)$ we obtain
\begin{equation}
(h|u_l)^2\le (h|h)
\end{equation}
and so we may write
\begin{equation}
\begin{split}
\chi^2 &\le \left[ (h|h) - (h|u)^2 \right]  \\
&= A^2\left[ (w|w) - (w|u)^2 \right] \\
&= A^2 \left[ 1 - ( 1 - ds^2 )^2 \right] \\
&\simeq 2A^2ds^2.
\end{split}
\end{equation}
Therefore, if $h=Aw+n$ where $w$ has a slight mismatch $ds^2=1-(w|u)$, then
$\chi^2$ has a non-central $\chi^2$ distribution with $\nu=2p-2$ degrees of
freedom and a non-central parameter $\lambda=2A^2ds^2$ and $A^2 =
\rho^2$.

A $\chi^2$ distribution with $\nu$ degrees of freedom has a mean of $\nu$
and a variance of $2\nu$ in Gaussian noise; hence one often considers the
quantity $\chi^2/\nu$ which would have a unit mean and a variance of two in
the presence of Gaussian noise alone.  A non central $\chi^2$ distribution
with $\nu$ degrees of freedom and non-central parameter $\lambda$ has a mean
of $\nu+\lambda$ and a variance of
$2(\nu+\lambda)\times[1+\lambda/(\nu+\lambda)]$.  (The factor in brackets in
the variance is always between 1 and 2, and is not really important for our
purposes.)  Thus, in this case, the quantity $\chi^2/(\nu+\lambda)$ has unit
mean and variance of between two and four.

For the reasons discussed above, a true signal will never be exactly matched
by one of our template waveforms, so we wish to conservatively modify the threshold
$\chi^2_\ast$ to allow for the case of a mismatched signal. While it is
possible to construct constant confidence thresholds on the non-central
$\chi^2$ distribution for various signal events, a crude (but perhaps
adequate) prescription is to threshold on the quantity
$\chi^2/(\nu+\lambda)$, which is roughly equivalent to thresholding on
$\chi^2/(p+A^2 ds^2)$.  Since real interferometer noise is not really
Gaussian, it is not important to use the exact result for the non-central
$\chi^2$ distribution, though this could certainly be done. This choice of
threshold is conservative as we would not reject signals more often than we
expect if using the true threshold.  In practice, we threshold on 
\begin{equation}
\chi^2 < \chi^2_\ast (p+\rho^2 \delta^2),
\label{eq:chisqthresholdtest}
\end{equation}
where $\rho$ is the signal-to-noise ratio of the signal and $\delta^2$ is a
parameter chosen to reflect the to be the largest amount mismatch that a true
signal may have with the template waveforms. Since we do not know all the
contributions to $\delta^2$ (in particular we do not accurately know the
contribution from errors in calibration), we set $\delta^2$ by Monte Carlo
techniques, which will be described in chapter \ref{ch:pipeline}.

\section{Trigger selection algorithm}
\label{s:maxoverchirp}

\begin{figure}[p]
\begin{center}
\includegraphics[angle=-90,width=\linewidth]{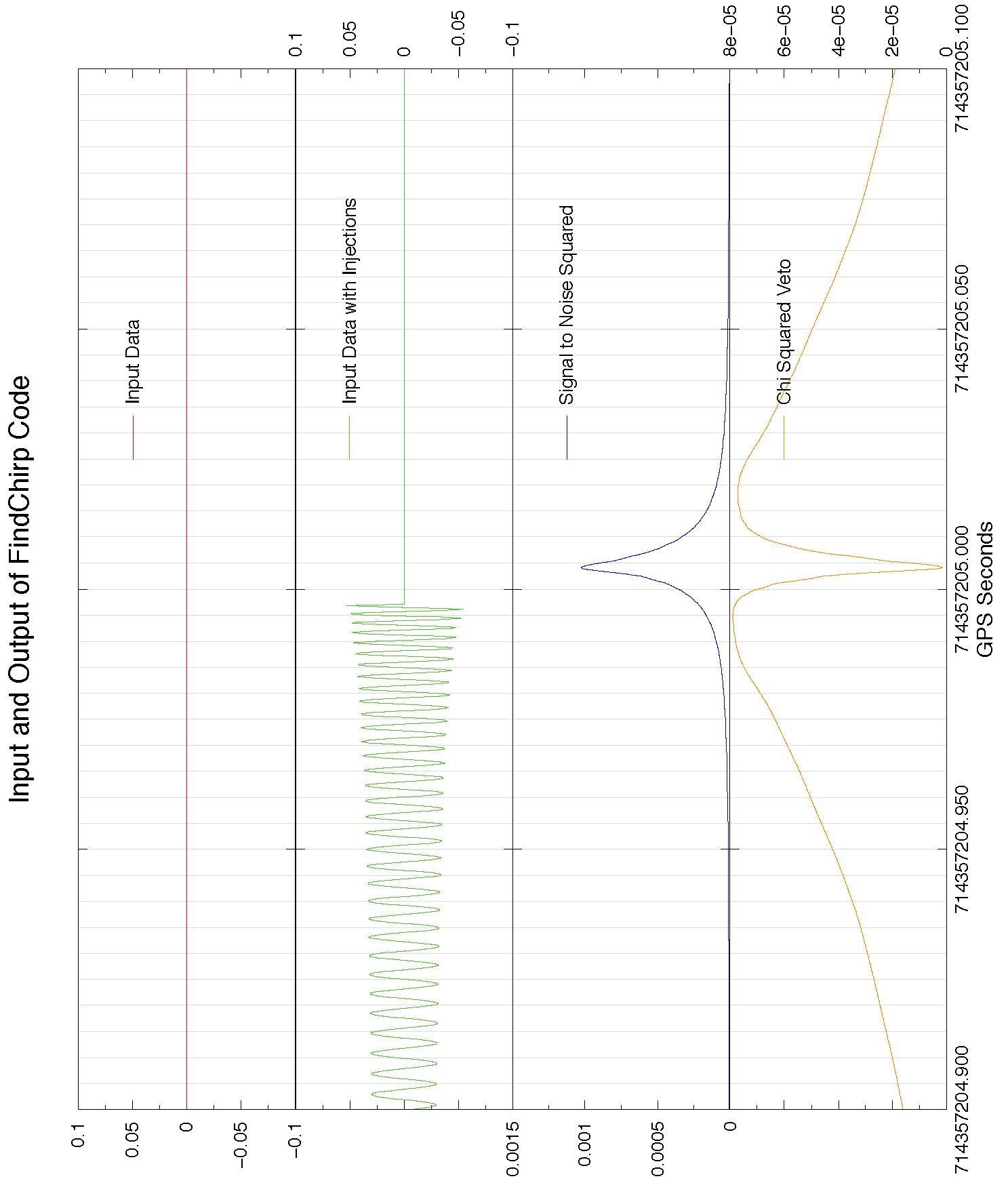}
\end{center}
\caption[Output of the Filtering Code for a Chirp In the Absence of Noise]{
\label{f:zero_inject_zoom}
Output time series from the filtering code
for an inspiral chirp in the absence of noise. A $(2.0,2.0)\,m_\odot$ inspiral
chirp is generated using the post$^2$-Newtonian time domain waveform
generation and injected into the data. This is filtered using the
post$^2$-Newtonian stationary phase waveform. The signal to noise squared and
$\chi^2$ time series are shown.  The signal to noise squared is a maximum at
the coalescence time of the \textit{template} inspiral signal. This occurs
slightly after the coalescence time of the injected signal. The difference in
coalescence times is due to the different methods of generating the chirp
signal.
}
\end{figure}

The object of the search algorithm is to generate a list of inspiral triggers.
A trigger is a time at which there may be a binary inspiral signal in the data
stream. The GPS time recorded in the trigger would correspond to the
coalescence time of an inspiral signal, which corresponds to the time at which
the signal-to-noise ratio squared is a maximum, as shown in figure
\ref{f:zero_inject_zoom}.

We have seen in sections \ref{s:wraparound} and \ref{ss:invspec} that if the
length of the chirp is $t_\mathrm{chirp}$ seconds and the length of the
inverse power spectrum is $t_\mathrm{PSD}$ seconds then we must ignore
$t_\mathrm{PSD}/2 + t_\mathrm{chirp}$ seconds of data at the beginning of the
data segment and $t_\mathrm{PSD}/2$ seconds of data at the end of the data
segment due to wrap-around corruption. To simplify the data management, we
ignore the first and last quarter of a segment; we test that $t_\mathrm{PSD}/2
+ t_\mathrm{chirp}$ is less that one quarter of a segment and generate an
error if it is not. The error informs the user that longer
data segments must be used to avoid corruption of the filter output.

\begin{figure}[p]
\begin{center}
\includegraphics[width=\linewidth]{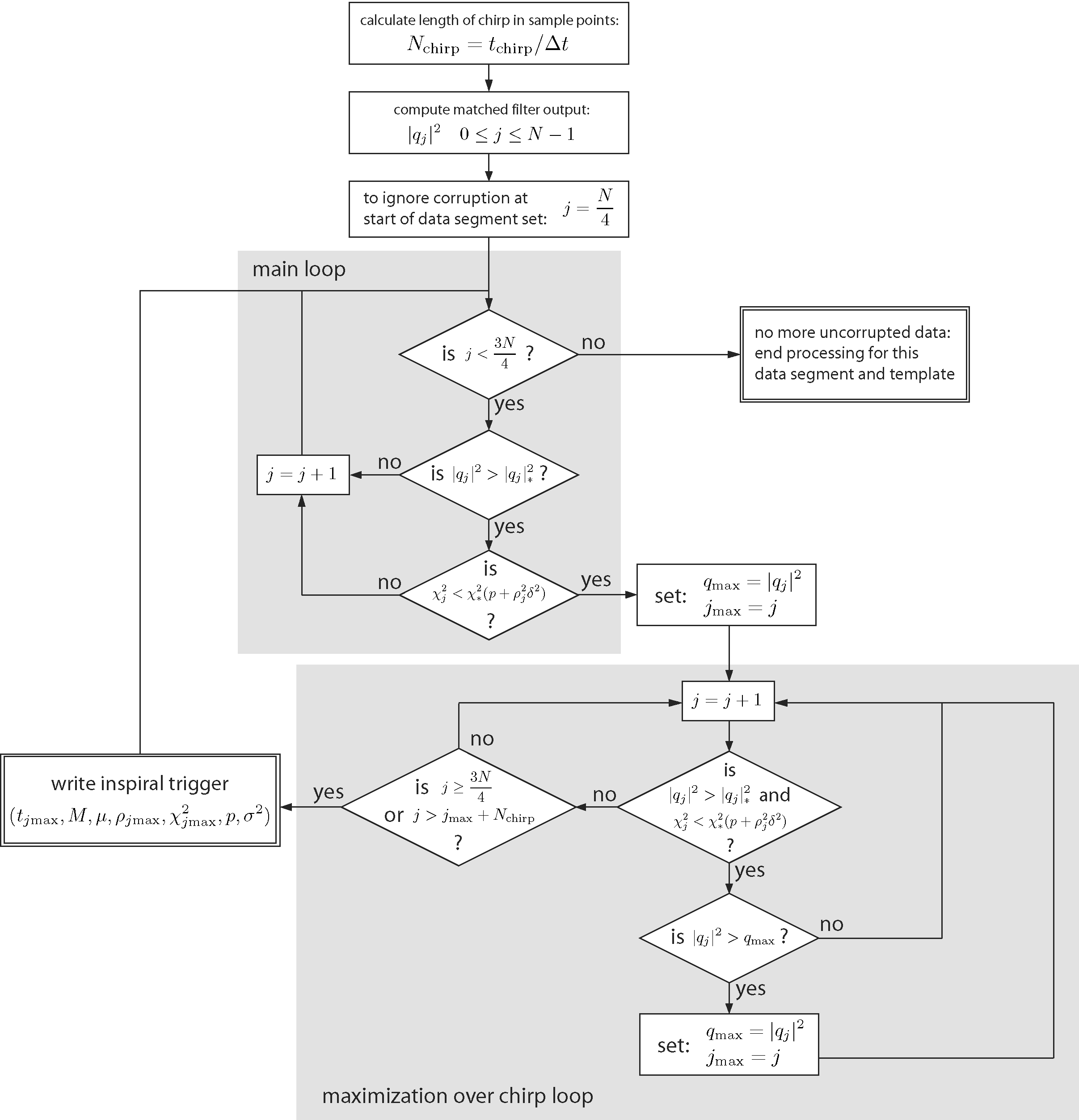}
\end{center}
\caption[Inspiral Trigger Selection Algorithm]{
\label{f:maxoverchirp}
The algorithm used to generate inspiral triggers. For a given inspiral
template we begin by calculating the length of the chirp and the filter
output. For a data segment of length $N$, the first and last $N/4$ points in
the segment may be corrupted due to FFT wrap-around, so we ignore them. For
the rest of the data segment, we step through the filter output looking for
times when the signal-to-noise and $\chi^2$ threshold are satisfied (the main
loop). If we find a point that passes the threshold tests, we label it
$j_\mathrm{max}$ and enter the maximization over chirp loop. This steps
through the data looking for the any larger values of $|q_j|^2$ within a chirp
length (given by $N_\mathrm{chirp}$) of the time $j_\mathrm{max}$. If a larger
value of $|q_j|^2$ is found, we reset $j_\mathrm{max}$ and keep looking for
any larger values. If no larger value is found within a chirp length (or we
reach the end of the uncorrupted data) we generate an inspiral trigger and
save its time, mass, signal-to-noise ratio, value of the $\chi^2$ veto, number
of $\chi^2$ bin ($p$) and the value of $\sigma^2$ for this data segment. 
}
\end{figure}

The signal-to-noise ratio $\rho^2$ of a trigger must exceed the threshold
$\rho^2_\ast$ and the $\chi^2$ statistic for the trigger must be less than the
threshold value $\chi^2_\ast/(p+\rho^2\delta^2)$.  When generating triggers,
we must consider the fact that the length of a data segment is greater than
the length of a chirp, so may be multiple chirps in a single segment.  We
could simply examine the time series for sample points where $\rho^2 >
\rho^2_\ast$, however it is likely that for a true signal there will be many
sample points above threshold corresponding to the same event.  Similarly, if
the data is noisy, we do not wish to generate a flood of events by considering
every sample point above threshold a unique event. We address this by a
trigger selection algorithm that we call \emph{maximization over a chirp}.
Figure \ref{f:maxoverchirp} shows the algorithm for constructing the list of
inspiral triggers.
It can be seen from the algorithm in figure \ref{f:maxoverchirp}  that
multiple triggers for the same template can be generated in one data segment.
The coalescence times for the different triggers must be separated by
\textit{at least} the length of the template waveform.  

The list of inspiral triggers is the final output of the filtering code. For
each trigger generated, we store the GPS time, mass parameters of the
template waveform, the signal-to-noise ratio, the value of the $\chi^2$ veto, the
effective distance $\mathcal{D}$ of the trigger in Mpc and the value of
$\sigma^2$.

The core of the inspiral analysis pipelines that we construct is the generation of
inspiral triggers. Once we have generated the inspiral triggers from the
matched filtering and $\chi^2$ code, we can test for coincidence
between multiple interferometers, examine environmental data and auxiliary
interferometer channels for associated artifacts, etc.  Construction of an
analysis pipeline is described in the next chapter.


\begin{table}[p]
\begin{center}
\begin{tabular}{l|c|c}
Random Noise Generator Seed& $\left\langle \rho^2(t) \right\rangle$ & $\mathrm{Var}( \rho^2(t) )$\\
\hline
$7$ & $2.0118$ & $4.0312$ \\
$15$ & $2.0059$ & $4.0196$ \\
$19$ & $1.9965$ & $3.9911$ \\
$43$ & $1.9998$ & $4.0023$ \\
$69$ & $1.9936$ & $3.9846$ \\
\end{tabular}
\end{center}
\caption[Mean and Variance of the Matched Filter Output for Gaussian Noise]{%
\label{t:normresults}
The mean and variance of the filter output $\rho^2(t)$ for five samples of
white, Gaussian noise with a constant power spectrum of $\ospsd =
2\varsigma^2\delta T$. The observed values agree with the expected value for the
mean and the variance, showing that the implementation of the matched
filter is correctly normalized.
}
\end{table}

\newpage

\Chapter{Detection Pipelines For Coalescing Binaries}
\label{ch:pipeline}

Chapter \ref{ch:findchirp} described the algorithms that we use to generate
inspiral triggers given an inspiral template and a \emph{single data segment}.  There
is more to searching for gravitational waves from binary inspiral than trigger
generation, however. To perform a search for a given class of sources in a
large quantity of interferometer data we construct a \emph{detection
pipeline}.  In section \ref{s:construction} we give an overview of the
the components used in a pipeline and how they fit together.
We then describe the building blocks of the pipeline in more detail. Section
\ref{s:dq} describes data quality cuts, which are used to discard 
data which is unsuitable for analysis. The application of trigger generation
to the data is explained in section \ref{s:pipetemplate}. The use of data from
multiple interferometers is described in section \ref{s:coincidence}. In
section \ref{s:vetoes} we show how data from the interferometer that does not
measure the gravitational wave signal can be used.  Finally in section
\ref{s:s2pipeline} we describe the pipeline that has been constructed to
search for binary neutron stars and binary black hole MACHOs in the S2 data.

\section{Construction of Inspiral Pipelines}
\label{s:construction}

\begin{figure}[p]
\begin{center}
\includegraphics[height=0.6\textheight]{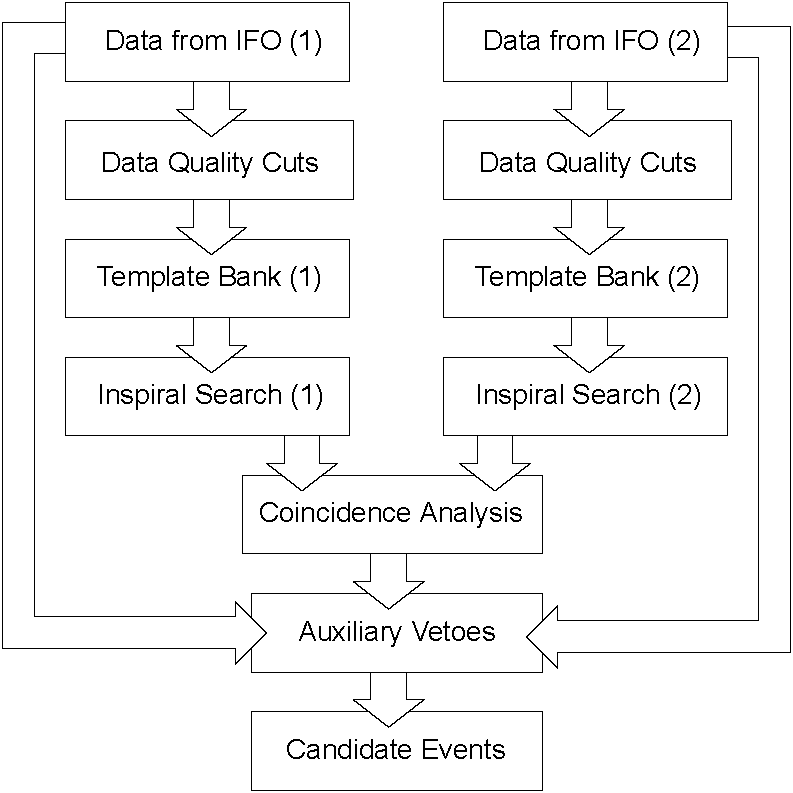}
\end{center}
\caption[Simple Pipeline to Search Data From Two Interferometers]{
\label{f:simple_pipe}
A simple pipeline used to search data from two interferometers for inspiral
signals. Raw data from interferometer, labeled $1$ and $2$, is recorded at the
observatories. Data quality cuts are then applied to the raw data to discard
times when the interferometer was not in a stable operating mode. Power
spectra generated from the data are used to generate a template bank for the
inspiral population being searched for. The template banks and interferometer
data are used to generate inspiral triggers for each interferometer. The
triggers for each interferometer are tested for coincidence, as a true
inspiral signal should be present in both interferometers at the same time, up
to the time it takes a gravitational wave to travel between the observatories.
Other coincidence tests, such as waveform parameter consistency, can be
applied at this stage. Transient noise sources may be detected in auxiliary
interferometer channels, for example seismometers. Such channels may be used to 
veto triggers that survive the coincidence analysis but are coincident with
the signature of noise in the auxiliary channel. Finally we obtain a sample of
candidate events for further investigation. Each step of the pipeline has many
parameters that can be tuned to minimize the false alarm and false dismissal 
rates.  
}
\end{figure}

A detection pipeline is a sequence of operations that starts with the raw data
from the interferometers and produces a list of \emph{candidate events}.
Figure \ref{f:simple_pipe} shows a simple pipeline to filter the data from a
pair of interferometers labeled IFO$1$ and IFO$2$.
We only use data that comes from interferometers in stable operation. Running
an interferometer is a complex process that requires human \emph{operators}
who are trained to \emph{lock} the interferometers. Locking the interferometer
is the process of bringing it from an uncontrolled state 
to a state where light is resonant in the interferometer.  From a
state in which the interferometer optics are freely swinging, the operators
manually align the optics of the interferometer using the interferometer
control systems. They then direct the automated lock acquisition
system\cite{Evans:thesis} to bring the Fabry-Perot cavities into resonance
with the beam splitter positioned so that the light at the anti-symmetric
port is a minimum. The recycling cavity is then brought into resonance and the
length sensing and control servo maintains the locked state by monitoring the
motion of the optics and adjusting their position accordingly.   In addition
to the operator, members of the LIGO Scientific Collaboration trained in the
operation of the interferometer are present at the observatory. The
collaboration member on duty is known as the as scientific monitor or
\emph{scimon}. Once the interferometer is locked, the operator and scimon
decide if the quality of the data being recorded is suitable to be flagged as
\emph{science mode data}. If the data is suitable it passes the first cut for
gravitational wave analysis and the operators and scimons continuously monitor
the data using various \emph{data monitoring tools}. Lock is lost
when the length sensing servo no longer has enough dynamic range to maintain
resonance in the interferometer. This is typically caused by large seismic
events which may be local to the observatory (e.g. liquid nitrogen tanks
creaking as the expand in the sun) or of global origin (e.g. a large
earthquake in China has caused loss of lock). Poor data quality in the
interferometer may also require a break in lock to remedy.  Continuous
locked operation has been maintained for up to $66.2$~hours in
the Hanford 4km interferometer. Seismic noise in the Livingston interferometer
limited the longest lock to $6.93$~hours during S2.

It is possible that the operators or scimons may make mistakes in deciding
that data should be flagged as science mode; they may forget to enable
calibration lines, for example. There may also be noise sources or transient
events in the data which make it unsuitable for analysis but which are not
easily detectable in the control room while the data is being taken. To
prevent such data from being used in an astrophysical analysis a list of
\emph{data quality cuts} is compiled. The manual selection of science mode
data may be considered the first data quality cut.  Additional tests of data
quality are described in section \ref{s:dq}.

The chirp signals from compact binary inspiral depend on the masses and spins
of the binary elements, as described in section \ref{s:inspiralgw}. Searching
for gravitational waves from BBHMACHOs requires signals which only depend on
the masses $m_1$ and $m_2$ of the binary.  The inspiral signals from BBHMACHOs
lie in some region of the template parameter space described by the component
masses $m_1$ and $m_2$. A single template is not sufficient to search for
signals from the population, as the template for a given pair of mass
parameters may not produce a high signal-to-noise ratio when used as a filter
to detect a signal with different mass parameters.  To search for signals from
a \emph{region} of parameter space we construct a \emph{bank} of inspiral
templates as described in section \ref{s:pipetemplate}. The template bank is
constructed to cover the parameter space in such a way that we do not discard
any signals from our target population.

We then use the bank of templates to filter the data for inspiral signals using
the matched filter and $\chi^2$ veto discussed in chapter \ref{ch:findchirp}.
This results in a list of \emph{inspiral triggers}. For the pipeline shown in
figure \ref{f:simple_pipe}, a bank of templates is generated used to filter
the data for inspiral signals for each interferometer. We describe how trigger
generation is used in a pipeline in section in detail in section
\ref{s:pipetemplate}.

One of the most powerful methods of rejecting false alarms is coincidence
between different interferometers. As described previously, there are three
LIGO interferometers which are operated simultaneously during science runs.
The H1 and H2 detectors are co-located at the LIGO Hanford Observatory and the
L1 detector is located at the LIGO Livingston Observatory. A true
gravitational wave should produce a signal in all operating detectors at the
same time, up to the time delay for the arrival of the wavefront of the
gravitational wave at the observatories.  We therefore require \emph{time
coincidence}, which demands that inspiral triggers be present in all operating
detectors simultaneously, with time offsets less than the light travel time
between detectors plus the measurement error of detection. While coincidence
is the most obvious use of multiple interferometers, other coincidence tests
may also be used, e.g. demanding consistency of the waveform parameters
between the triggers from two detectors, or consistency in the recovered
amplitude of the signal relative to the sensitivity of the detectors. We
describe these tests in section \ref{s:coincidence}.

While the goal during data acquisition is to ensure that the data recorded is
as stationary and Gaussian as possible, transient noise artifacts may still be
present in the data. For example, it has been seen that a person jumping up
and down in the control room at the observatory will cause a burst of noise in
the gravitational wave channel. There may also be occasional glitches in the
interferometer control systems that cause a transient in the gravitational
wave channel, despite the best efforts of the experimental team. To allow us
to distinguish such events from true gravitational wave signals, we record
several thousand data streams of auxiliary interferometer control channels and
physical environment monitor (PEM) channels. Auxiliary channels monitor the
state of the servo loops that control the interferometer and include
information about the pre-stabilized laser, the length sensing and control
system and the input and output optics.  PEM channels record the data from
devices such as seismometers, magnetometers and microphones placed in and
around the interferometer. These devices are designed to detect environmental
sources that may couple to signals in the gravitational wave channel. This
data can be used to construct \emph{vetoes} of the inspiral triggers if a
coupling can be identified between a noise source present in an auxiliary or
PEM channel and inspiral triggers in the gravitational wave channel.  We
demonstrate this process with examples in section \ref{s:vetoes}. 

The final step in constructing a pipeline is to turn the various elements
described above (data quality cuts, template bank generation, trigger
generation, coincidence and vetoing) into code that can be executed in an
automated way on a computing cluster. The execution of the code must ensure
that all the input data has been analyzed and the components of the pipeline
are executed in the correct sequence. We use a directed acyclic graph
(DAG) to describe the work flow of a pipeline.  For example, we may
construct construct a DAG to execute the simple pipeline in figure
\ref{f:simple_pipe} on all the data from L1 and H1 recorded in S2. The DAG
describing the work flow is submitted to a computing cluster via the Condor
high throughput computing system\cite{beowulfbook-condor}. The Condor DAG
manager executes the pipeline described in the DAG that we generate. This
process is described in more detail in section \ref{ss:dag}.

Implicit in the above discussion is that fact that there are many parameters
that must be set at each stage of the pipeline. For example: What data quality
cuts should we use? What signal-to-noise and $\chi^2$ thresholds should we use
when generating the inspiral triggers? What coincidence tests should we apply
and what should their parameters be? What auxiliary channels and PEM channels
should be used as vetoes, and how do we apply these vetoes to inspiral
triggers? Answering these questions is the key to turning a pipeline into a
full binary inspiral search; we call the process of selecting the parameters
\emph{tuning the pipeline}. In fact, pipeline tuning and construction of the
pipeline are not entirely separate. After constructing a pipeline and initial
tuning, we may decide to revisit the pipeline topology before performing
additional tuning.

When tuning the pipeline we may wish to minimize the \emph{false alarm rate},
i.e. minimize the number of candidate events that are not due to inspiral
signals.  We may simultaneously wish to minimize the false dismissal rate to
ensure that the pipeline does not discard triggers that are due to real
signals.  The false alarm rate can be studied by looking at candidate events
in the playground. The false dismissal rate can be studied by \emph{injecting}
signals into the data, that is generating a known inspiral signal and adding
it to the data before passing it through the pipeline.  Injection of signals
is described in section \ref{s:eff} and chapter \ref{ch:hardware}.

When tuning the pipeline for data that will be used to produce an upper limit,
we must ensure that we do not introduce statistical bias. A bias in the upper
limit could be introduced, for example, by selecting PEM channel events to
veto \emph{particular} inspiral triggers which were associated with the  PEM
events purely by chance. Clearly this could systematically eliminate true
inspiral events artificially in a way that would not be simulated in the
efficiency measurements described below. To do avoid such a possibility, we
select $10\%$ of all the data that we record as \emph{playground data}. The
data from GPS time $[t,t+600)$ is playground if 
\begin{equation}
t - 729273613 \equiv 0 \quad \mathrm{mod}(6370).
\end{equation}
Playground data is selected algorithmically to provide a representative sample
of the full data set. We are free to pursue whatever investigations we
wish on the playground data. Although we do not use this data in the upper limit
calculation, however we do not preclude the detection of a gravitational wave
signal in this data. We describe the process of tuning the S2 binary black
hole MACHO search in chapter \ref{ch:result}.

If we are using data from multiple interferometers we can measure the
\emph{background rate} of inspiral signals. We do this by introducing a time
shift into the data from different detectors before passing it through the
pipeline. If we assume that noise between the detectors is uncorrelated and
the time shift is sufficiently large, as described in section
\ref{s:background}, then any candidate events that survive the pipeline should
be due to noise alone and not astrophysical signals. By measuring the
background rate, we can measure the false alarm rate of the pipeline which can
be used for both tuning and the computation of the upper limit or detection
confidence.

\section{Data Quality Cuts}
\label{s:dq}

The theoretical matched filter is optimized for Gaussian data with a known,
noise spectrum that is stationary over the time scale of the data analyzed.
The filter therefore requires stable, well-characterized interferometer
performance. In practice, the interferometer performance is influenced by
optical alignment, servo control settings, and environmental conditions. The
list of science mode data provided by the operators can contain times when the
interferometer is not operating correctly. Unstable interferometer data can
produce false triggers that may survive both the $\chi^2$ test and
coincidence.  Data quality cut algorithms evaluate the interferometer data
over relatively long time intervals, using several different tests, or look
for specific behavior in the interferometer to exclude science mode data that
is unsuitable for analysis.  To decide if we should exclude science mode data
based on a particular data quality cut, we can examine the performance of the
inspiral code on data which is flagged as suspect by the cut. 

\subsection{Photodiode Saturation}
\label{ss:photodiode}

\begin{figure}[p]
\begin{center}
\begin{tabular}{cc}
\includegraphics[width=0.475\linewidth]{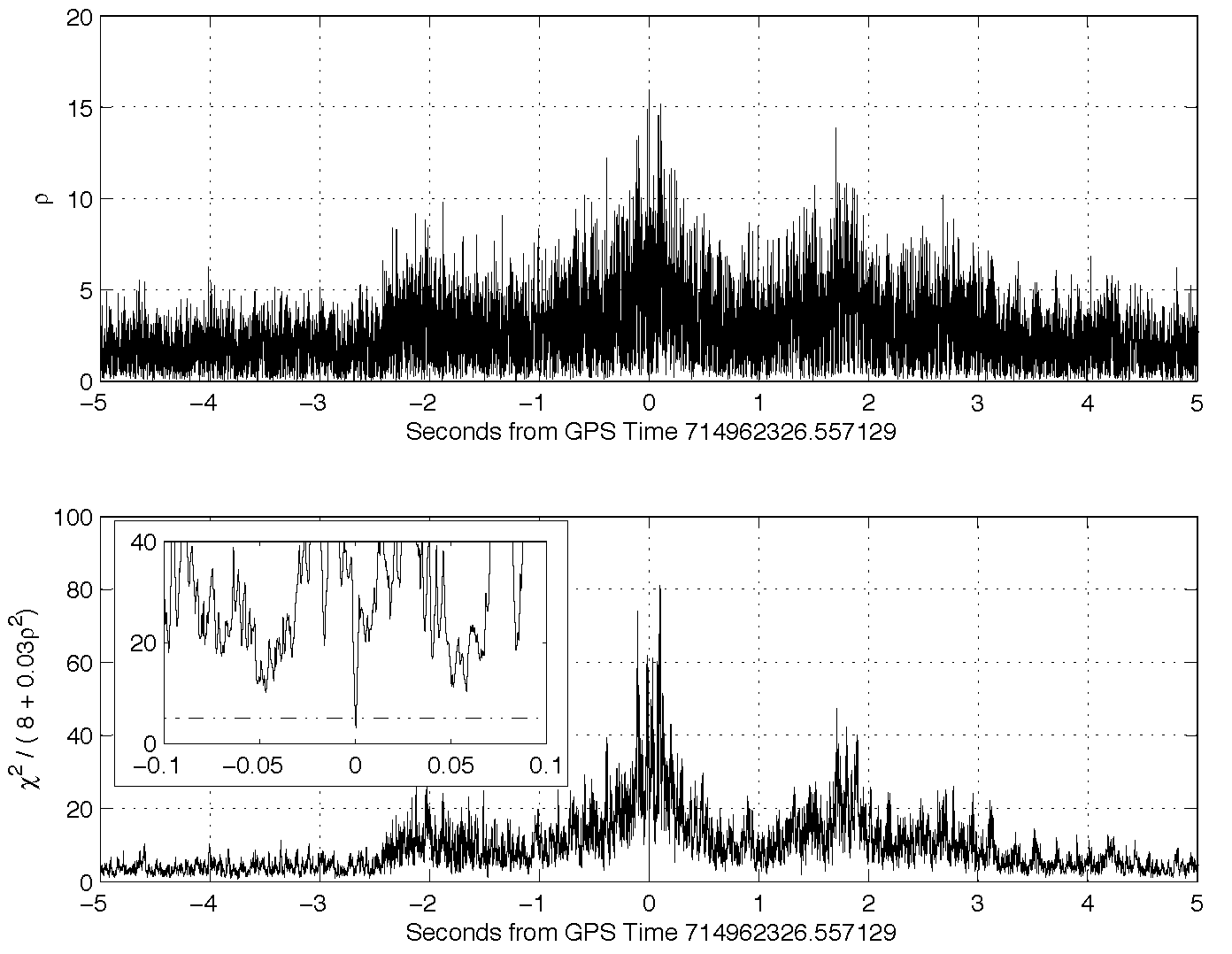} &
\includegraphics[width=0.475\linewidth]{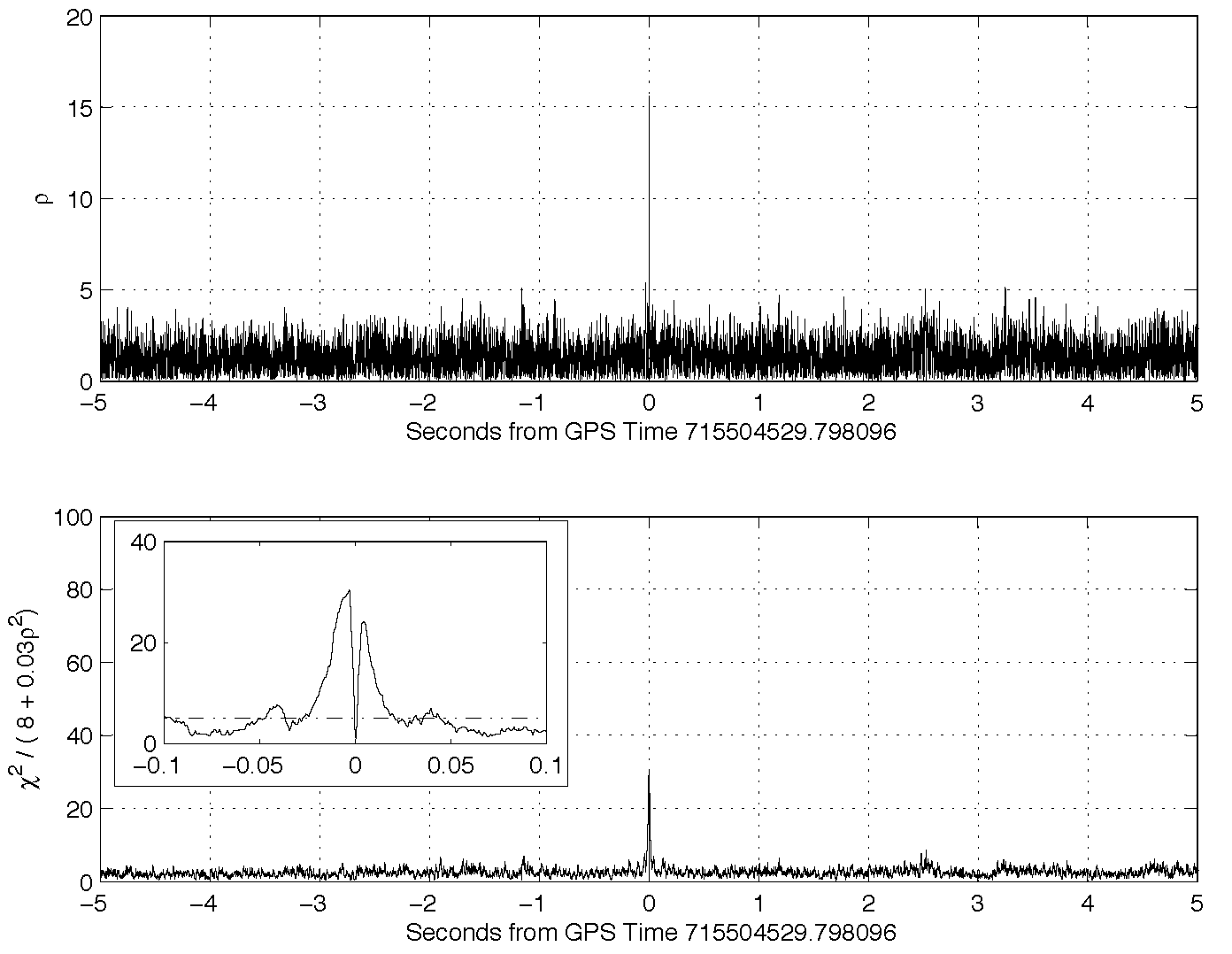}\\
\includegraphics[width=0.475\linewidth]{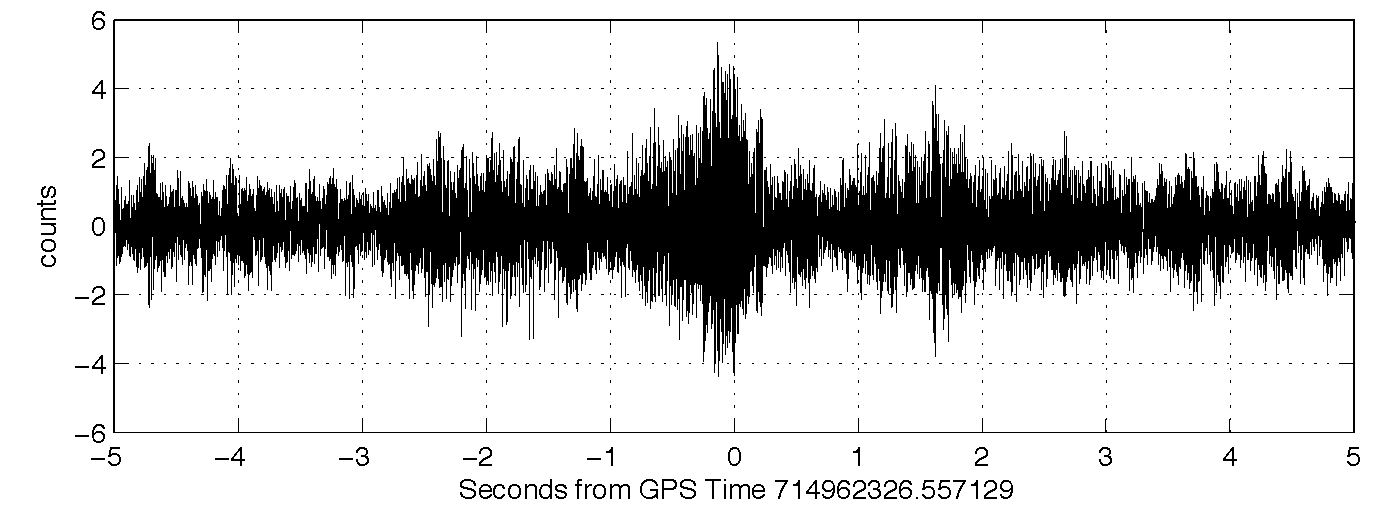} &
\includegraphics[width=0.475\linewidth]{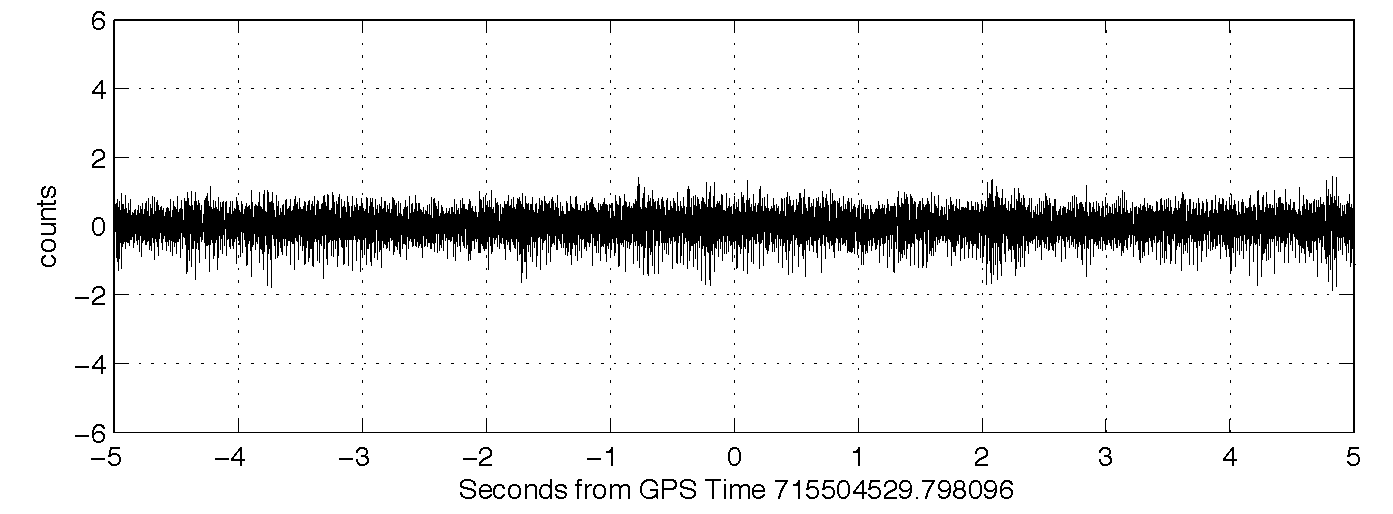}
\end{tabular}
\end{center}
\caption[Largest Inspiral Trigger Seen in the S1 Analysis]{
\label{f:s1loudest}
Left Panels: The largest signal-to-noise ratio candidate event seen during the
search of the LIGO S1 data. The top panel shows the signal-to-noise time
series, $\rho(t)$.  Notice that $\rho(t)$ is greater than the S1 threshold of
$6.5$ many times in a  $\sim 5$ second interval around the candidate event.
The center panel shows $\chi^2/ (p+ \delta^2 \rho^2)$ as a function of time
for the values of $\delta^2 = 0.03$ and $p = 8$ used in S1.  Notice $\chi^2 /
(p+ 0.03 \rho^2)$ is greater than the threshold of $5$ for $\sim 5$ seconds
around the candidate event,  but drops below this threshold right at the time
of maximum $\rho$.  The inset shows this more clearly for $\pm 0.1$ second
around the event where the threshold is indicated by a dot-dashed horizontal
line.  The bottom panel shows the time series for this candidate event after
applying a high-pass filter with a knee frequency of 200~Hz.  Notice the
bursting behavior which does not look like an inspiral chirp signal.  \break
Right Panels: A simulated injection into the L1 data.  This example was chosen
for comparison with the largest signal-to-noise ratio event shown in the left
panels since it similar in mass parameters, detected signal to noise and
$\chi^2$.   The instrument was behaving well at the time around the simulated
injection.  The top panel shows that $\rho(t) < 6.5$ except in close proximity
to the signal detection time.  The center panel shows $\chi^2/ (p+ 0.03
\rho^2)$ as a function of time.  Notice that it is much closer to threshold at
all times around the simulated injection; this contrasts dramatically with the
case of the candidate event shown in the left panels.  The inset shows this
more clearly for $\pm 0.1$ seconds around the injection.  The bottom panel
shows the time series for this simulated injection after applying a high-pass
filter with a knee frequency of 200~Hz.  The inspiral chirp signal is not
visible in the noisy detector output.
}
\end{figure}

Figure \ref{f:s1loudest} shows the signal-to-noise ratio and $\chi^2$ time
series of the loudest candidate event that was produced by the LIGO S1
inspiral search\cite{LIGOS1iul}. Also shown is the filter output for a simulated inspiral
with similar parameters that was injected into well behaved
interferometer data.  Notice that the time series of $\rho(t)$, $\chi^2(t)$
and the raw data are very noisy around the time of the S1 loudest candidate.
In contrast, the time series for the simulated signal is very clean. On
further investigation, it was determined that the photodiode that records the
light at the anti-symmetric port had saturated at the time of the S1 loudest
event. The system that converts the light into an electronic signal for the
length sensing and control servo had exceeded its dynamic range causing a
noise transient in the data.  The saturation is a symptom of poor
interferometer performance. Photodiode saturations are caused by large bursts
of noise in the gravitational wave channel which corrupt power spectral
estimation and matched filtering. A test was developed to monitor the
gravitational wave channel for photodiode saturations and this test has become
a data quality cut for current and future searches.

\subsection{Calibration Lines}
\label{ss:calcut}

A second example of a data quality cut is based on the presence of calibration
lines, which were described in section \ref{ss:calibration}. The calibration
lines track the response of the instrument to mirror movement or a
gravitational wave, which varies over time. Without the calibration lines, it
is not possible to construct an accurate response function. Since the inspiral
search needs correctly calibrated data, a simple data quality cut checks for
the presence of the calibration lines in the data. If they are absent, the
data is discarded.

\subsection{Data Quality Cuts Available in S2}
\label{ss:s2dq}

The full list of available data quality cuts and their meanings for S2 are
show in Table \ref{t:dqflags}. The table is divided into two sections,
mandatory and discretionary data quality cuts. Mandatory cuts represent
unrecoverable problems in data acquisition or calibration and so we must exclude
these times from the list of science mode segments. Discretionary data cuts
are optional for a particular search. For the inspiral search we decide
whether or not to use a cut based on the performance of the trigger generation
code in playground data when a particular data quality cut is active, as
described in section \ref{ss:s2dqselection}.  The times remaining after we
apply data quality cuts are called \emph{science segments}.

\section{Inspiral Trigger Generation}
\label{s:pipetemplate}

Chapter \ref{ch:findchirp} describes the algorithms that we use to determine
if an inspiral from a binary of masses ${m_1,m_2}$ is present in a single data
segment. The input to trigger generation is:
\begin{enumerate}
\item The template, $\tilde{h}_{ck}$, drawn from a \emph{template bank}.

\item The \emph{data segment} to be filtered, $\{v_j\}$ $j \in [0,N]$, where
$v_j$ is the raw (uncalibrated) interferometer output channel,
\texttt{LSC-AS\_Q}. A data segment is the unit of analysis for the filter and
is a subset of a science segment.

\item An average power spectral density, $S_v(|f_k|)$, of the channel 
\texttt{LSC-AS\_Q},

\item The instrumental response function, $R(f_k)$, which is required to
calibrate the data and the power spectrum. 
\end{enumerate}
In this section, we describe how these quantities are constructed and used to
generate inspiral triggers.

\subsection{Template Banks}
\label{ss:templatebank}

The matched filtering described in chapter \ref{ch:findchirp} has been used to
detect the inspiral waveforms from binary neutron stars in the mass range
$1\,M_\odot\le m_1, m_2\le 3\,M_\odot$ and binary black hole MACHOs in the
mass range $0.2\,M_\odot\le m_2, m_1\le 1\,M_\odot$. Since each mass pair
$\{m_1,m_2\}$ in the space produces a different waveform, we construct a {\em
template bank}, a discrete subset of the continuous family of waveforms that
belong to the parameter space. The placement of templates in the bank is
determined by the \emph{mismatch} of the bank, $\mathbb{M}$. The mismatch is
the fractional loss in signal to noise that results when an inspiral signal,
$s$, is not exactly correlated with a template in the bank, $h$. In terms of
the inner product defined in equation \ref{eq:innerproduct} of section
\ref{s:detectiontheory} it is
\begin{equation}
\mathbb{M} = 1 - \frac{(h|s)} {\sqrt{(h|h)(s|s)}}.
\end{equation}
If we consider the distribution of binaries to be uniform in space, then the
fraction of events lost due to the mismatch of the template from a population
is approximately $\mathbb{M}^3$, i.e the range is decreased by a factor of
$\mathbb{M}$. A mismatch of $3\%$, i.e. a $10\%$ loss of event rate, is
conventionally accepted as a reasonable goal for binary neutron stars. For
binary black hole MACHOs, we reduce the minimal match of the template bank to
$5\%$. This is due to the fact that the number of templates in the bank for a
given interferometer noise curve scales as approximately
$m_\mathrm{min}^{-8/3}$, where $m_\mathrm{min} = m_1 = m_2$ is the mass of the
lowest mass equal mass template in the bank\cite{Owen:1998dk}. If we lowering
the minimal match then the computational cost decreases as $\mathbb{M}^{-1}$.
The loss in signal-to-noise ratio is balanced by the fact that we are
searching for a population of binary black hole MACHOs in the galactic halo.
This population is far from homogeneous with almost all signals expected to
produce a signal-to-noise ratio far above threshold, so this $5\%$ loss in
signal-to-noise ration will hardly constitute any loss in observed event
rate.

The scaling in the number of templates as a function of the lower mass of the
bank parameter space is due to the fact that the number of templates required
to achieve a given minimal match is a function of the number of cycles that the
inspiral waveforms spends in the sensitive band of the interferometer. The
more cycles the matched filter correlates against, the greater its
discriminating power and so the loss in signal-to-noise ratio for a mismatched
template increases. A pair of inspiralling $1.4\,M_\odot$ neutron stars have
$347$ cycles in the S2 sensitive band (between $100$~Hz and $2048$~Hz),
compared to $1960$ cycles for a pair of $0.5\,M_\odot$ binary black hole
MACHOs; a pair of $0.1\,M_\odot$ binary black hole MACHOs have nearly
$28\,500$ cycles.

\begin{figure}[p]
\begin{tabular}{c}
\includegraphics[width=\linewidth]{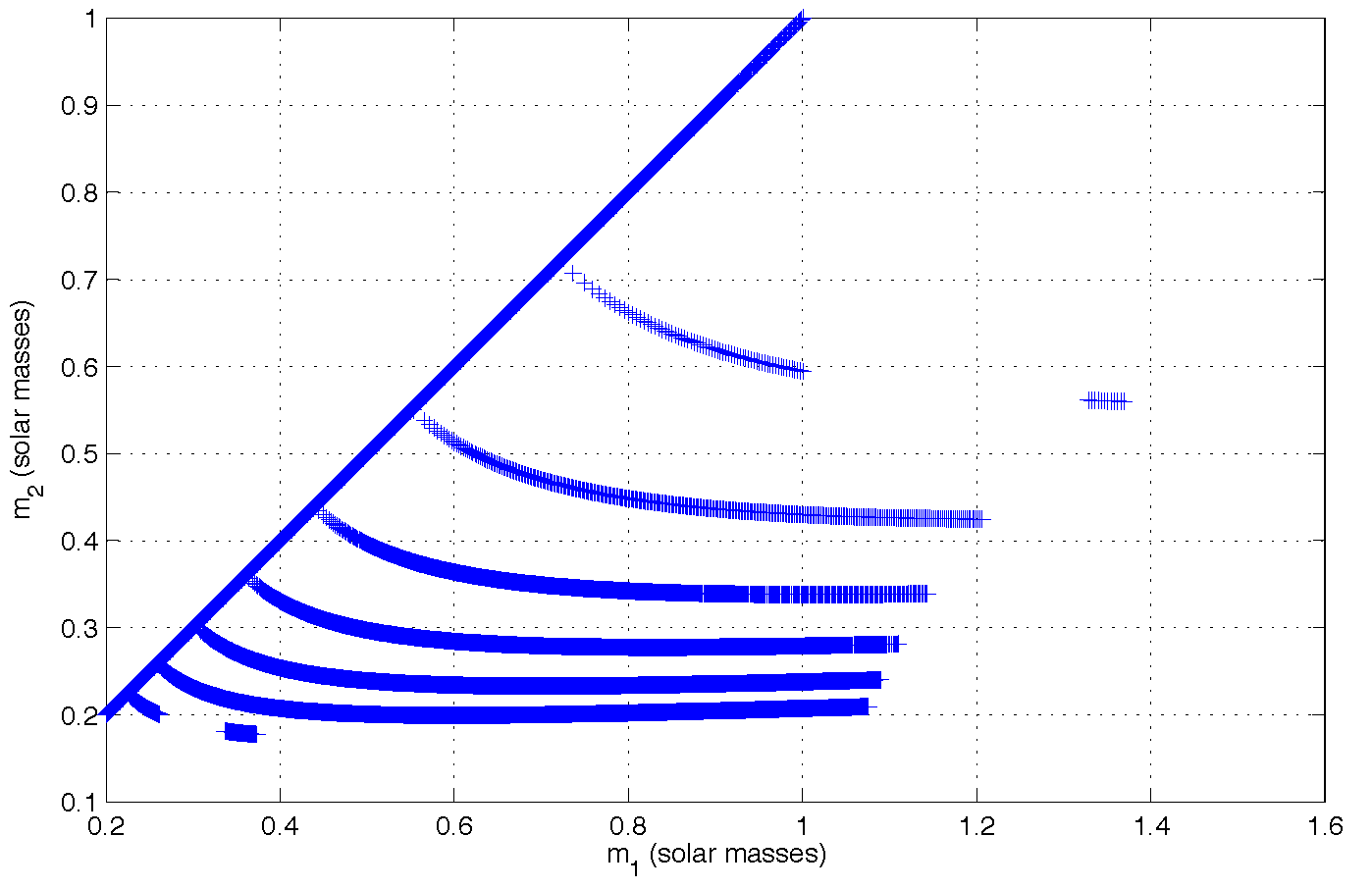}
\end{tabular}
\caption[Binary Black Hole MACHO Template Bank]{
\label{f:s2_banks}
The template bank required to cover the binary black hole MACHO parameter
space from $0.2$ to $1$ $M_\odot$ at a minimal match of $95\%$.  The template
bank is generated from the average power spectral density of a typical S2
analysis chunk (starting at GPS time 734256712.) The large number of
templates in the BBHMACHO bank is due to the lager number of cycles of the
MACHO templates in the sensitive band of the interferometer. The placement of
templates outside the mass parameter space is required to ensure that any
signal that lies in the space has a minimal match $> 0.95$.
}
\end{figure}

Since the number of templates is a function of the number of cycles of an
inspiral in the sensitive band of the interferometer, it also depends on the
shape of the power spectral density of the noise curve. In fact location of
the templates in the bank is also a function of the PSD, as described in
\cite{Owen:1998dk}. The power spectrum of the instrument changes over time and
so we must also change the template bank. We accomplish this by using the
power spectral density for an analysis chunk to generate a template bank that
is unique to that chunk. The PSD is calibrated and the template bank
generated. Figure \ref{f:s2_banks} shows binary neutron star and binary
black hole MACHO template banks generated for a typical stretch of S2 data.
The smallest and largest number of templates in a bank during S2 was $589$ and
$857$ binary neutron star templates and $11\,588$ and $17\,335$ binary black
hole MACHO templates.

\subsection{Data Management}
\label{ss:datamanagement}

Corruption due to the wrap-around of the matched filter means that not all the
time in a data segment can be searched for triggers. As described  in chapter
\ref{ch:findchirp}, we simplify the process of selecting uncorrupted data by
only searching for triggers in the signal-to-noise ratio, $\rho^2(t_j)$, when 
\begin{equation}
\frac{N}{4} \le j < \frac{3N}{4},
\end{equation}
where $N$ is the number of sample points in the data segment, and by demanding
that the amount of data corrupted is less than $N/4$ sample points. To ensure
that all data is analyzed we must overlap each data segment by $(N/2)\Delta 
t$~seconds. To compute an average power spectrum we require a sample of data
near the data segment being filtered so that a good estimate of the noise can
be obtained. We combine these two requirements by bundling several overlapping
data segments together in an \emph{analysis chunk}. The length of an analysis
chunk is bounded above by the amount of memory available in the computers
performing the filtering and bounded below by requiring a sufficiently large
number of segments in the computation of the average power spectrum. In the S2
pipeline, we construct analysis chunks of length $2048$~seconds from $15$
overlapped data segments of length $256$~seconds. The data segments are
overlapped by $128$~seconds, with the first and last $64$ seconds of data
segment ignored when searching for triggers. The analysis chunks themselves
are overlapped by $128$~seconds so that only the first and last $64$~seconds
of a science segment are not searched for inspiral triggers.

\begin{figure}[p]
\begin{center}
\includegraphics[width=\linewidth]{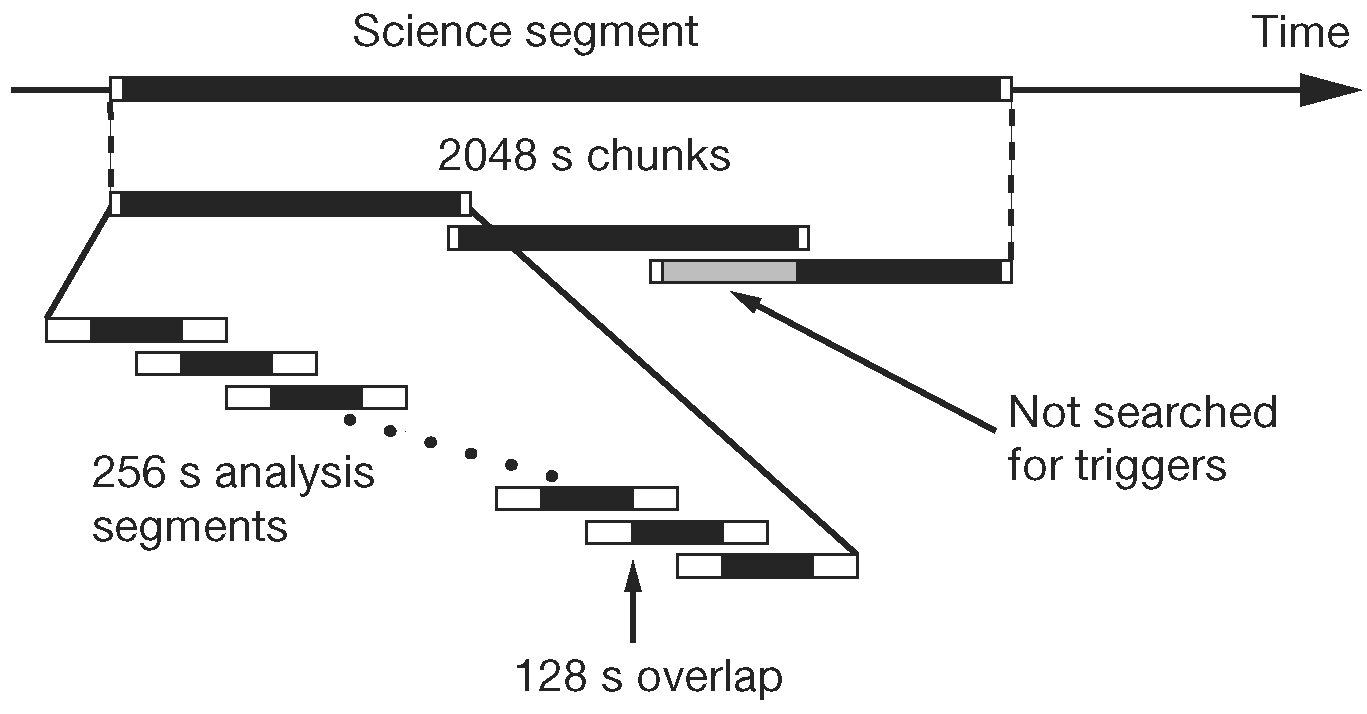}
\end{center}
\caption[Algorithm Used to Divide Science Segments into Data Analysis
Segments]{%
\label{f:s2_segments}
The algorithm used to divide science segments into data analysis segments.
Science segments are divided into $2048$~s chunks overlapped by $128$~s.
(Science segments shorter than $2048$~s are ignored.) An additional chunk with
a larger overlap is added to cover any remaining data at the end of a science
segment.  Each chunk is divided into $15$ analysis segments of length $256$~s
for filtering. The first and last $64$~s of each analysis segment is ignored,
so the segments overlap by $128$~s.  Areas shaded black are filtered for
triggers by the search pipeline. The gray area in the last chunk of the
science segment is not searched for triggers as this time is covered by the
preceding chunk, however this data is used in the PSD estimate for the final
chunk.
}
\end{figure}

Figure \ref{f:s2_segments} shows how analysis chunks are constructed from the
science segments in S2. The first analysis chunk is aligned with the start of
the science segment. Subsequent chunks overlap the previous one by
$128$~seconds. At the end of a science segment there is generally not enough
data to fit an entire analysis chunk but we cannot make the chunk shorter, as
we need all $15$ data segments to compute the average power spectrum. To solve
this problem, we align the end of the last chunk with the end of the science
segment and ignore any inspiral triggers generated for times that overlap the
previous chunk. 

\begin{figure}[p]
\begin{center}
\includegraphics[width=\linewidth]{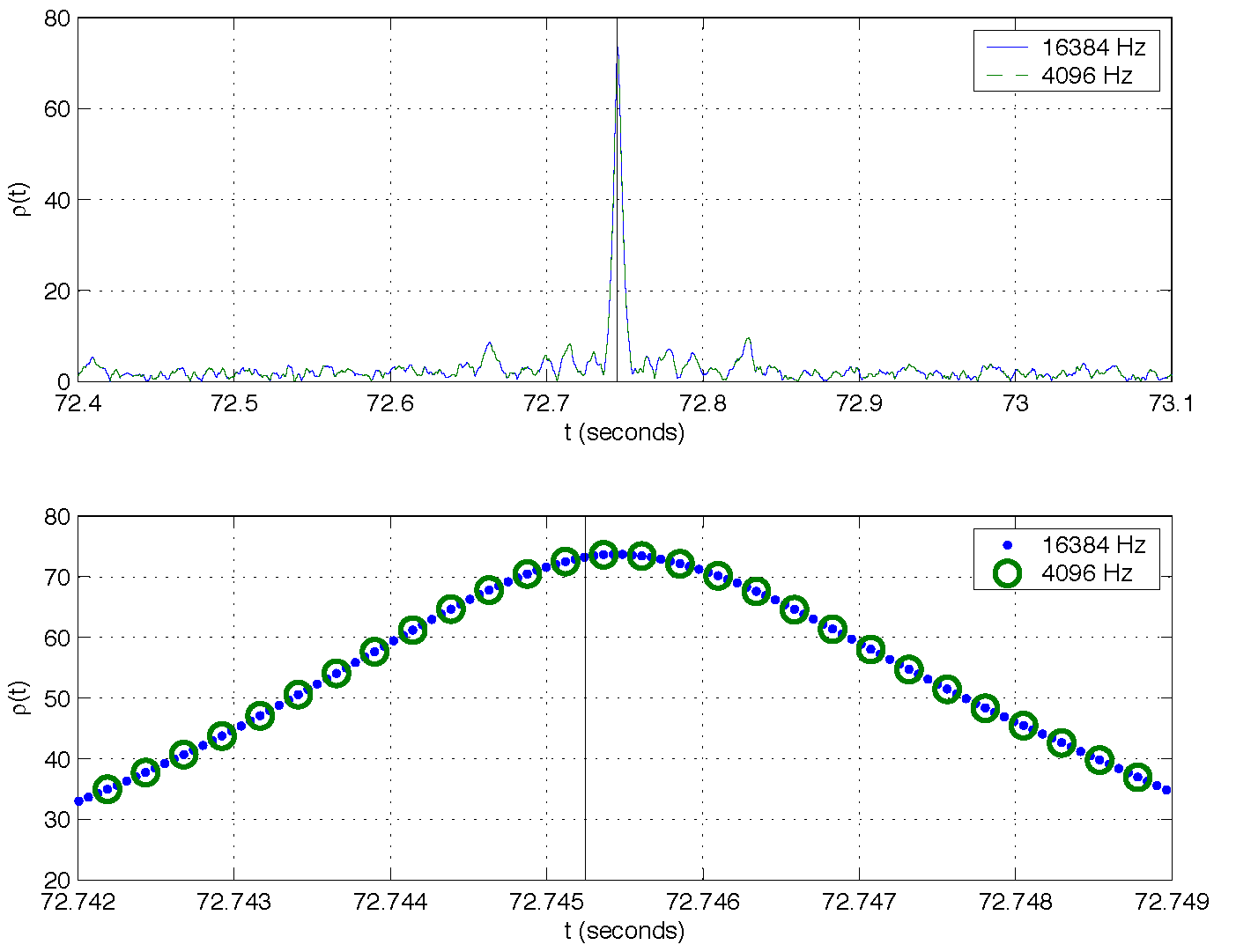}
\end{center}
\caption[Loss in Signal-to-noise Ratio Due to Resampling]{%
\label{f:snr_resample_loss}
The loss in signal-to-noise ratio for a $0.2,0.2\, M_\odot$ black hole
MACHO binary due to resampling. The inspiral waveform is generated at
$16\,384$~Hz and injected into data with a typical S2 noise curve. The end
time of the waveform is at $72.74525$~seconds, shown by the vertical line in
both plots. The top panned shows the signal-to-noise ratio for the data
segment when using data at the full bandwidth and data resampled to 
$4096$~Hz. The bottom panel shows the same data close to the end time of the
injection. The loss in signal-to-noise ratio for the inspiral trigger 
generated at $4096$~Hz is $0.1\%$.
}
\end{figure}

The interferometer records data at $16\,384$~Hz, and we down sample the
analysis chunk after reading it from disk to decrease the computational
resources required by the filtering code. We choose the new sample rate so
that the loss in signal-to-noise ratio due to the discrete time steps, 
$\Delta t$, is less than that due to the discrete choices of the template mass
parameters. It can be shown that for the initial LIGO noise curve, this is
true if the sample rate of the filtered data is greater than $\sim
2600$~Hz\cite{Owen:1998dk}. For simplicity, we chose sample rates that are
powers of two and so we resample the data to $4096$~Hz.  We do this by
applying a finite impulse response (FIR) low pass filter to remove power above
the Nyquist frequency of the desired sample rate, $2048$~Hz. The low passed
data is then decimated to the desired sample rate. Although the maximum
frequency of most of the BBHMACHO inspiral signals is greater than $2048$~Hz,
the loss of signal-to-noise ratio above this frequency is negligible as most
of the the signal-to-noise is accumulated at frequencies lower than $2048$~Hz,
as described in chapter \ref{ch:inspiral}. Figure \ref{f:snr_resample_loss} shows
the loss in signal-to-noise ratio of a BBHMACHO inspiral due to resampling.
The inspiral waveform of a pair of $0.2\,M_\odot$ black holes an effective
distance of $25$~kpc is generated at a sample rate of $16\,384$~Hz. The
maximum frequency of this waveform is $10\,112$~Hz. The waveform is injected
into raw (un-resampled) data with a typical S2 noise curve which is then
filtered at $16\,384$~Hz and $4096$~Hz. The maximum of the signal-to-noise
ratio is $\rho = 73.67$ for the raw data and $\rho = 73.58$ for the resampled
data, giving a difference in signal-to-noise ratio of $0.1\%$. This loss
combines the effects due to the discreteness of the resampled data and the
signal present above the Nyquist and is much less than the $5\%$ loss in
signal-to-noise ratio caused by the discrete nature of the template bank.

The interferometer data contains a large amount of power of seismic origin at
low frequencies.  This power is several orders of magnitude higher than the
noise in the sensitive frequency band of the interferometer. This power may
bleed across the frequency band when the data is Fourier transformed into the
frequency domain and dominate the true noise in the sensitive band of the
interferometer, In order to prevent this, we apply a Butterworth infinite
impulse response (IIR) high pass filter to the resampled data. The high pass
frequency and filter order are chosen so that power in the sensitive band of
the interferometer is not attenuated. The data is filtered forwards and
backwards through the filter to remove the dispersion of the filter.  The data
used to compute the power spectral estimate is windowed using a Hann window to
prevent power from line features in the spectrum (e.g. power line harmonics or
suspension wire resonances) from bleeding into adjacent frequency bands.
We also apply a low frequency cutoff in the frequency domain at a slightly
higher frequency than the time domain filter. The matched filter correlation
is not computed below this cutoff, so frequencies below it do not contribute
to the signal-to-noise ratio. 

A windowed copy of each of the $15$ data segment is used to construct the
average power spectral density used in the matched filter, as described in
section \ref{ss:psd}. Note that the data used in the matched filter is not
windowed; the windowed data is discarded once it has been used to compute the
power spectra.  The mean values over the analysis chunk of the calibration
parameters $\alpha$ and $\beta$ are used to construct the response $R(f_k)$,
as described in equation (\ref{eq:calibration}) of section \ref{ss:calibration}.
The same response function is used to calibrate the power spectral density and
all data segments in the chunk. 

A disadvantage to the above method of processing the input data is that we
require a science segment to be at least $2048$~seconds long. Any shorter
science segments are ignored, as there is not enough data to generate a power
spectral density. We note here that this lead to a significant amount of data
being discarded in the S2 analysis. As we will describe below, the S2 pipeline
requires the L1 interferometer to be operating in order to analyze the data.
L1 is the least stable of the interferometers as it is very sensitive to
seismic noise. High seismic noise can saturate the length sensing and control
servo and cause the loss of lock, which terminates a science segment.
Modification of the data management (e.g. construction of analysis chunks and
power spectral estimation) to allow us to use science segments shorter than
$2048$~seconds requires significant changes of the implementation of the
inspiral search code. It was not possible to implement and test these changes
within the time allowed to perform the S2 analysis. Fortunately, it is
expected that after the installation of addition seismic isolation at the
Livingston observatory, scheduled for completion in late 2004, the lengths of
science segments will be significantly increased and the number of short
segments discarded will decrease. Unfortunately data from the third science
run, S3, which was completed before the seismic upgrade, exhibits the problem
of short science segments, so a redesign of the filtering code may still be
required.

\subsection{Trigger Generation Parameters}
\label{ss:triggerparameters}

The process of template bank construction and inspiral trigger generation
relies on several parameters that can be tuned to minimize the false dismissal
or false alarm rate. We have touched on some of these already; in this section
we enumerate all of the tunable parameters for bank and inspiral trigger
generation.

Both template bank and inspiral trigger generation require a calibrated power
spectral density. In the inspiral trigger generation described above,
construction of the PSD is coupled to the length of the data segments and
analysis chunks. The following \emph{data conditioning} parameters are used to
construct the data segments, and so determine the characteristics of the PSD:
\begin{itemize}
\item Number of sample points in a data segment, $N$. This determines the
length of the segments correlated in the matched filter and the length of the
segments used in the PSD estimate. The number of points that subsequent data
segments are overlapped by is then $N_\mathrm{overlap} = N/2$.

\item Number of data segments in a chunk $N_\mathrm{segments}$. This sets the
number of data segments used in the PSD estimate and so the number of segments
in an analysis chunk.

\item Sample Rate, $1/\Delta t$. The sample rates used in LIGO data analysis
are integer powers of two Hertz.

\item Number of non-zero points in the square root of inverse power
spectrum  in the time domain, $N_\mathrm{invspectrunc}$. This
parameter was described in detail in section \ref{ss:invspec}.
\end{itemize}
We set $\Delta t\, N_\mathrm{invspectrunc} = 16$~seconds for the S2 analysis
based on the length of wraparound corruption allowed. A systematic study of
the value of $N_\mathrm{invspectrunc}$ has not been carried out for the S1
or S2 data, but is planned for future analysis.

Once the above parameters have be specified, it follows that the number of
sample points in an analysis chunk is
\begin{equation}
N_\mathrm{chunk} = 
\left(N_\mathrm{segments} - 1 \right) N_\mathrm{stride} + N
\end{equation}
where 
\begin{equation}
N_\mathrm{stride} = N - N_\mathrm{overlap}.
\end{equation}
The length of the analysis chunk, in seconds, is therefore
\begin{equation}
T_\mathrm{chunk} = \Delta t\, N_\mathrm{chunk} .
\end{equation}
The choice of these parameters is governed by the class of waveforms searched
for and the low frequency sensitivity of the interferometer. The longest
inspiral waveform in the template bank (which will be the smallest mass
template), is determined by the lowest sensitive frequency of the
interferometer. The sum of the length of the longest template and the length
of the inverse power spectrum must be shorter than the duration of the
signal-to-noise output, $\rho(t)$, that we ignore due to corruption. We
therefore require
\begin{equation}
\frac{1}{4} N \ge 2 N_\mathrm{invspectrunc} + N_\mathrm{longest}
\end{equation}
where $N_\mathrm{longest}$ is the number of points in the longest chirp.

Other data conditioning parameters control the cut-offs of the time
and frequency domain filters applied to the data. These are:
\begin{itemize}
\item The high pass filter cutoff, $f_\mathrm{hp}$, and high pass filter
order $O_\mathrm{hp}$. These parameters determine the shape of the IIR
Butterworth high pass filter applied to the analysis chunks before PSD
estimation and inspiral trigger generation.

\item The frequency domain low frequency cut off, $f_\mathrm{low}$. This
parameter allows us to exclude frequencies below a certain value from the
correlations in the matched filter and $\chi^2$ veto. A non-zero value of
$f_\mathrm{low}$ sets the value of the data in all frequency bins $k <
k_\mathrm{min} = f_\mathrm{low} / (N \Delta t)$ to zero which excludes this
data from the correlation. Note that $f_\mathrm{low} \ge f_\mathrm{hp}$ to
prevent data used in the correlation being attenuated by the high pass
filter.
\end{itemize}
During investigation of inspiral triggers in the S2 playground data, it was
discovered that many of the L1 inspiral triggers appeared to be the result of
non-stationary noise with frequency content around $70$~Hz.  An important
auxiliary channel, \texttt{L1:LSC-POB\_I}, proportional to the residual length
of the power recycling cavity, was found to have highly variable noise at
$70$~Hz.  There are understandable physical reasons for this, namely the power
recycling servo loop (for which \texttt{L1:LSC-POB\_I} is the error signal)
has a known instability around $70$~Hz, which often results in the appearance
of glitches in the detector output channel at around $70$~Hz.  As a
consequence, it was decided that to reduce sensitivity to these glitches the
high pass cut off should be set to $f_\mathrm{hp} = 100$~Hz with order
$O_\mathrm{hp} = 8$ and the low-frequency cutoff set to $f_\mathrm{low} =
100$~Hz.  This subsequently reduced the number of inspiral triggers
(presumably created by this problem); an inspection of artificial signals
injected into the interferometer revealed a very small loss of efficiency for
binary neutron star inspiral and BBHMACHO signal detection resulting from the
increase in the low frequency cutoff.

After we have defined the data conditioning parameters and a parameter space
for the search, the only remaining free parameter for template bank generation
is:
\begin{itemize}
\item The template bank mismatch, $\mathbb{M} \in [0,1)$. Given a value of
mismatch, $\mathbb{M}$, for a real signal that lies in the bank parameter
space the fractional loss in signal-to-noise ration should be no larger than
$\mathbb{M}$ when filtering with the signal against its exact waveform
compared to filtering the signal against the template in the bank with yields
the highest signal-to-noise ratio.
\end{itemize}
As described above, the value chosen for the S2 search is
$\mathbb{M}_\mathrm{BBHMACHO} = 5\%$.
Note that we do not truncate the power spectrum used for generating the
template bank, as there is no issue with wrap-around in the bank generation
algorithm.

For a given template we use matched filtering to construct the signal-to-noise
ratio, $\rho$, and search for times when this exceeds a threshold, $\rho >
\rho^\ast$. If the threshold $\rho_\ast$ is exceeded, we construct the
template based veto, $\chi^2$, with $p$ bins. Small values of $\chi^2$
indicate that the signal-to-noise was accumulated in a manner consistent with
an inspiral signal. If the value of the $\chi^2$ veto is below a threshold,
$\chi^2 < \chi^2_\ast (p + \delta^2 \rho^2)$, then an inspiral trigger is
recorded at the maximum value of $\{\rho| \chi^2 < \chi^2_\ast (p + \delta^2
\rho^2)\}$.  The parameters used available in trigger generation are:
\begin{itemize}
\item The signal to noise threshold, $\rho_\ast$.

\item The $\chi^2$ threshold, $\chi^2$.

\item The number of bins used in the $\chi^2$ veto, $p$.

\item The parameter $\delta^2$ used to account for this mismatch of a signal
and template in the $\chi^2$ veto, as described in section
\ref{ss:mismatchedchisq}.
\end{itemize}
Tuning of these trigger generation parameters is particular to the class of
search used. A detailed discussion of this tuning is for binary
black hole MACHOs is given in chapter \ref{ch:result}.

\section{Trigger Coincidence}
\label{s:coincidence}

Coincidence is a powerful tool for reducing the number of false event
candidates surviving a pipeline. The simplest test of coincidence is time
coincidence of triggers between two or more interferometers.  For a trigger to
be considered coincident in two interferometers, we demand that it is observed
in both interferometers within a temporal coincidence window $\delta t$. The
coincidence window must allow for the error in measurement of the time of the
trigger. It must also allow for the difference in gravitational time of
arrival if the interferometers are not located at the same observatory.  The
time difference between gravitational wave arrival time varies from
$0$~seconds, if the gravitational wave is propagating perpendicular to a line
joining the detectors, to $10$~ms if the gravitational wave is propagating
parallel to a line joining the detectors. The maximum time difference comes
from the time it takes a gravitational wave to propagate between the
observatories, given by $t = \frac{d}{c}$, where $d = 3002$~km is the
distance between the observatories and $c$ is the speed of light
(the propagation speed of gravitational waves).  Monte Carlo analysis with
simulated signals suggests that we cannot measure the time of the trigger to
an accuracy of less than $1$~ms. The time coincidence window is therefore
$\delta t = 1$~ms if the interferometers are located at the same observatory.
For coincidences between LHO and LLO triggers, we set $\delta t =
3000\,\mathrm{km} / c + 1\,\mathrm{ms} = 11\, \mathrm{ms}$. Any triggers
that fail this test are discarded. 

If a signal found in temporal coincidence is generated by real inspiral, it
should have the same waveform in both interferometers, up to issues of the
different detector antenna patterns yielding different combinations of $h_{+}$
and $h_{\times}$. This suggests that we could apply a waveform parameter test
to triggers candidate that survive the time coincidence test. We cannot
exactly extract the parameters of the waveform, however, since we filter the
interferometer data with a template bank which may not contain the true
waveform. In addition to this, the template banks will, in general, differ
between detectors and detector noise may cause error in the measurement of the
signal parameters, even if the template banks are the same. To account for
these sources of error, we can apply waveform parameter coincidence by
requiring that the two mass $m_1$ and $m_2$, of the templates are identical to
within an error of $\delta m$.

We now consider an amplitude cut on the signals. The Livingston and Hanford
detectors are not co-aligned. There is a slight misalignment of the detectors
due to the curvature of the earth and so the antenna patterns of the detectors
differ. This causes the measured amplitude and phase of a gravitational wave
to differ between the sites. In the extreme case, it is possible, for example,
for a binary to be completely undetectable by the L1 detector, but still
detectable by the H1 and H2 detectors. For a given inspiral trigger, we
measure the effective distance of the binary system. This is the distance at
which an optimally oriented binary would produce the observed signal-to-noise
ratio in a particular instrument---it is not the true distance of the binary.
Since the detectors have different antenna patterns they will report different
effective distances for the same gravitational wave.
Figure~\ref{f:gmst_dist_ratio} shows the ratio of effective distances between
the two LIGO observatories for the population of binary neutron stars
considered in the S2 analysis. The significant variation of the ratio of the
effective distances precludes using a naive test for amplitude coincidence. It
is possible to obtain information about sky position from time delay between
sites to construct a more complicated amplitude cut, but this has not be used
in the S2 analysis.

In the case of triggers from the H1 and H2 interferometers that are coincident
in time and mass, we can apply an amplitude cut that tests that the effective
distances of the triggers are coincident.  In this test we must allow for the
relative sensitivity of the detectors while allowing for error in the distance
measurement, as determined by Monte Carlo simulations. The amplitude cut for
triggers from H1 and H2 is given by 
\begin{equation}
\label{eq:eff_dist_test}
\frac{\left|\mathcal{D}_\mathrm{1} - \mathcal{D}_\mathrm{2}\right|}{D_\mathrm{1}}
< \frac{\epsilon}{\rho_\mathrm{2}} + \kappa,
\end{equation}
where $\mathcal{D}_1$ ($\mathcal{D}_2$) is the effective distance of the trigger in the first
(second) detector and $\rho_{2}$ is the signal-to-noise ratio of the trigger
in the second detector. $\epsilon$ and $\kappa$ are tunable parameters.
In order to disable the amplitude cut when comparing triggers from LLO and
LHO, we set $\kappa = 1000$.  When testing for triple coincident triggers we
accept triggers that are coincident in the L1 and H1 detectors that are
\emph{not} present in the H2 detector \emph{if} the effective distance of the
trigger is further than the maximum distance of H2 at the signal-to-noise
ratio threshold at the time of the candidate trigger.  Figure
\ref{f:coinc_test} summarizes the algorithm for the time, mass and distance
coincidence tests used in S2. 

We therefore have the following coincidence parameters that must be tuned for
the pipeline:
\begin{itemize}
\item The time coincidence window, $\delta t$, which is set to $1$~ms for
LHO-LHO coincidence and $11$~ms for LHO-LLO coincidence.

\item The mass coincidence window, $\delta m$.

\item The error on the measured effective distance due to the instrumental
noise, $\epsilon$, in the amplitude test.

\item The systematic error in measured effective distance, $\kappa$, in the
amplitude test.
\end{itemize}

If coincident triggers are found in H1 and H2, we can get an improved estimate
of the amplitude of the signal arriving at the Hanford site by coherently
combining the filter outputs from the two gravitational wave channels, 
\begin{equation}
\rho_H = \sqrt{ \frac{|z_{H1} + z_{H2}|^2}{\sigma_{H1}^2 +
     \sigma_{H2}^2} },
\label{eq:rhoH}
\end{equation}
where $z$ is the matched filter output given by equation (\ref{eq:zdef}).
In this combination, the more sensitive interferometer receives more weight in
the combined signal-to-noise ratio.  If a trigger is found in only one of the
Hanford interferometers, then $\rho_H$ is simply taken to be the $\rho$ from
that interferometer.

Finally, we cluster the coincident triggers over a $4$~second time interval.
Clustering is performed so that a noise transient that may cause several
templates to trigger within a small window is only counted as a single event
in the data sample.

\section{Auxiliary and Environmental Channel Vetoes}
\label{s:vetoes}

\begin{figure}[p]
\vspace{5pt}
\begin{center}
\begin{tabular}{cc}
(i) & (ii) \\
\includegraphics[width=0.475\linewidth]{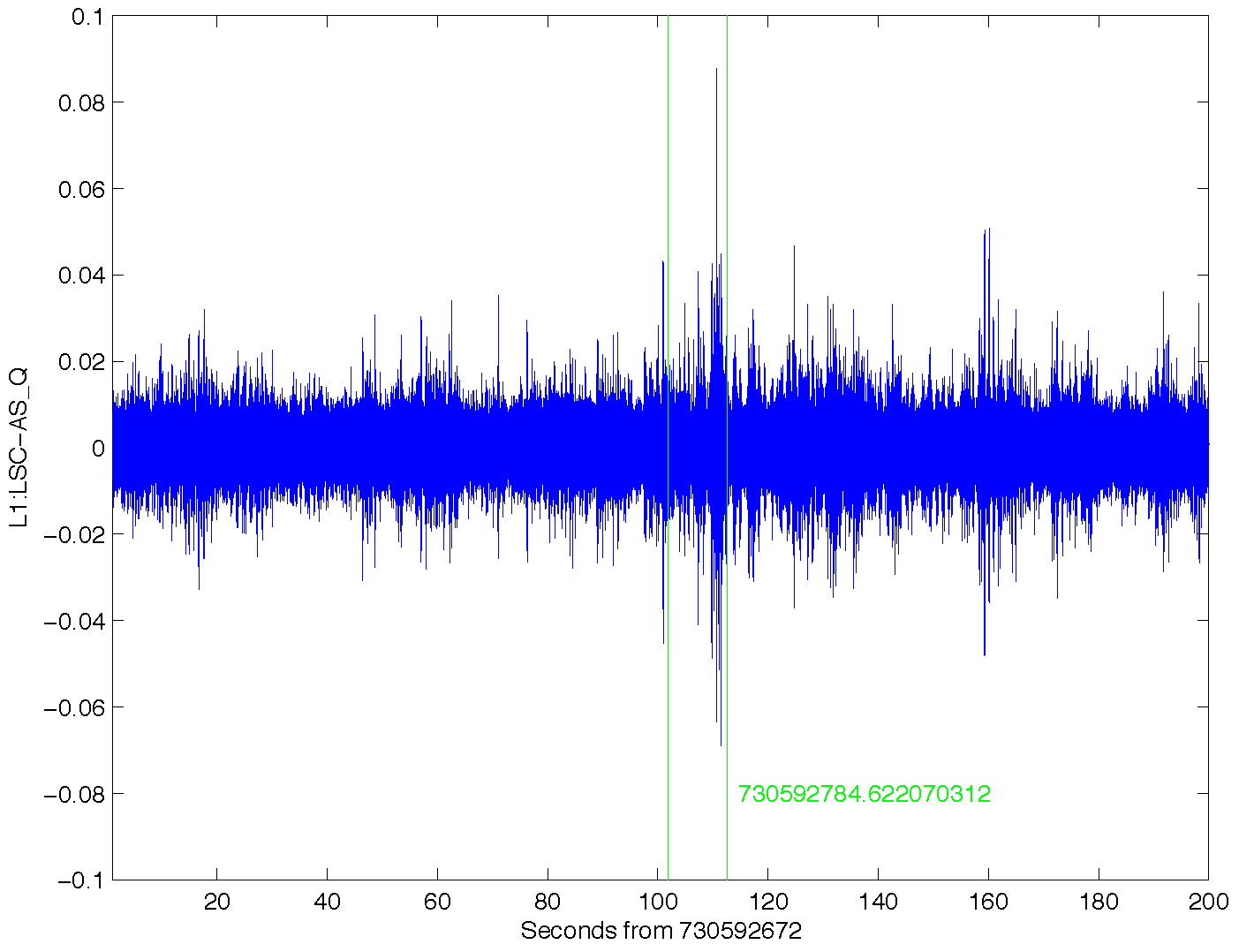} &
\includegraphics[width=0.475\linewidth]{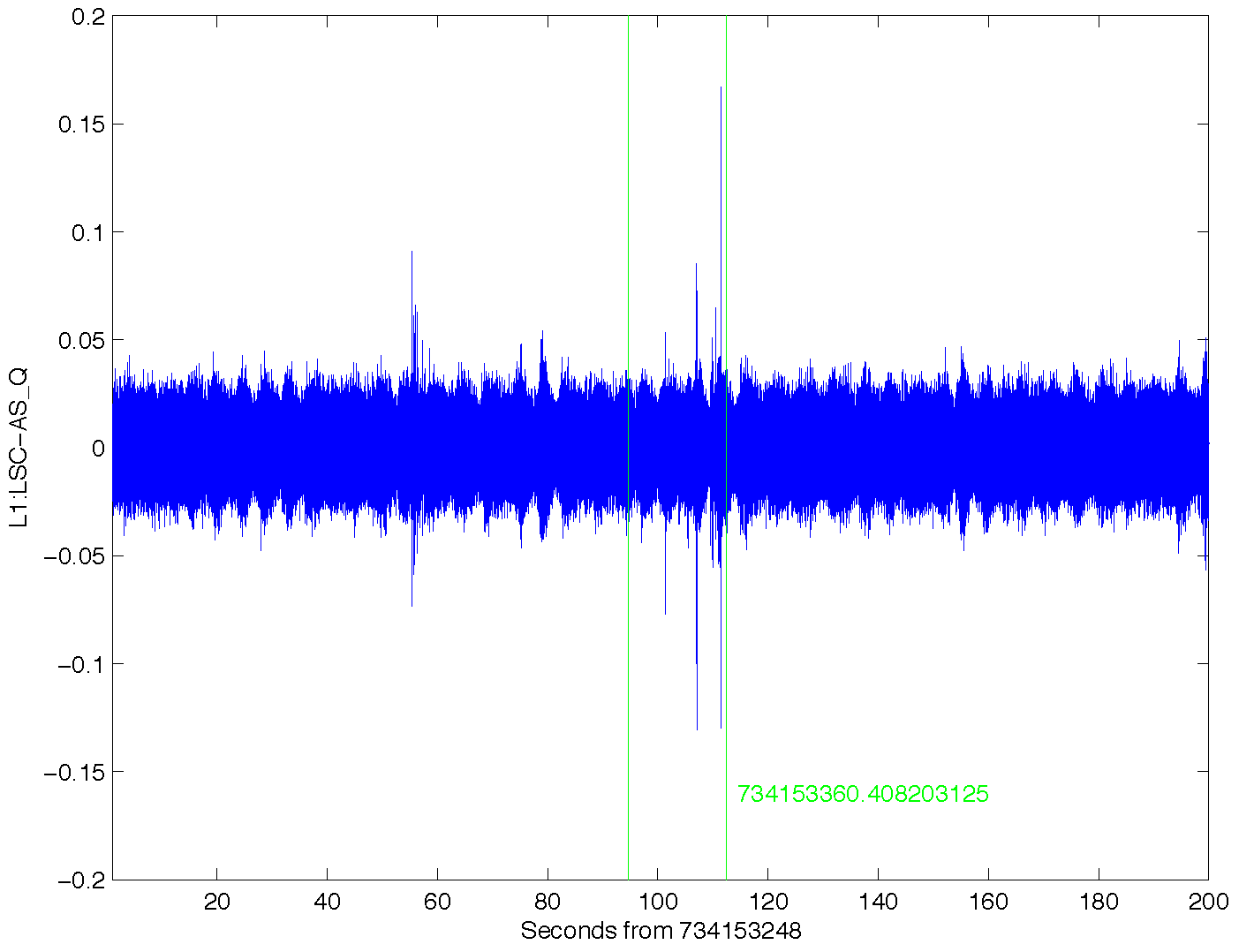}\\
(iii) & (iv) \\
\includegraphics[width=0.475\linewidth]{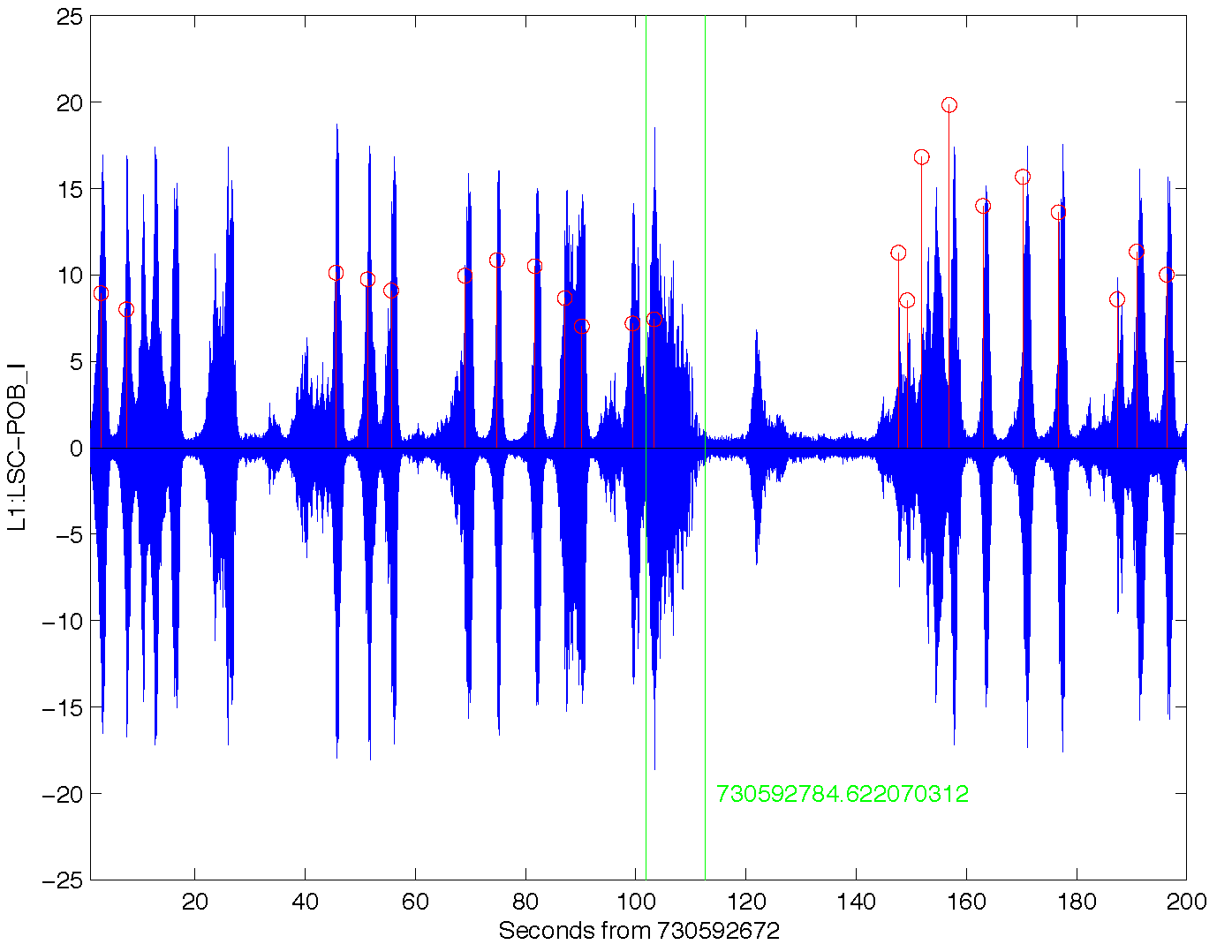} &
\includegraphics[width=0.475\linewidth]{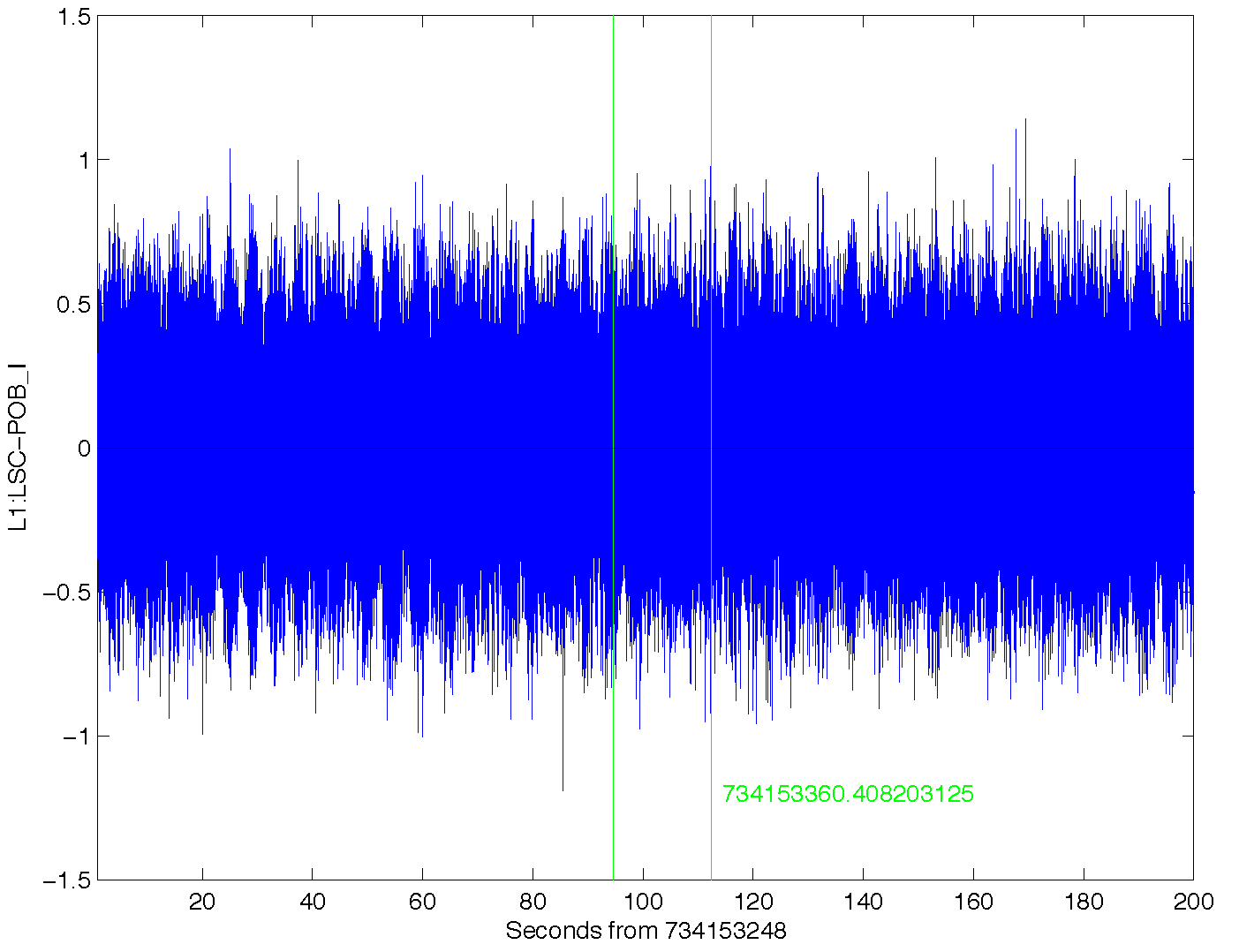}\\
\end{tabular}
\end{center}
\caption[Auxiliary Channel Veto Investigation of Two Candidate Triggers]{%
\label{f:vetoes}
An auxiliary channel veto investigation of two candidate triggers. Panel (i)
shows the gravitational wave channel, L1:LSC-AS\_Q, high passes above 100 Hz
for a an inspiral trigger at GPS time 730592784 with a signal-to-noise ratio
$\rho = 10.6$. The vertical line shows the time of the inspiral trigger.
Panels (iii) shows the auxiliary interferometer channel L1:LSC-POB\_I high
passed above 70 Hz. The vertical lines with circles show the location of
glitchMon triggers produced by the noise in L1:LSC-POB\_I. By excluding
inspiral triggers within a time window of these glitchMon triggers, we can
reduce the number of false event candidates in the pipeline. For contrast,
panel (ii) shows an gravitational wave channel high passed above 100 Hz for
an inspiral trigger at GPS time 734153360. This trigger has a similar
signal-to-noise ratio, $\rho = 10.9$, as the trigger in panel (i). For this
trigger there does not seem to be a correlated noise source in the auxiliary
channel L1:LSC-POB\_I shown high passed above 70 Hz in panel (iv).  
}
\end{figure}

In addition to data quality cuts, another method to exclude false alarms is to
look for signatures in environmental monitoring channels and auxiliary
interferometer channels which would indicate an external disturbance or
instrumental glitches. This allows us to {\it veto} any triggers recorded at
that time.  Auxiliary interferometer channels (which monitor the light in the
interferometer at points other than the antisymmetric port---where a
gravitational wave would be most evident) are examined, with the aim being to
look for correlations between glitches found in the readouts of these channels
and inspiral event triggers found in the playground data.  By doing so, we are
capable of identifying instrumental artifacts that directly affect the light
that is measured in the gravitational wave channel, so these vetoes are
potentially very powerful. Figure \ref{f:vetoes} demonstrates the the use of
auxiliary channels to identify the source of an inspiral trigger in the
gravity wave channel.

When choosing vetoes, we must consider the possibility that a gravitational
wave itself could produce the observed glitches in the auxiliary channel due
to some physical or electronic coupling.  This possibility was tested by means
of hardware injections, in which a simulated inspiral signal is injected into
the interferometer by physically moving one of the end mirrors of the
interferometer. Hardware injections allow us to establish a limit on the
effect that a true signal would have on the auxiliary channels.  Only those
channels that were unaffected by the hardware injections were considered
``safe'' for use as potential veto channels. The process of testing veto
safety with hardware injections is described in more detail in chapter
\ref{ch:hardware}.

We used a computer program, {\it glitchMon}\cite{glitchMon}, to examine the
data and identify large amplitude transient signals in auxiliary channels.
Numerous channels, with various filters and threshold settings, were examined
and which produced a list of times when the glitches occurred. The glitch
event list was compared with times generated by triggers from the inspiral
search (Note that these studies were all conducted on the playground
data.)  A time window around a glitch was defined, and any inspiral event
within this window was rejected. One can associate the veto with inspiral
event candidates and evaluate a veto efficiency (percentage of inspiral events
eliminated), use percentage (percentage of veto triggers which veto at least
one inspiral event), and dead-time (percentage of science-data time eliminated
by the veto). A ``good'' veto will have a large veto efficiency and use
percentage with a small dead time suggesting that it is well correlated with
events in the gravitational wave channel that produce inspiral triggers.
Followup studies are performed on such candidate vetoes to determine the
physical coupling between the auxiliary channel and the gravitational wave
channel. If a sound coupling mechanism is found, then the veto will be used.

Tuning of the vetoes for binary black hole MACHOs is described
in chapter \ref{ch:result}.

\section{Background Estimation}
\label{s:background}

Since we restrict the S2 analysis to coincident data and require that at least
two of the interferometers must be located at different observatories, we may
measure a background rate for our analysis.  We estimate the background rate
by introducing an artificial time offset, or {lag}, $\mathbb{T}$ to the
triggers coming from the Livingston detector relative to the Hanford detector.
We call this ``sliding the triggers by $\mathbb{T}$.'' After generating
triggers for each interferometer, we slide the triggers from the LHO
interferometers relative to the LLO interferometer and look for coincidences
between the offset and zero lag triggers.  The triggers which emerge from the
end of the pipeline are then considered a single trial representative of an
output from a search if no signals are present in the data.   By choosing a
lag of more than 20~ms, we ensure that a true gravitational wave will not be
coincident in the time-shifted data streams.  In fact, we use lags longer than
this to avoid correlation issues; the minimum lag is $17$~seconds.  Note that we do
not time-shift the two Hanford detectors relative to one another since there
may be real correlations due to environmental disturbances.  If the times of
background triggers are not correlated in the two interferometers then the
background rate can be measured; we assume that there is no such correlation
between LHO and LLO triggers.

\section{Detection Efficiency}
\label{s:eff}

In absence of detection, we will construct an upper limit on event rate.  To
do this we need to measure the \emph{detection efficiency}, $\varepsilon$, of
the analysis pipeline to our population. This is the fraction of true signals
from a population that would produce triggers at the end of the pipeline. A
Monte Carlo method is used to measure this efficiency. We simulate a
population of binary neutron stars and \emph{inject} signals from that
population into the data from all three LIGO interferometers. The injection is
performed in software by generating an inspiral waveform and adding it to
interferometer data immediately after the raw data is read from disk. We
inject the actual waveform that would be detected in a given interferometer
accounting for both the masses, orientation, polarization, sky position and
distance of the binary, the antenna pattern and calibration of the
interferometer into which this signal is injected.  The effectiveness of
software injections for measuring the response of the instrument to an
inspiral signal is validated against hardware injections where an inspiral
signal is added to the interferometer control servo during operation to
produce the same output signal as a real gravitational wave. This validation
is described in chapter \ref{ch:hardware}. The data with injections is run
through the full analysis pipeline to produce a list of inspiral triggers.
We may combine the signal-to-noise rations from coincident triggers from
several interferometers into a single \emph{coherent} signal-to-noise ratio,
\begin{equation}
\hat{\rho} = f(\rho_\mathrm{L1},\rho_\mathrm{H})
\end{equation}
where the form of $f$ is chosen based on studies of the playground and
background triggers. We can then construct a final threshold,
$\hat{\rho}_\ast$, on triggers that survive the pipeline. The detection
efficiency, $\varepsilon(\hat{\rho})$, is the ratio of the number of signals
with $\hat{\rho} > \hat{\rho}_\ast$  to the number of injected signals.

\section{The S2 Data Analysis Pipeline}
\label{s:s2pipeline}

In this section we describe the pipeline constructed to search the S2 data for
gravitational waves from inspiralling binary neutron stars and binary black
hole MACHOs. The data quality cuts used are common to both searches and are
described in section \ref{ss:s2dqselection}. The detection of a
gravitational-wave inspiral signal in the S2 data would (at the least) require
triggers in both L1 and one or more of the Hanford instruments with consistent
arrival times (separated by less than the light travel time between the
detectors) and waveform parameters. During the S2 run, the three LIGO
detectors had substantially different sensitivities, as can be seen from
figure \ref{f:s2noisecurve}. The sensitivity of the L1 detector was greater
than those of the Hanford detectors throughout the run. Since the orientations
of the LIGO interferometers are similar, we expect that signals of
astrophysical origin detected in the Hanford interferometers will most often
be also detectable in the L1 interferometer.  We use this and the requirement
that a signal be detected in both the Livingston and at least one of the
Hanford interferometers to construct a {\em triggered search} pipeline. 

\subsection{Selection of Data Quality Cuts for S2}
\label{ss:s2dqselection}

Playground data from each of the three interferometers was analyzed separately
producing a list of inspiral triggers from each interferometer. Only the
mandatory data quality cuts were used to exclude time from the science mode
segments. Each interferometer was filtered separately using template banks
particular to that interferometer. No coincidence was applied between
interferometers; data quality cuts were tested independently on the three
lists of inspiral triggers produced. Table \ref{t:s2dqresults} shows the the
correlation of inspiral triggers with a particular data quality cut for
triggers of different signal to noise. When selecting the data quality cuts we
must be aware of three constraints. The first is that the data quality cuts
are based on data from the gravitational wave channel so it is important to
ensure that a data quality cut is not triggered by a real signal in the data.
For this reason we always use caution when selecting a cut base on noise in
AS\_Q.\@ The second constraint is that we do not wish to exclude large amounts
of data from the analysis.  Finally we base our choice on advice from the
experimental team. A member of the experimental team may decide that a
cut should be used, even if it does not correlate with inspiral triggers, as
any detection made in such a time could not be trusted. Table
\ref{t:s2dqchoice} shows the final choice of discretionary data quality cuts
and the reasons for them.

\subsection{A triggered search pipeline}
\label{ss:triggeredsearch}

\begin{figure}[p]
\vspace{5pt}
\begin{center}
\includegraphics[width=\textwidth]{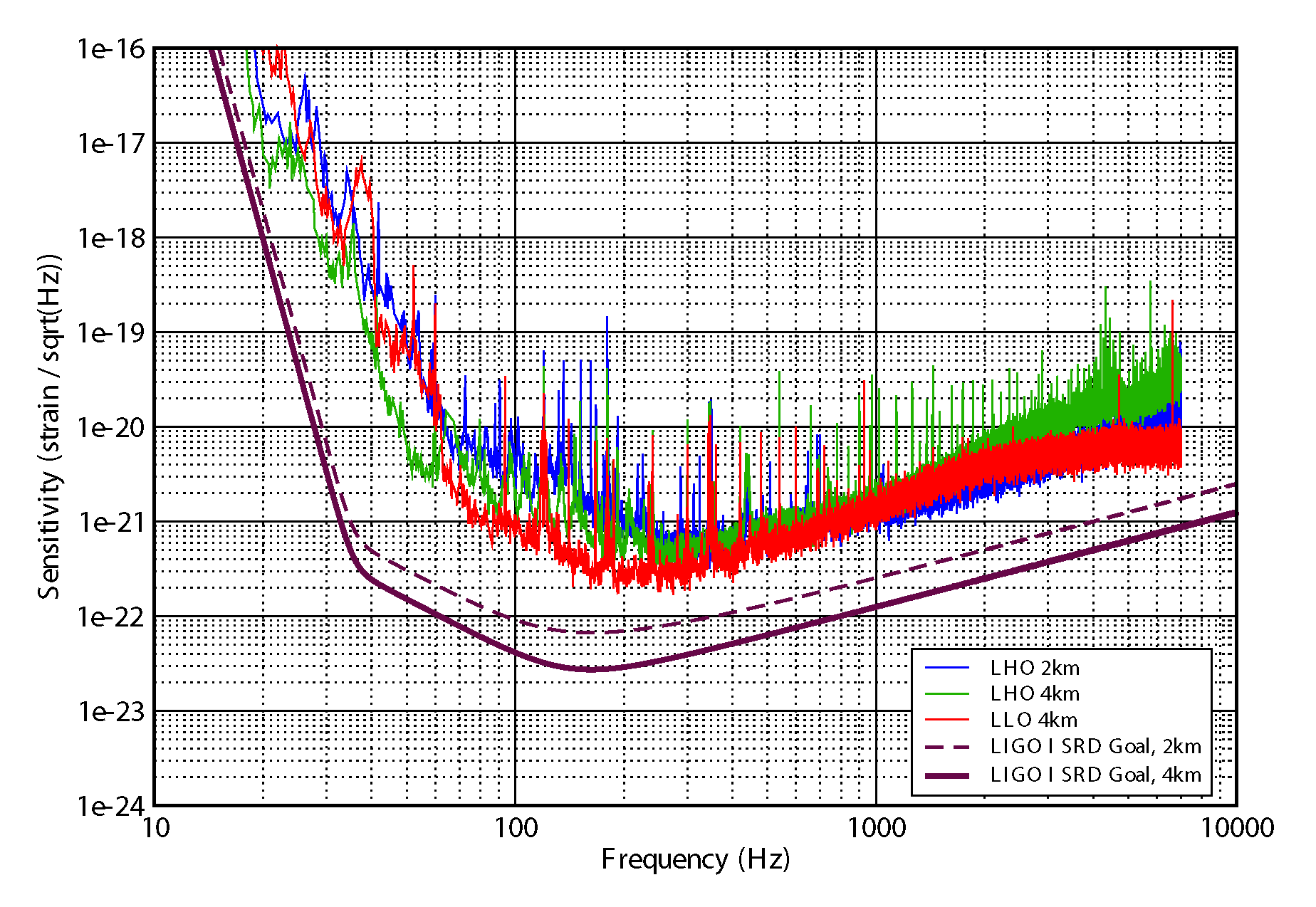}    
\end{center}
\caption[Sensitivity of LIGO Interferometers During S2]{%
\label{f:s2noisecurve}
Typical sensitivities of the three LIGO interferometers during the second LIGO
science run shown as strain amplitude spectral density,
$\tilde{h}/\sqrt{\mathrm{Hz}}$. The smooth solid curve shows the design
sensitivity (SRD Goal) of the $4$~km interferometers and the smooth dashed
curve shows the design sensitivity of the $2$~km interferometer.
}
\end{figure}

During the S2 run, the three LIGO detectors had substantially different
sensitivities, as can be seen from figure \ref{f:s2noisecurve}. The Livingston
interferometer is more sensitive than either of the Hanford interferometers.
We use this and the requirement that a signal be detected in both the
Livingston and at least one of the Hanford interferometers to construct a {\em
triggered search} pipeline, summarized in Fig.~\ref{f:pipeline}. We search for
inspiral triggers in the most sensitive interferometer (L1), and only when a
trigger is found in this interferometer do we search for a coincident trigger
in the less sensitive interferometers. This approach reduces the computational
power necessary to perform the search.

The power spectral density (PSD) of the noise in the Livingston detector is
estimated independently for each L1 chunk that is coincident with operation of
a Hanford detector (denoted $\mathrm{L1} \cap (\mathrm{H1} \cup
\mathrm{H2})$).  The PSD is used to lay out a template bank for filtering that
chunk, according to the parameters for mass ranges and minimal
match\cite{Owen:1998dk}. The data from the L1 interferometer for the chunk is
then filtered, using that bank, with a signal-to-noise threshold
$\rho_{\mathrm{L}}^\ast$ and $\chi^2$ veto threshold $\chi^2_{\ast\mathrm{L}}$
to produce a list of triggers as described in section~\ref{s:pipetemplate}.
For each chunk in the Hanford interferometers, a \emph{triggered bank} is
created by adding a template if it produced at least one trigger in L1 during
the time of the Hanford chunk.  This is used to filter the data from the
Hanford interferometers with signal-to-noise and $\chi^2$ thresholds specific
to the interferometer, giving a total of six thresholds that may be tuned.
For times when only the H2 interferometer is operating in coincidence with L1
(denoted $\mathrm{L1} \cap (\mathrm{H2} - \mathrm{H1})$) the triggered bank is
used to filter the H2 chunks that overlap with L1 data; these triggers are
used to test for L1-H2 double coincidence.  All H1 data that overlaps with L1
data (denoted $\mathrm{L1} \cap \mathrm{H1}$) is filtered using the triggered
bank for that chunk. For H1 triggers produced during times when all three
interferometers are operating, a second triggered bank is produced for each H2
chunk by adding a template if it produced at least one trigger found in
coincidence in L1 and H1 during the time of the H2 chunk and the H2 chunk is
filtered with this bank.  These triggers are used to search for triple
coincident triggers in H2.  The remaining triggers from H1 when H2 is not
available are used to search for L1-H1 double coincident triggers.

\subsection{A directed acyclic graph (DAG) for the S2 pipeline}
\label{ss:dag}

\begin{figure}[p]
\begin{center}
\hspace*{-0.2in}\includegraphics[width=0.6\textheight]{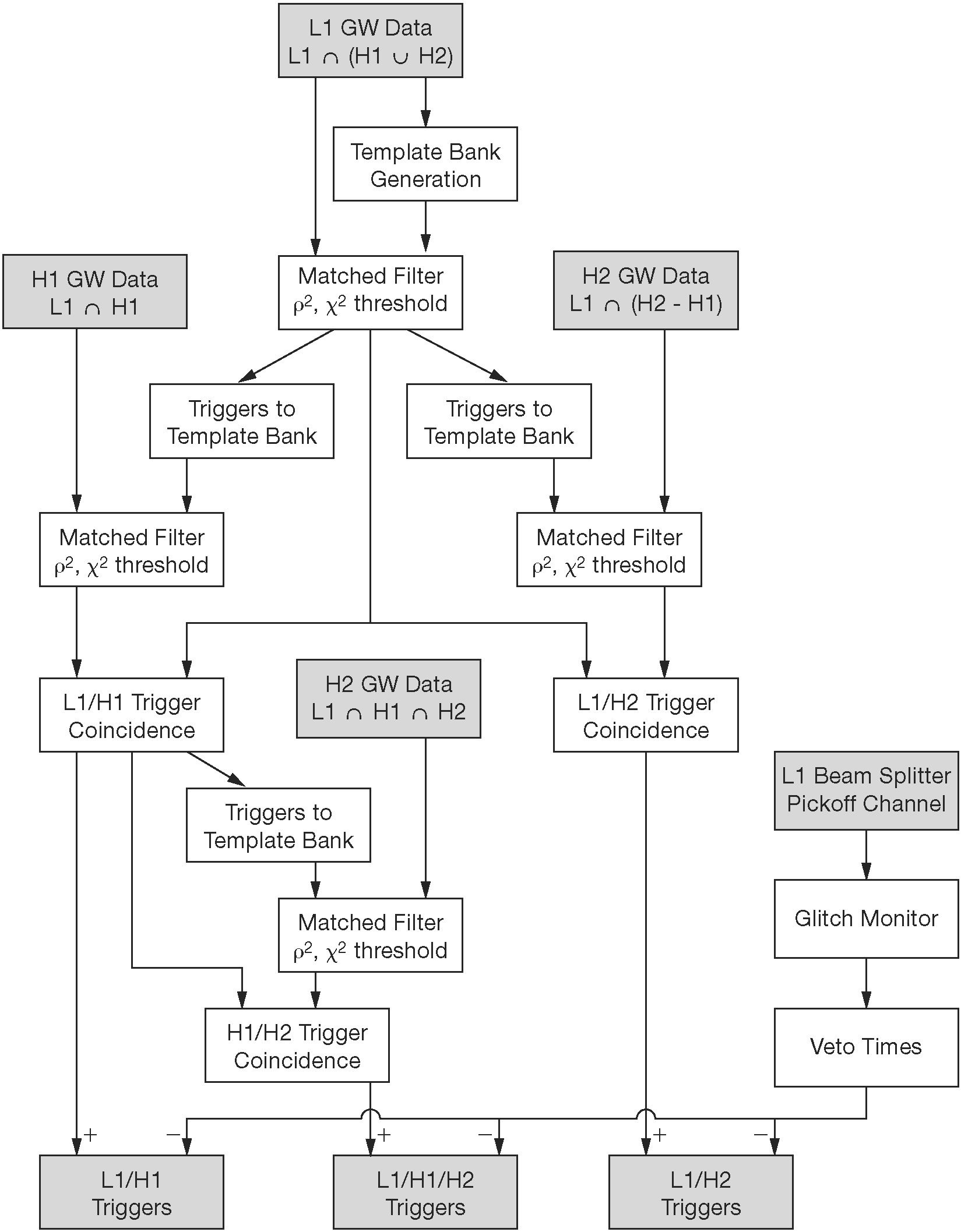}
\end{center}
\caption[Structure of the S2 Triggered Search Pipeline]{%
\label{f:pipeline}
The inspiral analysis pipeline used to determine the reported upper
limit. $\mathrm{L1} \cap (\mathrm{H1} \cup \mathrm{H2})$ indicates times when
the L1 interferometer was operating in coincidence with one or both of the
Hanford interferometers. $\mathrm{L1} \cap \mathrm{H1}$ indicates times when
the L1 interferometer was operating in coincidence with the H1 interferometer.
$\mathrm{L1} \cap (\mathrm{H2} - \mathrm{H1})$ indicates times when the L1
interferometer was operating in coincidence with only the H2 interferometer.
The outputs of the search pipeline are triggers that belong to one of the
two double coincident data sets or to the triple coincident data set.}
\end{figure}

In this section we demonstrate how the S2 pipeline in figure \ref{f:pipeline}
can be abstracted into a DAG to execute the analysis. We illustrate the
construction of the DAG with the short list of science segments shown in table
\ref{t:fakesegslist}. For simplicity, we only describe the construction of the
DAG for zero time lag data. The DAG we construct filters more than the
absolute minimum amount of data needed to cover all the double and triple
coincident data, but since we were not computationally limited during S2, we
chose simplicity over the maximum amount of optimization that could have used.

\begin{figure}[p]
\begin{center}
\hspace*{-0.2in}\includegraphics[width=\linewidth]{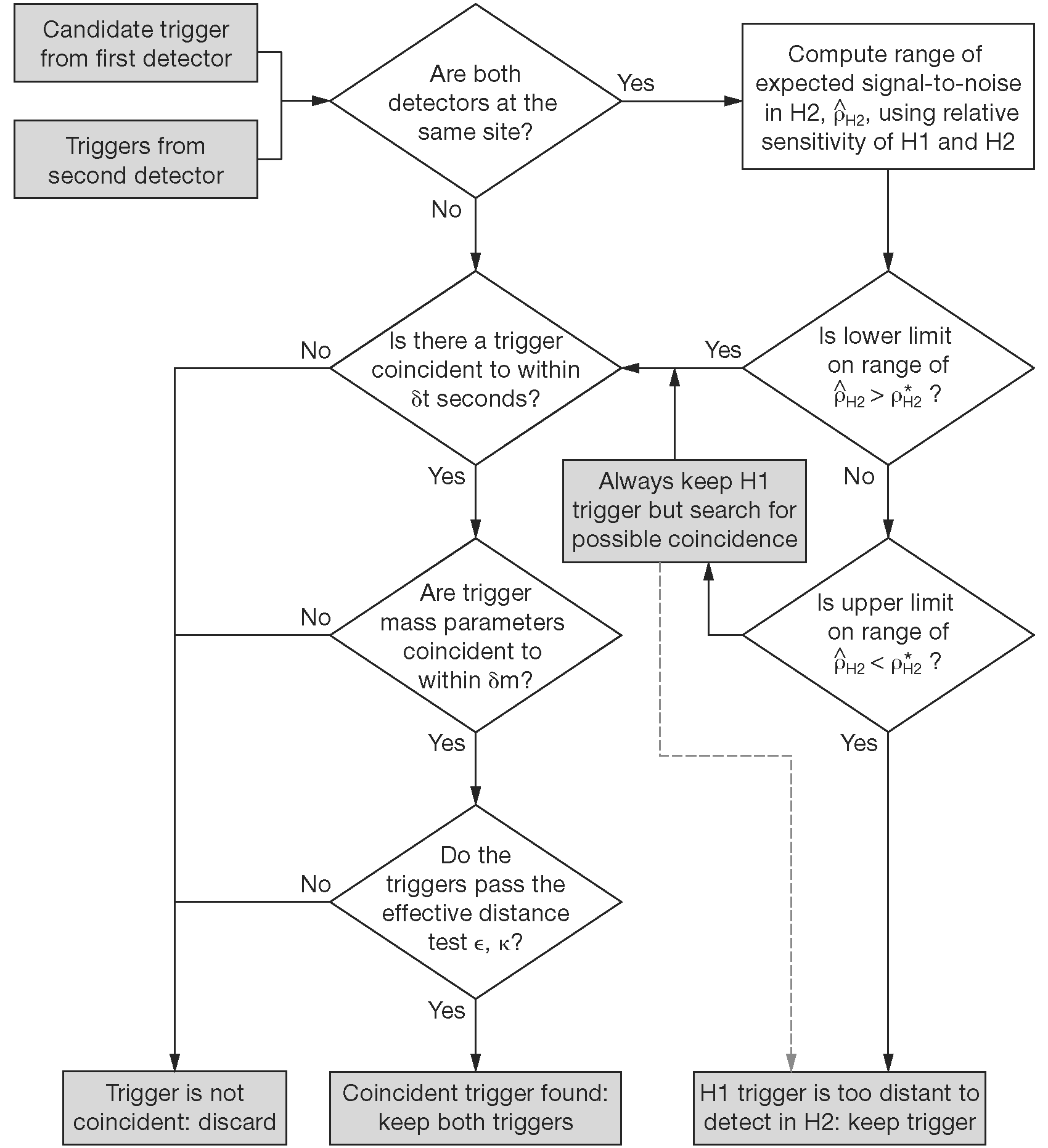}
\end{center}
\caption[Trigger Coincidence Test]{%
\label{f:coinc_test}
The test to decide if a trigger in the first detector has a coincident 
trigger in the second detector. If detectors are at different sites, time and
mass coincidence are demanded. The effective distance cut is disabled by 
setting $\kappa = 1000$. If the detectors are at the same site, we ask if the
maximum distance to which H2 can see at the signal-to-noise threshold
$\rho_\mathrm{H2}^\ast$ is greater than the distance of the H1 trigger,
allowing for errors in the measurement of the trigger distance. If this is the
case, we demand time, mass and effective distance coincidence.  If distance to
which H2 can see overlaps the error in measured distance of the H1 trigger, we
search for a trigger in H2, but always keep the H1 trigger even if no
coincident trigger is found. If the minimum of the error in measured distance
of the H1 trigger is greater than the maximum distance to which H2 can detect
a trigger we keep the H1 trigger without searching for coincidence.}
\end{figure}

A DAG consists of \emph{nodes} and \emph{edges}. The nodes are the programs
which are executed to perform the inspiral search pipeline. In the S2
pipeline, the possible nodes of the DAG are:
\begin{enumerate}
\item\textsc{datafind} locates data for a specified time interval on the
compute cluster and creates a file containing the paths to the input data that
other programs can read.

\item\textsc{tmpltbank} generates an average power spectral density for a
chunk and computes a template bank for a given region of mass parameter space
and minimal match.

\item\textsc{inspiral} filters an analysis chunk using a template bank and
generates inspiral triggers for further analysis.

\item\textsc{trigtotmplt} generated a triggered template bank from the
output of the inspiral code.

\item\textsc{inca} (INspiral Coincidence Analysis) implements the
coincidence analysis described in section \ref{s:coincidence} and figure
\ref{f:coinc_test}.
\end{enumerate}
The edges in the DAG define the relations between programs; these relations
are determined in terms of \emph{parents} and \emph{children}, hence the
directed nature of the DAG. A node in the DAG will not be executed until all
of its parents have been successfully executed. There is no limit to the
number of parents a node can have; it may be zero or many. In order for the
DAG to be acyclic, no node can be a child of any node that depends on the
execution of that node. By definition, there must be at least one node in the
DAG with no parents. This node is executed first, followed by any other nodes
who have no parents or whose parents have previously executed.  The
construction of a DAG allows us to ensure that inspiral triggers for two
interferometers have been generated before looking for coincidence between the
triggers, for example.

The S2 DAG is generated by a program called the \emph{pipeline script}, which
is an implementation of the logic of the S2 pipeline in the Python programming
language. The pipeline script takes as input the list of science segments for
each interferometer, with data quality cuts applied. The script reads in all
science segments longer than $2048$~seconds and divides them into \emph{master
analysis chunks}. If there is data at the end of a science segment that is
shorter than $2048$~seconds, the chunk is overlapped with the previous one, so
that the chunk ends at the end of the science segment. An option named
\verb|trig-start-time| is set and passed to the inspiral code. No triggers are
generated before this time and so no triggers are duplicated between chunks.
For example, the first L1 science segment in the fake segment list in table
\ref{t:fakesegslist} starts at GPS time 730000000 and ends at GPS time
730010000. It is divided into the following master chunks:
\begin{verbatim}
<AnalysisChunk: start 730000000, end 730002048>
<AnalysisChunk: start 730001920, end 730003968> 
<AnalysisChunk: start 730003840, end 730005888> 
<AnalysisChunk: start 730005760, end 730007808> 
<AnalysisChunk: start 730007680, end 730009728> 
<AnalysisChunk: start 730007952, end 730010000, trig_start 730009664>
\end{verbatim}
Although the script generates all the master chunks for a given
interferometer, not all of them will be filtered. Only those that overlap with
double or triple coincident data are used for analysis.  The master analysis
chunks are constructed for L1, H1 and H2 separately by reading in the three
science segment files. The full list of master chunks for the fake segments is
written to a log file.

The pipeline script next computes the disjoint regions of double and triple
coincident data to be searched for triggers. 64 seconds is subtracted from the
start and end of each science segment (since this data is not searched for
triggers) and the script performs the correct intersection and unions of the
science segments from each interferometer to generate the following segments
containing the times of science mode data to search:
\begin{verbatim}
Writing 2 L1/H1 double coincident segments 
<ScienceSegment: start 730007936, end 730009936, dur 2000>
<ScienceSegment: start 731001064, end 731002436, dur 1372>
total time 3372 seconds 

Writing 2 L1/H2 double coincident segments 
<ScienceSegment: start 730002564, end 730004064, dur 1500>
<ScienceSegment: start 731004564, end 731005936, dur 1372>
total time 2872 seconds 
 
Writing 2 L1/H1/H2 triple coincident segments 
<ScienceSegment: start 730004064, end 730007936, dur 3872>
<ScienceSegment: start 732000064, end 732002936, dur 2872>
total time 6744 seconds
\end{verbatim}
The GPS start and end times are given for each segment to be searched for
triggers.  The script uses this list of science data to decide which master
analysis chunks need to be filtered. All L1 master chunks that overlap with H1
or H2 science data to be searched are filtered. An L1 template bank is
generated for each master chunk and the L1 data is filtered using this bank.
This produces two intermediate data products for each master chunk, which are
stored as XML data. The intermediate data products are the template bank file,
\verb|L1-TMPLTBANK-730000000-2048.xml|, and the inspiral trigger file,
\verb|L1-INSPIRAL-730000000-2048.xml|. The GPS time in the filename
corresponds to the start time of the master chunk filtered and the number
before the \verb|.xml| file extension is the length of the master chunk.

All H2 master chunks that overlap with the L1/H2 double coincident data to
filter are then analyzed. For each H2 master chunk, a triggered template bank
is generated from L1 triggers between the start and end time of the H2 master
chunk. The triggered bank file generated is called
\verb|H2-TRIGBANK_L1-730002500-2048.xml|, where the GPS time corresponds to
start time of the master H2 chunk to filter. All L1 master chunks that overlap
with the H2 master chunk are used as input to the triggered bank generation to
ensure that all necessary templates are filtered.  The H2 master chunks are
filtered using the triggered template bank for that master chunk to produce H2
triggers in files named \verb|H2-INSPIRAL_L1-730002500-2048.xml|. The GPS
start time in the file name is the start time of the H2 master chunk.

All H1 master chunks that overlap with either the L1/H1 double coincident data
or the L1/H1/H2 triple coincident data are filtered. The bank and trigger
generation is similar to the L1/H2 double coincident case. The triggered
template bank is stored in a file names
\verb|H1-TRIGBANK_L1-730004000-2048.xml| and the triggers in a file named
\verb|H1-INSPIRAL_L1-730004000-2048.xml| where the GPS time in the file name
is the GPS start time of the H1 master chunk. The H2 master chunks that
overlap with the L1/H1/H2 triple coincident data are described below.

For each L1/H1 double coincident segments to search, an inca process is run to
perform the coincidence test. The input to inca is all L1 and H1 master chunks
that overlap the segment to search. The GPS start and stop times passed to
inca are the start and stop times of the double coincident segment to search.
The output is a file names \verb|H1-INCA_L1H1-730007936-2000.xml|. The GPS
start time in the file name is the start time of the double coincident
segment.  A similar procedure is followed for each L1/H2 double coincident
segment to search. The output files from inca are names
\verb|H2-INCA_L1H2-731004564-1372.xml|, and so on.

For each L1/H1/H2 triple coincident segment, an inca process is run to create
the L1/H1 coincident triggers for this segment. The input files are all L1 and
H1 master chunks that overlap with the segment. The start and end times to
inca are the start and end times of the segment. This creates a file named
\begin{verbatim}
H1-INCA_L1H1T-730004064-3872.xml
\end{verbatim}
where the GPS start time and duration in the file name are those of the triple
coincident segment to search.  For coincidence between L1 and an LHO
interferometer, we only check for time, $dt$, and mass, $dm$, coincidence.
The parameter $\kappa$ in the effective distance cut is set to $1000$, so the
amplitude cut is disabled.

For each H2 master chunk that overlaps with triple coincident data, a triggered
template bank is generated. The input file to the triggered bank generation is
the inca file for the segment to filter that contains the master chunk. The
start and end times of the triggered bank generation are the start and end
times of the master chunk. This creates a file called
\verb|H2-TRIGBANK_L1H1-730004420-2048.xml|.  The H2 master chunk is filtered
through the inspiral code to produce a trigger file
\verb|H2-INSPIRAL_L1H1-730004420-2048.xml|.
 
For each triple coincident segment to filter, and inca is run between the H1
triggers from the L1H1T inca and the H2 triggers produced by the inspiral
code. The input files are the H1 inca file \verb|H1-INCA_L1H1T-730004064-3872.xml|
and all H2 master chunk inspiral files that overlap with this interval. The
coincidence is performed as follows:
\begin{enumerate}
\item For each H1 trigger compute the effective distance of the trigger minus
$\kappa$ times the effective distance (this is the lower bound on the error
allowed in effective distance).

\item Compute the maximum range of H2 for the trigger mass.

\item If the lower bound on the H1 trigger is further away than can be seen in
H2, keep the trigger.

\item If the lower bound on the effective distance of the H1 trigger is less
than the range of H2, but the upper bound is greater, keep the trigger in H1.
If a H2 trigger is found within the interval, store it as well.

\item If upper bound on the distance of the H1 trigger is less than the range
of H2, check for coincidence. A coincidence check is performed in $\delta t$, $\delta m$,
$\epsilon$ and $\kappa$. If there is no coincident trigger discard the H1 trigger.
\end{enumerate}
This coincidence step creates two files
\begin{verbatim}
H1-INCA_L1H1H2-730004064-3872.xml
H2-INCA_L1H1H2-730004064-3872.xml
\end{verbatim}
where the GPS start time and duration of the files are the start and duration
of the triple coincident segment.  The L1/H1 coincidence step is then executed
again to discard any L1 triggers coincident with a H1 triggered that has been
discard by H2. The input to the inca are the files
\begin{verbatim}
L1-INCA_L1H1T-730004064-3872.xml
H1-INCA_L1H1H2-730004064-3872.xml
\end{verbatim}
and the output is the files
\begin{verbatim}
L1-INCA_L1H1H2-730004064-3872.xml
H1-INCA_L1H1H2-730004064-3872.xml.
\end{verbatim}
The H1 input file is overwritten as it is identical to the H1 output file.
Finally, we obtain the data products of the search which contain the candidate
trigger found by the S2 pipeline in these fake segments. The for the fake
segments described here, the final output files will be:

\subsubsection*{Double Coincident L1/H1 Data}
\begin{verbatim}
L1-INCA_L1H1-730007936-2000.xml L1-INCA_L1H1-731001064-1372.xml 
H1-INCA_L1H1-730007936-2000.xml H1-INCA_L1H1-731001064-1372.xml
\end{verbatim}

\subsubsection*{Double Coincident L1/H2 Data}
\begin{verbatim}
L1-INCA_L1H2-730002564-1500.xml L1-INCA_L1H2-731004564-1372.xml 
H2-INCA_L1H2-730002564-1500.xml H2-INCA_L1H2-731004564-1372.xml
\end{verbatim}

\subsubsection*{Triple Coincident L1/H1/H2 Data}
\begin{verbatim}
L1-INCA_L1H1H2-730004064-3872.xml L1-INCA_L1H1H2-732000064-2872.xml 
H1-INCA_L1H1H2-730004064-3872.xml H1-INCA_L1H1H2-732000064-2872.xml 
H2-INCA_L1H1H2-730004064-3872.xml H2-INCA_L1H1H2-732000064-2872.xml
\end{verbatim}

As the size of the input science segment files increase, so the number of
nodes and vertices in the DAG increases, however the algorithm for generating
the DAG remains the same.

\begin{table}[p]
\begin{center}
\begin{tabular}{ll}
Mandatory Data Quality Cut & Description \\\hline
OUTSIDE\_S2   &Data is outside of official S2 time interval \\
MISSING\_RAW  &Raw data is missing \\
DAQ\_DROPOUT       & Dropout in data acquisition system \\
MISSING\_RDS       & Data is unavailable for analysis \\
INVALID\_TIMING    & Timing believed to be unreliable \\
CALIB\_LINE\_NO\_RDS\_V03 & Problem with accessing data for calibration \\
DAQ\_REBOOT               & One or more data acquisition system rebooted \\
INVALID\_CALIB\_LINE  & Problem with calibration line strength \\
NO\_CALIB                 & Calibration line turned off \\
LOW\_CALIB                & Calibration line strength too low \\
\hline\hline
\\
Discretionary Data Quality Cut & Description \\\hline
MICH\_FILT    &One or more Michelson control loop \\
&filters was not in its nominal state \\
AS\_PD\_SATURATION & Antisymmetric port photodiode saturated \\
ASQ\_LARGEP2P      & Large peak-to-peak range seen in AS\_Q \\
&at end of lock \\
NONSTAND\_CTRLS    & Non-standard controls affecting calibration\\
&and couplings \\
ASQ\_OUTLIER\_CLUSTER     & Cluster of large AS\_Q outliers \\
&in short time interval \\
ASQ\_OUTLIER\_CORRELATED  & Large ASQ outliers correlated with \\
&outliers in auxiliary IFO channel \\
ASQ\_LOWBAND\_OUTLIER     & High noise below 100 Hz in AS\_Q \\
ASQ\_UPPERBAND\_OUTLIER   & High noise in 100-7000 Hz in AS\_Q \\
\hline\hline
\end{tabular}
\end{center}
\caption[Data Quality Cuts Available During S2]{%
\label{t:dqflags}
Data quality cuts available during the S2 science run and their meanings. 
Some data quality flags monitor human error in the operation of the
instrument that make the data unsuitable for analysis, such as NO\_CALIB and
NONSTAND\_CTRLS. Others cuts identify hardware or software failures in the
operation of the instrument, for example MISSING\_RAW and DAQ\_REBOOT.
Additional cuts like AS\_PD\_SATURATION and ASQ\_UPPERBAND\_OUTLIER monitor
the gravitational wave channel, AS\_Q, for unusable data.
}
\end{table}

\begin{figure}[p]
\vspace{5pt}
\begin{center}
\includegraphics[width=\linewidth]{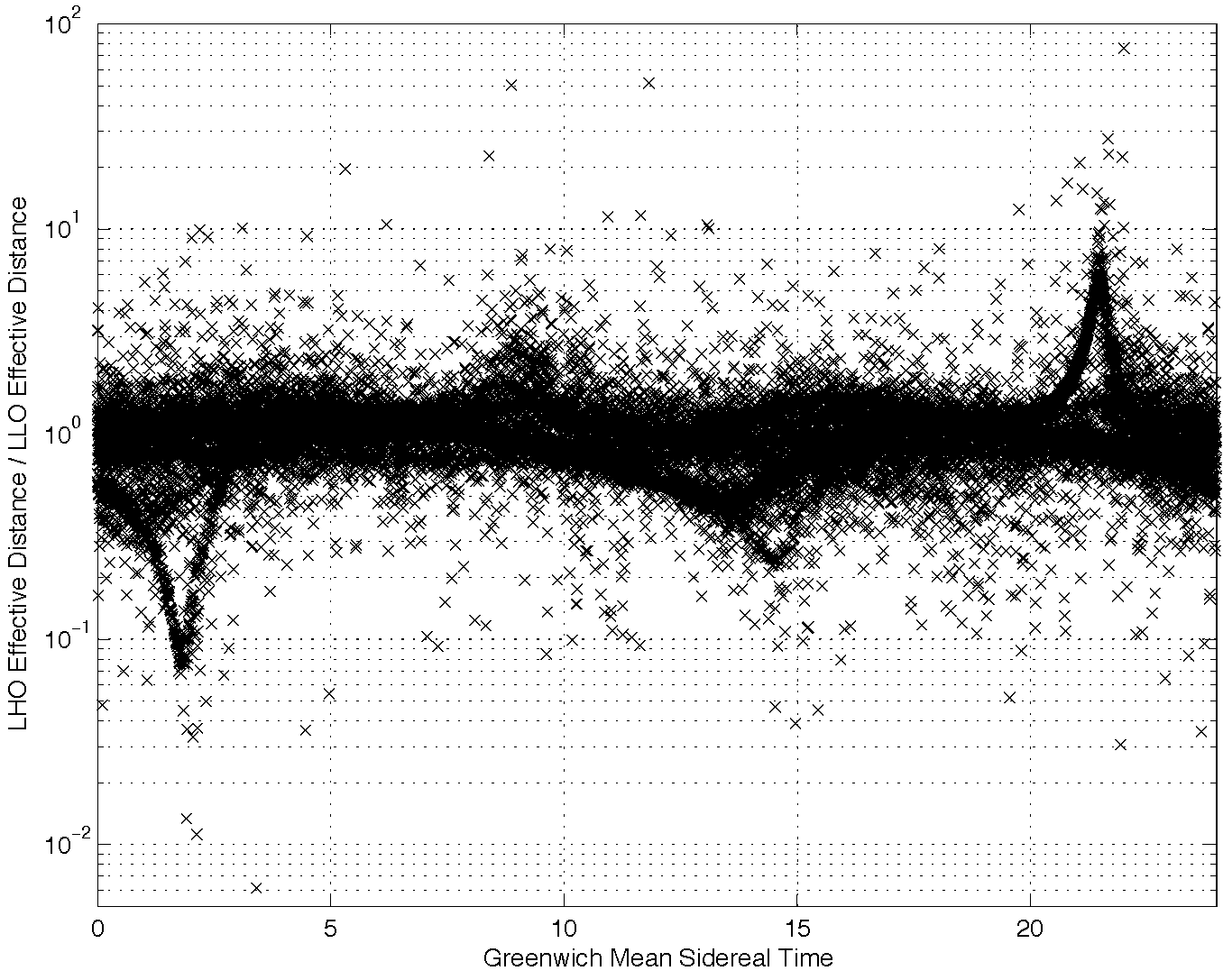}    
\end{center}
\caption[Ratio of Effective Distance of Injected Signals Between Observatories]{%
\label{f:gmst_dist_ratio}
The ratio of the known effective distance of an injected signal in the Hanford
Observatory (LHO) to the known effective distance of an injected signal in the
Livingston Observatory (LLO) as a function of Greenwich Mean Sidereal Time.
The slight misalignment of the interferometers at the two different
observatories due to the curvature of the earth causes the antenna pattern of
the detectors to differ. As a result the distance at which a binary system
appears is different in each detector, even in the absence of noise.  The
ratio of effective distances can be significant, so this precludes the use of
an amplitude cut when testing for inspiral trigger coincidence between
different observatories.
}
\end{figure}

\begin{table}[p]
\begin{center}
\begin{tabular}{lcccccccc}
H1 Data Quality Cut      &$T_\mathrm{total}$&$T_\mathrm{play}$&$T_\mathrm{done}$&$\rho>8$&$\rho>10$&$\rho>12$ \\
(1)                      &(2)     &(3)    &(4)    &(5)     &(6)     &(7)\\\hline
ASQ\_LOWBAND\_OUTLIER    &  14741 &  1990 &  1536 &   625  &  178   &   2 \\
ASQ\_OUTLIER\_CLUSTER    &  20407 &  1800 &  1800 &     0  &    0   &   0 \\
ASQ\_OUTLIER\_CORRELATED &   3126 &   558 &   456 &   390  &  167   &   2 \\
ASQ\_UPPERBAND\_OUTLIER  &  22817 &  1876 &  1876 & 15435  &10159   &7574 \\
AS\_PD\_SATURATION       &     72 &     5 &     0 &     0  &    0   &   0 \\
MICH\_FILT               & 118807 & 11400 & 11400 &  4443  & 3922   &3185 \\
\hline\hline
\\
H2 Data Quality Cut      &$T\mathrm{total}$&$T_\mathrm{play}$&$T_\mathrm{done}$&$\rho>8$&$\rho>10$&$\rho>12$ \\
(1)                      &(2)     &(3)    &(4)    &(5)     &(6)     &(7)  \\\hline
AS\_PD\_SATURATION        &    4   &   0   &   0  &    0  &    0   &   0 \\
MICH\_FILT                &64368   &6570   &5648  & 1294  &  164   &   7 \\
\hline\hline
\\
L1 Data Quality Cut      &$T\mathrm{total}$&$T_\mathrm{play}$&$T_\mathrm{done}$&$\rho>8$&$\rho>10$&$\rho>12$ \\
(1)                      &(2)     &(3)    &(4)    &(5)     &(6)     &(7)  \\\hline
ASQ\_LARGEP2P            &   2699 &   380 &     0  &    0 &    0  &    0 \\
ASQ\_OUTLIER\_CORRELATED &    840 &    60 &    60  &    0 &    0  &    0 \\
AS\_PD\_SATURATION       &    646 &    61 &    10  &  813 &  119  &    6 \\
MICH\_FILT               & 203539 & 21696 & 17794  & 6393 &  497  &   32 \\
NONSTAND\_CTRLS          &   4020 &   843 &    18  &    0 &    0  &    0 \\
\end{tabular}
\end{center}
\caption[Inspiral Triggers Generated at Times of Data Quality Cuts]{
\label{t:s2dqresults}
The table shows the inspiral triggers generate from science mode data with the
mandatory data quality cuts applied. For each discretionary data quality cut
applied a given interferometer (1), the amount of time that would be excluded
from the total science mode data by the cut is given (2). Since we tune data
quality cuts on playground data, the amount of playground time excluded is
also shown (3) and the amount of playground data analyzed for triggers (4).
These may differ for reasons explained in section \ref{ss:datamanagement}. The
number inspiral triggers generated when a particular data quality cut is
active is shown for different signal-to-noise thresholds (5--7). To generate
the triggers, interferometer data was high passed above $50$~Hz in the time
domain and a low frequency cutoff of $70$~Hz was applied to frequency domain.
Template banks were generated with a minimal match of $0.97$ and the
signal-to-noise threshold for the matched filter was set to $\rho_\ast = 8$. A
$\chi^2$ veto with $8$ bins applied with a threshold of $\chi^2 < 20 (8 +
0.03^2 \rho^2)$. 
}
\end{table}

\begin{table}[p]
\begin{center}
\begin{tabular}{ll}
Discretionary Data Quality Cut  & Applied \\\hline\hline
MICH\_FILT                 & No \\
\multicolumn{2}{l}{\parbox{\linewidth}{\footnotesize The cut would exclude a
large number of triggers, but would reduce the amount of data in the search
significantly. It was decided not to apply this cut and to try and exclude
false triggers from these times by a combination of coincidence, vetoes and
reducing the $\chi^2$ threshold.}}\\
\\
AS\_PD\_SATURATION        & Yes \\
\multicolumn{2}{l}{\parbox{\linewidth}{\footnotesize Clear correlation with
inspiral triggers with large signal-to-noise ratios in L1 and the study
described in section \ref{ss:photodiode} suggest that this should be used. The
lack of correlated trigger in H1 was due to the fact that the playground did
not sample any times with photodiode situations.\baselineskip=14pt}}\\
\\
ASQ\_LARGEP2P             & No \\
\multicolumn{2}{l}{\parbox{\linewidth}{\footnotesize A loud inspiral signal
could trigger this cut, so it is unsafe for use.\baselineskip=14pt}} \\
\\
NONSTAND\_CTRLS           & Yes \\
\multicolumn{2}{l}{\parbox{\linewidth}{\footnotesize Advice from experimental
team advised that detections made during this time could not be
trusted.\baselineskip=14pt}} \\
\\
ASQ\_OUTLIER\_CLUSTER     & No \\
\multicolumn{2}{l}{\parbox{\linewidth}{\footnotesize Not 
well correlated with inspiral triggers. \baselineskip=14pt}} \\
\\
ASQ\_OUTLIER\_CORRELATED  & No \\
\multicolumn{2}{l}{\parbox{\linewidth}{\footnotesize Not 
well correlated with inspiral triggers.\baselineskip=14pt}} \\
\\
ASQ\_LOWBAND\_OUTLIER     & No \\
\multicolumn{2}{l}{\parbox{\linewidth}{\footnotesize Not 
well correlated with inspiral triggers.\baselineskip=14pt}} \\
\\
ASQ\_UPPERBAND\_OUTLIER   & Yes \\
\multicolumn{2}{l}{\parbox{\linewidth}{\footnotesize Times with high upper band
noise in H1 are clearly correlated with high signal-to-noise ratio triggers.
In order to prevent the cut from begin triggered by real signals we also
require that the cut is on for more that $180$ seconds. The longest inspiral
signal in the S2 analysis is $52$ seconds.\baselineskip=14pt}}\\
\end{tabular}
\end{center}
\caption[Choice of Data Quality Cuts in S2]{%
\label{t:s2dqchoice}
The final selection and justification of discretionary data quality cuts for
the S2 binary neutron star and binary black hole MACHO searches.}
\end{table}

\begin{table}[p]
\begin{center}
\begin{tabular}{llll}
Interferometer&Start&End&Duration\\
\hline
L1 &  730000000 &730010000 &  10000  \\
L1 &  731001000 &731006000 &   5000  \\
L1 &  732000000 &732003000 &   3000  \\
\hline
H1 &  730004000 &730013000 &   8000  \\
H1 &  731000000 &731002500 &   2500  \\
H1 &  732000000 &732003000 &   3000  \\
\hline
H2 &  730002500 &730008000 &   5500  \\
H2 &  731004500 &731007500 &   2500  \\
H2 &  732000000 &732003000 &   3000
\end{tabular}
\end{center}
\caption[Fake Science Segments Used to Test DAG Generation]{
\label{t:fakesegslist}
The fake science segments used to construct the DAG shown in figure
\ref{f:fake_segs_dag}.
}
\end{table}

\begin{sidewaysfigure}[p]
\begin{center}
\hspace*{-0.2in}\includegraphics[width=\linewidth]{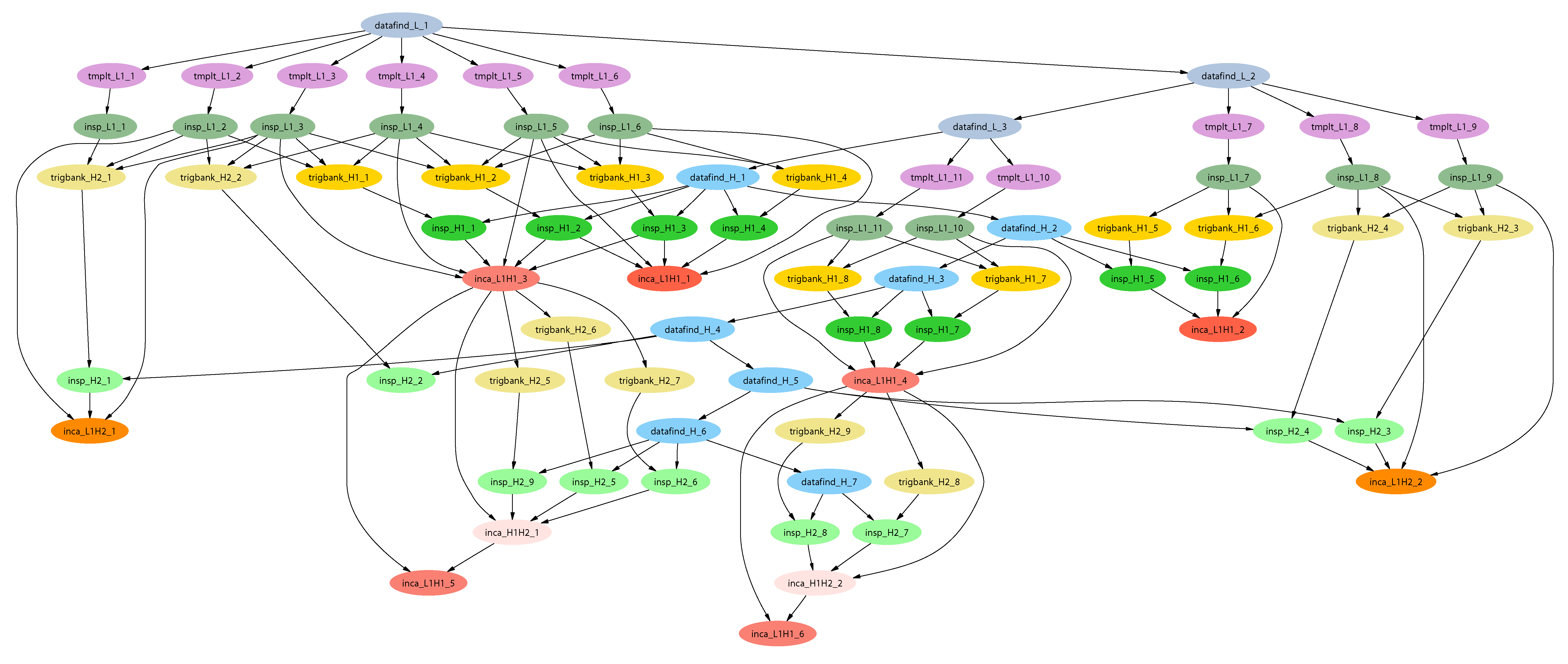}
\end{center}
\caption[DAG Generated from Fake Segments]{%
\label{f:fake_segs_dag}
The DAG generated from the pipeline shown in figure \ref{f:pipeline} and the
fake science segment list described in section \ref{ss:dag}. The figure shows
the structure of the DAG with all job dependencies needed to execute the S2
pipeline. Note that it appears that there are several LHO master chunks
analyzed that do not need to be filtered for a zero time lag.  These are added
to the DAG to ensure that all the data necessary for a background estimation
with a maximum time side of 500 seconds is analyzed.
}
\end{sidewaysfigure}

\Chapter{Hardware Signal Injections}
\label{ch:hardware}

\begin{figure}[p]
  \vspace{5pt}
  \begin{flushright}
    \includegraphics[width=\textwidth]{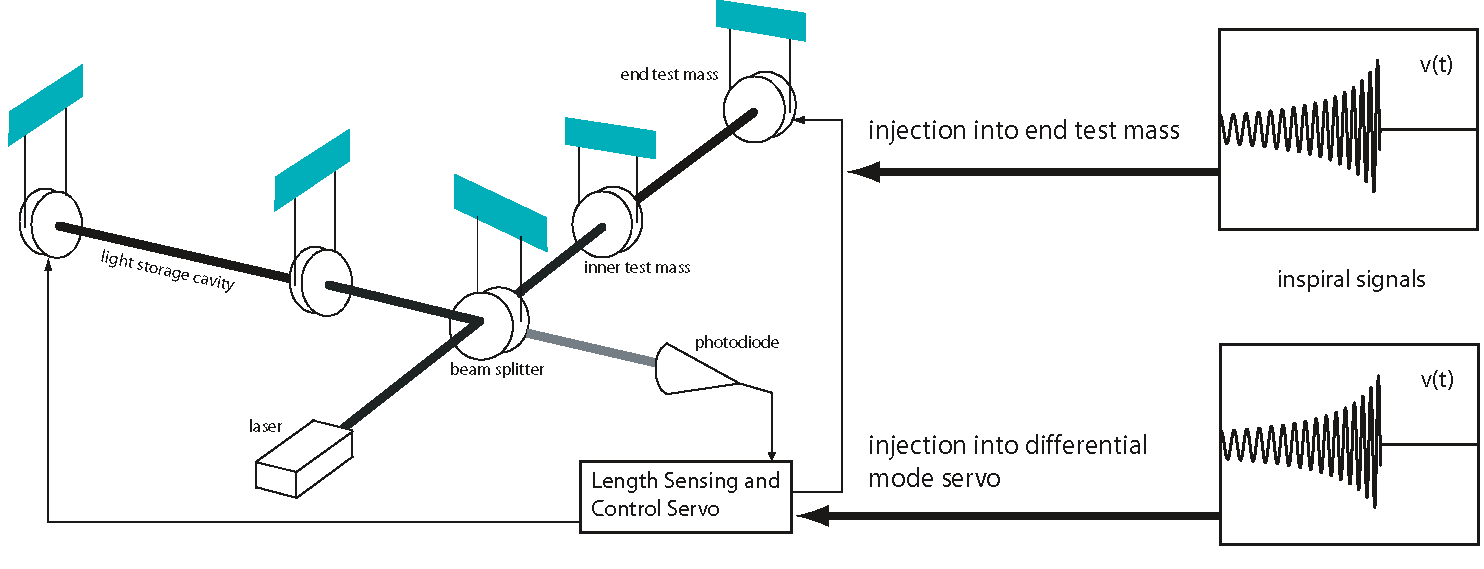}    
  \end{flushright}
  \caption[Schematic of LIGO Interferometer Showing Injection Points]{%
\label{f:ifo_inj}
  A schematic diagram of the LIGO interferometer showing the injection points
  used in S1 hardware injections. Inspiral signals were injected either
  directly into the end test mass drive of one arm or into the differential
  mode servo, and this into both arms. Care was taken to ensure that the
  correct transfer function, $T(f)$, was used in each case.
  }
\end{figure}

Gravitational radiation incident on the LIGO interferometers from an
inspiralling binary will cause the test masses to move relative to each other.
This produces a differential change in length of the arms as described in
section \ref{s:effect}.  \emph{Injection} is the process of adding a waveform to
interferometer data to simulate the presence of a signal in the noise. We use
injections to measure the performance of the binary inspiral analysis
pipeline as described in section \ref{s:eff}.  \emph{Software injections},
which add a simulated signal to the data after it has been recorded, are used
for efficiency measurements. Since they performed \emph{a posteriori} the
interferometer is not affected while it is recording data.  Alternatively, a
simulated signal can be added to the interferometer control system to make the
instrument behave as if an inspiral signal is present.  The interferometer
Length Sensing and Control system has excitation points which allow arbitrary
signals to be added into the servo control loops or to the drives that control
the motion of the mirrors\cite{LIGOS1instpaper}.  We call this \emph{hardware
injection}; the data recorded from the instrument contains the simulated
signal. Figure \ref{f:ifo_inj} shows the hardware injection points on a
schematic diagram of the interferometer and length sensing and control loop.

Analysis of hardware injections allows us to ensure that the analysis pipeline
is sensitive to real inspiral signals and validates the software injections
used to test the pipeline efficiency.  In order to perform an accurate upper
limit analysis for binary inspirals, we must measure the efficiency of our
pipeline. That is, we inject a known number of signals into the pipeline and
determine the fraction of these detected.  Injecting signals into the
interferometer for the duration of a run is not practical and would
contaminate the data, so we use the analysis software to inject inspiral
signals into the data.  By comparing software and hardware injections we
confirm that software injections are adequate to measure the efficiency of the
upper limit pipeline.

Hardware injections provide a very complete method of testing the inspiral
detection pipeline. By recovering the physical parameters of an injected
signal, we test our understanding of all aspects of the pipeline, including
the instrumental calibration, the filtering algorithm and veto safety. We
injected inspiral signals immediately after the first LIGO science run (S1) in
September 2002. The resulting data was analyzed using the
software tools used to search for real signals.  In this chapter, we describe
the results of analysis of the S1 hardware injections. The analysis pipeline
used in S1 differs from that used in S2\cite{LIGOS1iul}. Here we are examining
the response of the filtering code to the hardware injections, however, and so
the differences between the S1 and S2 pipelines are unimportant.

\section{Injection of the Inspiral Signals}
\label{s:injecting}

To inject the signals, we generate the interferometer strain $h(t)$ produced
by an inspiralling binary using the restricted second order post-Newtonian
approximation in the time domain\cite{Blanchet:1996pi}.  The LSC calibration group
supplies a transfer function $T(f)$ which allows us to construct a signal
$g(t)$ that produces the desired strain when it is injected into the
interferometer.  The transfer function $T(f)$ should be identical to the
actuation function $A(f)$ described in section \ref{ss:calibration}, however
in S1 this was simplified to contain only the pendulum response of the mirrors,
given by 
\begin{equation}
T(f) = \frac{L}{C}\frac{f^2}{f_0^2}
\end{equation}
where $L$ is the length of the interferometer, $C$ is the calibration of the
excitation point in nm/count and $f_0$ is the pendulum frequency of the test
mass. Damping is neglected as it is unimportant in the LIGO frequency band.
The code used to generate the hardware injections is the same as that used
for software injections; only the transfer function used to generate the
injected signal differs since we are injecting into the control signal $g$
rather than the error signal $v$.

During S1, we injected signals corresponding to an optimally oriented binary.
Injections of a $1.4\,M_\odot$ inspiralling binary at distances from $10$
kpc to $80$ kpc were used to test the neutron star analysis.  We also injected
signals from a $1.4,\,4.0\,M_\odot$ binary and several $1.4,1.4\,M_\odot$
binaries at closer distances.  These signals were injected into the
differential mode servo and directly into an end test mass drive.

\section{Detection of the Injected Signals}
\label{s:detection}

\begin{figure}[p]
  \vspace{5pt}
  \begin{flushright}
    \includegraphics[width=\textwidth]{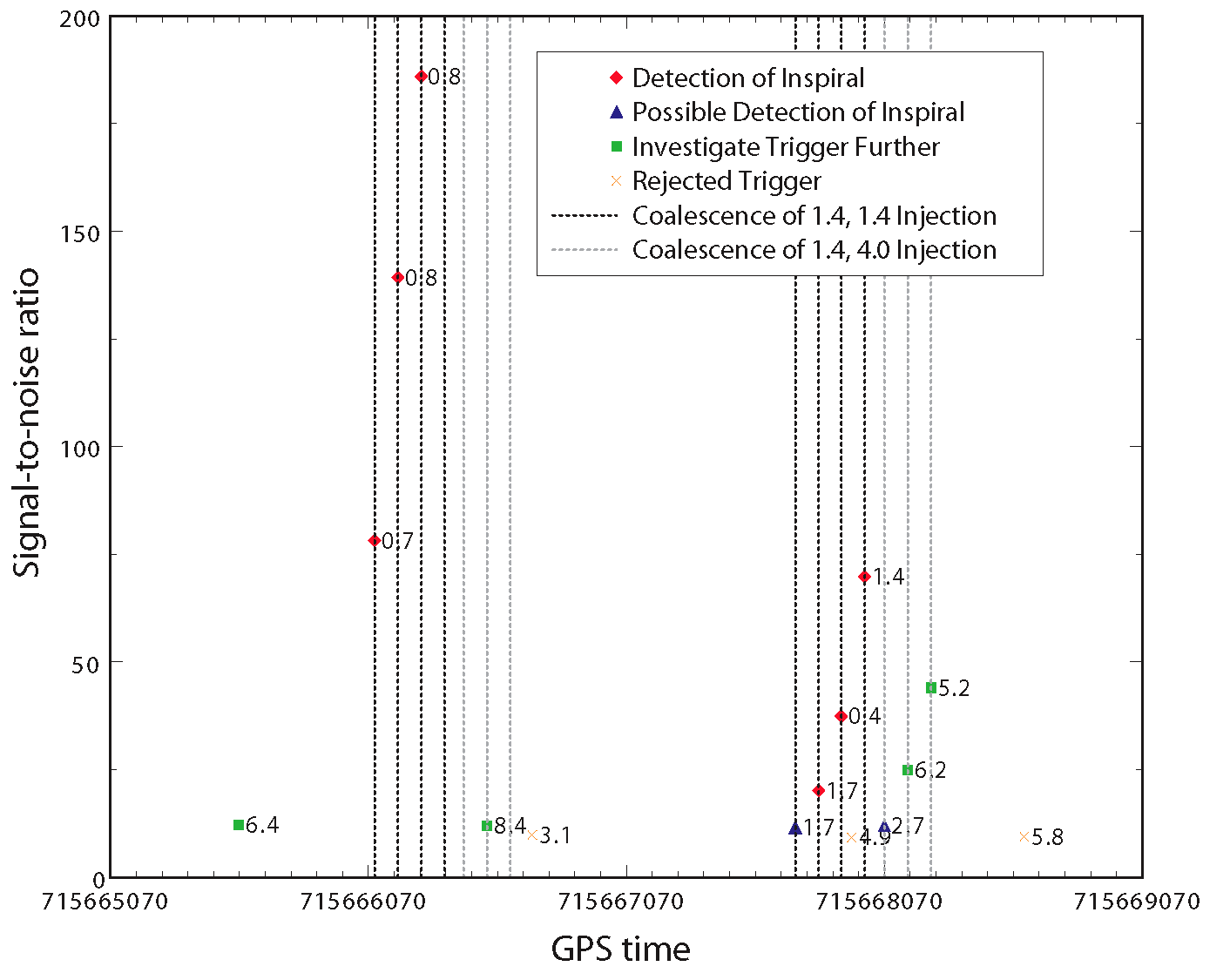}    
  \end{flushright}
  \caption[Candidate Events from Hardware Injections]{%
\label{f:inj_snr}
The candidate events generated by processing 4000 seconds of data from the
Livingston 4 km interferometer through the S1 analysis pipeline.  This data
included two sets of injections; the known coalescence times are indicated by
the dashed vertical lines. The signal-to-noise ratio is plotted and the value
of the $\chi^2$ veto is shown next to the candidate event.
  }
\end{figure}

Figure \ref{f:inj_snr} shows the events generated by processing 4000 seconds
of data from the Livingston 4 km interferometer (L1) on 10 September 2002
during the post-run hardware injections.

The first set of injections were large amplitude signals used to verify the
inspirals were being correctly injected. We ignore these and concentrate on
the second set, which were at more appropriate distances.  We only consider
the 1.4 solar mass inspiral injections, as the 1.4,4.0 injection lies
outside the template bank space used in the S1 binary neutron star analysis.

It can be seen that all of the hardware injections are identified as candidate
events since they have high signal-to-noise ratios and values of the $\chi^2$
test lower than $5$, which was the threshold used in the S1 analysis
pipeline\cite{LIGOS1iul}. Some of the 1.4,4.0 injections are also flagged
for further investigation as they cause templates inside the bank to ring, but
have high $\chi^2$ values as they are not exactly matched.

Since we know the exact coalescence time of the injected waveform, we can
compare this with the value reported by the search code and ensure that the
search code is reporting the correct time.  The known and measured parameters
for the second set of $1.4,1.4\ M_\odot$ injections are shown in table
\ref{t:triggers}. The raw data is resampled to $4096$ Hz before being
filtered. For each of the signals injected, we were able to detect the
coalescence time of the injection to within one sample point of the correct
value at $4096$ Hz, which is consistent with the expected statistical error
and confirms that the pipeline has not introduced any distortion of the
signals.

\section{Checking the Instrumental Calibration}
\label{s:calibration}

Calibration measurements of the interferometers were performed before and
after the run; these are the reference calibrations. In general, the
calibration changes due to changes in the alignment on time scales of minutes.
This variation is encoded in the parameter $\alpha$ which is
monitored using a sinusoidal signal injected into the
detector (see section \ref{ss:calibration}). $\alpha$ is used as input
to the data analysis pipeline and varied between $0.4$ and $1.4$ during S1.
Data in S1 was analyzed in 256 second segments.  For each 256 seconds of data
starting at time $t_0$, we construct the calibration, $R(f;t_0)$ by using
$\alpha(t_0)$ and a reference calibration.  $R(f;t_0)$ is then used to
calibrate the 256 seconds of data.

\begin{figure}[p]
  \vspace{5pt}
  \begin{flushright}
    \includegraphics[width=\textwidth]{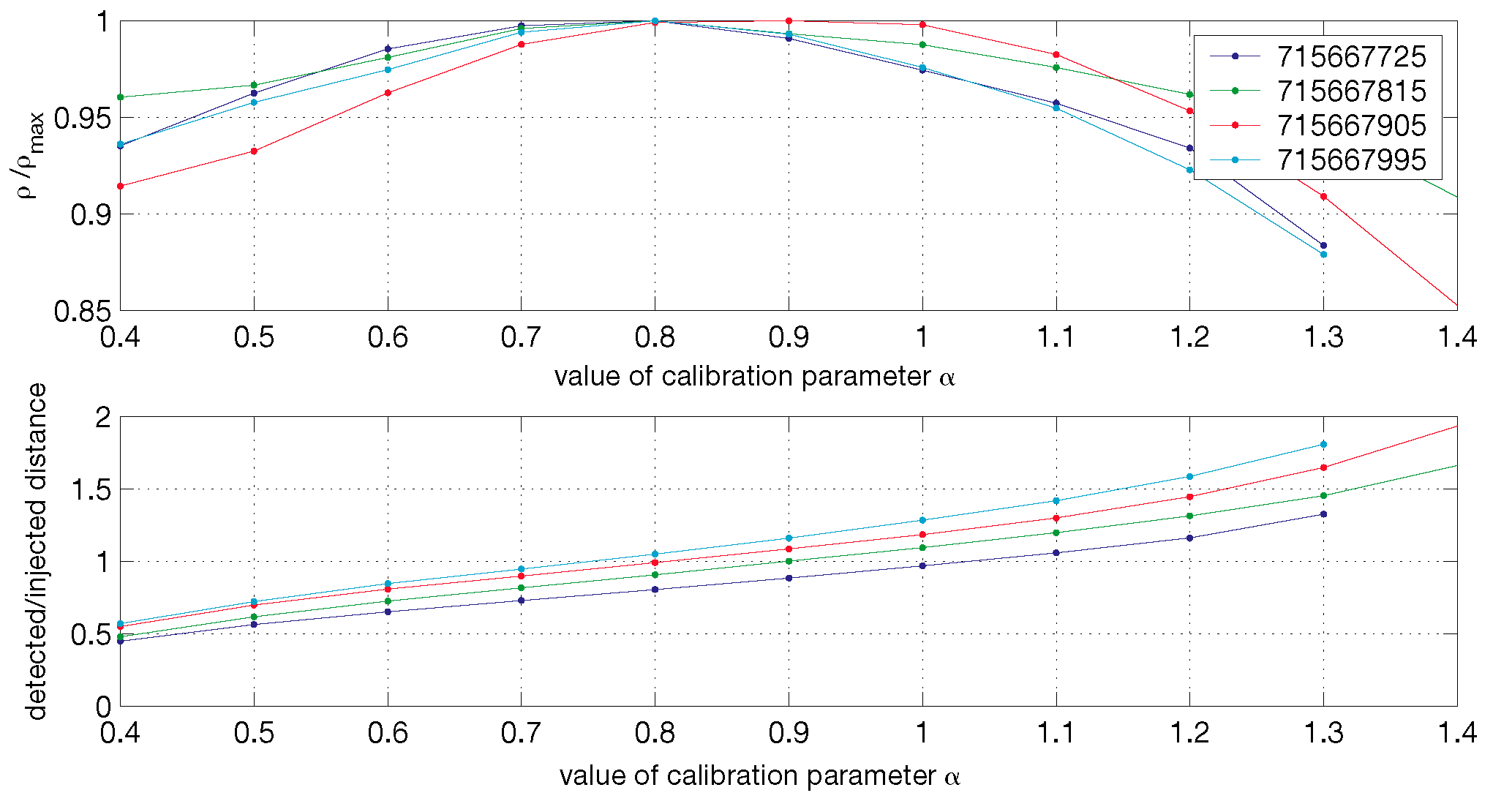}    
  \end{flushright}
  \caption[Study of Calibration Using Hardware Injections]{%
\label{f:calibration}
  Each curve corresponds to a hardware injection at the given GPS time. We
  re-analyze each injection with different calibrations to show how the
  detected quantities vary with $\alpha$. The upper plot shows the ratio of
  signal-to-noise ratio, $\rho$, to its maximum value, $\rho_{\mathrm{max}}$.
  The lower plot shows the ratio of the detected distance to the known distance of
  the hardware injection.
  }
\end{figure}

Figure \ref{f:calibration} shows a set of injections into the Livingston
interferometer analyzed with different calibrations generated by varying the
value of $\alpha$. We expect that the signal-to-noise varies quadratically and
the effective distance varies linearly with changes in $\alpha$\cite{Allen:1996}.
This is confirmed by the injections.  There is no single value of $\alpha$
that gives the correct effective distance for all the injections; this is
consistent with the estimated systematic errors in the calibration.
Unfortunately the calibration line was not present during the time the
hardware injections were performed, so we cannot directly compare a measured
calibration with the result of the injections.

\section{Safety of Vetoes}
\label{s:safety}

\begin{figure}[p]
  \vspace{5pt}
  \begin{flushright}
    \includegraphics[width=\textwidth]{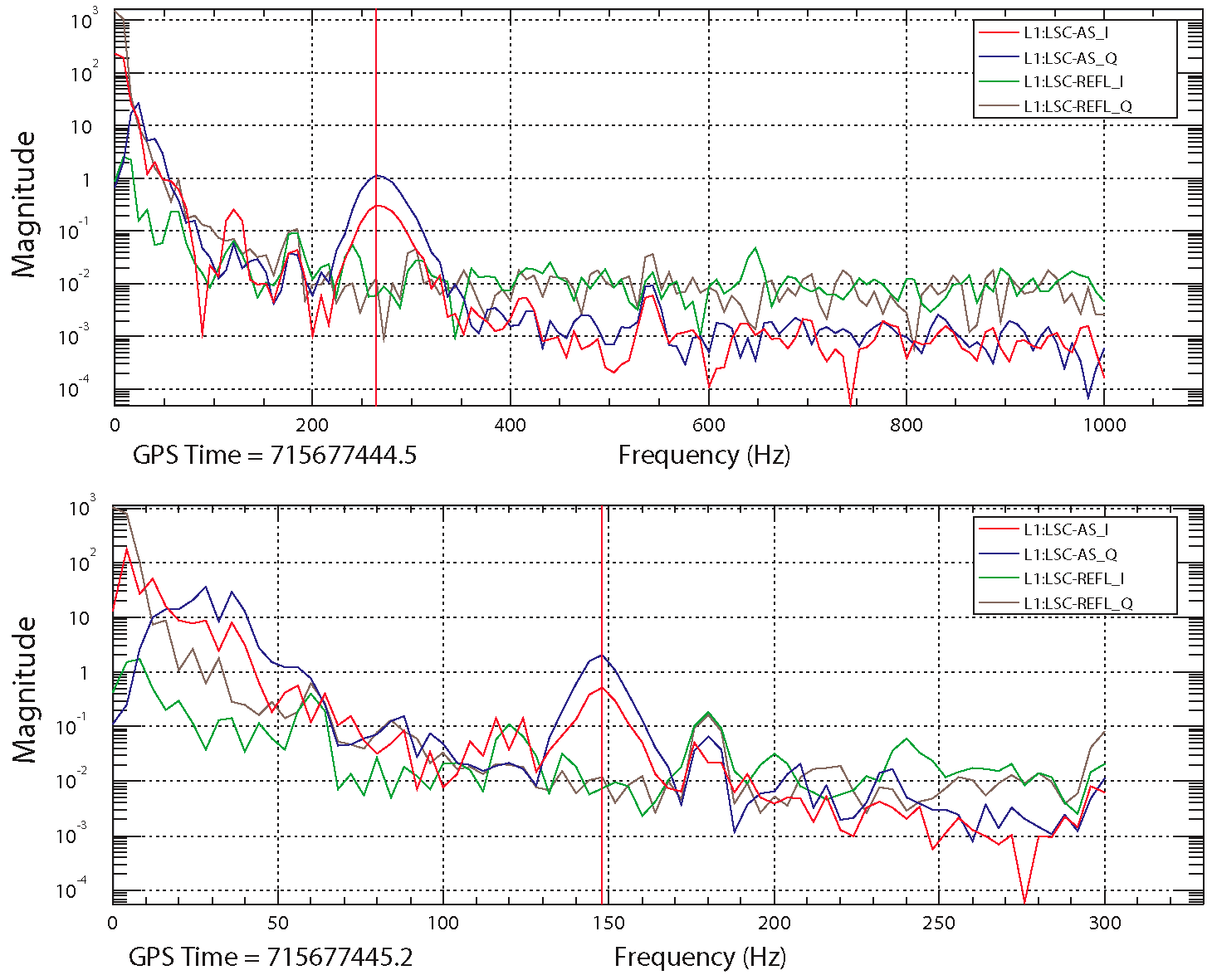}    
  \end{flushright}
  \caption[Study of Veto Safety Using Hardware Injections]{%
\label{f:veto_safe}
  Power spectra of the gravitational wave channel LSC-AS\_Q and the auxiliary
  channels LSC-AS\_I, LSC-REFL\_I and LSC-REFL\_Q during a 
  hardware injection. The broad peak in the spectrum is the inspiral signal
  and the two power spectra taken at subsequent times show it sweeping across
  the band as the frequency of the inspiral signal increases with time.
  }
\end{figure}

During construction of the the inspiral pipeline we considered using inspiral
triggers found in auxiliary interferometer channels as vetoes on triggers in
the gravitational wave channel. Concern was raised that a real inspiral signal
may couple between these channels and a real signal may be inadvertently
vetoed.  To check this, we examined coupling between the channels at the time
of an injection.  Figure \ref{f:veto_safe} shows the power spectra of the
gravitational wave channel, LSC-AS\_Q, and the auxiliary channels that we
considered using as vetoes during S1: LSC-AS\_I, LSC-REFL\_I and
LSC-REFL\_Q.  The injected inspiral signal can clearly be seen coupling
to the auxiliary channel LSC-AS\_I, but there is no obvious coupling
between the injected signal and LSC-REFL\_I or LSC-REFL\_Q. This
led us to discard LSC-AS\_I as a possible veto channel in the S1
analysis. Similar studies have been performed for the S2 data when auxiliary
channels are proposed as veto channels.

\newpage

\begin{table}[p]
  \begin{flushright}
  \begin{tabular}{l|l|c|c}
  End time of Injection&End Time of Detection&$\rho$&$\chi^2$\\
  \hline
  $04:35:12.424928$ & $04:35:12.424927$ & $11.623546$ & $1.653222$ \\
  $04:36:42.424928$ & $04:36:42.425171$ & $20.230101$ & $1.671016$ \\
  $04:38:12.424928$ & $04:38:12.424927$ & $37.488770$ & $0.443966$ \\
  $04:39:42.424928$ & $04:39:42.424927$ & $69.815262$ & $1.375486$ \\
  \end{tabular}
  \end{flushright}
  \caption[Hardware Injections Found by the Analysis Pipeline]{%
\label{t:triggers}
  Hardware injection events found by the inspiral analysis pipeline. End time
  of injection is the known end time of the injected signal and end time of
  detection is the end time of the signal as reported by the analysis
  pipeline. Times are Universal Time (UTC) on 10 September 2002. The values of
  signal-to-noise ratio $\rho$ and $\chi^2$ veto are given for each event.
  }
\end{table}

\Chapter{The Rate of Binary Black Hole MACHO Inspirals in the Halo}
\label{ch:result}

In this chapter, we present the results of a search for gravitational waves
from the inspiral of binary black hole MACHOs in data from the second LIGO
science run (called S2).  The goal of the search is the
detection of the gravitational waves. In the absence of a detection, however, we
place an upper limit on the rate of inspiralling BBHMACHOs. This
limit may be compared to the predicted rate of $5 \times 10^{-2} \times
2^{\pm 1}$ discussed in chapter \ref{ch:macho}.  

Analysis of the full S2 data set for gravitational waves from inspiralling
binary black hole MACHOs is complete and the result of this search will appear
in \cite{S2Macho:2004}. Since this result is currently embargoed pending LIGO
Scientific Collaboration internal review, we instead present the result of the
search on the playground data. No gravitational waves from BBHMACHO inspirals
were found in the playground, so in section \ref{s:s2upperlimit} we compute an
upper limit on the rate of binary black hole MACHO inspirals in the playground
data.  Although this result is statistically biased, as it is computed from
data used to tune the pipeline, it allows us to make a reasonable prediction
of the upper limit available using the full S2 data and assuming no BBHMACHO
signals are detected in the full data set.

In section \ref{s:s2run} we describe the data sample used in the analysis.
Section \ref{s:s2tuning} describes how the parameters of the search listed in
the previous chapter were tuned on the playground data. Section \ref{s:monte}
described the Monte Carlo simulations used to measure the efficiency of the
pipeline. In section \ref{s:s2background} we describe the background observed
in the S2 data.

\section{The Second LIGO Science Run}
\label{s:s2run}

All three LIGO detectors operated during the second science run, referred to
as S2, which lasted for 59 days (1415 hours) from February 14 to April 14,
2003.  Although the detectors were manned by operators and scientific monitors
around the clock, the amount of data flagged for scientific analysis was
limited by environmental factors (especially high ground motion at LLO and
strong winds at LHO), occasional equipment failures, and periodic special
investigations.  The total amount of science data obtained was 536 hours for
L1, 1044 hours for H1, and 822 hours for H2.

\begin{figure}[p]
\begin{center}
\includegraphics[width=0.75\linewidth]{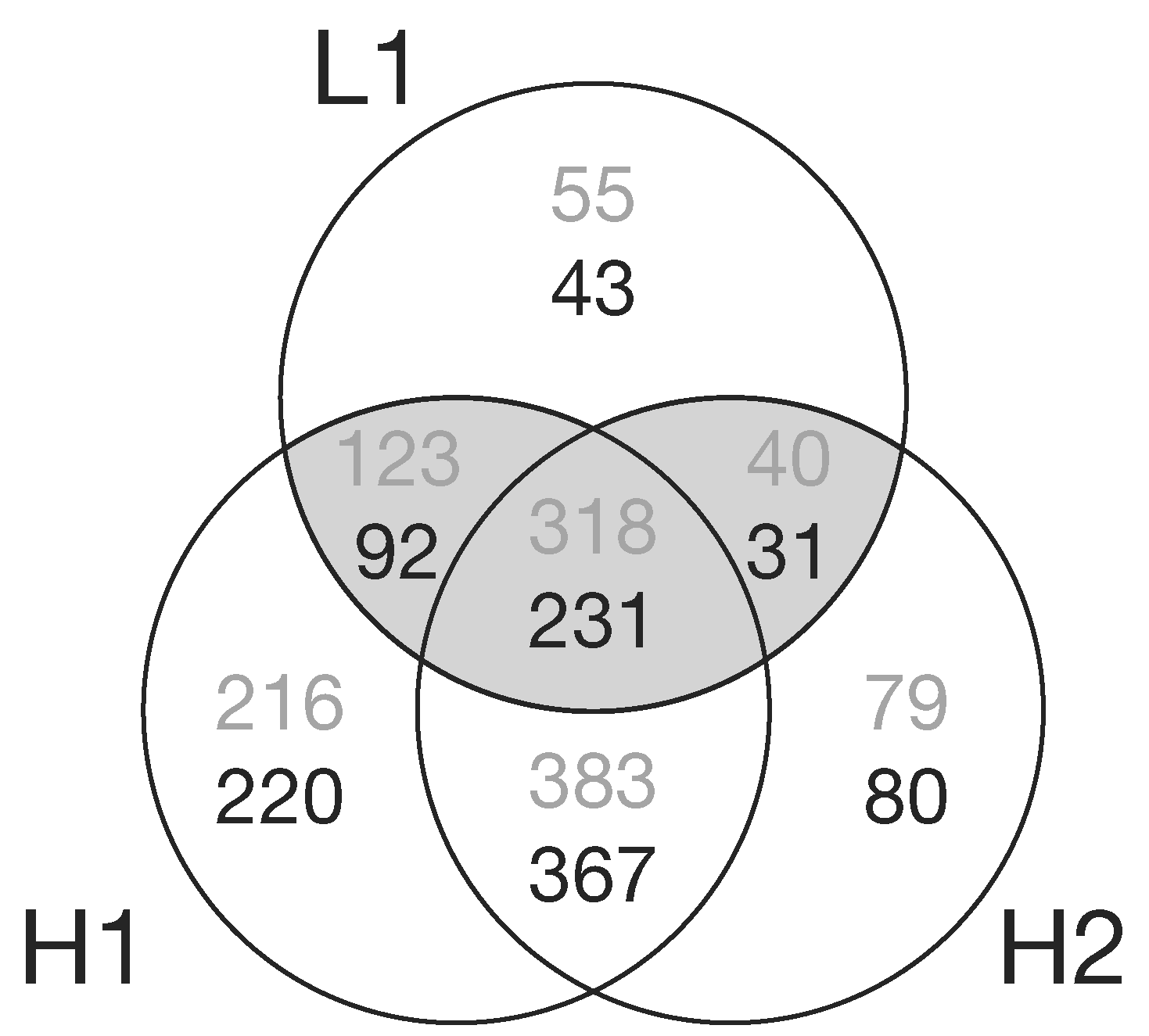}
\end{center}
\caption[Amounts of Single and Coincident Interferometer Data in S2]{%
\label{f:S2times}%
The Venn diagram shows the number of hours that each detector combination was
operational during the S2 run.  The upper number gives the amount of time the
specific instruments were operational.  The lower number gives the total
non-playground time which was searched for inspiral triggers.  The shaded
region corresponds to the data used in the S2 MACHO search.}
\end{figure}

The analysis described in this thesis uses data collected while the LLO
detector was operating at the same time as one or both of the LHO detectors in
order to make use of the triggered search pipeline.  Science mode data during
which both H1 and H2 were operating but L1 was not, amounting to 383 hours,
was not used in this analysis because of concerns about possible
environmentally-induced correlations between the data streams of these two
co-located detectors. This data set, as well as data collected while only one
of the LIGO detectors was in science mode, will be combined with data from the
third LIGO science run in a future analysis. Figure~\ref{f:S2times} shows a
breakdown by interferometer of the data recorded during S2. The data used in
this search is indicated by the shaded region.

\section{Tuning the Analysis Pipeline}
\label{s:s2tuning}

The entire analysis pipeline was explored first using the playground data set
in order to tune the the various thresholds and other parameters. The goal of
tuning the pipeline is to maximize the efficiency of the pipeline to detection
of gravitational waves from binary inspirals without producing an excessive
rate of spurious candidate events. In the absence of a detection, a pipeline
with a high efficiency and low false alarm rate allows us to set the best
upper limit. It should be noted, however, that our primary motivation is to
enable reliable detection of gravitational waves. The efficiency is measured
by Monte Carlo simulations in which signals from the hypothetical population
are added to the data and then sought. This approach accounts for any
systematic error associated with the methods used in our pipeline.  Note that
another factor in the tuning of the pipeline are the available computational
resources. We would like to be able to complete the search in less time than
the length of the data being analyzed, so that real-time searches are possible
when the interferometers are taking continuous data. For this reason, certain
tuning decisions are based on the computational efficiency of the pipeline;
these decisions will be clearly identified below.

Prior to commencing the binary black hole MACHO search, a search for
inspiralling binary neutron stars (BNS) was conducted on the S2 data using the
pipeline described in chapter~\ref{ch:pipeline}\cite{LIGOS2iul}. The mechanics
of the BNS search are very similar to those described here, except that the
template bank covers binaries with $1.0\,M_\odot < m_1, m_2 < 3.0\,M_\odot$,
where $m_1$ and $m_2$ are the masses of each object in the binary. Since the
BNS and binary black hole MACHO searches are very similar, and share the same
playground data, we may use the parameters of the BNS search (which were tuned
on the playground data) as a starting point for tuning the binary black hole
MACHO search. 

There are two sets of parameters that we are able to tune in the pipeline: (i)
the single interferometer parameters which are used in the matched filter and
$\chi^2$ veto to generate inspiral triggers in each interferometer, and (ii)
the coincidence parameters used to determine if triggers from two
interferometers are coincident. The single interferometer parameters include
the signal-to-noise threshold $\rho^\ast$, the number of frequency sub-bands in
the $\chi^2$ statistic $p$, the $\chi^2$ cut threshold $\Xi^\ast$,  and the
coefficient on the signal-to-noise dependence of the $\chi^2$ cut, i.e.
$\delta^2$ in equation (\ref{eq:chisqthresholdtest}). These are tuned on a
per-interferometer basis, although some of the values chosen are common to two
or even three detectors.  The coincidence parameters are the time coincidence
window $\delta t$ for triggers, the mass parameter coincidence window $\delta
m$ and the effective distance cut parameters $\epsilon$ and $\kappa$ in
equation.~(\ref{eq:eff_dist_test}).  Due to the nature of the triggered search
pipeline, parameter tuning was carried out in two stages. We first tuned the
single interferometer parameters for the primary detector (L1).  We then used
the triggered template banks (generated from the L1 triggers) to explore the
single interferometer parameters for the less sensitive Hanford detectors.
Finally the parameters of the coincidence test were tuned.

\subsection{Template Bank Generation}
\label{ss:tunebank}

Recall that the number of templates needed to cover a given region of
parameter space at a specified minimal match fluctuates as the shape of the
noise power spectrum changes. The greater the sensitivity at low frequencies,
relative to higher frequencies, the larger the template bank (see section
\ref{ss:templatebank}). The computational resources available for the MACHO
search are limited and the computational cost is proportional number of
templates in the bank. We therefore tuned the template bank parameters to
allow the search to be completed within the available resources.

\begin{figure}[p]
\begin{center}
\includegraphics[width=\linewidth]{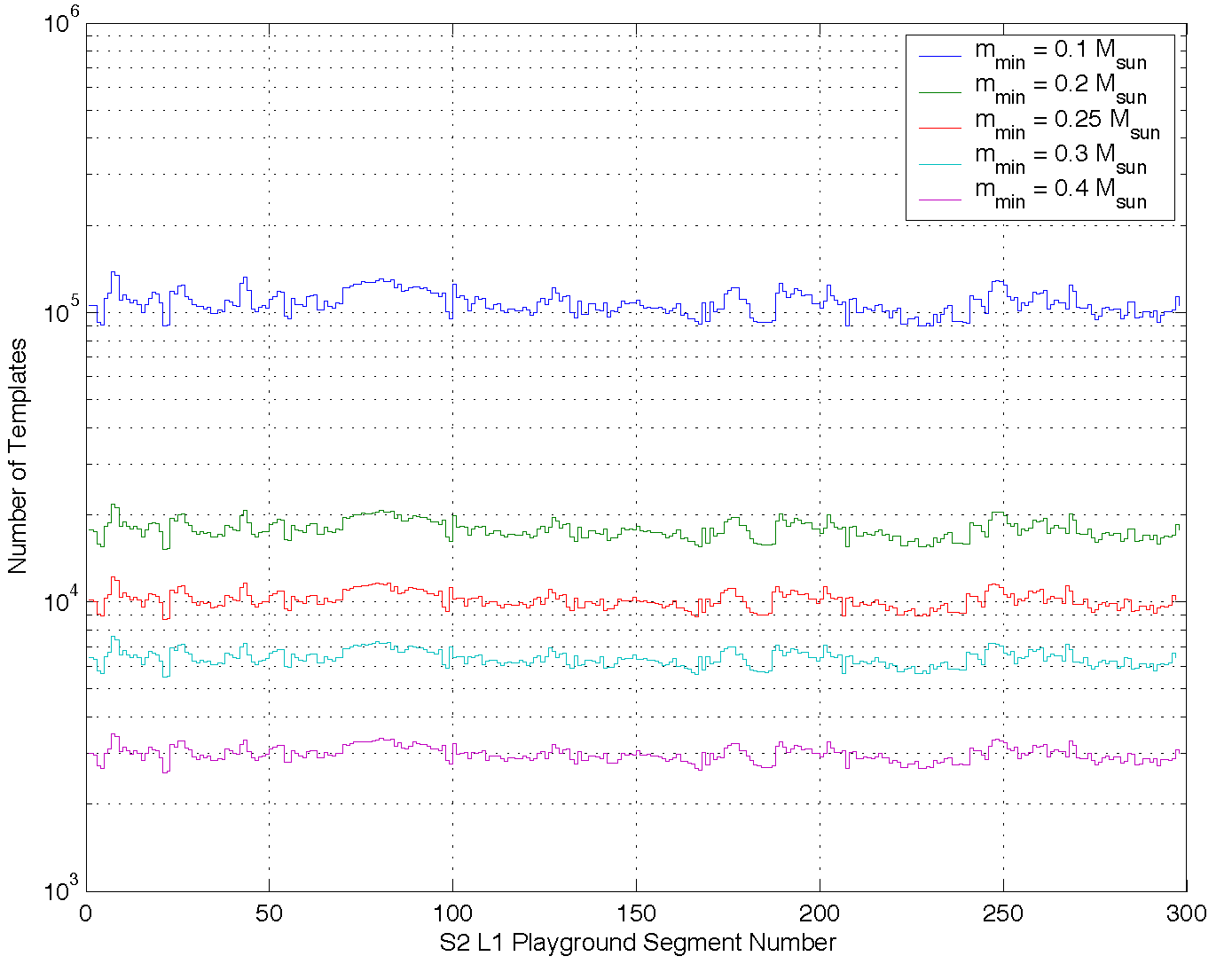}
\end{center}
\caption[MACHO Template Bank Size for Various Lower Masses]{%
\label{f:bank_size}%
The plot shows the size of the template bank, generated with a minimal match
of $97\%$, for various values of $m_\mathrm{min}$. As described in section
\ref{ss:templatebank} the size of the template bank is proportional to
$m_\mathrm{min}^{-8/3}$, where $m_\mathrm{min}$ is the mass parameter of the
smallest equal mass binary in the template bank. Using these data, it was
decided that the lowest mass accessible was $m_\mathrm{min} = 0.2\,M_\odot$,
given the available computational resources.}
\end{figure}

Due to the algorithm used to construct the template bank\cite{Owen:1998dk}, the
smallest mass template in the bank will be the equal mass binary
$(m_\mathrm{min},m_\mathrm{min})$, where $m_\mathrm{min}$ is the (user
specified) minimum binary component mass. Figure~\ref{f:bank_size} shows the
size of the template bank necessary to cover each playground analysis chunk at
a minimal match of $0.97\%$ for several values of $m_\mathrm{min}$.  The
maximum binary component mass in each case is $1\,M_\odot$, so the largest
mass binary in the bank parameter space is $(1,1)\,M_\odot$. For fixed
$m_\mathrm{min}$ the number of templates remains reasonably constant over the
course of the S2 run, but there is a large variation in the number of templates
required as a function of lower mass. The scaling of template number
as a function of lower mass is consistent with that described in
\cite{Owen:1998dk}.

As described in chapter \ref{ch:macho}, the MACHO mass range measured by
microlensing is $0.15\,M_\odot$ to $0.9\,M_\odot$ at $95\%$ confidence. It
would therefore be desirable for the binary black hole MACHO inspiral search
to cover a region of mass parameter space slightly larger than this, say
$0.1\,M_\odot$ to $1.0\,M_\odot$.  It can be seen, however, that almost an
order of magnitude more templates are needed to decrease the lower boundary of
the mass parameter space from $0.2\,M_\odot$ to $0.1\,M_\odot$.  Therefore,
given the computational resources available for the BBHMACHO search, the
lowest mass template in the bank was set to $0.2\,M_\odot$. Note that for the
final search, the match of the template bank was also lowered to $0.95\%$ to
further decrease the number of templates to an average of $14\,179$ per
analysis chunk over the S2 run.  The latter choice is be justified by the
sensitivity to galactic binary black hole MACHOs in S2, as will be seen below.
The size of the inspiral template bank used is shown in
figure~\ref{f:inspiral_summary}.
 
\subsection{Interferometer Sensitivity and Signal-to-noise Thresholds}
\label{ss:snrthreshold}

\begin{figure}[p]
\begin{center}
\includegraphics[width=\linewidth]{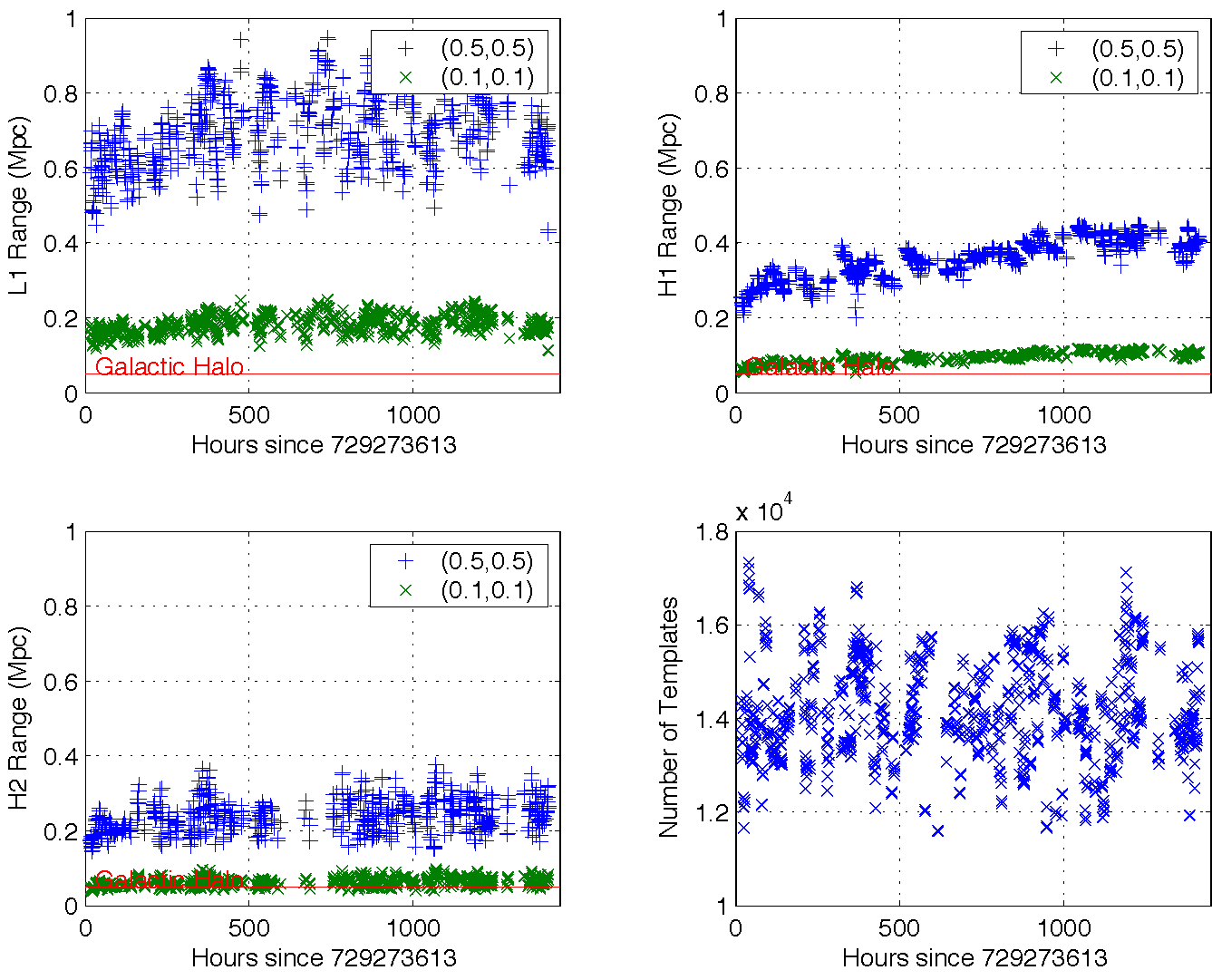}
\end{center}
\caption[S2 Interferometer Ranges and Template Bank Size]{%
\label{f:inspiral_summary}%
The bottom right plot shows the variation in the size of the MACHO search
template bank over the course of the S2 run. As described in the text, the
template bank is generated using L1 data to cover a region of parameter space
from $0.2\,M_\odot$ to $1.0\,M_\odot$ (component mass) at $95\%$ minimal
match. This template bank is used to filter the L1 data in the triggered
search pipeline. The other three plots show the variation in distance to which
the three LIGO interferometers can see an optimally oriented binary at
signal-to-noise ratio 8 over the S2 run. Since this is a function of the
masses of the binary, this range is shown for a $(0.1,0.1)\,M_\odot$ and a
$(0.5,0.5)\,M_\odot$ binary.
}
\end{figure}

The noise power spectrum also determines the sensitivity of the interferometer
to binary inspirals. We can quantify the sensitivity in terms of the distance
to which we can see an optimally oriented binary inspiral at a given
signal-to-noise ratio. This is the maximum distance at which the
interferometer can detect a binary (at this signal-to-noise ratio), since the
gravitational wave strain in the interferometer is a maximum when the binary
is optimally oriented.  The maximum inspiral ranges for an optimally oriented
binary at signal-to-noise ratio $\rho^\ast = 8$ are shown in
figure~\ref{f:inspiral_summary}.  The distance to which we can detect an
optimally oriented binary is also a function of the mass of the binary,
scaling as $\mu^{1/2}M^{1/3}$; figure~\ref{f:inspiral_summary} shows the
ranges for a $(0.5,0.5)\,M_\odot$ and a $(0.1,0.1)\,M_\odot$ binary.

Notice that there are no times when either of the LHO interferometers are more
sensitive than the LLO interferometer and so demanding triggers are always
present in the most sensitive interferometer means that they are required to
be found in L1. In fact the LLO interferometer has a significantly larger
range than either of the LHO interferometers, at times being sensitive to
BBHMACHO inspirals in Andromeda at around $0.7$~Mpc.  Since we require
coincidence between L1 and one of the LHO interferometers to make a detection,
however, we are restricted to a search for BBHMACHOs in the Galactic halo.  

Based on the sensitivity plots shown in figure~\ref{f:inspiral_summary}, we
set the signal-to-noise threshold to $7$ in all three interferometers; we
justify this as follows. All three interferometers are sensitive to optimally
oriented inspirals with $\rho^\ast \ge 8$ at distances greater than the size
of the Galactic halo. A binary black hole MACHO in the Galaxy may have an
unfavorable orientation, however, causing it to appear at a large effective
distance.  For this reason, we want to set the signal-to-noise ratio threshold
as low as possible without producing an excessive false alarm rate. Lowering
the signal-to-noise threshold has a computational impact on our search:
when the signal-to-noise ratio for a template crosses threshold, we
perform the $\chi^2$ veto which requires $p$ additional complex inverse FFTs,
where $p$ is the number of frequency bins used in the veto. If we set the
signal-to-noise threshold too low, we may exceed the available computational
resources due to the extra operations required to perform the $\chi^2$ veto.
In fact this is what happens with the S2 binary black hole MACHO search since
the template banks are so large. Whereas in the S2 BNS search we were able to
lower the signal-to-noise threshold to $6$, the BBHMACHO search is limited to
a signal-to-noise threshold to $7$.

\subsection{Tuning the $\chi^2$ Veto Parameters}
\label{ss:chisqtuning}


Recall from section~\ref{ss:mismatchedchisq} that the $\chi^2$ veto thresholds
on
\begin{equation}
\chi^2 < \chi^2_\ast (p+\rho^2 \delta^2),
\label{eq:chisqthreshold2}
\end{equation}
where $\rho$ is the signal-to-noise ratio of the signal and $\delta^2$ is a
parameter chosen to be reflect the largest expected mismatch that a true
signal will have with the templates in the bank. The initial parameters used
for the $\chi^2$ veto, based on tuning of the BNS search, were $p = 15$ and
$\delta^2 = 0.04$ with the threshold set to $\chi^2_\ast(\mathrm{L1}) = 5.0$ in
the L1 interferometer and $\chi^2_\ast(\mathrm{H1}) = \chi^2_\ast(\mathrm{H2}) =
12.5$ in the LHO interferometers.

\begin{figure}[p]
\begin{center}
\includegraphics[width=\textwidth]{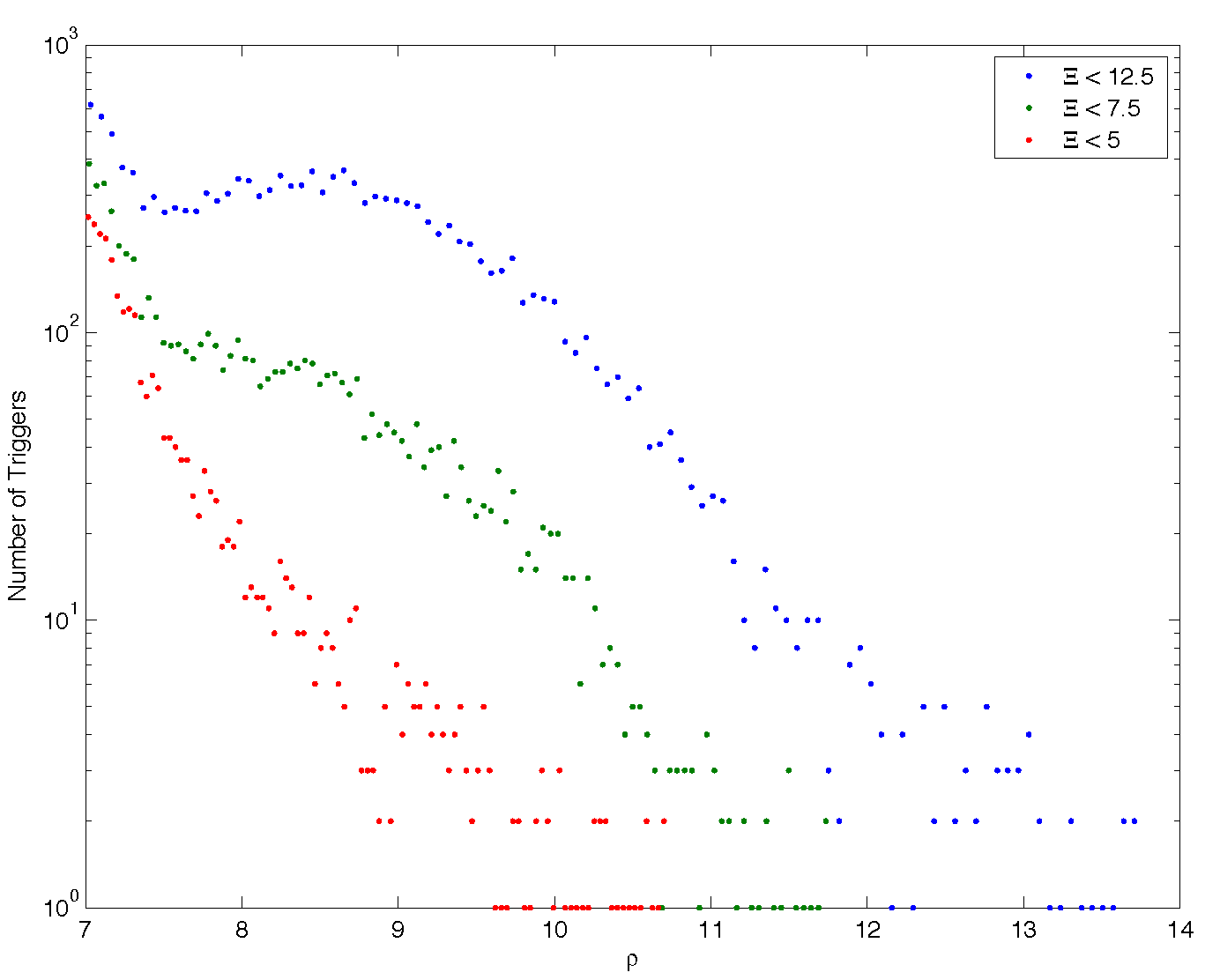}
\end{center}
\caption[Tuning the $\chi^2$ Veto for H1]{%
\label{f:h1_chiqsq_tuning}%
The figure shows a histogram of all the triggers generated from the H1 data
using the triggered search (i.e. no coincidence with L1 or H2 has been applied
to the triggers). The signal-to-noise threshold is $\rho_\ast = 7$ and the
parameters of the $\chi^2$ veto are $p = 15, \delta^2 = 0.04, \Xi = 12.5$, as
in the S2 binary neutron star search. If the interferometer data is Gaussian,
then we would expect the histogram to be monotonically decreasing with
increasing signal-to-noise ratio; however, there is a pronounced ``hump'' in the
histogram at $\rho\approx 9$ suggesting some non-Gaussian feature in the data.
By lowering the value of $\Xi$ to $5$, we can remove this feature from the
histogram, but we must be careful in doing so that we do not reduce the
detection efficiency of the pipeline.
}
\end{figure}

\begin{figure}[p]
\begin{center}
\includegraphics[width=\textwidth]{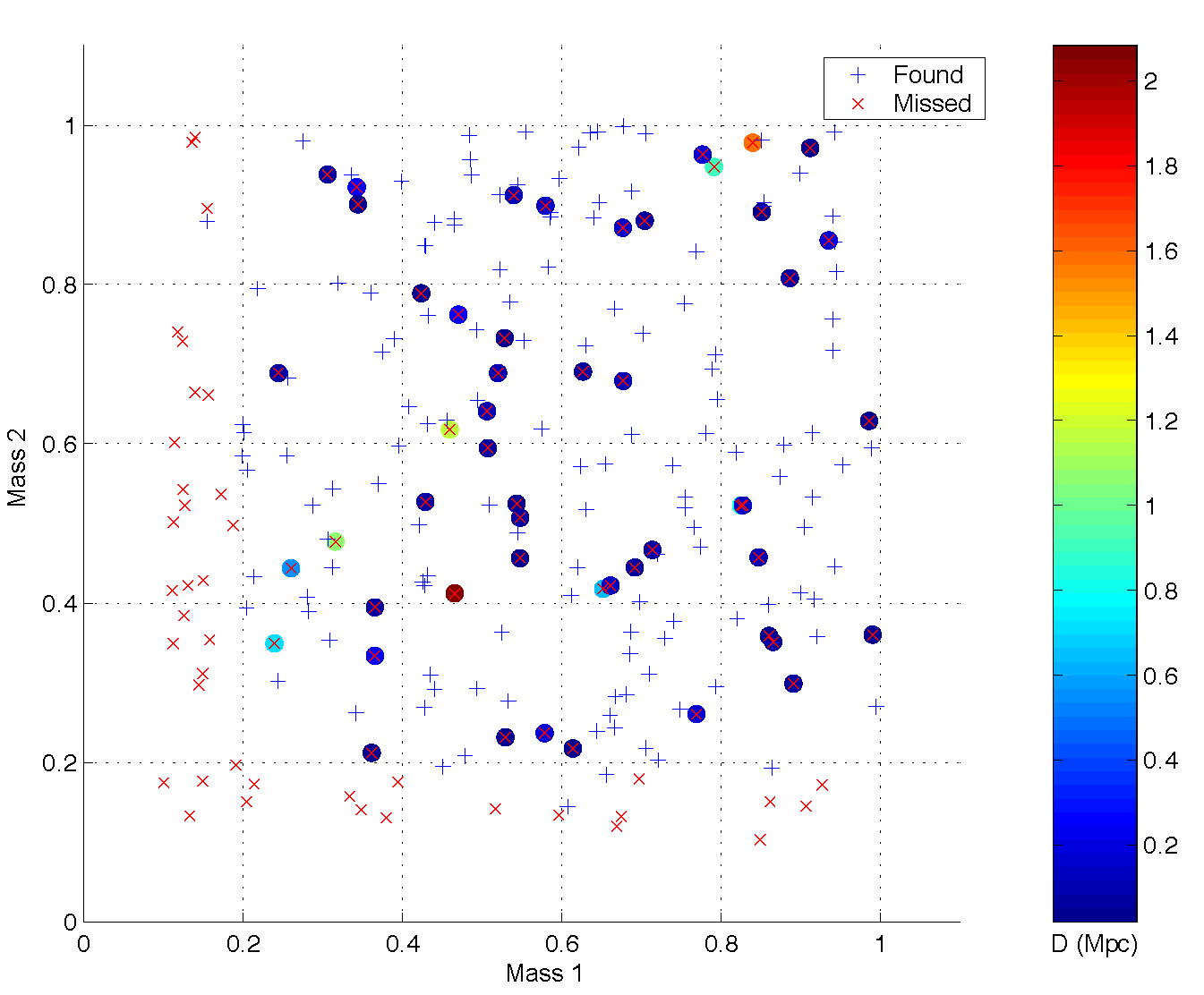}
\end{center}
\caption[Found and Missed H1 Injections for $\Xi = 12.5 \delta^2 = 0.04$]{%
\label{f:h1_missed_tuning}%
The figure shows the results of a small Monte Carlo simulation used to test
the detection efficiency of the pipeline using H1 triggers (i.e. no
coincidence with L1 or H2 has been applied to the triggers). The
signal-to-noise threshold is $\rho_\ast = 7$ and the parameters of the
$\chi^2$ veto are $p = 15, \delta^2 = 0.04, \Xi = 12.5$. Found injections are
shown with a $+$, missed injections are shown with a $\times$ and the masses
of the injection are shown as the $x$ and $y$ coordinates. We would expect to
miss any injections with a mass component below $0.2\,M_\odot$ due to the
coverage of the template bank; however injections in the region inside the
bank should be detected, unless they are at an effective distance larger than
the range of the interferometer. The missed injections that we would expect to
find are color coded according to the effective distance at which they are
injected.  Several injections are missed as they are at a large effective
distance (e.g.  the injection at $(0.48,0.42)\,M_\odot$); however there are
may missed injections at distances $< 200$~kpc which should be detectable in
the H1 data (e.g. the injection at $(0.88,0.91)\,M_\odot$). Investigation of
the missed injections showed they had large values of signal-to-noise ratio,
but were vetoed by the $\chi^2$ test. This suggests that the search parameters
used must be re-tuned to increase the detection efficiency.
}
\end{figure}

\begin{figure}[p]
\begin{center}
\includegraphics[width=0.7\textwidth]{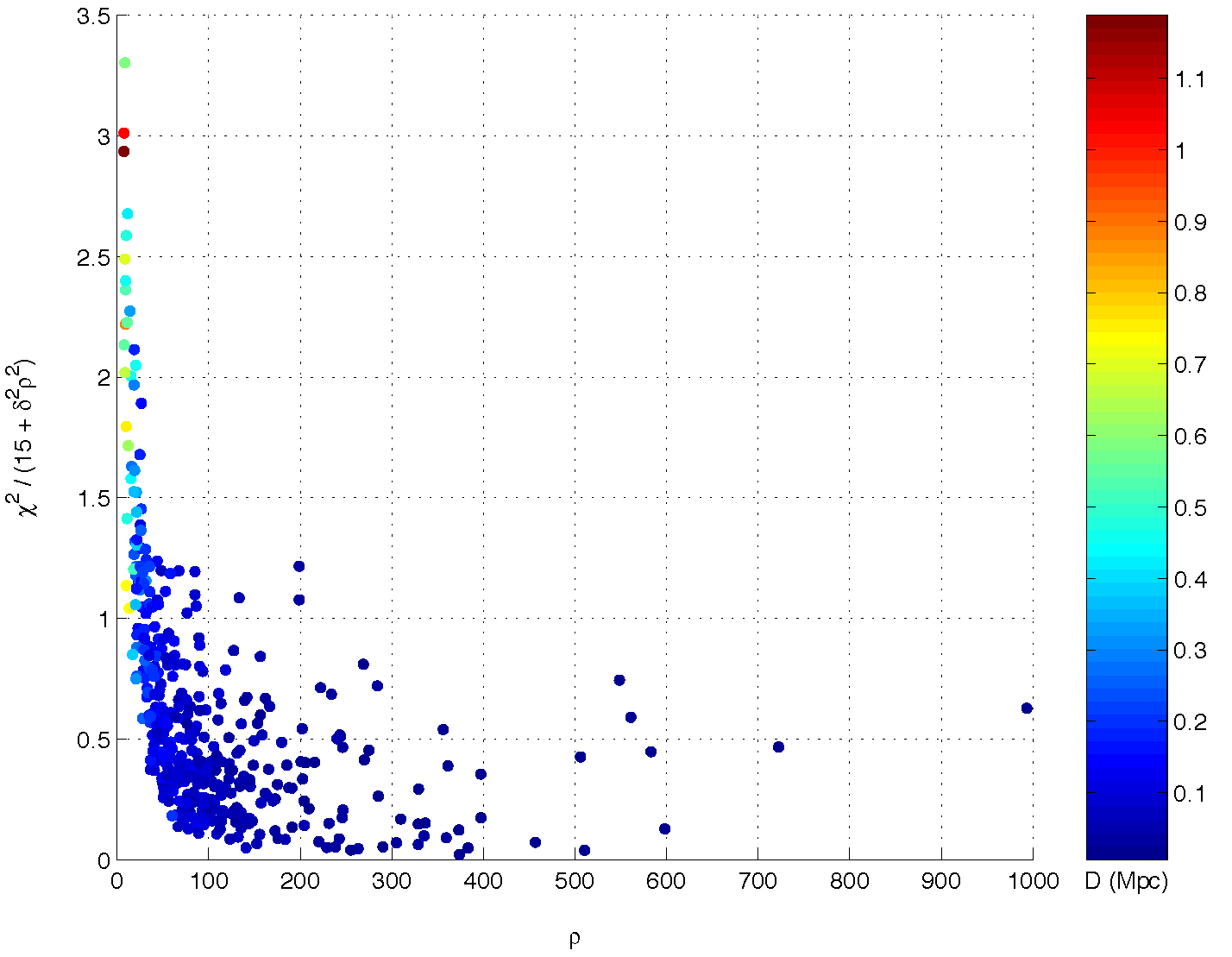}\\
\includegraphics[width=0.7\textwidth]{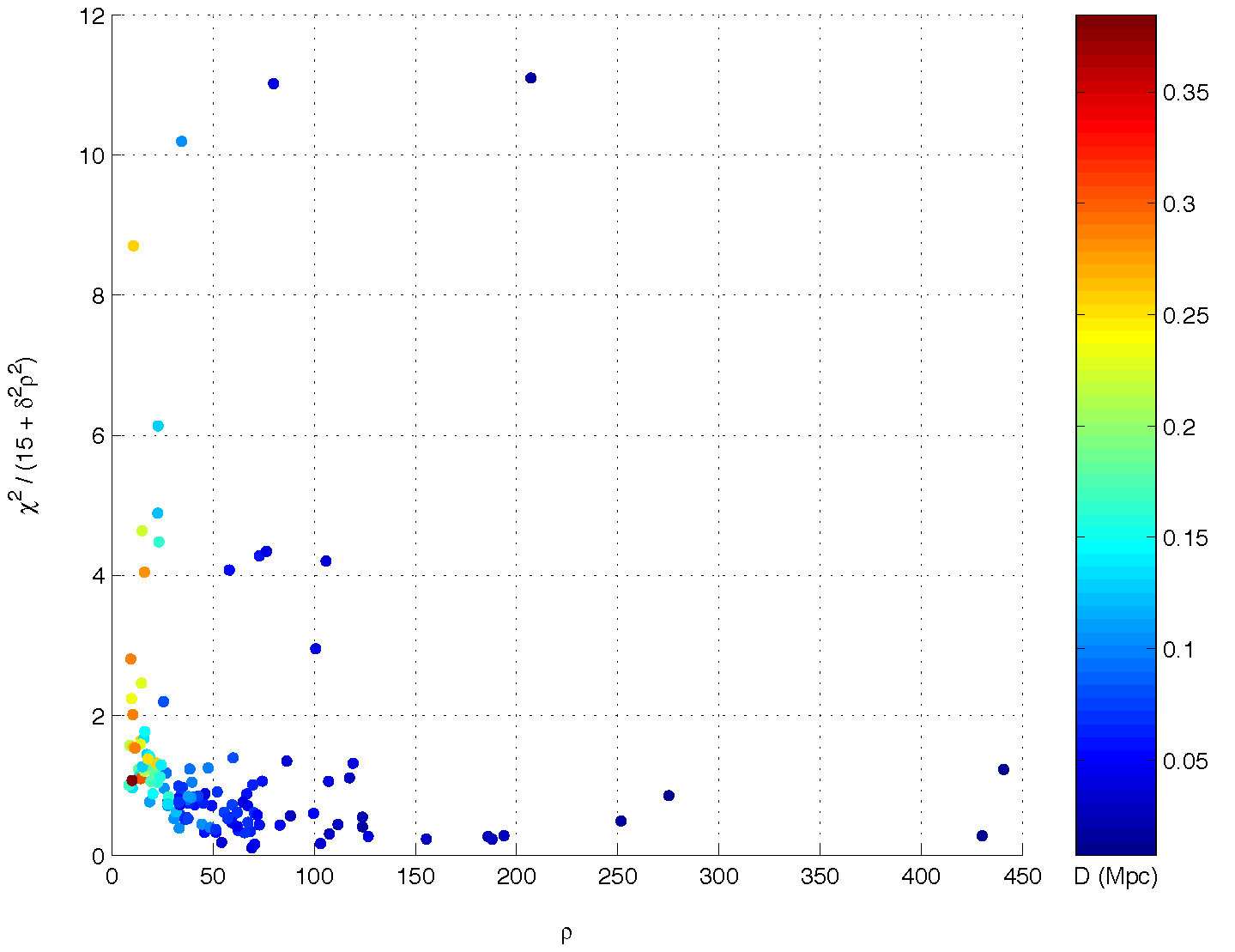}
\end{center}
\caption[Illustration of Tuning Based on Injections]{%
\label{f:h1_inj_xi_tuning}%
The plots in this figure show the values of $\rho$ and $\Xi = \chi^2/(15 +
\delta^2\rho^2)$ for the inspiral triggers corresponding to injected signals
found by the triggered search pipeline. The upper plot shows L1 triggers and
the lower plot shows H1 triggers (which correspond to the found injections of
figure \ref{f:h1_missed_tuning}). No coincidence has been applied to the H1
triggers at this stage; however they are generated using template banks
produced from L1 triggers. The color of each trigger shows the effective
distance at which it was injected.  Both plots are generated with a
signal-to-noise threshold of $\rho_\ast = 7$, and the parameters of the
$\chi^2$ veto were $p = 15, \delta^2 = 0.04$ and $\Xi_\mathrm{L1} = 12.5,
\Xi_\mathrm{L1} = 5.0$, values chosen based on the tuning of the S2 binary
neutron star search. It can be seen that, at a given signal-to-noise ratio,
the H1 triggers typically have higher values of $\Xi$ than the L1 triggers.
This is due to the a larger mismatch between the injected signal and the
templates in the H1 triggered bank.
}
\end{figure}

\begin{figure}[p]
\begin{center}
\includegraphics[width=0.7\textwidth]{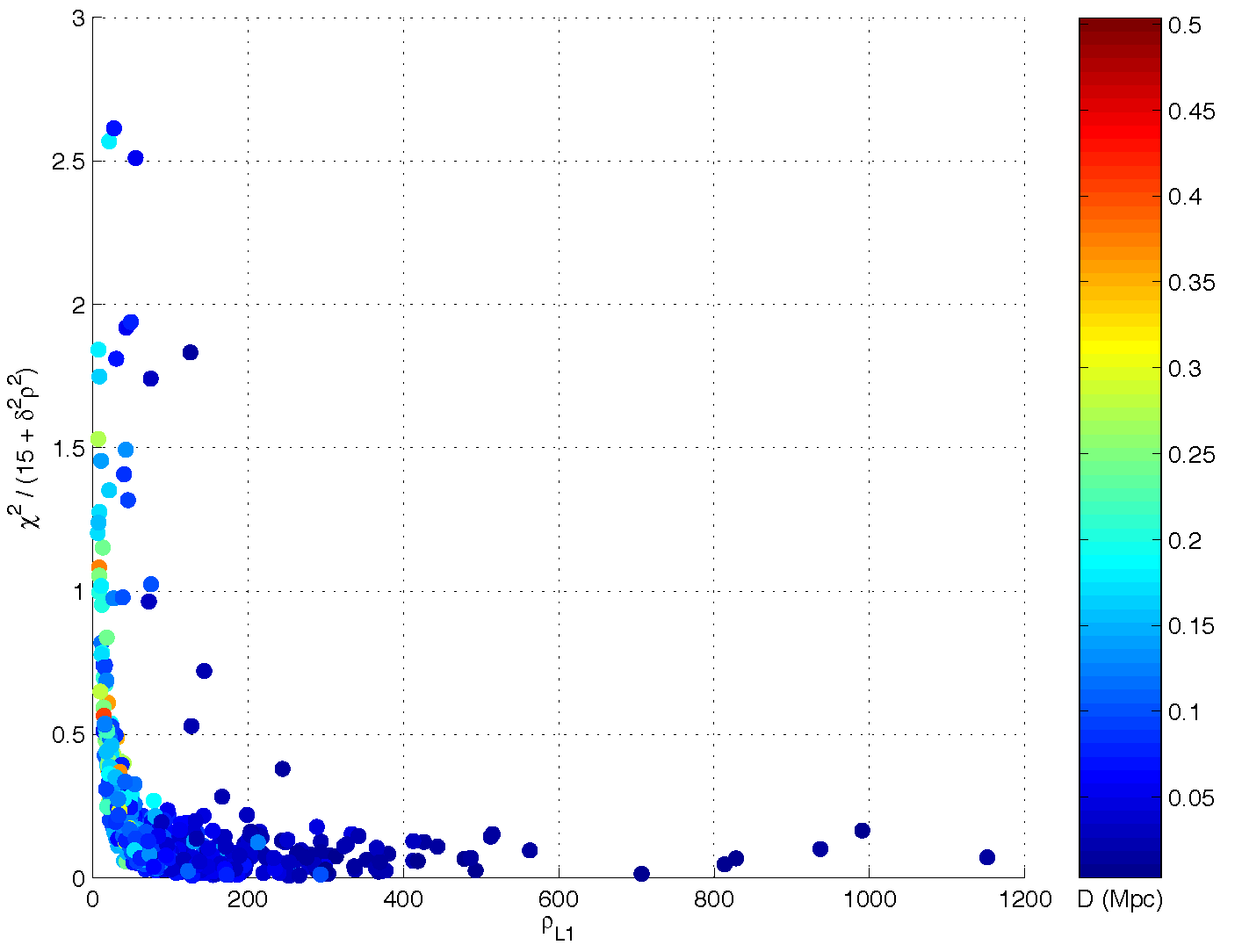}\\
\includegraphics[width=0.7\textwidth]{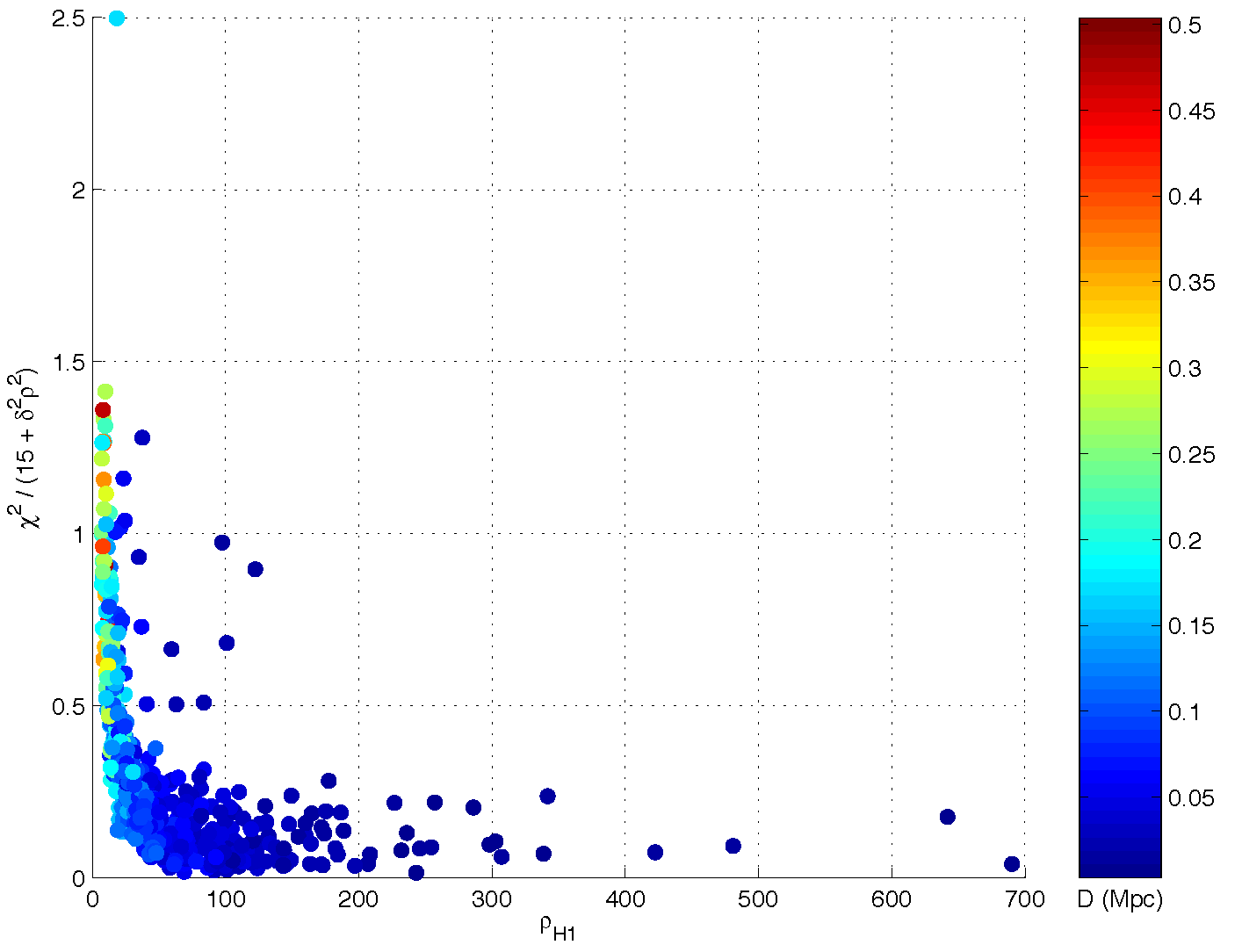}
\end{center}
\caption[Signal-to-noise ratio and $\chi^2$-veto for Injected Signals]{%
\label{f:h1_inj_xi_final}%
The plots in this figure should show the observed values of $\rho$ and $\Xi
= \chi^2/(15 + \delta^2\rho^2)$ for the inspiral triggers corresponding to
injected signals found by the triggered search pipeline using the final set of
parameters chosen. The upper plot shows L1 triggers and the lower plot shows
H1 triggers. No coincidence has been applied to the H1 triggers at this stage;
however they are generated using template banks produced from L1 triggers.
These plots should be compared to those shown in
figure~\ref{f:h1_inj_xi_tuning}. By tuning the value of $\delta^2$ to $0.2$,
it can be seen that much lower values of $\Xi$ are obtained for the H1
injections. This suggests that we could further reduce the threshold
$\Xi_\ast$, although this was not done as no coincident triggers were found in
the playground data and a looser value of $\delta$ allowed us to probe the
region slightly outside the template bank parameter space.
}
\end{figure}

Figure \ref{f:h1_chiqsq_tuning} illustrates tuning of the $\chi^2$ veto on the
H1 playground triggers. If the interferometer noise is Gaussian, then the
square of the signal-to-noise ratio should be $\chi^2$ distributed with two
degrees of freedom. This means that the histogram of the signal-to-noise ratio
of the triggers should be a monotonically decreasing function of $\rho$. We
can see from figure~\ref{f:h1_chiqsq_tuning}, however, that there is an excess
in the number of triggers with $\rho \approx 8.5$; this suggests some
non-Gaussian behavior in the data that we would like to remove. It can be seen
that decreasing the threshold $\chi^2_\ast$ to $5$ removes this hump in the
distribution and decreases the signal-to-noise ratio of the loudest event from
$\rho = 13.8$ to $\rho = 10.7$. This suggests that lowering the $\chi^2$
threshold is desirable.  Figure~\ref{f:h1_missed_tuning}, which shows the
results of a small Monte Carlo simulation, demonstrates the danger of making
such decisions without reference to the detection efficiency, however. For
each simulated signal added to the data, the figure shows whether or not it was
detected in the H1 data using the initial choice of $\chi^2$ veto parameters
($\delta^2 = 0.04$, $\chi^2_\ast(\mathrm{H1}) = 12.5$). Several injections are
missed at effective distances well within the range of the H1 interferometer.
Follow up investigations of these missed triggers show that, although they
have very large values of signal-to-noise $\rho \sim 10^2 - 10^3$, they are
missed because they fail the $\chi^2$ veto. This is caused by the mismatch
between the signal and the template.  The results of this study imply that we
should loosen the $\chi^2$ veto, in contradiction to the results suggested by
figure~\ref{f:h1_chiqsq_tuning}.  Notice, however, that the excess of H1
triggers occurs at low values of $\rho$ and the missed injections are at
higher values of $\rho$. Figure~\ref{f:h1_inj_xi_tuning} shows the values of
$\chi^2/(p+\rho^2\delta^2)$ and $\rho$ for the detected L1 injections and the
detected H1 injections before coincidence is applied. These triggers are taken
from the same Monte Carlo simulation as the triggers shown in
figure~\ref{f:h1_missed_tuning}. The values of $\chi^2/(p+\rho^2\delta^2)$
observed for the detected H1 injections are significantly higher than those
for the L1 injections. This suggests that we should increase the parameter
$\delta^2$, which has the effect of decreasing the value of
$\chi^2/(p+\rho^2\delta^2)$ for triggers with high signal-to-noise ratios. If
we increase $\delta^2$ we may be able to decrease the value of
$\chi^2/(p+\rho^2\delta^2)$ to remove the excess of triggers in H1, without
adversely affecting the pipeline detection efficiency. After several
iterations, the values $\delta^2 = 0.4$, $\chi^2_\ast(\mathrm{L1}) = 3.1$,
$\chi^2_\ast(\mathrm{H1}) = 5.0$ and $\chi^2_\ast(\mathrm{H2}) = 10.0$ were
chosen.  Figure~\ref{f:h1_inj_xi_final} shows the values of $\Xi$ and $\rho$
for the detected L1 injections and the detected H1 injections with these new
parameters. Notice that the values of $\chi^2/(p+\rho^2\delta^2)$ for the loud
signal-to-noise triggers are considerably lower than before. It appears from
the results in figure~\ref{f:h1_inj_xi_final} that it would be possible to
reduce $\chi^2_\ast(\mathrm{H1})$ further to $2.51$, without loss of
efficiency. No coincident triggers survived the pipeline in the playground
data with a threshold of $\chi^2_\ast(\mathrm{H1}) = 5.0$, however, so it was
decided not to reduce this threshold further.

\subsection{Coincidence Parameter Tuning}
\label{ss:coinc_tuning}

After the single interferometer parameters had been selected, the coincidence
parameters were tuned using the triggers from the single interferometers.  As
described in section~\ref{ch:hardware}, the coalescence time of an inspiral
signal can be measured to within $\le1$~ms. The light travel time between
observatories is $10$~ms, so $\delta t$ was chosen to be $1$~ms for LHO-LHO
coincidence and $11$~ms for LHO-LLO coincidence. The mass coincidence
parameter was initially chosen to be $\delta m = 0.03$, however testing with
the binary neutron star search showed that this could be set to $\delta m =
0.0$ ({\it i.e.}\ requiring the triggers in each interferometer to be found
with the exact same template) without loss of efficiency.

Having tuned the time and mass parameters, we tune the effective
distance parameters $\kappa$ and $\epsilon$. Initial estimates of $\epsilon =
2$ and $\kappa = 0.2$ were used for testing, however it was discovered that
many injections were missed using these thresholds to test for LLO-LHO
coincidence. This is due to the fact that the detectors are slightly
misaligned, so the ratio of effective distance of a trigger between the two
observatories can be large for a significant fraction of the population, as
shown in Fig.~\ref{f:gmst_dist_ratio}. As a result, we disabled the
effective distance cut for triggers generated at different observatories.
A study of simulated signals injected into H1 and H2 interferometers, both
located at the LIGO Hanford Observatory, suggested using values
of $\epsilon_{HH} = 2$ and $\kappa_{HH} = 0.5$. Note
that, as described above, we demand that an L1/H1 trigger pass the H1/H2
coincidence test if the effective distance of the trigger in H1 is within the
maximum range of the H2 detector at threshold.

\section{Results of Injection Monte Carlo Simulations}
\label{s:monte}

\begin{figure}[p]
\begin{center}
\includegraphics[width=0.7\textwidth]{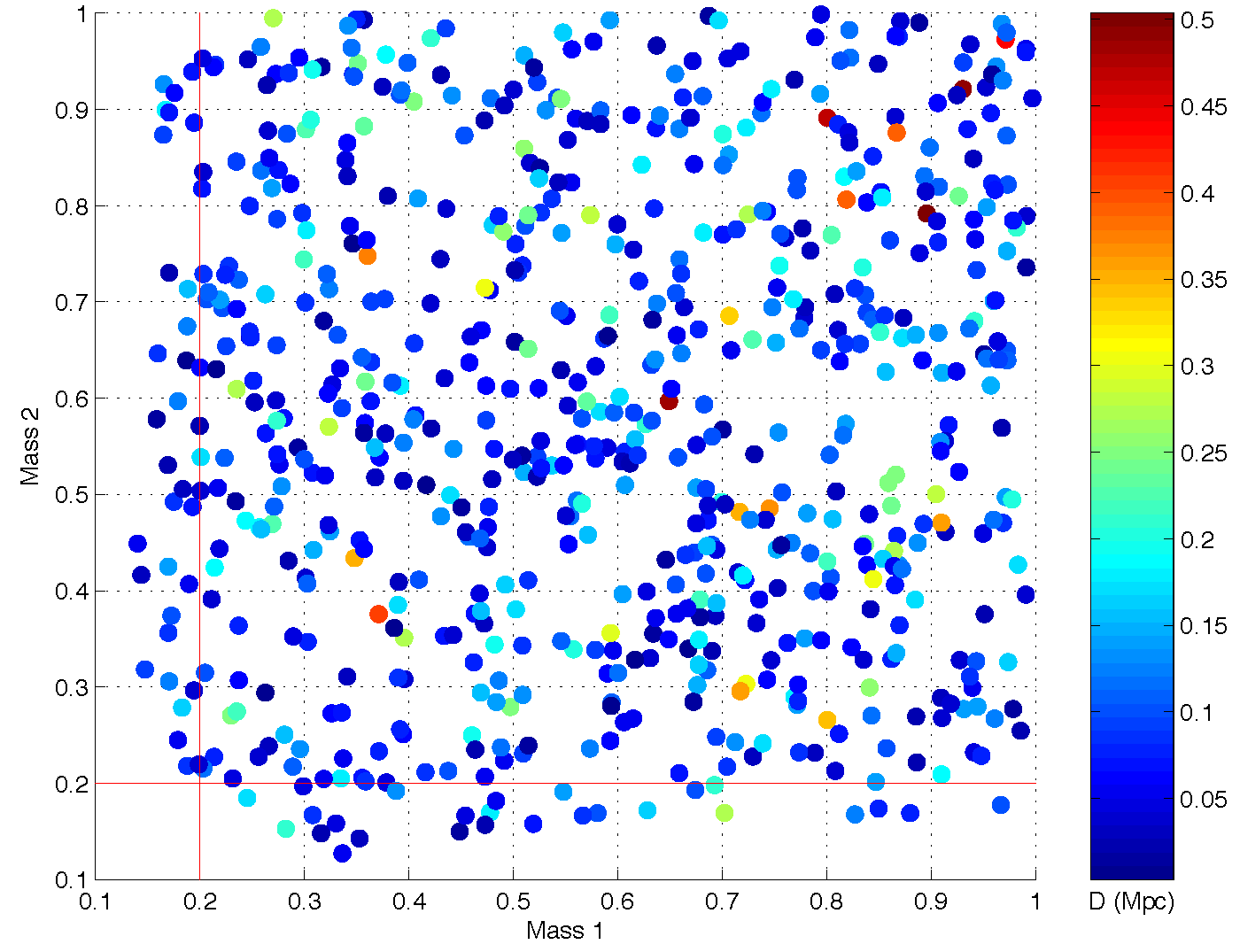}\\
\includegraphics[width=0.7\textwidth]{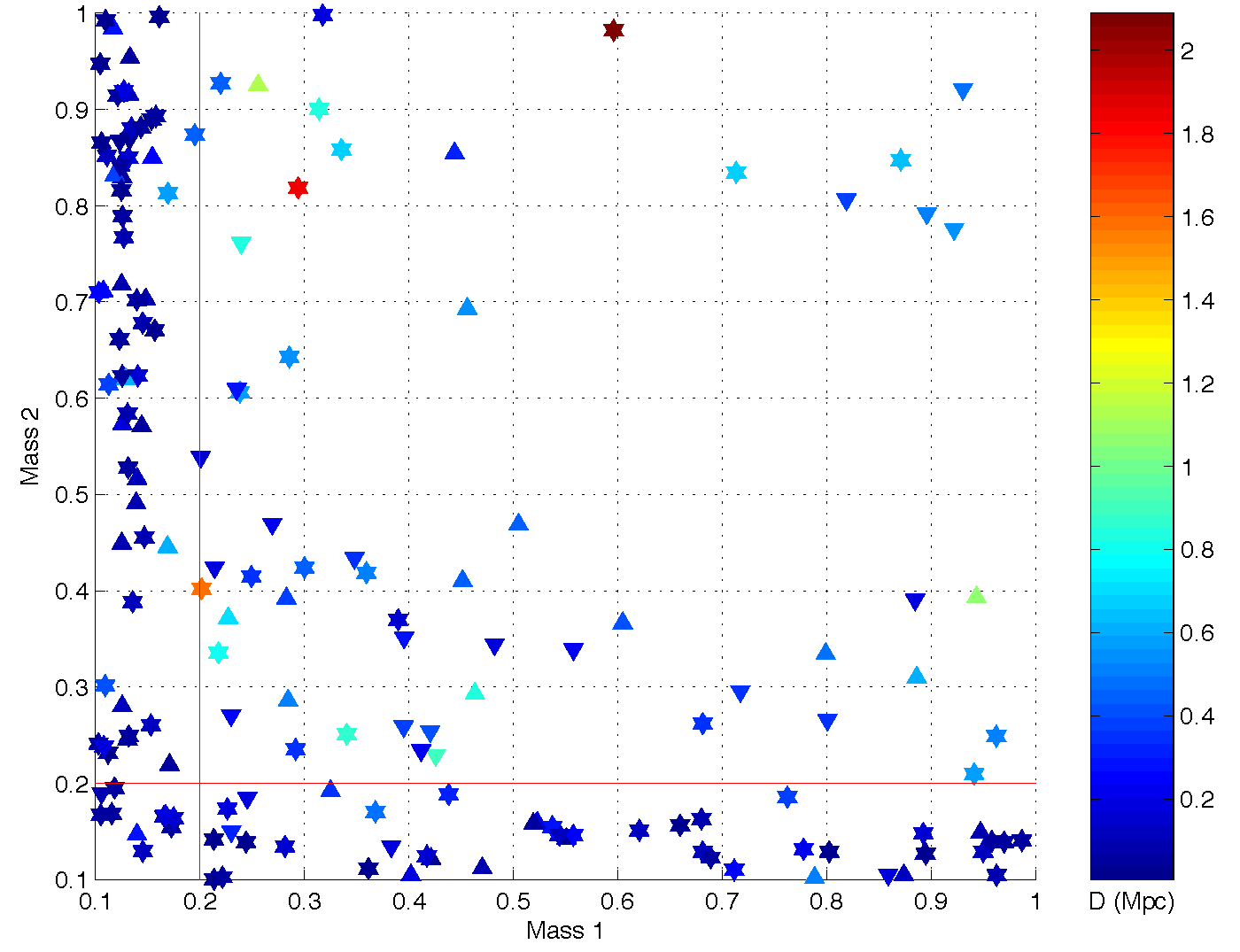}
\end{center}
\caption[Injections Detected and Missed by Monte Carlo Simulation]{%
\label{f:m1m2_found_missed}%
This figure shows the results of the Monte Carlo simulation used to measure
the efficiency of the pipeline once parameter tuning had been completed; these
detected triggers have survived all threshold and coincidence tests.  The
injections that are detected are shown as circles on the upper plot.  The
lower plot shows the injections that were not detected: stars correspond to
missed injections in the triple coincident data, upward pointing triangles to
the L1-H1 data and downward pointing triangles to the L1-H2 data. The $x$ and
$y$ coordinates are the mass parameters $m_1$ and $m_2$ of each injection,
respectively. The color of each injection represents the effective distance in
the Hanford interferometers at which the waveform was injected (since the LHO
interferometers limit the sensitivity of the search).  The horizontal and
vertical red lines show the edge of the template bank parameter space. 
}
\end{figure}

\begin{figure}[p]
\begin{center}
\includegraphics[width=0.7\textwidth]{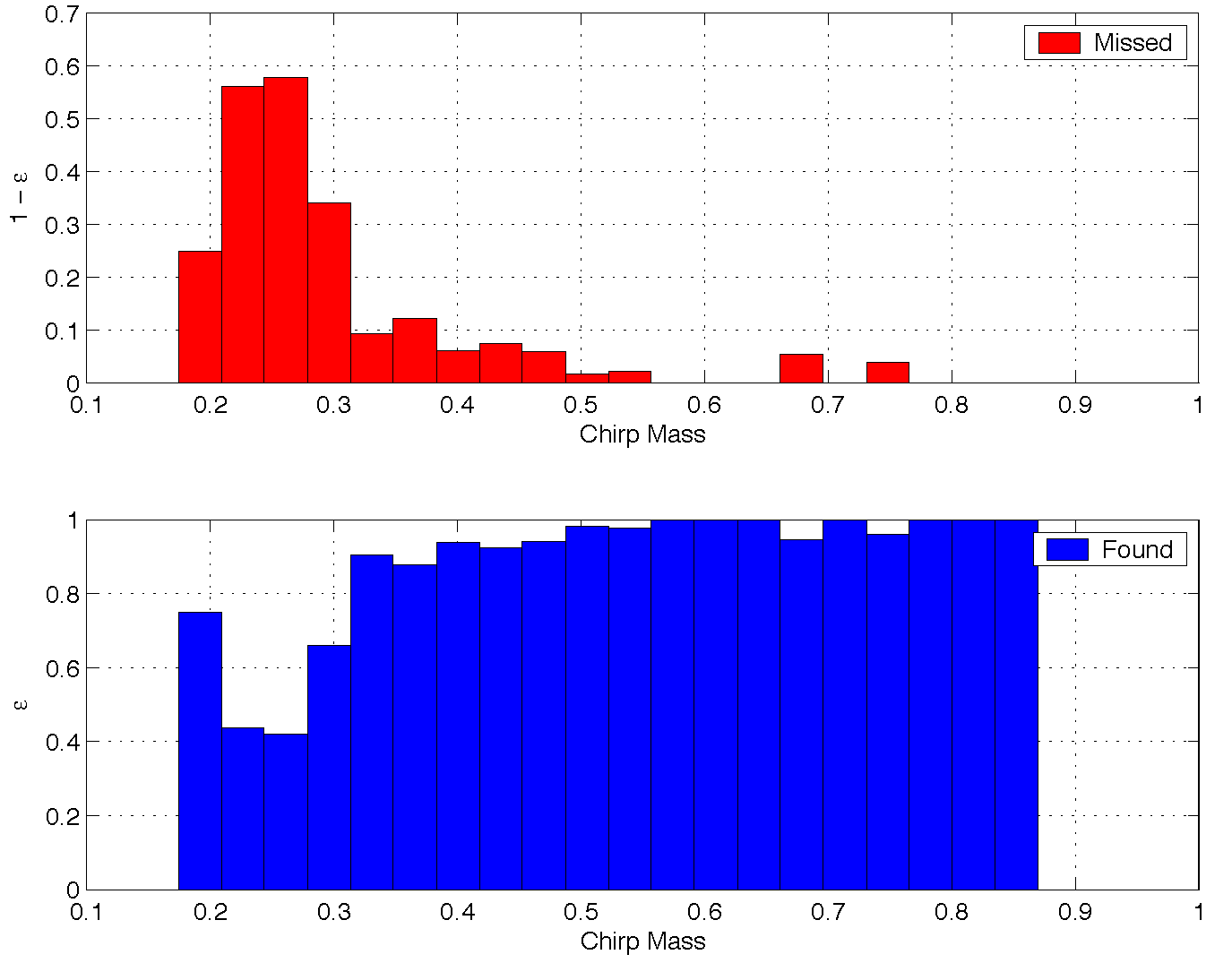} \\
\includegraphics[width=0.7\textwidth]{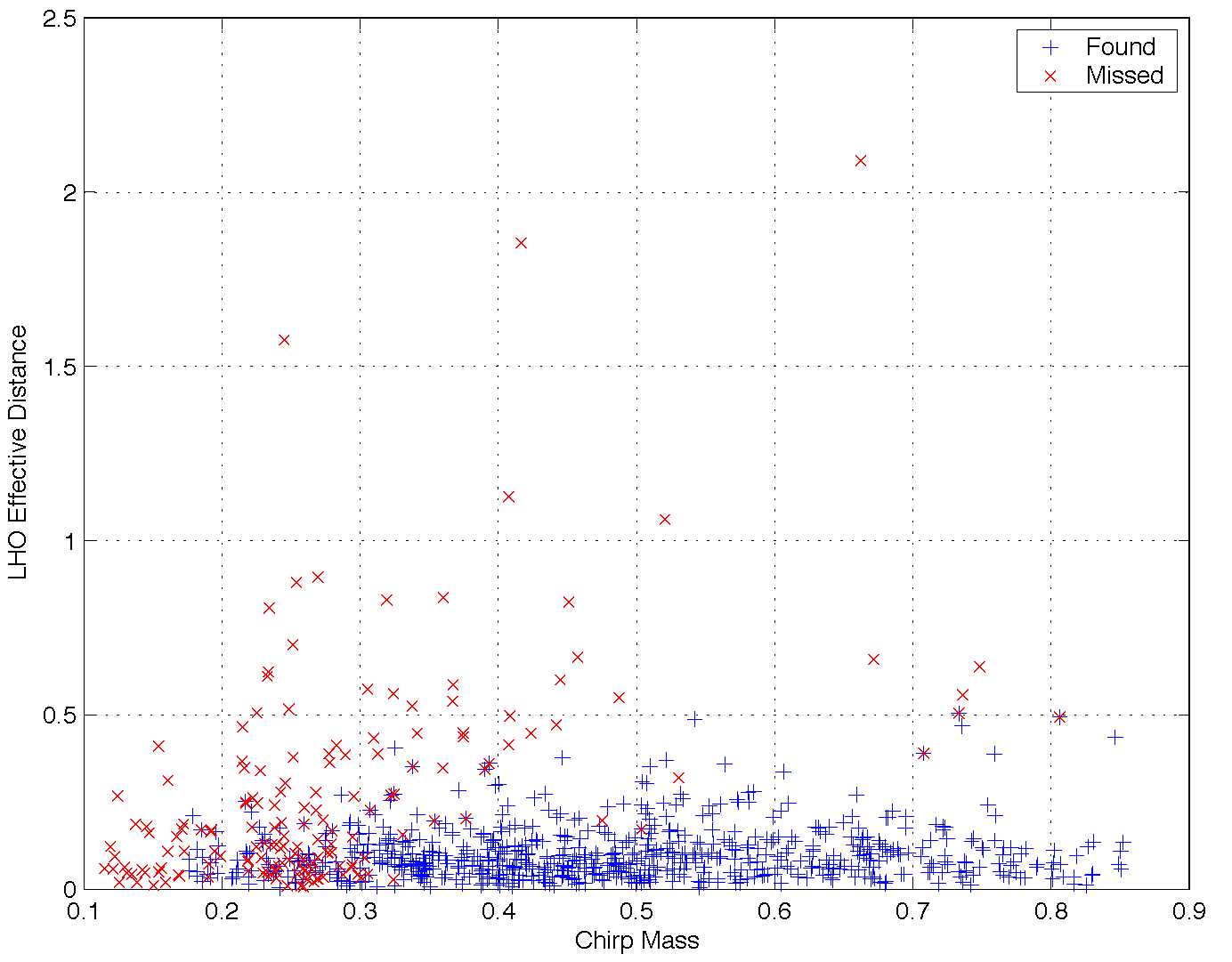}
\end{center}
\caption[Search Efficiency as a Function of Chirp Mass]{%
\label{f:mchirp_eff}%
The upper two plots show the efficiency of the pipeline $\varepsilon$ and the
loss of the search $1-\varepsilon$ as a function of the injected signal chirp
mass $\mathcal{M}$ measured by the Monte Carlo Simulation. The lower plot
shows the chirp mass and effective distance in LHO of the injections used to
measure the pipeline efficiency; detected injections are shown with a $+$ and
missed injections are shown with a $\times$. It can be seen that the
efficiency of the pipeline is unity very close to unity for high values of
$\mathcal{M}$ and falls as the chirp mass decreases. There appears to be an
anomalously large value of $\varepsilon$ at $\mathcal{M}\approx 0.18$, however
it can be seen from the lower plot that there were comparatively few
injections at this chirp mass, so this may be an effect of small number
statistics. Further Monte Carlo simulations will be able to test this
hypothesis.
}
\end{figure}

The final parameter values chosen are shown in table~\ref{t:ifo_params}. Once
fixed, an injection Monte Carlo was performed to measure the efficiency of the
search pipeline. For each playground interval in the S2 data, an inspiral
waveform was generated using the population model described in
section~\ref{s:bbhmachopopulation}. These signals has masses between
$0.1\,M_\odot$ and $1.0\,M_\odot$. Each inspiral signal was injected into the
data at a random time during a unique playground interval and the data
analyzed through the full pipeline with the final set of parameters. Four
separate simulation runs were performed giving a total of 849 injections in
the analyzed playground data.  Figure~\ref{f:m1m2_found_missed} shows the
results of this simulation in the $(m_1,m_2)$ plane. The figure shows which of
the simulated signals were detected and which were missed by the full pipeline
in the double and triple coincident playground data. Also shown in the figure
is the effective distance at which these signals were injected in the LHO
interferometers, since it is generally the less sensitive detector that limits
the detection efficiency.  Although the template bank only coves the region
above and to the right of the red lines at $0.2\,M_\odot$, the upper plot
shows that some signals are detected with component masses between $\sim
0.15$--$0.20\,M_\odot$. This is a direct result of increasing $\delta^2$ in
the $\chi^2$ veto, equation (\ref{eq:chisqthreshold2}), allowing loud, but
slightly mismatched signals to be detected. The lower plot in
figure~\ref{f:m1m2_found_missed} shows the injections that are missed by
the pipeline. Injections missed from triple, L1-H1 double and L1-H2 double
coincident data are shown with a star, an upward triangle and a downward
triangle respectively. These missed signals are color coded with the injected
effective distance in the LHO detectors. Injections are
only missed when their effective distance is comparable or greater than the
ranges for the LHO detectors shown in figure~\ref{f:inspiral_summary}; there
are no anomalous missed injections.  Figure~\ref{f:mchirp_eff} shows the
efficiency of the search as a function of chirp mass. As expected, given the
strength of the BBHMACHO signals and the sensitivity of the detectors, the
efficiency is $\varepsilon \sim 1$ for $\mathcal{M} >0.35$ with the small loss
of efficiency coming from systems with an unfavorable orientation. The
efficiency drops for $\mathcal{M}$ below $0.35$ due to the combined effect
of the signals becoming weaker as the chirp mass decreases and falling outside
the region of good template bank coverage.  Notice that there appears to be an
anomalous value of $\varepsilon$ at $\mathcal{M} \approx 0.2$; this appears to
be associated with a dearth of injections in this mass range.  Large scale
Monte Carlos, with many more injections, are currently being performed to
explore this region of parameters space and investigate this anomaly in the
full S2 data set.

\begin{figure}[p]
\begin{center}
\includegraphics[width=\textwidth]{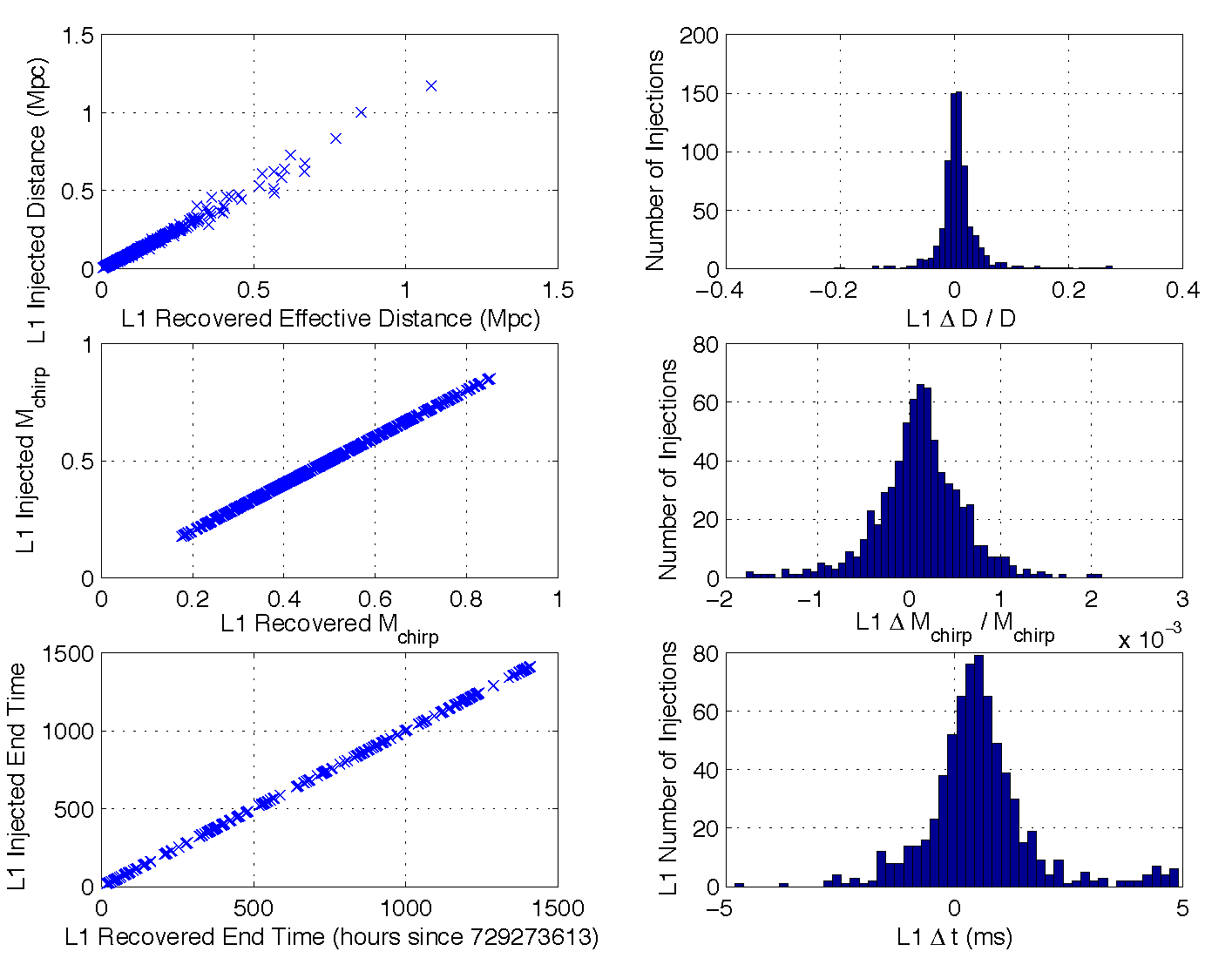}
\end{center}
\caption[Measurement accuracy of L1 Injection Parameters]{%
\label{f:l1_param_error}%
The panels in this figure compare the measured values of effective distance,
chirp mass and end time with the known values for the injected waveforms in
the Monte Carlo simulation in the L1 detector.
}
\end{figure}

\begin{figure}[p]
\begin{center}
\includegraphics[width=\textwidth]{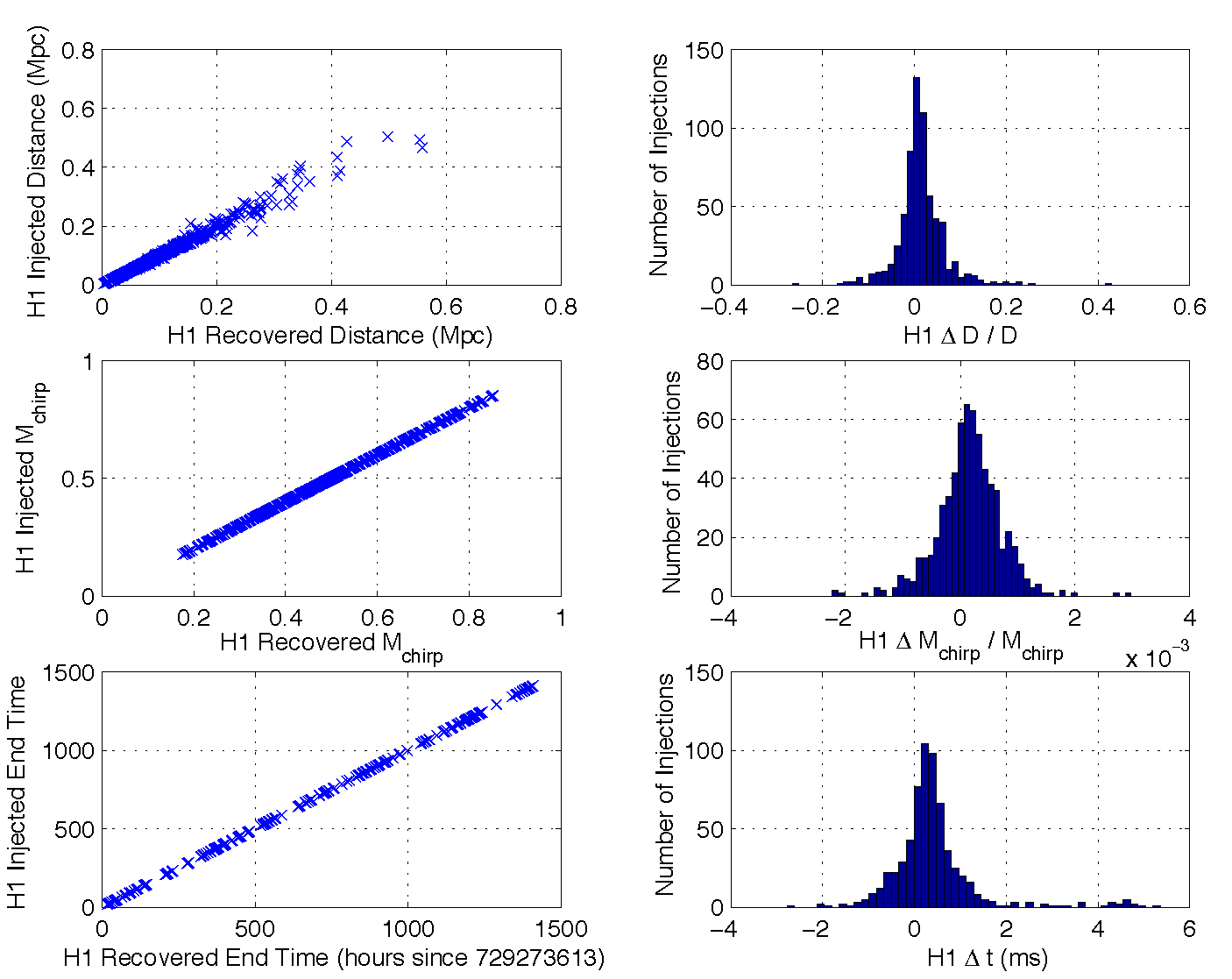}
\end{center}
\caption[Measurement accuracy of H1 Injection Parameters]{%
\label{f:h1_param_error}%
The panels in this figure compare the measured values of effective distance,
chirp mass and end time with the known values for the injected waveforms in
the Monte Carlo simulation in the H1 detector.
}
\end{figure}

\begin{figure}[p]
\begin{center}
\includegraphics[width=\textwidth]{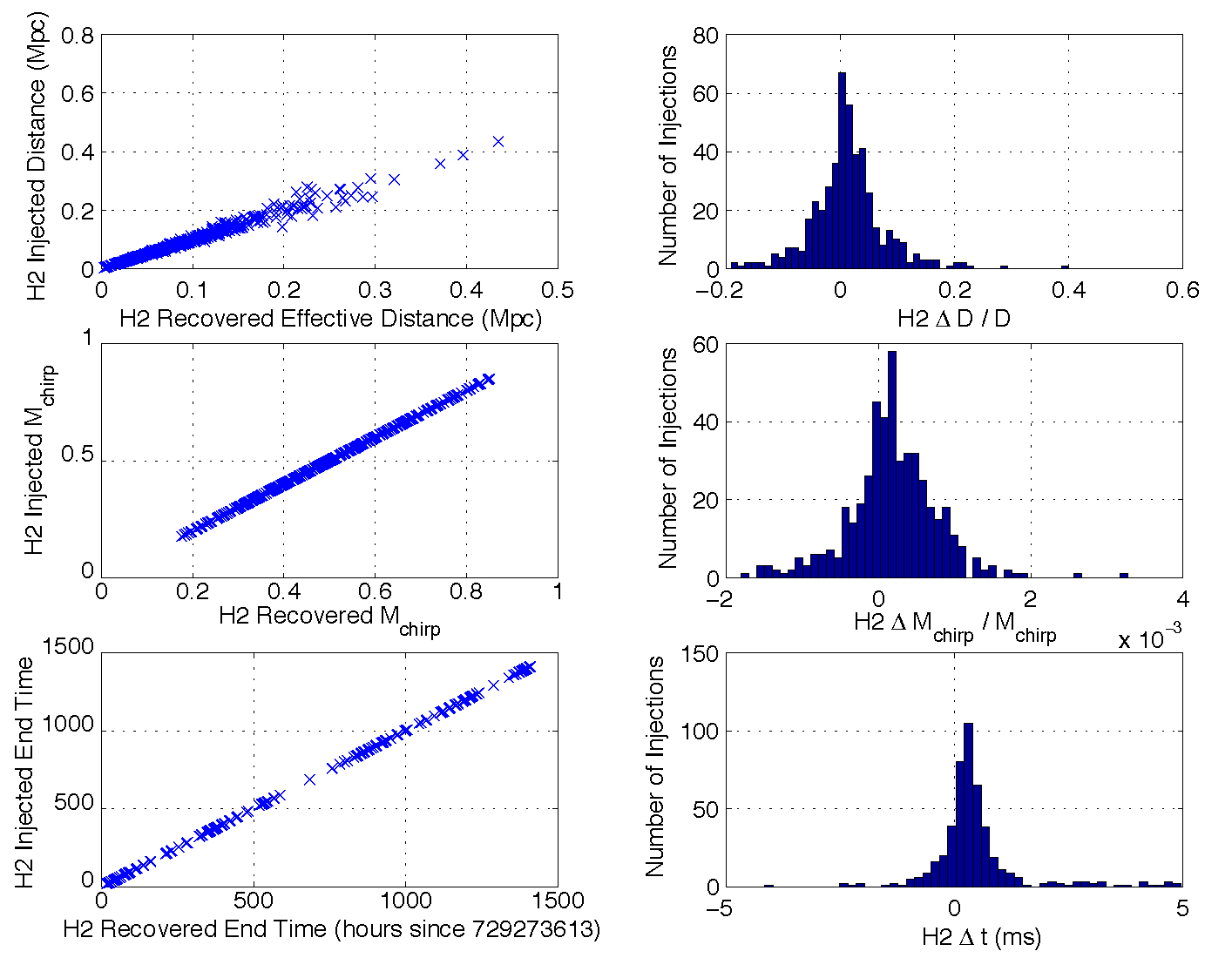}
\end{center}
\caption[Measurement accuracy of H2 Injections Parameters]{%
\label{f:h2_param_error}%
The panels in this figure compare the measured values of effective distance,
chirp mass and end time with the known values for the injected waveforms in
the Monte Carlo simulation in the H2 detector.
}
\end{figure}

A comparison of the inspiral parameters recovered by the pipeline and the
known injection parameters from signals injected in the Monte Carlo simulation
are shown for L1, H1 and H2 in figures \ref{f:l1_param_error},
\ref{f:h1_param_error} and \ref{f:h2_param_error} respectively. It can be seen
that the effective distance recorded by the search code is unbiased in all
cases, and can typically be recovered to an accuracy of $\sim 10\%$. This is
comparable to the $10\%$ distance uncertainty to nearby galaxies.  Using the
measured coalescence phase, effective distance and difference in time of
arrive at the two detectors, it is possible to gain information about the
location of the signals, however a comprehensive study of this has not been
performed for the S2 data. For all interferometers, the chirp mass is
recovered extremely well with an accuracy of $0.1\%$. This consistent with the
results quoted in \cite{Cutler:1994} and is encouraging for the parameter
measurement in the case of a detection. Although there appears to be a bias in
the measure end time of the signal, this is due to the fact that the current
implementation of the filtering code measures the end time of the template,
not the coalescence time of the binary. The injected signals are generated
with a time domain waveform generator\cite{LAL} and recovered with the
stationary phase waveforms described in chapter \ref{ch:findchirp}. There is a
slight difference in the frequency at which these waveforms terminate: the
time domain waveforms are terminated when the post-Newtonian phase evolution
of equation (\ref{eq:biwwphase}) can no longer be evolved and the stationary
phase waveforms are terminated when they reach the gravitational wave
frequency of a test particle in the innermost stable circular orbit of
Schwarzschild spacetime. As a result of this there is a small mass dependent
offset between the end times of the waveform as recorded by the injection code
and the filtering code. This will not affect the time coincidence test,
however, as we demand that the templates have the same mass parameters
in all the detectors. As a result this time offset will be identical between
detectors and we can apply coincidence. Changes to the filtering code are
planed to remove this offset in time measurement.

\section{Background Estimation}
\label{s:s2background}

\begin{figure}[p]
\begin{center}
\includegraphics[width=\textwidth]{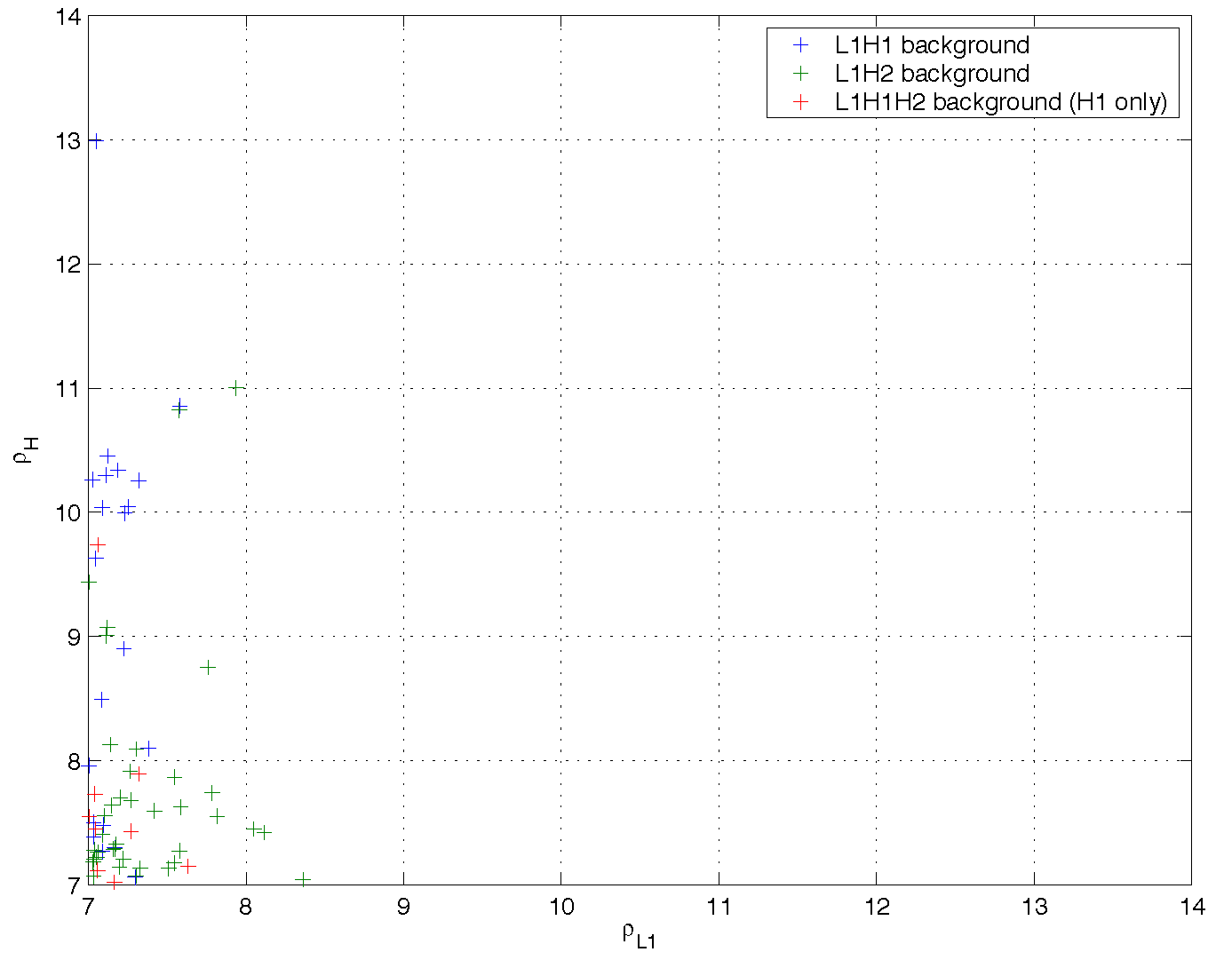}
\end{center}
\caption[Background Triggers from 20 Time Slides]{%
\label{f:bkg}%
The signal-to-noise ratio of the background triggers produced by 20
time-slides. No triple coincident background triggers were observed.
The colors are color coded depending on whether they were found in the triple,
L1-H1 double of L1-H1 double triggers coincident data. No background triggers 
were found coincident in all three detectors, so the triggers from the triple
coincident data set are from L1-H1 coincidence only.
}
\end{figure}

\begin{figure}[p]
\begin{center}
\includegraphics[width=0.7\textwidth]{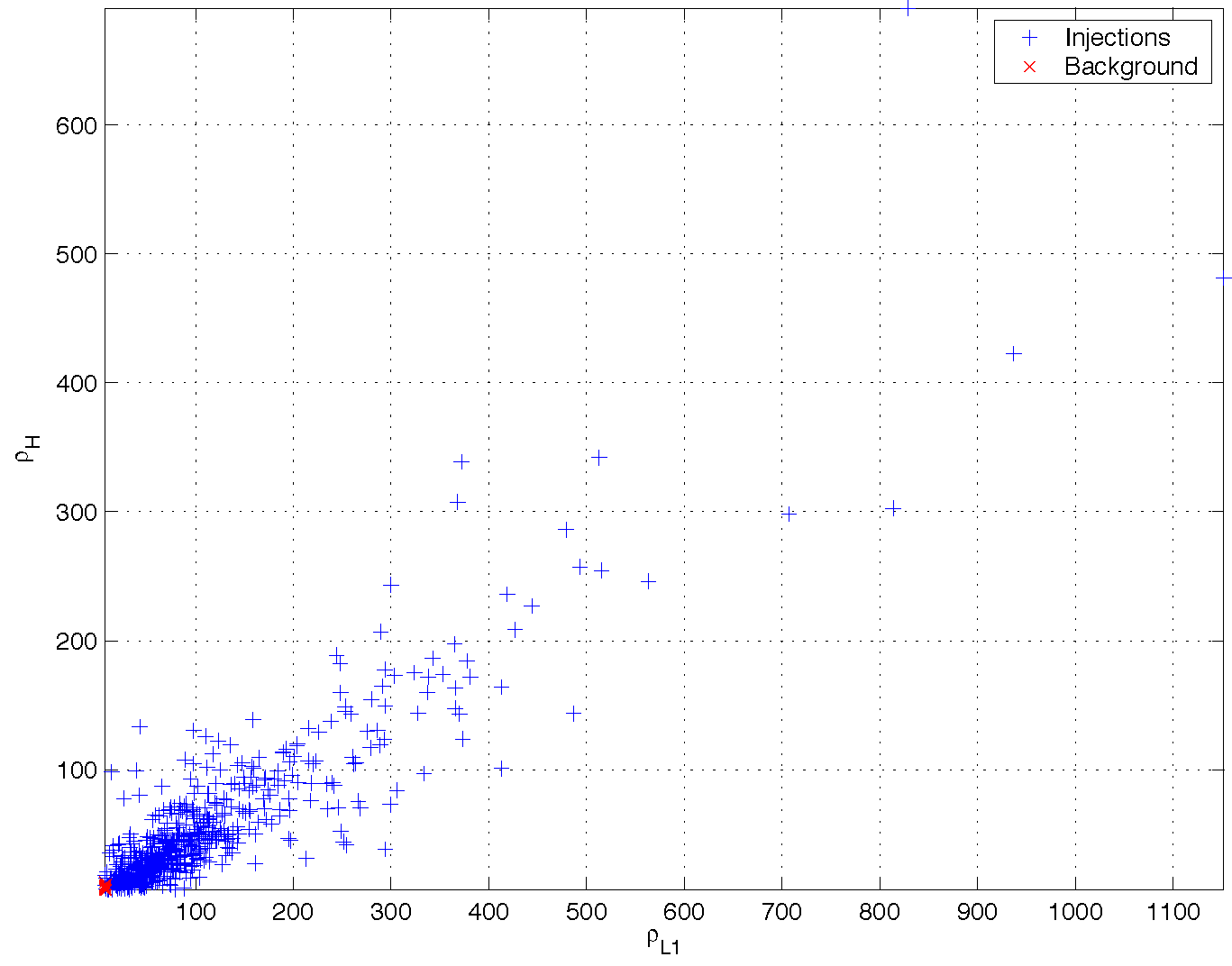}\\
\includegraphics[width=0.7\textwidth]{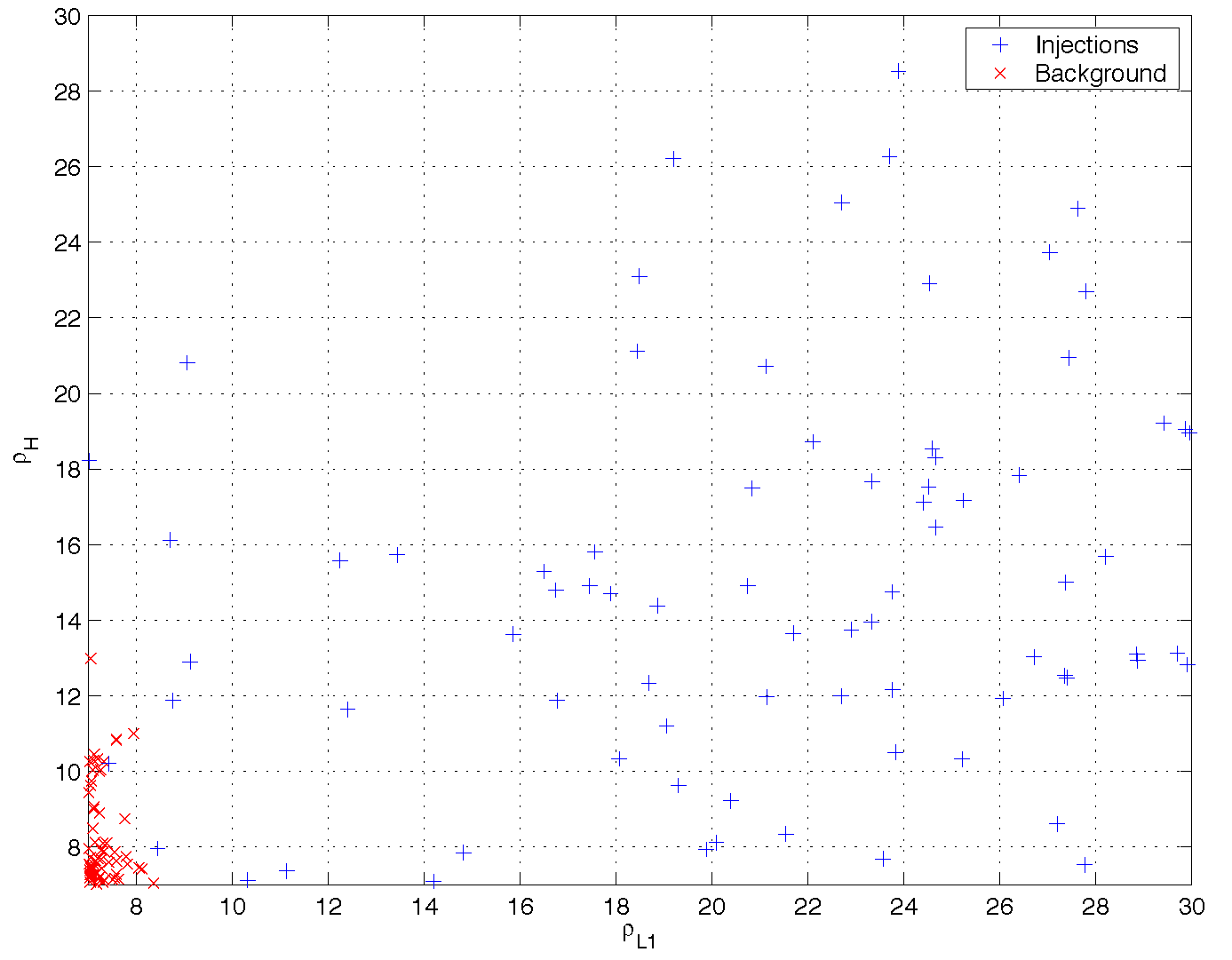}
\end{center}
\caption[Comparison of Background Triggers and Injections]{%
\label{f:bkg_inj}%
The plots in this figure compare the signal-to-noise ratios of the background
triggers to those of the triggers corresponding to software injections from the
Monte Carlo simulation. Notice in the upper plot that the signal-to-noise
ratio of detected injections in L1 is a factor of $\sim 2$ higher than the
signal-to-noise ratio in the LHO detectors, due to the greater sensitivity of
L1. The lower plot shows a magnification of the low signal-to-noise ratio
region. It can be seen that background triggers generally have a larger
signal-to-noise ratio in the LHO detectors, suggesting the coherent statistic
described in the text that gives greater weight to the L1 signal-to-noise
ratio.
}
\end{figure}

We estimate the background rate for this search by introducing an artificial
time offset, or {lag}, $\Delta t$ in the triggers coming from the Livingston
detector relative to the Hanford detector.  The time-lag triggers are then fed
into subsequent steps of the pipeline.  The triggers that emerge from the end
of the pipeline are considered a single trial representative of an output from
a search if no signals are present in the data.   By choosing a lag of more
than 20~ms, we ensure that a true gravitational wave will not be coincident in
the time-shifted data streams.  We do not time-shift the two Hanford detectors
relative to one another since there may be real correlations due to
environmental disturbances.  If the times of background triggers are not
correlated at the sites, then the background rate can be measured. A total of
20 time-lags were analyzed to estimate the background. Note that the time lags
use all the data and are not restricted to playground. The resulting
distribution of time-lag triggers in the
$(\rho_{\mathrm{H}},\rho_{\mathrm{L}})$ plane is shown in figure~\ref{f:bkg};
the distribution of background triggers and injected signals are compared in
figure~\ref{f:bkg_inj}. It can be seen that the signal-to-noise ratios of
background triggers are higher in the LHO interferometers than in the L1
interferometer, whereas the signal-to-noise ratios of injections are louder in the
L1 interferometer. This distribution suggested the form of a ``coherent''
signal-to-noise ratio $\hat{\rho}$ which gives a factor of 2 more significance
to the signal-to-noise ratio in L1 compared to the signal-to-noise ratio in
L1. Based on the studies of the background and injections, we chose a
combined signal-to-noise statistic
\begin{equation}
\hat{\rho}^2 = \rho^2_{\mathrm{L1}} +  \frac{\rho_{\mathrm{H}}^2 }{ 4 }.
\label{eq:combinedsnr}
\end{equation}

\section{Upper Limit on BBHMACHO Inspiral in the S2 Playground Data}
\label{s:s2upperlimit}

After the data quality cuts, discarding science segments with durations
shorter than 2048 sec, and application of the instrumental veto in L1, a
total of {35.2} hours of playground data remained; {22.0} hours of
triple-detector data,  {10.2} hours of L1-H1 and {3.97} hours of L1-H2. The
analysis of the playground data produced no double or triple coincident
inspiral candidates.

To determine an upper limit on the event rate we use the \emph{loudest event
statistic}\cite{loudestGWDAW03} which uses the detection efficiency at the
signal-to-noise ratio of the loudest trigger surviving the pipeline to
determine an upper limit on the rate. If no triggers survive the pipeline, we
use the signal-to-noise threshold as the loudest event.  Suppose the
population of sources produces Poisson-distributed events with a rate
$\mathcal{R}$ per year per Milky Way Equivalent Galaxy (MWEG) and
$N_G(\rho^\ast)$ is the number of MWEGs to which the search is sensitive at
$\rho \geq \rho^\ast$.   Then the probability of observing an inspiral signal
with $\rho > \rho^\ast$, given some rate $\mathcal{R}$ and some observation
time $T$, is
\begin{equation}
P(\rho>\rho^\ast;{\mathcal{R}}) = 1 - e^{-{\mathcal{R}}T N_G(\rho^\ast)}.
\label{eq:foreground-poisson}
\end{equation}
A trigger can arise from either an inspiral signal in the data or from
background.   If $P_b$ denotes the probability that all background triggers
have signal-to-noise ratio less than $\rho^\ast$,  then the probability of observing either an
inspiral signal or a background trigger with $\rho > \rho^\ast$ is given by
\begin{equation}
P(\rho>\rho^\ast;{\mathcal{R}},b) = 1 - P_b e^{-{\mathcal{R}}TN_G(\rho^\ast)}.
\label{eq:joint-dist}
\end{equation}
Given the probability $P_b$, the total observation time $T$, and the number of
Milky Way equivalent galaxies $N_{{G}}$ to which the search is sensitive, we
find that the rate of binary black hole MACHO inspirals per MWEG is
\begin{equation}
  \mathcal{R}_{90\%} = \frac{2.303+\ln P_b}{T N_{G}(\rho^\ast)}
\end{equation}
with 90\% confidence. This is a frequentist upper limit on the rate.
For ${\mathcal{R}}>{\mathcal{R}}_{90\%}$, there is more than $90\%$
probability that at least one event would be observed with SNR greater
than $\rho_{\text{max}}$. 

Since no coincident events were observed in the playground data, we determine
the rate by measuring the efficiency of the pipeline at the (combined)
signal-to-noise threshold
\begin{equation}
\hat{\rho}^2_\mathrm{max} =  7^2 + \frac{7^2}{4} = 61.25.
\label{eq:threshold}
\end{equation}
This is a conservative limit on the rate. If we lowered the
signal-to-noise thresholds until we observed the loudest trigger, then this
trigger will have a signal-to-noise ratio less than the value given in equation
(\ref{eq:threshold}); the measured efficiency will be greater than or equal to
that used here so the rate will be less than or equal to which we
quoted here. Furthermore, since we only have a small number of time lags, we
neglect the background term $P_b$. Dropping this term will give a conservative
value for upper limit\cite{loudestGWDAW03}.

If we restrict the mass parameters of the injected signals to those with a
component in the range $0.15$ to $1.0\,M_\odot$, the mass range suggested by
microlensing observations, we find that efficiency of the search pipeline at
$\hat{\rho}^\ast = \sqrt{61.25}$ is 
\begin{equation}
\varepsilon\left( \hat{\rho}^\ast = \sqrt{61.25} \right) = \frac{ 692 } { 756
} = 0.915.
\end{equation}
The observation time for the playground data is $T = 35.2\,\mathrm{hours} = 4
\times 10^{-3}\,\mathrm{yr}$ and so the upper limit on the rate of binary
black hole MACHO inspirals in the playground data is
\begin{equation}
\mathcal{R}_{90\%} = \frac{2.303}{0.915 \times 4 \times 10^{-3}} = 
627\,\mathrm{yr}^{-1}\,\mathrm{MWEG}^{-1}.
\end{equation}

The amount of non-playground data in the full S2 data, again discarding
science segments with durations shorter than 2048 sec,  and application of the
instrumental veto in L1, there is a total of 345 hours of non-playground data;
225 hours of triple-detector data, 90 hours of L1-H1 and 30 hours of L1-H2.
Assuming that the signal-to-noise ratio of the loudest event in the full data
is comparable to $\hat{\rho}^2 = 61.25$, which is suggested by the background
triggers, we may estimate the achievable upper limit as
\begin{equation}
\mathcal{R}_{90\%} = \frac{2.303}{0.915 \times 3.9 \times 10^{-2}} = 
64\,\mathrm{yr}^{-1}\,\mathrm{MWEG}^{-1}.
\end{equation}
This upper limit is three orders of magnitude larger than the upper bound on
the rate of $R = 0.05\times2^{\pm 1}\,\mathrm{yr}^{-1}\,\mathrm{MWEG}^{-1}$,
so we are not yet able to constrain the fraction of the Galactic halo in
BBHMACHOs. The sensitivity of the interferometers during the S2 search is
roughly an order of magnitude from design sensitivity, however. For a
$(0.5,0.5)\,M_\odot$ binary, at design sensitivity the interferometers will be
sensitive to $\sim 50$~MWEG, so in one year of data taking, assuming no
detections have been made, an upper limit on rate of BBHMACHO inspirals of $R
= 3\times10^{-2}\,\mathrm{yr}^{-1}\,\mathrm{MWEG}^{-1}$ are possible (assuming
$P_b = 0.5$) which will significantly impact the theoretical rate estimates.
Additionally, since the range of the search scales as a function of the mass,
the possibility for detecting more massive BBHMACHO binaries increases as more
galaxies become accessible and the rate can be further constrained.


\newpage

\begin{table}[p]
\begin{tabular}{cllr}
Parameter & Description  & value \\
\hline 
$f_\mathrm{hp}$ & High Pass Filter Frequency  & $100$~Hz \\
$O_\mathrm{hp}$ & High Pass Filter Order  & $100$~Hz \\
$f_\mathrm{low}$ & Low Frequency Cutoff  & $100$~Hz \\
$m_\mathrm{min}$ & Template bank lower component mass  & $0.2\,M_\odot$ \\
$m_\mathrm{max}$ & Template bank upper component mass  & $1.0\,M_\odot$ \\
$\mathbb{M}$ & L1 template bank minimal match  & 0.95 \\
$\rho^\ast_\mathrm{L1}$ & L1 signal-to-noise ratio threshold & 7.0 \\
$\Xi^\ast_\mathrm{L1}$ & L1 $\chi^2$ veto threshold & 3.1 \\
$\rho^\ast_\mathrm{H1}$ & H1 signal-to-noise ratio threshold & 7.0 \\
$\Xi^\ast_\mathrm{H1}$ & H1 $\chi^2$ veto threshold & 5.0 \\
$\rho^\ast_\mathrm{H2}$ & H2 signal-to-noise ratio threshold & 7.0 \\
$\Xi^\ast_\mathrm{H2}$ & H2 $\chi^2$ veto threshold & 10.0 \\
$p$ & Number of bins in $\chi^2$ veto & 15 \\
$\delta^2$ & $\chi^2$ veto mismatch parameter & 0.2 \\
$\delta m$ & Trigger mass coincidence parameter & 0.0 \\
$\delta t_\mathrm{HH}$ & H1-H2 trigger time coincidence parameter & 0.001 s \\
$\delta t_\mathrm{LH}$ & L1-H1, L1-H2 trigger time coincidence parameter & 0.011 s \\
$\kappa_{HH}$ & H1-H2 trigger amplitude coincidence parameter & 0.5 \\
$\kappa_{LH}$ & L1-H1, L1-H2 trigger amplitude coincidence parameter & 1000.0 \\
$\epsilon$ & Trigger amplitude coincidence parameter & 2.0
\end{tabular}
\caption[Pipeline Parameters used in S2 BBHMACHO Search]{%
\label{t:ifo_params}%
A complete list of the parameters that were selected at the various
stages of the pipeline. These values are justified in the text.
}
\end{table}

\Chapter{Conclusion}
\label{ch:conclusion}

\begin{figure}[p]
\vspace{5pt}
\begin{center}
\includegraphics[width=\textwidth]{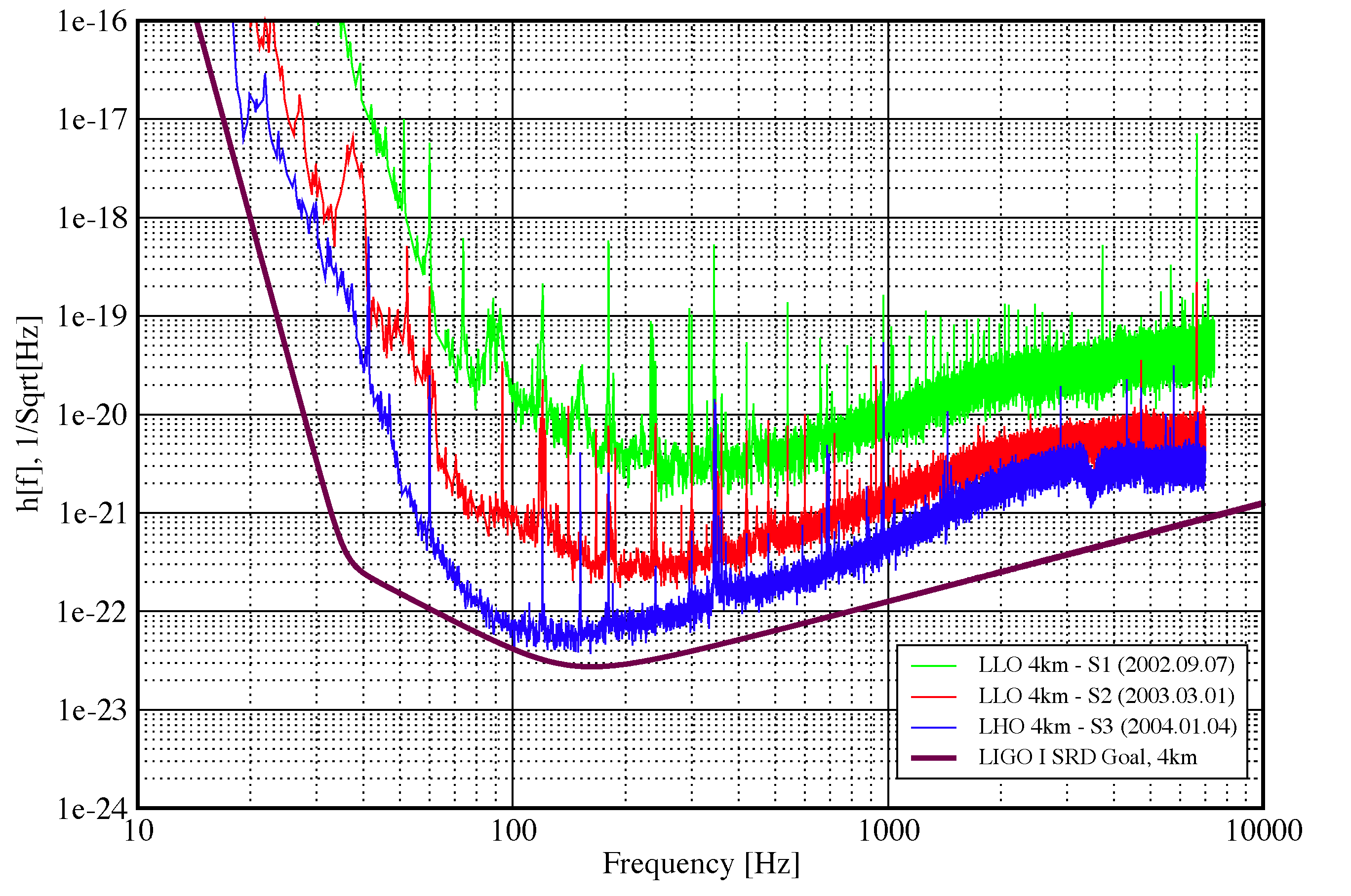}
\end{center}
\caption[Comparison of Best LIGO Interferometer Sensitivity]{%
\label{f:s3strain}
Comparison of the best sensitivities of the LIGO interferometers between
science runs. The solid curve shows the design sensitivity for the $4$~km
interferometers: the LHO $4$~km is only a factor of $\sim 2$ away from design
at $100$~Hz during S3.
}
\end{figure}

Although the upper limit that we have placed on the rate of binary black hole
MACHO inspirals in the galaxy is lower than the upper bound of the predicted
rates, the LIGO interferometers were not at design sensitivity when the S2
data was taken. At present, the sensitivities of the instruments are
significantly better than during S2, as can be seen from
figure~\ref{f:s3strain}, and progress on reducing noise in the interferometers
continues apace.  The increase in detector sensitivity makes a larger volume
of the Universe accessible to searches for binary inspirals. In addition to
this, the amount of data is also increasing as the interferometers become more
stable.

These improvements in the instruments will increase the chance of detecting
gravitational waves from binary inspirals. If the rates of binary black hole
MACHO coalescence are truly as high as predicted, then initial LIGO would
stand an excellent chance of detecting an inspiral. The first detection of
gravitational waves will be a major scientific breakthrough and will yield and
enormous amount of scientific information, particularly if the detection came
from a binary black hole MACHO. The length of binary black hole MACHO
inspirals in the sensitive band of the interferometer will allow extremely
accurate parameter estimation as well as tests of post-Newtonian theory. For
systems with total mass greater than $\sim 0.64\,\mathrm{M}_\odot$ LIGO will
be sensitive to the coalescence of the binary and will be able to study the
strong gravitational field effects when two binary black holes merge. When
this is coupled with the accurate parameter estimation available from the
earlier part of the waveform, the inspiral of a binary black hole MACHO could
be an excellent laboratory for General Relativity.  A detection would also
impact the studies of halo dark matter and early universe physics, providing a
MACHO component to the halo and suggesting that primordial black holes do
indeed form in the universe.

In the absence of detection, the improvements in detector sensitivity will
dramatically improve the upper limits placed on the rate of binary black hole
MACHO inspirals. Once these rates are below the predicted rates, we may begin
to use observations from gravitational wave interferometers to constrain the
fraction of galactic halos in the form of primordial black hole MACHOs. While
this may not be as significant as a detection, it will still be of interest to
the astrophysical community.

\newpage 

\clearpage
\bibliographystyle{unsrt}
\bibliography{references}
\addcontentsline{toc}{chapter}{\numberline {Bibliography}}

\clearpage
\birthplacedate{Nottingham, United Kingdom \>\>January 25, 1976}
\collegewherewhen{%
\>University of Newcastle Upon Tyne \>\>1994--1999, \>M.Math.\\
\>\uwm	\>\>1999--2004, \>Ph.D.}

\newpage
\null\vskip1in%
\begin{center}
{\Large\bf Curriculum Vitae}
\end{center}
\vskip 2em
\begin{tabbing}
\tabset
Title of Dissertation\\
\>Searching for Gravitational Radiation from Binary Black Hole MACHOs \\
\>in the Galactic Halo
\end{tabbing}
\vskip 1em

\begin{startvita}
\end{startvita}

\renewenvironment{thebibliography}[1]%
  {\begin{list}{\labelenumi\hss}%
     {\usecounter{enumi}\setlength{\labelwidth}{3em}%
      \setlength{\leftmargin}{5em}}}%
  {\end{list}}
\renewcommand{\bibitem}[1]{\item\label{#1}\relax}%
\renewcommand{\theenumi}{\arabic{enumi}}%
\begin{publications}
\putbib[papers]
\end{publications}

\begin{honorarysocieties}
2003 \> UWM Chancellor's Graduate Student Fellowship\\
2003 \> UWM Dissertator Fellowship\\
2002 \> Physics Graduate Student Trust Fund Award\\
2002 \> UWM Chancellor's Graduate Student Fellowship\\
2002 \> UWM Graduate School Fellowship\\
2001 \> Papastamatiou Scholarship\\
1999 \> Institute of Mathematics and Its Applications Prize\\
1997 \> Stroud Book Prize for Theoretical Physics
\end{honorarysocieties}

\finishvita
\end{document}